\documentclass[11pt]{article}
     
\usepackage[margin=1.in]{geometry}     

\usepackage{amsmath,amsthm,amsfonts,amssymb,amscd}
\usepackage[mathscr]{euscript}
\usepackage[dvipsnames]{xcolor}
	\usepackage{soul}
\usepackage[subfigure]{tocloft}

	\usepackage[toc,titletoc]{appendix}
\usepackage[square,numbers]{natbib}
	\usepackage{etoolbox,xstring}
	\makeatletter
	\patchcmd{\NAT@test}{\else\NAT@nm}{\else\NAT@nmfmt{\NAT@nm}}{}{}
	\let\NAT@up\scshape
	\makeatother
	\makeatletter
	\patchcmd{\NAT@test}{\else\NAT@nm}{\else\NAT@nmfmt{\NAT@nm}}{}{}
	\renewcommand{\NAT@nmfmt}{\expandafter\aliNAT@nmfmt\expandafter}
	\newcommand\aliNAT@nmfmt[1]{{%
  	\noexpandarg
  	\def~{}%
  	\edef\temp#1\edef\temp{\detokenize\expandafter{\temp}}%
 	 \begingroup\edef\x{\endgroup
   	 \noexpand\StrSubstitute{\temp}{\detokenize{etal}}}\x
    	{\textnormal{et\nobreakspace al}}[\tempa]%
  	\textsc{\tempa}}}
	\makeatother
\usepackage[T1]{fontenc}
\usepackage{titlesec}
	\titleformat{\section}{\large\bfseries\scshape}{\thesection.}{0.5em}{\centering}
	\titleformat{\subsection}{\normalsize\bfseries}{\thesubsection.}{0.5em}{\centering\textit}
	\titleformat{\subsubsection}[runin]{\normalsize\bfseries}{\thesubsubsection.}{0.5em}{}
\usepackage{fancyhdr}
\usepackage{bm}
\usepackage{bbm}
\usepackage{array}
	\newcolumntype{P}[1]{>{\centering\arraybackslash}p{#1}}		
	\newcolumntype{M}[1]{>{\centering\arraybackslash}m{#1}}
	\usepackage{multirow}
	\usepackage{booktabs}
	\usepackage{diagbox}
\usepackage{float}
\usepackage{tikz}
	\usetikzlibrary{decorations.pathreplacing,angles,quotes,patterns}
\usepackage{graphicx}
	\graphicspath {{figures/}}
\usepackage[hang]{subfigure}
	\usepackage{wrapfig}
	\usepackage[font=footnotesize,labelfont=bf]{caption}
\usepackage{mathtools}
\usepackage{stmaryrd}
\usepackage{MnSymbol}
\usepackage{algorithm}
\usepackage{enumitem}
\usepackage{hyperref}
	\hypersetup{colorlinks=True, linkcolor=blue, urlcolor=blue, citecolor=blue}
\usepackage{cleveref}

\theoremstyle{plain}

\theoremstyle{definition}

\newtheorem{remark}{\scshape Remark}

\newcommand{\sgn}{\operatorname{sgn}}
\newcommand{\pvint}{\operatorname{P}\hspace{-0.5em}\int}
\def\p{\partial}
\def\bbR{\mathbb{R} }

\def\ft #1{{\widehat{#1}}}

\def\divv{\operatorname{div}}
\providecommand{\e}[1]{\ensuremath{\times 10^{-#1}}}

\title{\Large\textbf{\textsc{A  multiscale model for Rayleigh-Taylor and 
Richtmyer-Meshkov instabilities}}}
\author{
 {\normalsize \textbf{\textsc{Raaghav Ramani}}}
 \vspace{-.05 in}
\\{\small Department of Mathematics}
\vspace{-.05 in}
\\{\small University of California}
\vspace{-.05 in}
\\{\small Davis, CA 95616 USA}
\vspace{-.05 in}
\\{\small\textit{\url{rramani@math.ucdavis.edu}}} \and
 {\normalsize \textbf{\textsc{Steve Shkoller}}}
 \vspace{-.05 in}
\\{\small Department of Mathematics}
\vspace{-.05 in}
\\{\small University of California}
\vspace{-.05 in}
\\{\small Davis, CA 95616 USA}
\vspace{-.05 in}
\\{\small\textit{\url{shkoller@math.ucdavis.edu}}}
}

\date{\today}

\begin{document}

\maketitle

\begin{abstract}
We develop a novel multiscale model of interface motion for the Rayleigh-Taylor instability (RTI) and Richtmyer-Meshkov instability (RMI) for two-dimensional, inviscid, 
compressible flows with vorticity, which yields a fast-running numerical algorithm that produces both qualitatively and quantitatively
similar results to a resolved gas dynamics code, while running approximately two orders  of magnitude  (in time)
 faster.   Our multiscale model is founded upon a new
 compressible-incompressible decomposition of the velocity field $u=v+w$.     The incompressible component $w$ of the velocity  is also irrotational and is solved using
 a new asymptotic model of the Birkhoff-Rott singular integral formulation of the incompressible Euler equations, which reduces
 the problem to one spatial dimension.  This asymptotic model, called the higher-order $z$-model, is derived using
 small nonlocality as the asymptotic parameter, allows for interface turn-over and roll-up,  and yields a significant simplification for the equation describing the evolution of the amplitude of vorticity.
 This incompressible component  $w$ of the velocity controls the small scale structures of the interface and can be solved efficiently on fine grids.  Meanwhile, 
 the compressible component of the velocity $v$  remains continuous near contact discontinuities and can be computed on relatively coarse grids, while receiving subgrid scale 
 information from $w$.  We first validate the incompressible higher-order $z$-model by comparison with classical RTI experiments as well as full point vortex simulations.  We
 then consider both the RTI and the RMI problems for our multiscale model of compressible flow with vorticity, and show excellent agreement with our high-resolution gas dynamics solutions.
\end{abstract}

{\small
\tableofcontents}

\section{Introduction}

The instability that occurs when an interface separating two fluids of different densities is perturbed and  
subjected to an acceleration force is a fundamental problem in fluid mechanics. 
The Rayleigh-Taylor instability (RTI) \cite{Rayleigh1882,Taylor1950} 
occurs when the lighter fluid is accelerated towards the heavier fluid (under the action of gravity, for instance).
 The Richtmyer-Meshkov instability (RMI) 
\cite{Richtmyer1960,Meshkov1969} is initiated by the 
passage of a shock wave across the perturbed interface separating the two fluids. In either case, 
perturbations of the interface initially grow according to the linear theory, before the system enters the 
nonlinear regime, in which the light fluid \emph{bubbles} into the heavy fluid, while the heavy fluid 
\emph{spikes} into the light fluid. The velocity of the resulting flow is discontinuous at the material interface (or contact
discontintuity), which
initiates the Kelvin-Helmholtz instability (KHI) \cite{Kelvin1871,Helmholtz1868}. This causes the 
interface to \emph{roll up} 
into complex vortical structures, and eventually leads to turbulent mixing. Each of these
instabilities arises in numerous important applications, including in astrophysics \cite{HesterEtAl1996}, 
inertial confinement fusion \cite{BettiEtAl1998}, and ocean mixing \cite{Smyth2012}. We refer the reader 
to the works \cite{Sharp1984,Kull1991,Brouillette2002} and the references therein for further details. 

The fundamental mathematical model for the RT and RM instabilities is the Euler 
system of hydrodynamics equations, consisting of the conservation of mass, momentum, and energy.  The  mathematical analysis
of the Euler equations is extremely challenging due to the ill-posed nature of the equations in the absence of
 of stabilizing mechanisms such as surface tension or viscosity, with the RTI and RMI causing growth of perturbations at 
 the smallest scales available. The highly unstable nature of both the RTI and RMI also poses significant 
  difficulties for numerical methods, and the development of algorithms to  study these instabilities has been the subject of 
intensive research over the last several decades
\cite{Daly1967,BaMeOr1980,GlimmEtAl1986,Tryggvason1988,GlimmEtAl1990,DimonteEtAl2004,
AlmgrenEtAl2010,LaScSo2007b}, and continues to remain a challenge.

 As the linear theory shows, the highest frequency perturbations of the 
interface have the largest growth rates; numerical solutions thus often suffer from the development 
of spurious small scale structure \cite{LiWe2003}, which does not appear to agree with laboratory 
experiments \cite{WaNiJa2001}. Numerical methods with a large amount of implicit 
diffusion suppress these small scale eddies, but, in doing so, prevent the development of the KHI mixing 
zones. 

Moreover, even when numerical schemes can be manipulated into producing better solutions 
\cite{RaReSh2019b}, simulations can be prohibitively (computationally) expensive.  
Direct Numerical Simulations (DNS) solve the complete governing equations with exact 
physical parameters and sufficient resolution to represent all the scales of the flow \cite{MoMa1998}, but 
the requirement that all the spatial and temporal scales be numerically resolved results in 
overwhelmingly expensive calculations, both in terms of computational runtime, as well as other basic 
computational resources, such as memory. As such, simulations are currently generally 
limited to small Reynolds 
number flows and simple geometries \cite{VeMa2007}.
These observations indicate a great need for fast algorithms that can be used to accurately 
predict the RTI and RMI mixing layers and associated growth rates. 

In this work, we  develop a \emph{multiscale} model for 
interface evolution during RTI and RMI for two-dimensional, inviscid, compressible flow with vorticity. 
 Our multiscale model is founded upon a new
 compressible-incompressible decomposition of the velocity field $u=v+w$, which is, in turn, based
 upon a two-phase elliptic system of Hodge type \cite{ChSh2017}.    The incompressible component $w$ of the velocity  is also irrotational and is solved using
 a new asymptotic model of the Birkhoff-Rott singular-integral formulation of the incompressible Euler equations, which already reduces
 the problem to one spatial dimension.  This asymptotic model, called the {\it higher-order $z$-model}, is derived using
 small nonlocality as the asymptotic parameter, allows for interface turn-over and roll up,  and yields a significant simplification for the equation describing the evolution of the amplitude of vorticity.
 This incompressible component  $w$ of the velocity controls the small scale structures of the interface and can be solved efficiently on fine grids.  Meanwhile, 
 the compressible component of the velocity $v$ field remains smooth near contact discontinuities and can be computed on relatively coarse grids, while receiving subgrid scale 
 information from $w$.

Specifically, our higher-order $z$-model, approximates the Birkhoff-Rott  (BR)
equations \cite{CoCoGa2010}
of interface evolution in two-dimensional multiphase incompressible and irrotational flow, which are, in turn, a reduction of
the incompressible and irrotational Euler equation to one-dimensional evolution for the parameterization of the interface 
$z( \alpha , t)=\left(z_1( \alpha,t), z_2( \alpha ,t)\right)$ and the amplitude of vorticity $\varpi( \alpha , t)$.
The original (low-order) $z$-model was derived  by \citet{GrSh2017} using an asymptotic expansion in a small 
\emph{non-locality} parameter, and its main advantage over the full BR evolution  is a drastic simplification of the 
dynamics for the  amplitude of vorticity $\varpi$. Without this simplification, the BR dynamics of $\varpi$
is nonlinear, non-local, and is in fact a Fredholm integro-differential equation of the second kind.
Numerical methods for this type of equation are thus often quite complex and computationally expensive. On the 
other hand, the $\varpi$ dynamics given by the $z$-model allow us to implement an 
extremely simple numerical method which avoids costly upwinding and iterative procedures 
\cite{BaBe2004,Sohn2004a}; in particular, we use a simple Fourier collocation method
to evolve $\varpi$.  For the evolution of the interface $z$, the original (low-order) $z$-model of \cite{GrSh2017}
used a local equation, while our new (high-order) $z$-model instead uses 
 Krasny's $\delta$-desingularization 
\cite{Krasny1986b} of the singular integral kernel.   The solution of $z$ and $\varpi$ then provide us with the
incompressible velocity field $w$ via an efficient kernel computation.     The compressible velocity field $v$ is solved
on a very coarse grid (while receiving small-scale information from $w$) using a very simple WENO scheme together with a nonlinear, spacetime smooth, artificial viscosity
method termed the $C$-method \cite{ReSeSh2012,RaReSh2019a,RaReSh2019b}.

We first validate our incompressible and irrotational (high-order) $z$-model by
performing a number of numerical experiments, including both the single-mode and 
multi-mode RTI, to demonstrate the accuracy of the $z$-model and its numerical implementation.  
The computed $z$-model solutions are compared with observations from laboratory experiments 
\cite{WaNiJa2001,Read1984,Youngs1989}, and are shown to achieve very
similar growth rates of the bubbles and spikes, as well as the mixing layer. We additionally compare our 
$z$-model solutions with ``reference'' solutions computed using a sophisticated numerical method for 
the complete Birkhoff-Rott equations \cite{Sohn2004a}, and show that the two solutions are in excellent 
agreement, thereby demonstrating the validity of the $z$-model. Moreover, our 
simplified model equations allow for a 
numerical computation that is a factor of \emph{at least} 75 times faster than the reference solution 
calculation. 
\textcolor{black}{
We also compare the simple Krasny desingularization used in the numerical implementation of the 
$z$-model with two other  higher-order regularizations that smooth the singular integral kernel via convolution with Gaussian-type
functions which satisfy certain moment conditions. We demonstrate that all three numerical methods for smoothing the singular integral
produce similar numerical solutions (in a sense to be made precise below); as will be shown,  these solutions   are in reasonable agreement in the asymptotic limit as the 
mesh spacing ${\scriptstyle\Delta} \alpha$ and viscosity parameter $\delta$ converge to zero. 
}

Then, we use our {multiscale} model, designed to simulate interface evolution in compressible
flows with vorticity.   As we noted above,  we decompose $u=v+w$, where $w$ is both 
divergence-free and curl-free, but has a  discontinuity in its tangential component across the contact discontinuity, 
while $v$ is continuous across the contact, but is forced by the bulk 
compression and vorticity of the fluid. By analogy with turbulence models, such as 
large-eddy simulation (LES) \cite{MeSa2006}, Reynolds-averaged Navier-Stokes (RANS), and 
Lagrangian-averaged Navier-Stokes (LANS-$\alpha$) \cite{MoKoShMa2003}, our 
{multiscale model} involves the decomposition of the flow into a part which can be solved on 
a coarse grid (the mean flow), and a part which must be solved on a fine grid (the sub-grid scale fluctuations).  
The novelty of our approach is that, for the RTI and RMI, the fine grid coincides with the interface itself, and is 
thus one-dimensional. This means that fine structures can be simulated with much less 
computational expense than is required for fully two-dimensional calculations on similarly fine meshes. 

We describe a simple Eulerian-Lagrangian algorithm for our multiscale model that couples 
the equations on a coarse two-dimensional mesh with the equations on the high resolution one-dimensional 
interface. For modeling the RMI, a modified set of equations is used, in which we account for both the effects of 
shock-contact interaction, as well as the classical Taylor ``frozen turbulence'' hypothesis \cite{Taylor1938}. 
We then discuss the numerical implementation of the algorithm, which uses our incompressible $z$-model,
as well as simple interpolation and integral-kernel calculation techniques. 
A number of
numerical experiments for the RTI and RMI are performed
to demonstrate the efficacy of our multiscale model and algorithm for compressible flows with vorticity. In particular, we show that our algorithm 
produces solutions that agree both qualitatively and quantitatively with  (relatively) high-resolution reference solutions. \textcolor{black}{We again perform some basic convergence studies, and find good agreement between the 
multiscale solutions and high-resolution reference solutions in the limit as the 
interfacial mesh spacing ${\scriptstyle\Delta} \alpha$ and desingularization parameter $\delta$ converge to zero.} 
 Moreover,  the run times of our multiscale algorithm
are two orders of magnitude (or more) faster than  those of the corresponding high-resolution reference solutions.

\vspace{.1in}
\noindent
{\bf Outline of the paper.}   \Cref{sec:preliminaries} is devoted to the  notation and definitions that will be used throughout the paper. 
In  \Cref{sec:Euler}, we introduce the full system of Euler equations for compressible flow, followed by the
incompressible and irrotational simplification.  For the latter, we explain how those equations can be solved
using the Birkhoff-Rott \emph{singular integral-kernel} equations for the interface parameterization and 
amplitude of vorticity.   We then describe our asymptotic (in nonlocality) $z$-model.
 We next consider the full compressible Euler equations as a two-phase 
 elliptic system for the velocity, and derive  a novel compressible-incompressible decomposition 
 of the velocity. This decomposition is the foundation of our multiscale model and algorithm. 
 
 In \Cref{z-model-numerical-implementation}, we consider the numerical implementation of the 
 incompressible
 $z$-model. A simple numerical method is introduced, and results for several numerical experiments are 
 shown, including comparisons with laboratory experiments, theoretical predictions and models, and 
 benchmark numerical simulations. We then present,  in \Cref{sec:multi-scale-algorithm},  our multiscale 
 model and algorithms for the compressible RTI and RMI, and give
  details about their numerical implementations. 
  
 Our multiscale  algorithm is   then applied to  two RTI and two RMI test problems
 \Cref{sec:multi-scale_numerical_simulations}, and compared against both high-resolution simulations and low-resolution simulations.
 Finally,  our conclusions are in  \Cref{sec:conclusion}.    Two short sections of the Appendix are provided: the first concerns mesh refinement studies
 for the multiscale algorithm and the second summarizes our numerical method for gas dynamics.

\section{Preliminaries} \label{sec:preliminaries}
\subsection{Some notation and definitions}
\subsubsection{Derivatives} We write
$$
\partial_i f=\frac{\partial f}{\partial x_i}\ \text{ for } i=1,2\,, \ \  \partial_t f=\frac{\partial f}{\partial t} \,, \ \
\nabla=(\p_{\!1}, \p_2)\,,    \ \ \nabla^\perp=(- \p_2, \p_{\!1}) \,,
$$
and for a vector $F$,
$$
\operatorname{div} F=  \nabla \cdot F  \ \text{ and } \ \operatorname{curl} F = \nabla ^\perp\cdot F \,.
$$
The Laplace operator is defined as $ \Delta = \p_{\!1}^2 + \p_2^2$.   Given a transport velocity $u(x,t)$, we shall denote the material
derivative $\p_t + u \cdot \nabla $ by $\tfrac{\mathrm{D}}{\mathrm{D}t} $.

\subsubsection{Fourier series}\label{sec::Fourier}  Let $ \mathbb{T} _L $ denote the interval $[-L/2,L/2]$.
If  $f:\mathbb{T}_L \to \bbR$ is a square-integrable $L$-periodic function,  then it has the Fourier series representation
$f(\alpha) = \sum\limits_{k=- \infty }^ \infty  \ft{f}(k) e^{\frac{2 \pi i k \alpha}{L}}$ for all $\alpha \in \mathbb{T}_L$, where the complex Fourier coefficients
are defined by $\mathcal{F}\{f\}(k) \equiv  \ft{f}(k) = {\dfrac{1}{L}} \int_{\mathbb{T}_{L}}{}\, f(\alpha)
e^{-\frac{2 \pi i k \alpha}{L}} \, \mathrm{d} \alpha$. 
We have the following standard identity:
\begin{equation} \label{ft-derivative}
\mathcal{F}\{\partial^n_{\alpha} f\}(k) = \left(\frac{2 i\pi}{L} k \right)^n \ft{f}(k)\,,
\end{equation}
where $\partial_\alpha = \frac{\partial}{\partial \alpha}$.  
We shall sometimes write $\ft{f}_k$ for $\ft{f}(k)$.

\subsubsection{Principal value integral}

The \emph{principal value integral} of a function 
$f: \mathbb{R} \to \mathbb{R} $ is defined as 
\begin{equation} \label{eq:pv}
\pvint_{\mathbb{R}} f(\beta) \,\mathrm{d}\beta \coloneqq \lim_{\varepsilon \to 0^+ } \int_{(-1/\varepsilon,-\varepsilon)\cup(\varepsilon,1/\varepsilon)} f(\beta) \,\mathrm{d}\beta\,.
\end{equation}

\subsubsection{Hilbert transform}   
The \emph{Hilbert transform} of a function $f: \mathbb{R} \to \mathbb{R}$ is 
defined as 
\begin{equation}\label{hilbert-r}
\mathcal{H} f (\alpha) = \frac{1}{\pi} \pvint_{\mathbb{R}} \frac{f(\beta)}{\alpha - \beta}\, \mathrm{d}\beta \,.
\end{equation}

If $f$ is an $L$-periodic function on $\mathbb{T}_L$, then 
\begin{equation}\label{hilbert-periodic}
\mathcal{H}f(\alpha) = \frac{1}{L} \pvint_{\mathbb{T}_L} \frac{f(\beta)}{\tan(\tfrac{\pi}{L}(\alpha - \beta))} \,\mathrm{d}\beta \,.
\end{equation}
Equivalently, using the Fourier representation, the Hilbert transform $ \mathcal{H} $ can be defined as
\begin{align}\label{Hilbert}
\ft{\mathcal{H} f}(k)=-i\text{sgn}(k) \ft{f}(k) \,.
\end{align}
In particular, we note that
$\mathcal{H} ^2=-1$.

\subsubsection{Discrete operators in Fourier space}
Let $\alpha \in \mathbb{T}_L$. We discretize the parameter $\alpha$ with $N+1 = 2^r + 1$ nodes, 
$$
\alpha_k = -L/2 + (k-1) {\scriptstyle\Delta} \alpha \,,
$$
with ${\scriptstyle\Delta} \alpha = L/N$. 
Given an $L$-periodic function $f(\alpha)$, we denote by $f_k = f(\alpha_k)$ the function $f$ evaluated at 
a point $\alpha_k \in \mathbb{T}_L$ 
Let $\tilde{\mathcal{F}}$ and $\tilde{\mathcal{F}}^{-1}$ denote the discrete Fourier and inverse Fourier 
transforms, respectively, defined for sequences $(f_k)$ of length $N=2^r$ by 
$$
\tilde{\mathcal{F}} \left\{ f_k \right\}_m = \sum_{l=1}^{N} f_l \cdot e^{-\frac{2 i \pi}{N} (m-1) (l-1) } \quad \text{ and } \quad \tilde{\mathcal{F}}^{-1} \left\{ \ft{f}_m \right\}_k = \frac{1}{N}\sum_{l=1}^{N} \ft{f}_l \cdot e^{\frac{2 i \pi}{N} (k-1)(l-1) } \,.
$$
We define the discrete Fourier operators $(H_k)$, $(D_k)$, $(D^2_k) \in \mathbb{C}^N$ as
\begin{align}
H_k &= \begin{cases}
		0 & \text{if } k=1 \,, \\ 
		-i & \text{if } 2 \leq k \leq (N+1)/2 \,, \\
		i & \text{if } k > (N+1)/2  \,,
		\end{cases}  \label{discrete-hilbert} \\
D_k &= \begin{cases}
		\frac{2  \pi i}{L} (k-1) & \text{if } k < (N+1)/2  \,,\\
		0 & \text{if } k = (N+1)/2 \,, \\ 
		-\frac{2 \pi i}{L} (N-k) & \text{if } k > (N+1)/2 \,,
		\end{cases} \label{discrete-partial} \\
D^2_k &= \begin{cases}
		-\left(\frac{2 \pi}{L} \right)^2 (k-1)^2 & \text{if } k \leq (N+1)/2\,, \\
		-\left(\frac{2 \pi}{L} \right)^2 (N-k)^2 & \text{if } k > (N+1)/2\,. 
		\end{cases} \label{discrete-partial-squared}
\end{align}
Formula \eqref{discrete-hilbert} is the discrete Hilbert transform in Fourier space, while  
\eqref{discrete-partial} and \eqref{discrete-partial-squared} are the discrete derivative operators 
$\partial_\alpha$ and $\partial^2_{\alpha}$, respectively, in Fourier space.

\subsection{Computational platform and code optimization}

All of the numerical simulations conducted in this work were run on a Macbook Pro laptop using a 
2.4 GHz Intel Core i5 processor with 8 GB of RAM. The operating system is macOS High Sierra 10.13.6, and 
the GFortran F90 compiler is used. 

The codes for the numerical methods described in the paper are implemented in the same 
programming framework, but are not otherwise specially optimized (apart from a specific 
calculation described in the paper). The same input, output, and timing routines are used in all of the codes. 
This consistency allows for a reliable comparison of the different algorithms and their associated 
imposed computational burden. 

\section{The Euler equations} \label{sec:Euler}

\subsection{The compressible Euler equations} 
The fundamental mathematical  model for the motion of an inviscid two-dimensional fluid is given by the
compressible Euler equations:
\begin{subequations}
\label{euler2d}
\begin{align}
\partial_t \rho + \divv  (\rho u) &= 0 \,, \label{euler2d-eqn1} \\
\partial_t (\rho u) + \divv  (\rho u \otimes u) + \nabla p + \rho g e_2 &= 0\,,  \label{euler2d-eqn2}\\ 
\partial_t E + \divv  (u(E+p)) + \rho g u_2 &=0 \,, \label{euler2d-eqn3}
\end{align}
\end{subequations}
where $\otimes$ denotes the tensor product,  and $\divv M$ denotes the row-wise divergence of a matrix $M$. The velocity vector is  ${u} = (u_1\,,u_2)$ with horizontal component $u_1$ and vertical
component $u_2$, $\rho>0$ is the fluid density (assumed strictly positive),  $E$ denotes the energy, and $p$ is the pressure defined by an
equation of state.  
These equations are, in fact, the basic 
conservation laws of fluid dynamics: \eqref{euler2d-eqn1} is conservation of mass, \eqref{euler2d-eqn2} is conservation of linear momentum, and \eqref{euler2d-eqn3} is conservation of energy.

The system \eqref{euler2d} can be written in classical conservation-law form as the Cauchy problem
\begin{subequations}
\label{Euler-2d}
\begin{alignat}{2}
\partial_t {\bf{U}}(x,t)+ \partial_{x_1} {\bf{F}}({\bf{U}}(x,t)) + \partial_{x_2} {\bf{G}}({\bf{U}}(x,t)) = {\bf H}(x,t),& && \ \ \  {x} \in \mathbb{R}^2 \,,   t > 0,  \label{Euler-2d-motion} \\
{\bf{U}}(x,0)  = { \bf{U}}_0({x}),& && \ \ \ {x} \in \mathbb{R}^2 \,,   t = 0,
\end{alignat}
\end{subequations}
where the 4-vector $\bf{U}$ and the \emph{flux functions} ${\bf{F}}({\bf{U}})$ and ${\bf{G}}({\bf{U}})$ are defined as
\begin{equation}
{\bf U} = \left ( \begin{array}{c} \rho \\ \rho u_1 \\ \rho u_2 \\ E \end{array} \right ) \quad \text{ and } \quad {\bf F}({\bf U}) = \begin{pmatrix}
		\rho u_1 \\
		\rho u_1^2+p \\
		\rho u_1 u_2 \\
		u_1(E+p)
		\end{pmatrix}\, \quad \text{ and } \quad 
\bf{G}(\bf{U}) = \begin{pmatrix}
		\rho u_2 \\
		\rho u_1 u_2 \\
		\rho u_2^2 + p \\
		u_2(E+p)
		\end{pmatrix}\,. 
\end{equation}
The space coordinate is $x=(x_1,x_2)$, with $x_1$ denoting the horizontal component, $x_2$ denoting the vertical component, and $t\ge 0$ denoting time.
 The function $\bf H$ denotes the forcing function due
 to gravity, and so will be given as 
${\bf H} = (0\,,0\,,-\rho g\,,-\rho g u_2)$, where $g$ is a gravitational acceleration constant. 
The pressure $p$ is  defined by the ideal gas law, 
\begin{equation}\label{pres-eqn}
p = (\gamma-1)\left(E - \frac{1}{2}\rho | {u} |^2 \right)\,,
\end{equation}
where $\gamma$ is the adiabatic constant, which we will assume takes the value $\gamma=1.4$, unless
otherwise stated. 
We also define the specific internal energy per unit mass of the 
fluid as $e=p/(\rho(\gamma-1))$.    Once the initial data $u_0(x)$, $\rho_0(x)$, $E_0(x)$ are specified, solutions of \eqref{Euler-2d} provide the velocity, density, and energy
for each instant of time for which the solution exits.

\subsection{The incompressible and irrotational Euler equations}  In the absence of sound waves, 
the system \eqref{euler2d} can be simplified to model incompressible flows.  The incompressible Euler equations are written as
\begin{subequations}
\label{euler2d_incompressible}
\begin{alignat}{2}
\rho \left[ \partial_t {w}(x,t)+ (w \cdot \nabla) w \right] + \nabla p + g \rho e_2 = 0,& && \ \ \  {x} \in \mathbb{R}^2 \,,   t > 0,   \label{euler-inc-eqn1}  \\
\divv w = 0,& && \ \ \  {x} \in \mathbb{R}^2 \,,   t \ge 0,   \label{euler-inc-eqn2} \\
 {w}(x,0)  = {w}_0({x}),& && \ \ \ {x} \in \mathbb{R}^2 \,,   t = 0\,,  \label{euler-inc-eqn3}
\end{alignat}
\end{subequations} 
where $w=(w_1,w_2)$ denotes a divergence-free velocity vector field, 
the density $\rho$ is assumed to be a constant (or piecewise constant as we shall consider below), and the pressure $p$ is a Lagrange multiplier
which enforces the incompressibility constraint \eqref{euler-inc-eqn2}.   
We define the two-dimensional vorticity function $ \omega = \operatorname{curl} w$, where 
$$
\operatorname{curl} w := \nabla ^\perp  \cdot w = \p_1 w_2 - \p_2 w_1\,.
$$
Computing the $\operatorname{curl}$ of \eqref{euler-inc-eqn1} and using 
\eqref{euler-inc-eqn2} shows that the two-dimensional vorticity is transported by
incompressible flows,
$$
\partial_t \omega + w \cdot \nabla \omega =0 \,,
$$
and hence if the initial velocity $w_0(x)$ is chosen to be irrotational such that $\omega_0(x) =0$, then
 $\omega(x,t)=0$ for all time $t$ for which the solution exists.  Thus,
for such data, we supplement (\ref{euler2d_incompressible}) with
\begin{alignat}{2}
\operatorname{curl}  w = 0,& && \ \ \  {x} \in \mathbb{R}^2 \,,   t \ge 0 \,.  \tag{\ref*{euler2d_incompressible}d}  
\label{euler-inc-eqn4}
\end{alignat}

\subsection{The compressible Euler equations as a two-phase hyperbolic system}
\begin{figure}[H]
\centering
\begin{tikzpicture}
\draw [line width=0.75mm,blue] plot [smooth,tension=1] coordinates {(-1,-1) (0,-0.6) (1,1) (2,-1.5) (3,1) (3.5,0) (4.5,0.5) (4.0,-1) (5.25,-0.25) (5.9,-0.85) (6.5,-1.0)};
\node at (2.0,-2.0) {$\Omega^{-}(t)$};
\node at (2.0,1.) {$\Omega^{+}(t)$};
\node at (-0.25,-0.25) {$\Gamma(t)$};
\node at (4.25,1.0) {$\rho^{+}$, $u^{+}$, $E^{+}$};
\node at (4.25,-1.5) {$\rho^{-}$, $u^{-}$, $E^{-}$};
\draw [->, line width=0.35mm] (5.5,-0.25) -- (6.0,0.5);
\draw [->, line width=0.35mm] (5.5,-0.25) -- (6.25,-0.75);
\node at (6.15,0.65) {\small $n$};
\node at (6.45,-0.65) {\small $\tau$};
\end{tikzpicture}
\vspace{-.1 in}
\caption{An example of a time-dependent contact discontinuity $\Gamma(t)$ separating the two fluid regions $\Omega^+(t)$ and $\Omega^-(t)$.}\label{fig1}
\end{figure}
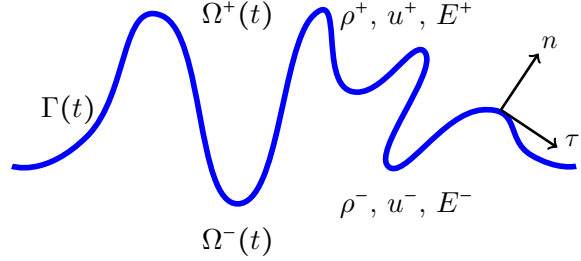

%
%
%
%
We are particularly interested in two-dimensional discontinuous solutions of the Euler equations 
\eqref{Euler-2d}  which propagate curves of discontinuity, whose evolution is determined by the
Rankine-Hugoniot conditions (see, for example, \cite{Dafermos2016}).   Specifically, 
our focus is on two-dimensional  solutions ${\bf U} = (\rho, \rho u_1, \rho u_2, E)$ to \eqref{Euler-2d} which have  jump
discontinuities across a time-dependent, space-periodic material interface $\Gamma(t)$ (see Figure \ref{fig1}).
The two-dimensional fluid domain is written as
$$\mathbb{R}^2 = \Omega^+(t) \cup \Omega^-(t) \cup \Gamma(t)\,,$$
where $\Omega^+(t)$ denotes the time-dependent open 
domain lying above $\Gamma(t)$, while $\Omega^-(t)$ denotes the open domain lying below $\Gamma(t)$. 
We let $n(\cdot , t)$ denote the unit normal vector to $\Gamma(t)$ pointing into $\Omega^+(t)$ and let $\tau(\cdot ,t)$ denote the unit tangent vector to $\Gamma(t)$, so that the
pair $(\tau,n)$ denote a right-handed basis.
We
denote by ${\bf{U}}^+$ the solution in the domain $\Omega^+(t)$ and  by ${\bf U}^-$,  the solution in $\Omega^-(t)$.    The jump of a function ${\bf{U}}$ across $\Gamma(t)$ is denoted
by
$$
 \llbracket {\bf{U}} \rrbracket = {\bf{U}}^+ - {\bf{U}}^-  \text{ on } \Gamma(t) \,.
$$

The Rankine-Hugoniot conditions relate the speed of propagation ${\sigma}(t)$ of the curve of discontinuity 
$\Gamma(t)$ with the jump discontinuity in the variables ${\bf U} = (\rho, \rho u_1, \rho u_2, E)$ via the relation
\begin{subequations}
\label{RH-conditions2}
\begin{align}
{\sigma} \llbracket{\rho} \rrbracket &= \llbracket \rho u \cdot n \rrbracket\,, \label{RH-1} \\
{\sigma} \llbracket{\rho u} \rrbracket &= \llbracket (\rho u \cdot n)u + p n \rrbracket\,, \label{RH-2} \\
{\sigma} \llbracket{E} \rrbracket &= \llbracket (E+p)u \cdot n \rrbracket\,, \label{RH-3}
\end{align}
\end{subequations}
which represent, respectively,  the conservation of mass, linear momentum, and energy across the discontinuity. Notice that
\eqref{RH-2} admits solutions with $\llbracket p \rrbracket = 0$ and 
$\llbracket (\sigma - u \cdot n) \rho u\rrbracket = 0$, and that the latter condition is satisfied if  
$u^{\pm} \cdot n = \sigma$. Such discontinuities are known as 
\emph{contact discontinuities}, in which case the interface $\Gamma(t)$ is
 transported by the fluid velocity $u$,  and  the pressure $p$ is continuous across
 $\Gamma(t)$. Contact discontinuities are the class of two-dimensional discontinuous solutions solving the  following  coupled {\it two-phase} 
 system of hyperbolic equations:
\begin{subequations}
\label{ceuler}
\begin{alignat}{2}
\partial_t \rho^\pm + \divv  (\rho u^\pm ) &= 0,  \qquad && \text{ in } \Omega^\pm(t)\,,  \label{ceuler-a} \\
\partial_t (\rho^\pm u^\pm) + \divv (\rho^\pm u^\pm \otimes u^\pm) + \nabla p^\pm + \rho g e_2 &= 0,  \qquad && \text{ in } \Omega^\pm(t)\,,   \label{ceuler-b}   \\ 
\partial_t E^\pm + \divv (u^\pm(E^\pm+p^\pm)) + \rho g u^\pm_2 &=0,  \qquad && \text{ in } \Omega^\pm(t) \,, \\
\llbracket {u} \cdot n \rrbracket  &= 0, \qquad && \text{ on } \Gamma(t) \,, \\
 \llbracket {u} \cdot \tau \rrbracket &\neq 0, \qquad && \text{ on } \Gamma(t) \,, \\
  \llbracket p  \rrbracket &= 0, \qquad && \text{ on } \Gamma(t)  \,, \\
\left(u^\pm(x,0), \rho^\pm(x,0), E^\pm(x,0), \Gamma(0)\right) &= \left(u^\pm_0, \rho^\pm_0, E^\pm_0, \Gamma_0\right), \qquad && \text{ at } t=0  \,.  \label{ceuler-g}
\end{alignat}
\end{subequations}
As already noted the interface $\Gamma(t)$ is transported by the velocity $u$ and we will make the dynamics of $\Gamma(t)$ precise, once we introduce
a parameterization for $\Gamma(t)$.
The tangential velocity jump discontinuity is the primary mechanism that initiates the Kelvin-Helmholtz instability . 
The densities $\rho^\pm$ and the energies $E^\pm$ are, in general, also discontinuous across $\Gamma(t)$.   The initial data is specified in \eqref{ceuler-g}.


\subsection{The two-phase incompressible and irrotational Euler equations}
The incompressible and irrotational Euler equations for two-phase flow are written as
\begin{subequations}
\label{ieuler}
\begin{alignat}{2}
\rho^\pm \left( \partial_t {w}^{\pm} + {w}^{\pm} \cdot \nabla {w}^{\pm} \right) + \nabla p^{\pm} + \rho^{\pm} g e_2 & = 0 \qquad && \text{ in } \Omega^\pm(t)\,, \\
 \operatorname{curl} w= \operatorname{div} w &= 0  \qquad && \text{ in } \Omega^\pm(t)\,, \\ 
\llbracket {w} \cdot n \rrbracket  &= 0 \qquad && \text{ on } \Gamma(t) \,, \\
 \llbracket {w} \cdot \tau \rrbracket &\neq 0\qquad && \text{ on } \Gamma(t) \,, \\
  \llbracket p  \rrbracket &= 0\qquad && \text{ on } \Gamma(t)  \,,  \label{ieuler-e} \\
\left(w^\pm(x,0), \Gamma(0)\right) &= \left(w^\pm_0, \Gamma_0\right) \qquad && \text{ at } t=0  \,,
\end{alignat}
\end{subequations}
and $\rho^+$ and $\rho^-$ are constant in each phase.   Again, the interface $\Gamma(t)$ is transported by the velocity $w$, and for incompressible
and irrotational flows, the interface $\Gamma(t)$ is called a \emph{vortex sheet}, because the vorticity is restricted to the one-dimensional interface as a measure,
as will be made precise.

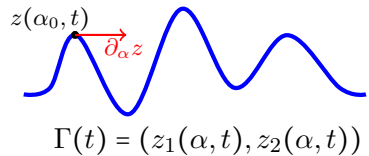
\begin{figure}[H]
\centering
\begin{tikzpicture}[scale=0.35]
    \draw (16,-2) node { $\Gamma(t)= \left( z_1( \alpha , t), z_2(\alpha ,t)\right)$}; 
 \draw[color=blue,ultra thick] plot[smooth,tension=.6] coordinates{(9,-0.25) ( 10,0) (11,2) (13,-1) (15, 3) (17, 0) (19, 2) (21,0) (22,-0.25) };
  \draw (11,2) node { $.$}; 
 \draw[color=black,ultra thick] (10.95,2.01) circle (0.7mm);
  \draw[color=red,thick,->] (10.95,2.01) -- (12.9,2.01);
 \draw (10.,2.6) node { \footnotesize{$ z( \alpha_0 , t)$}}; 
  \draw[color=red] (12.8,1.5) node { \footnotesize{$\partial_ \alpha  z$}}; 
\end{tikzpicture} 
\vspace{-.1 in}
\caption{The curve $\Gamma(t)$ is parameterized by $z( \alpha ,t)=(z_1( \alpha ,t), z_2( \alpha ,t))$.  A tangent vector at a point $z( \alpha , t)$ on $\Gamma(t)$ is
given by $\partial_ \alpha z( \alpha , t)$. }\label{fig2}
\end{figure}

 Incompressibility and irrotationality of the flow allow for a reduction of the 
system \eqref{ieuler} to a coupled
system of evolution equations in one space dimension.   We let $\mathbb{T}_L = [-L/2\,,L/2]$ denote a (periodic) interval of length $L$, and
 introduce a {\it parameterization} of the interface $\Gamma(t)$  by a mapping $z: \mathbb{T}_L  \to \mathbb{R}^2  $, so that for each $ \alpha$ in  $\mathbb{T}_L $,
the vector $z( \alpha , t) = {\left(z_1( \alpha ,t), z_2(\alpha , t)\right)}$ represents a point on the interface $\Gamma(t)$.
Moreover, for any $ \alpha_0$,
 the vector $\partial_ \alpha z( \alpha_0 , t)$ is tangent to $\Gamma(t)$ at the point $z( \alpha_0 ,t)$, and 
 $\tau = \partial_ \alpha z/ | \partial_ \alpha z|$ is the unit tangent vector at that point (as shown in Figure \ref{fig2}).
  Now, since $\Gamma(t)$ moves with speed $w \cdot n$, it follows that
 $\partial_t z ( \alpha , t)  = \left[w( z( \alpha , t),t)\cdot n\right] n$ and that the tangential motion of the interface $\partial_t z \cdot \partial_ \alpha z$ has no constraints at all.
The dynamics of the interface $\Gamma(t)$ are governed by the evolution equation
\begin{equation}\label{z-eqn}
\partial_t z(\alpha,t) = w(z(\alpha,t),t) \,.
\end{equation}

The vorticity $\omega$ vanishes in each of $\Omega^{\pm}(t)$, and is in fact a \emph{measure} 
supported on $\Gamma(t)$, written as
$$
\omega = \varpi \delta_{\Gamma(t)}\,,
$$
where $\delta_{\Gamma(t)}$ is the Dirac delta distribution supported on $\Gamma(t)$, and the 
function $\varpi$ is the \emph{amplitude of vorticity} along $\Gamma(t)$.
More precisely, if $\varphi$ is any smooth test function with compact support in $ \mathbb{R}^2  $,
then 
$$
\langle \omega \,, \varphi \rangle = \int_{\mathbb{R}} \varpi(\beta,t) \varphi(z(\beta,t)) \, \mathrm{d}\beta\,. 
$$
The amplitude of vorticity $\varpi$ may be computed in terms of the jump in the velocity as
\begin{equation*}
\begin{split}
\int_{\mathbb{R}} \varpi(\beta,t) \varphi(z(\beta,t)) \, \mathrm{d}\beta = \langle \omega \,, \varphi \rangle & \coloneqq - \int_{\Omega} w \cdot \nabla^{\perp} \varphi \,\mathrm{d}x  \\
& = \int_{\Omega^{+} \cup \Omega^{-}} \varphi \nabla^{\perp} \cdot w \,\mathrm{d}x - \int_{\Gamma} \llbracket w \cdot \tau \rrbracket \varphi \,\mathrm{d}S \,.
\end{split}
\end{equation*}
Since the vorticity $\omega = \nabla^{\perp} \cdot w = 0$  in 
$\Omega^{+}(t) \cup \Omega^{-}(t)$,  it  then follows that 
$$
\int_{\mathbb{T}_L} \varpi(\beta,t) \varphi(z(\beta,t)) \, \mathrm{d}\beta = - \int_{\Gamma} \llbracket w \cdot \tau \rrbracket \varphi \,\mathrm{d}S = - \int_{\mathbb{R}} \llbracket w \cdot \tau \rrbracket \varphi(z(\beta,t)) |\partial_{\alpha} z(\beta,t) |  \,\mathrm{d} \beta
$$
for any smooth test function $\varphi $ with compact support in $\mathbb{R}^2  $, which implies that 
$$
\varpi = - \llbracket w \cdot \tau \rrbracket | \partial_{\alpha} z | = - \llbracket w \cdot \partial_{\alpha} z \rrbracket \,.
$$

Due to the fact that the flow is both irrotational and incompressible, there exist scalar \emph{stream functions}
$\psi^{\pm}(x,t)$ such that $\Delta \psi^{\pm} = 0$ in $\Omega^{\pm}(t)$ and 
$w^{\pm} = \nabla^{\perp} \psi^{\pm}$.

Following \cite{Rott1956,Birkhoff1962}, we next
reduce  \eqref{ieuler} to a system of coupled evolutionary integro-differential equations  in one space dimension. 
The incompressible and irrotational velocity $w$ can be reconstructed from the vorticity measure $\varpi$ using the well-known 
Biot-Savart kernel $\mathcal{K}_{\mathbb{R}}(x)$, which is an integral representation for 
$\nabla^{\perp} \Delta^{-1}$ in $\mathbb{R}^2$. The kernel is defined by
\begin{equation}\label{biot-savart-r2}
\mathcal{K}_{\mathbb{R}}(x) = \frac{x^{\perp}}{2\pi |x|^2} = \frac{1}{2\pi} \left( \frac{-x_2}{x_1^2 + x_2^2} \,, \frac{x_1}{x_1^2 + x_2^2}\right)\,.
\end{equation}
Away from the interface, the velocity $w$ is then given as 
\begin{equation}\label{BS-velocity-R}
w(x,t) = \pvint_{\mathbb{R}} \mathcal{K}_{\mathbb{R}}(x-z(\beta,t)) \varpi(\beta,t) \,\mathrm{d} \beta\,, 
\end{equation}
for $x \in \Omega^{+}(t) \cup \Omega^{-}(t)$, where we recall that the integral is to be understood in the 
principal value sense \eqref{eq:pv}. At the interface $\Gamma(t)$, the velocity 
$w|_{\Gamma(t)}$ is defined to 
be the average $(w^+ + w^-)/2$ on $\Gamma(t)$. The Plemelj formulae give
$$
w^{\pm}(z(\alpha,t),t) = \pvint_{\mathbb{R}} \mathcal{K}_{\mathbb{R}}(z(\alpha,t)-z(\beta,t)) \varpi(\beta,t) \,\mathrm{d} \beta \pm \frac{1}{2} \frac{\varpi(\alpha,t)}{|\partial_{\alpha} z(\alpha,t)|} \tau \,, 
$$
from which it follows that
\begin{equation}\nonumber
w(z(\alpha,t),t) = \pvint_{\mathbb{R}} \mathcal{K}_{\mathbb{R}}(z(\alpha,t)-z(\beta,t)) \varpi(\beta,t) \,\mathrm{d} \beta\,.
\end{equation}
This integral is over the real line.  For horizontally periodic flows, 
the integral can be summed over the periodic images to yield an integral over a single
period, with the kernel given by
\begin{equation}\label{biot-savart-torus}
\mathcal{K}_{\mathbb{T}_L}(x) = \frac{\left(-\sinh(2\pi x_2/L)\,, \sin(2\pi x_1/L) \right)}{2L\left(\cosh(2 \pi x_2/L)-\cos(2 \pi x_1 /L)\right)} \,,
\end{equation}
where $\mathbb{T}_L = [-L/2\,,L/2]$ denotes the periodic interval with period $L$.   Hence,
\begin{equation}\label{wBR}
w(z(\alpha,t),t) = \pvint_{\mathbb{T}_L } \mathcal{K}_{\mathbb{T}_L }(z(\alpha,t)-z(\beta,t)) \varpi(\beta,t) \,\mathrm{d} \beta\,.
\end{equation}

Together with \eqref{z-eqn}, 
the system is closed by determining the evolution equation for the amplitude of vorticity $\varpi$. 
A lengthy computation \cite{CoCoGa2010,GrSh2017} using the Bernoulli equation, the Plemelj formulae, 
and \eqref{ieuler-e} provides the dynamics for $\varpi$; together with \eqref{z-eqn} and \eqref{wBR}, we obtain the following
coupled system:
\begin{subequations}
\label{zeuler}
\begin{align} 
\p_t z( \alpha , t) &= \pvint_{\mathbb{T}_L} \mathcal{K}_{\mathbb{T}_L}(z(\alpha,t)-z(\beta,t)) \varpi(\beta,t) \,\mathrm{d}\beta \,,  \label{zeuler-a} \\
\partial_t \varpi (\alpha,t) &= -\partial_{\alpha} \left[ \frac{A}{4 \pi^2} \left| \pvint_{\mathbb{R}} \varpi(\beta,t) \frac{(z(\alpha,t)-z(\beta,t))^{\perp}}{|z(\alpha,t)-z(\beta,t)|^2}\,\mathrm{d}\beta \right|^2 - \frac{A}{4} \frac{\varpi(\alpha,t)^2}{|\partial_{\alpha}z(\alpha,t)|^2} - 2Agz_2 \right]   \label{zeuler-b} \\ 
& \qquad \qquad+ \frac{A}{\pi} \partial_t \left[   \pvint_{\mathbb{R}} \varpi(\beta,t) \frac{(z(\alpha,t)-z(\beta,t))^{\perp}}{|z(\alpha,t)-z(\beta,t)|^2} \cdot \partial_{\alpha}z(\alpha,t)\,\mathrm{d}\beta \right] \,, \nonumber
\end{align}
\end{subequations}
where 
$$A = (\rho^{+} - \rho^{-})/(\rho^{+} + \rho^{-})$$ is the Atwood number.   These equations are solved for $ \alpha \in \mathbb{T}  _L$ and $t>0$.   The coupled equations 
\eqref{zeuler}  are the incompressible
and irrotational Euler equations, reduced to a one-dimensional problem for the three unknowns $(z_1,z_2, \varpi)$. 

\textcolor{black}{The analysis of the BR system \eqref{zeuler} is difficult, due to 
the presence of the Kelvin-Helmholtz instability. 
Linear stability analysis yields perturbation solutions with arbitrarily large growth rates, so that the 
problem is ill-posed in the sense of Hadamard \cite{Ebin1988,KaLe2005}. \citet{Delort1991} proved
existence of  global weak solutions for initial data that is a signed vorticity measure (concentrated on the interface); 
see also \cite{Majda1993,EvMu1994,Xin2001}. Uniqueness of these solutions has not been proved,
and there is evidence to suggest that such solutions are, in fact, not unique 
\cite{Sze2011,Pullin1989,MaMaZh1994,Zheng2006}.
}

\subsection{An asymptotic model for incompressible interface motion: the $z$-model}\label{subsec:z-model}
As we will explain in Section \ref{z-model-numerical-implementation}, the numerical solution of the system \eqref{zeuler} can be computationally expensive and difficult to implement. Moreover, the equations are
sufficiently complex that, in many cases, the dynamics of solutions is extremely difficult to analyze. As such, 
there has been a sustained effort to 
develop  \emph{model} equations that can suitably approximate the Euler 
equations in certain asymptotic 
regimes. For water waves (i.e. $A=-1$), there are a number of such equations (see, for example, \cite{AuGrShWi2019,CrSu1993} and references therein), and
for the two-fluid case (i.e. $-1 < A < 1$), a number of 
modal models have been proposed for the evolution of the interface, such as the models of 
\cite{Goncharov2002} and \cite{Haan1991}; we refer the reader to \citet{Zhou2017} for an extensive 
review of the subject.

The fundamental difficulty is the nonlocal nature of the singular integral equations \eqref{zeuler}, 
in which the dynamics at a point on the interface require information at all other points on the interface. 
By developing a new asymptotic procedure in which $z$ and $\varpi$ are expanded in a small non-locality parameter, \citet{GrSh2017}
obtained model equations, approximating the solution to \eqref{zeuler}, which allow for interface turn-over and place no constraints on the steepness of the
interface.   These localized equations are 
\begin{subequations}
\label{zmodel0}
\begin{align} 
\partial_t z( \alpha ,t)& = {\frac{1}{2}} \mathcal{H} \varpi( \alpha , t) \frac{\p_ \alpha z^\perp( \alpha ,t)}{|\partial_{\alpha} z (\alpha,t)|^2} \,, \label{zmodel0-a} \\
\partial_t \varpi(\alpha,t) &= - \partial_{\alpha} \left[ \frac{A}{2|\partial_{\alpha} z (\alpha,t)|^2}  \mathcal{H} \left( \varpi(\alpha,t) \mathcal{H} \varpi(\alpha,t) \right) - 2Ag z_2(\alpha,t) \right]\,,
\end{align} 
\end{subequations} 
where $\mathcal{H}$ denotes the Hilbert transform, defined in \eqref{Hilbert}. 
The equations \eqref{zmodel0} are called the (lower-order) $z$-model.

A number of numerical experiments of the (lower-order) $z$-model were performed in \cite{GrSh2017}, which demonstrated very good agreement with experimental  data and theoretical 
predictions of interface growth, but the localized nature of the evolution for $z( \alpha ,t)$ in \eqref{zmodel0-a} can inhibit the initiation of Kelvin-Helmholtz roll-up.   On the
other hand, the fundamental challenge in simulating the Euler system \eqref{zeuler} stems from the evolution equation for $\varpi$.    As such we introduce the {\it 
higher-order} $z$-model as the following system:
\begin{subequations}
\label{zmodel}
\begin{align}
\partial_t z(\alpha,t) &= \pvint_{\mathbb{T}_L} \mathcal{K}_{\mathbb{T}_L}(z(\alpha,t)-z(\beta,t)) \varpi(\beta,t) \,\mathrm{d}\beta \,, \label{BR-eqn}\\
\partial_t \varpi(\alpha,t) &= - \partial_{\alpha} \left[ \frac{A}{2|\partial_{\alpha} z (\alpha,t)|^2}  \mathcal{H} \left( \varpi(\alpha,t) \mathcal{H} \varpi(\alpha,t) \right) - 2Ag z_2(\alpha,t) \right] \label{wbar-eqn}\,,
\end{align}
\end{subequations}
in which the asymptotic model for $\varpi$ evolution is coupled to the integral equation for $z$.

\subsection{The Euler equations as a two-phase elliptic system for velocity}  
We now reformulate the full compressible Euler equations \eqref{ceuler} as a two-phase elliptic system for
the  compressible velocity vector $u$.   
As we have already stated, by using the parameterization $z( \alpha , t)$ for the interface $\Gamma(t)$,  
the dynamics of the interface $\Gamma(t)$ are governed by the evolution equation
$\partial_t z(\alpha,t) = u(z(\alpha,t),t)$, 
and from the definition of  the amplitude of vorticity  $\varpi$, we have the following jump conditions for the velocity:
$$
 \llbracket u \cdot \tau \rrbracket = -     \tfrac{\varpi}{| \partial_{\alpha} z |} \text{ and }   \llbracket u \cdot n \rrbracket  =0 \,.
$$

Next, from equation \eqref{ceuler-a}, we have that
$$
F^\pm:= \divv u^\pm = - \frac{  \tfrac{\mathrm{D}\rho^\pm}{\mathrm{D}t}}{ \rho^\pm} \,.
$$
Letting the operator $ \operatorname{curl} $ act on \eqref{ceuler-b}, and setting $\omega= \operatorname{curl} u$, we find
that $\omega$ is the solution of
$$
\tfrac{\mathrm{D} \omega}{\mathrm{D} t} + \omega \divv u = -  \tfrac{\nabla \rho \cdot \nabla^{\perp} p}{\rho^2}  \,.
$$
Thus, given $z$, $\varpi$, $F^\pm$, and $ \omega^\pm$, we can reconstruct $u^\pm$ by solving the following two-phase elliptic system:
\begin{subequations}
\label{celliptic}
\begin{alignat}{2}
\divv u^\pm&= F^\pm,  \qquad && \text{ in } \Omega^\pm(t)\,, \\
\operatorname{curl} u^\pm&= \omega^\pm,  \qquad && \text{ in } \Omega^\pm(t)\,, \\ 
\llbracket {u} \cdot n \rrbracket  &= 0 ,\qquad && \text{ on } \Gamma(t) \,, \\
 \llbracket {u} \cdot \tau \rrbracket &= -     \tfrac{\varpi}{| \partial_{\alpha} z |}, \qquad && \text{ on } \Gamma(t) \,. 
\end{alignat}
\end{subequations}
For the two-dimensional geometry that we are considering, the system \eqref{celliptic} is uniquely solvable  \cite{ChSh2017}, and thus $u$ obtained from \eqref{celliptic}
is the velocity field solving the compressible Euler equations \eqref{ceuler}.

\subsection{A compressible-incompressible decomposition of the Euler equations}\label{sec:comp-incomp}  
We substitute the additive decomposition 
$u= v+w $ 
into the two-phase elliptic system \eqref{celliptic} and define
$v^\pm$ and $w^\pm$ to be the solutions of
\begin{subequations}
\label{elliptic_vw}
\begin{alignat}{5}
\divv  v^\pm &= F^\pm &&\text{in}\quad \Omega^\pm(t) \,,  \qquad \qquad \qquad && \divv w^\pm &&= 0&&\text{in}\quad \Omega^\pm(t)\,, \\
\operatorname{curl}  v^\pm &= \omega^\pm &&\text{in}\quad \Omega^\pm(t) \,, \qquad \qquad\qquad  &&\operatorname{curl}  w^\pm &&= 0 &&\text{in}\quad \Omega^\pm(t)\,,\\
 \llbracket v \cdot n \rrbracket&= 0 \qquad &&\text{on}\quad \Gamma(t) \,,\qquad \qquad \qquad  &&  \llbracket w \cdot n  \rrbracket&&= 0 \qquad &&\text{on}\quad \Gamma(t)\,,\\ 
 \llbracket v \cdot \tau\rrbracket&= 0  && \text{on}\quad \Gamma(t) \,, \qquad \qquad \qquad && \llbracket w \cdot \tau  \rrbracket&&= - \tfrac{\varpi}{ |\p_{\alpha} z|  }   \quad && \text{on}\quad \Gamma(t) \,,
\end{alignat}
\end{subequations}
together with
\begin{align}
\p_t z( \alpha , t) & = v(z( \alpha ,t),t) + w(z( \alpha ,t),t) \,,  \tag{\ref*{elliptic_vw}e} \label{elliptic_vw-e} \\
 \varpi( \alpha , t) & = - \llbracket {w(z( \alpha,t),t)  \cdot \partial_{\alpha} z( \alpha , t)} \rrbracket  \,.  \tag{\ref*{elliptic_vw}f}  \label{elliptic_vw-f}
\end{align} 

The velocity $w$ is incompressible and irrotational,  but has a discontinuity in its tangential component, while the velocity $v$ is continuous and is forced by
the bulk compression and vorticity of the fluid.   There are a number of different ways to find velocities $v$ and $w$ such that $u$ solves the full compressible Euler
equations \eqref{ceuler}.   We shall  simultaneously solve for the pair $(v,w)$ as the solution of the following system:
\begin{subequations}
\label{w-eqn}
\begin{alignat}{2}
\bar \rho^\pm \left( \partial_t {w}^{\pm} + {w}^{\pm} \cdot \nabla {w}^{\pm} \right) + \nabla p^{\pm} + \bar\rho^{\pm} g e_2 & = 0 \qquad && \text{ in } \Omega^\pm(t)\,, \\
 \operatorname{curl} w^\pm= \operatorname{div} w^\pm &= 0  \qquad && \text{ on } \Omega^\pm(t)\,, \\ 
\llbracket {w} \cdot n \rrbracket  &= 0 \qquad && \text{ on } \Gamma(t) \,, \\
 \llbracket {w} \cdot \tau \rrbracket &= -     \tfrac{\varpi}{| \partial_{\alpha} z |}\qquad && \text{ on } \Gamma(t) \,, \\
  \llbracket p  \rrbracket &= 0\qquad && \text{ on } \Gamma(t)  \,, \\
\left(w^\pm(x,0), z( \alpha ,0)\right) &= \left(0, z_0( \alpha )\right) \qquad && \text{ at } t=0  \,,
\end{alignat}
\end{subequations}
where $z_0(\alpha )$ is the initial data for the parameterization of the interface,  the initial amplitude of vorticity is computed as
$$
\varpi( \alpha , 0)  = - \llbracket u(z_0( \alpha),0)  \cdot \partial_{\alpha} z_0( \alpha) \rrbracket  \,,
$$
and the density functions $\bar\rho^\pm$ are constants given by $\bar\rho^\pm = \rho_0^\pm|_{\Gamma_0}$.
This is coupled to 
\begin{subequations}
\label{v-eqn}
\begin{alignat}{2}
\partial_t \rho^\pm + \operatorname{div}  (\rho^\pm v^\pm ) &= - \operatorname{div} (\rho^\pm w^\pm)  \qquad && \text{ in } \Omega^\pm(t)\,,  \label{v-eqn-a} \\
\partial_t (\rho^\pm v^\pm) + \operatorname{div} (\rho^\pm v^\pm \otimes v^\pm) + \nabla p^\pm + \rho^\pm g e_2 &= -\p_t (\rho^\pm w^\pm) \nonumber  \\
& \hspace{-1.5 in} - \operatorname{div} \left( \rho^\pm(w^\pm \otimes w^\pm + w^\pm \otimes v^\pm + v^\pm \otimes w^\pm ) \right)
\qquad && \text{ on } \Omega^\pm(t)\,,  \label{v-eqn-b} \\ 
\partial_t E^\pm + \operatorname{div} (v^\pm(E^\pm+p^\pm)) + \rho^\pm g v^\pm_2 &= - \operatorname{div} ( w^\pm(E^\pm+p^\pm)) -g\rho^\pm w^\pm_2 \qquad && \text{ on } \Omega^\pm(t) \,, \\
\llbracket {v} \rrbracket  &= 0 \qquad && \text{ on } \Gamma(t) \,, \\
  \llbracket p  \rrbracket &= 0\qquad && \text{ on } \Gamma(t)  \,, \\
\left(v^\pm(x,0), \rho^\pm(x,0), E^\pm(x,0), z( \alpha ,0)\right) &= \left(u^\pm_0, \rho^\pm_0, E^\pm_0, z_0( \alpha )\right) \qquad && \text{ at } t=0  \,,
\end{alignat}
\end{subequations}
together with \eqref{elliptic_vw-e} and \eqref{elliptic_vw-f}.    This decomposition of the flow into velocities $v$ and $w$ provides a natural setting for a multiscale model of compressible 
interface evolution.   In particular, we shall develop a two-scale solution strategy, in which \eqref{w-eqn} is solved over small scales using our higher-order incompressible
 $z$-model \eqref{zmodel}, and \eqref{v-eqn} is solved over large scales.
 
An equivalent formulation for this $(v,w)$ system is given by replacing \eqref{elliptic_vw-e},
\eqref{elliptic_vw-f}, and \eqref{w-eqn} with
 \begin{subequations}
\label{zeuler2}
\begin{align} 
\p_t z( \alpha , t) &= \pvint_{\mathbb{T}_L} \mathcal{K}_{\mathbb{T}_L}(z(\alpha,t)-z(\beta,t)) \varpi(\beta,t) \,\mathrm{d}\beta + v(z( \alpha , t),t) \,, \\
\partial_t \varpi (\alpha,t) &= -\partial_{\alpha} \left[ \frac{A}{4 \pi^2} \left| \pvint_{\mathbb{R}} \varpi(\beta,t) \frac{(z(\alpha,t)-z(\beta,t))^{\perp}}{|z(\alpha,t)-z(\beta,t)|^2}\,\mathrm{d}\beta \right|^2 - \frac{A}{4} \frac{\varpi(\alpha,t)^2}{|\partial_{\alpha}z(\alpha,t)|^2} - 2Agz_2 \right] \nonumber \\ 
& \qquad \qquad+ \frac{A}{\pi} \partial_t \left[   \pvint_{\mathbb{R}} \varpi(\beta,t) \frac{(z(\alpha,t)-z(\beta,t))^{\perp}}{|z(\alpha,t)-z(\beta,t)|^2} \cdot \partial_{\alpha}z(\alpha,t)\,\mathrm{d}\beta \right] \,, \label{zeuler2-2}
\end{align}
\end{subequations}
 and coupling these equations with \eqref{v-eqn}.   Then $w$ is computed using \eqref{BS-velocity-R}.   
 This will be the basis for our multiscale modeling approach.

 \textcolor{black}{We note that both the two-dimensional compressible and incompressible Euler equations are ill-posed in Sobolev spaces for most vortex sheet initial 
 data.\footnote{ \textcolor{black}{ The two-dimensional compressible Euler equations are 
 weakly well-posed in Sobolev spaces if the
 the initial vorticity measure (tangential jump of velocity along the interface) is sufficiently large relative to the Mach number of the flow
 \cite{CoSe2007,CoMo2004}}} 
    On the other hand, these systems of equations become well-posed in Sobolev spaces if either
bulk viscosity \cite{Den2007}  or surface tension along the contact discontinuity \cite{ChCoSh2007} is added.  As such, any 
state-of-the-art high-order numerical discretization of the compressible Euler equations 
uses some form of regularization  to remove small-scale instability and oscillations (see the review paper \cite{LiWe2003}).
In order to generate high-resolution reference solutions for comparison with our multiscale algorithm, we  too rely on a regularization scheme
that employs  a new type of  anisotropic artificial viscosity operator which is described  in \Cref{C-method-numerical-implementation}.  This anisotropic
operator adds nonlinear viscosity only in directions tangential to the 
 evolving front while adding virtually  zero viscosity in the direction normal to the interface.  This approach
ensures that that the contact discontinuity does  not become too smeared (which indeed occurs for more traditional isotropic artificial 
viscosity operators).}
 
 \textcolor{black}{In deriving the 
multiscale  decomposition, we return to the inviscid Euler setting in which all regularization is removed.   The 
inviscid velocity $u$ is decomposed into  $v$ and $w$, which are evolved by the equations
  \eqref{v-eqn} and  \eqref{zeuler2}.    As such $v$ and $w$ are again governed by inviscid systems, and in particular, $w$
is governed by the inviscid incompressible and irrotational Euler equations, which are once again ill-posed, while $v$ no longer has
a discontinuity and does not suffer from the same small-scale instabilities as the original Euler system it was derived from.     Now,
the $w$ system must be numerically regularized, but before we do so, we make a significant simplification for the dynamics for the
amplitude of vorticity $\varpi$.  The equation \eqref{zeuler2-2} is a highly nonlinear and nonlocal equation.  We replace this complicated evolution with our asymptotic
$z$-model \eqref{wbar-eqn}. 
This  is a Burgers-type equation which leads to shock formation from  initial data consisting of  small perturbations of equilibrium.  This
Burgers-like equation can be stabilized in the same way as the one-dimensional  Burgers equation, and the use
of artificial viscosity is the most natural (and efficient) method for this purpose.
As shown in \cite{AuGrShWi2019}, the $\varpi$ equation \eqref{wbar-eqn} (with and without regularization) is locally well-posed for analytic  data; for
such  data (with periodic boundary conditions), there exists a limit of zero artificial viscosity.}

\textcolor{black}{
On the other hand,  since the foundational work of \citet{ChBe1973}, it is very natural to 
use the vortex blob method (to be described below in \Cref{subsec-z-model-smoothing})
to solve for $z_1$ and $z_2$.   While there are many other choices for regularizing the $z$-model, using this combination of artificial viscosity for
$\varpi$ and vortex blobs for $z_1$ and $z_2$ produces an efficient and stable algorithm which allows for convergence of the large-scale structures
of the flow (such as bubble and spike locations), as will be demonstrated below.
}
  
  \textcolor{black}{
An alternative approach to our multiscale decomposition might have been to decompose  solutions
of the compressible Navier-Stokes equations into large-scale and small-scale velocities, but special (and very restrictive) interface conditions would
then be required to keep the interface sharp, while the standard interface conditions would instead enforce continuity of the velocity across the
interface.   While  vortex methods for viscous flows \cite{CoKo2000} have been developed by \citet{Chorin1973,Chorin1980} using the random walk method and by
 \citet{Degond1989} using  weighted particle methods, we instead rely upon a 
 simple artificial viscosity scheme restricted to
 the interface for our multiscale algorithm which is 
computationally   less  expensive than the vortex methods for viscous flows.}

\section{Numerical implementation of the $z$-model}\label{z-model-numerical-implementation}   
Our multiscale model will rely on a fast-running numerical implementation of the higher-order $z$-model \eqref{zmodel}.   In this section, we explain the method,
and perform  some classical numerical experiments to demonstrate the efficacy of our scheme.

\subsection{A regularization of the incompressible $z$-model}\label{subsec-z-model-smoothing}

A simple method for approximating the singular integral on the right-hand side of \eqref{BR-eqn} is to use a 
standard trapezoidal quadrature rule. This is the original \emph{point vortex method} of 
\citet{Rosenhead1931}. \textcolor{black}{Unfortunately, as demonstrated in 
\cite{Krasny1986a,ChBe1973}, solutions computed 
using the point vortex method often suffer from irregular point vortex motion due to small 
perturbation errors introduced by round-off error\footnote{Few digits of precision can also be regularizing, 
as shown by the calculations of \citet{Rosenhead1931}.}.
Such irregular motion is then  amplified by the Kelvin-Helmholtz instability.}
Moreover, this irregular motion persists as the mesh is refined, and is in fact initiated at earlier times as 
the number of nodes increases.

In \cite{Krasny1986b,Krasny1987}, 
the equation \eqref{BR-eqn} is \emph{desingularized} by smoothing the singular kernel 
$\mathcal{K}_{\mathbb{T}_L}$ to yield the desingularized kernel 
\begin{equation}\label{smoothed-kernel}
\mathcal{K}^\delta_{\mathbb{T}_L}(x) = \frac{\left(-\sinh(2\pi x_2/L)\,, \sin(2\pi x_1/L) \right)}{2L\left(\delta^2 + \cosh(2 \pi x_2/L)-\cos(2 \pi x_1 /L)\right)} \,.
\end{equation}
with $\delta \in \mathbb{R}$ some constant. This yields a more numerically stable set of equations to which 
the standard trapezoidal quadrature rule can be applied. Computational evidence \cite{Krasny1986b} 
suggests that this approximation converges beyond the singularity time if the mesh is refined and the 
smoothing parameter $\delta$ is decreased, in the appropriate order. 
In our numerical experiments, we have found that the 
scaling 
\begin{equation}\label{delta-scaling}
\delta^2 = | {\scriptstyle\Delta} \alpha \log{{\scriptstyle\Delta} \alpha} | \cdot \tilde{\delta}^2 \,,
\end{equation}
with $\tilde{\delta}$ a 
constant, yields 
stable solutions with increasing amounts of roll-up as ${\scriptstyle\Delta} \alpha \to 0$. 
\textcolor{black}{The details of how the parameter $\tilde{\delta}$ is chosen are provided 
in \S \ref{subsubsec-delta-tests}, in which we provide an example of the procedure applied to 
a KHI test problem.}

\textcolor{black}{The full-space version of the desingularized kernel \eqref{smoothed-kernel} is given by 
\begin{equation}\label{smoothed-kernel-r}
\mathcal{K}^\delta_{\mathbb{R}}(x) = \frac{1}{2 \pi} \frac{x^{\perp}}{ \left( |x|^2 + \delta^2  \right) } \,.
\end{equation}
We will make use of \eqref{smoothed-kernel-r} in the numerical experiments in \Cref{z-model-numerical-study}.}
  

\textcolor{black}{
Convergence of the point vortex and vortex blob methods for smooth flows is proved in 
\cite{Hald1979,BeMa1982,GoHoLo1990}.
For vortex sheets, where the initial data is not smooth, 
\citet{CaLo1989} proved global existence of analytic solutions from arbitrary analytic initial data for 
the desingularized equations in the case $A=0$. They also proved short time convergence of the vortex blob
method as the desingularization parameter $\delta$, mesh size, and time-step converge to zero.
When the sheet is analytic, the error due to the desingularization is 
$\mathcal{O}(\delta)$ \cite{CaLo1989}, assuming round-off errors are sufficiently 
small. 
Convergence for weak solutions was proved by \citet{LiuXin1995} in the case that the vorticity measure 
is of distinguished sign (and the Atwood number vanishes, $A=0$).}

\textcolor{black}{Let us note that the full-space desingularized kernel \eqref{smoothed-kernel-r} satisfies the 
sufficient conditions of the theorem of \citet{LiuXin1995}, whereas the periodic desingularized kernel 
\eqref{smoothed-kernel} does not satisfy the assumptions. Consequently, it is not known whether the solutions 
to the system induced by the regularized kernel \eqref{smoothed-kernel} converge to a weak solution of the
incompressible Euler system. 
Nonetheless, there is numerical evidence to suggest that the numerical solution does indeed converge
 \cite{Krasny1986b,BaPh2006}.}

Next, we turn to the evolution equation for the amplitude of vorticity $\varpi$. 
The nonlinearity in \eqref{wbar-eqn} often results in the development of steep profiles of the variable 
$\varpi$, analogous to the formation of shocks for solutions to nonlinear conservation laws. The shock 
formation in $\varpi$ generally occurs at late times, and is followed by the 
roll-up of the vortex sheet. One can handle this phenomenon by using shock-capturing methods 
(see for instance \citet{Sohn2004a}). However, we use the simplest possible technique, namely a linear 
artificial viscosity operator $\mu \partial^2_{\alpha}$ with $\mu \geq 0$, to \emph{smear} the shock 
over a small number of cells and thereby stabilize the solution.

We therefore consider the following \emph{regularization} of the higher-order $z$-model \eqref{zmodel} as follows:
\begin{subequations}
\label{BR-system-regular}
\begin{align}
\partial_t z(\alpha,t) &= \int_{\mathbb{T}_L} \mathcal{K}^\delta_{\mathbb{T}_L}(z(\alpha,t)-z(\beta,t)) \varpi(\beta,t) \,\mathrm{d}\beta \,, \label{BR-eqn-regular}\\
\partial_t \varpi(\alpha,t) &= - \partial_{\alpha} \left[ \frac{A}{2|\partial_{\alpha} z (\alpha,t)|^2}  \mathcal{H} \left( \varpi(\alpha,t) \mathcal{H} \varpi(\alpha,t) \right) - 2Ag z_2(\alpha,t) \right] \label{wbar-eqn-regular} + \mu \partial^2_{\alpha} \varpi(\alpha,t)\,.
\end{align}
\end{subequations}
When $\delta = 0$, the Cauchy principal value of the integral in \eqref{BR-eqn-regular} must 
be taken, while for $\delta > 0$, the integral is proper and the equations are then a regularized 
approximation to periodic vortex sheet evolution. 

\subsubsection{Other regularizations of the singular kernel \eqref{biot-savart-torus}}

\textcolor{black}{For the purposes of comparison with the Krasny approximation, we consider two other desingularized 
kernels that approximate the singular kernels \eqref{biot-savart-torus} or \eqref{biot-savart-r2}. 
These kernels were derived by 
\citet{BaBe2004} (see also \cite{BeMa1985}), and are of the form
\begin{equation}\label{kernel-baker}
\mathcal{K}^{\delta}_i(x) = \mathcal{K}(x) ( 1 + g_i ( \eta_\delta ) )\,,
\end{equation}
where the subscript $i$ is related to the number of moment conditions satisfied by the kernel 
(see \cite{BaBe2004} for the details). More precisely, we shall consider $g_i$ for $i=1,3$, in which case we 
have
\begin{subequations}
\begin{align*}
 g_1(\eta_\delta) &= - \exp(- \eta_\delta^2)  \,, \\
g_3(\eta_\delta) &= (-1 + 2 \eta_\delta^2) \exp(- \eta_\delta^2) \,, 
\end{align*}
\end{subequations}
Here, $\mathcal{K}(x)$ refers to either the $\mathbb{R}^2$ kernel \eqref{biot-savart-r2} or the periodic 
kernel \eqref{biot-savart-torus}. In the former case, the variable $\eta_ \delta $ is given by $\eta_ \delta  = |x|/\delta$, while in 
the latter case, $\eta_ \delta ^2 = 2 \left( \cosh(x_2) - \cos(x_1) \right) / \delta^2$. 
}

\textcolor{black}{Formula \eqref{kernel-baker} is indeed a desingularization due to the fact that the functions 
$g_i$ satisfy $1 + g_i(\eta_ \delta ) = \mathcal{O}(\eta_ \delta ^2)$ as $\eta_ \delta  \to 0$, whereas the singular kernel 
$\mathcal{K}(x)$ has an $\mathcal{O}(1/x)$  type singularity 
at the origin, so that the denominator is cancelled and 
$\mathcal{K}^\delta_i(x)$ is a smooth function of $x$. We note that the exponential decay of 
\eqref{kernel-baker} as $|\eta_ \delta | \to \infty$ is in contrast to the slower
algebraic decay of the Krasny desingularization \eqref{smoothed-kernel}.}

\textcolor{black}{The kernels $\mathcal{K}^\delta_i$ satisfy the sufficient conditions of the theorem of 
\citet{LiuXin1995}, and consequently, solutions to the regularized system converge
as $\delta \to 0$, when the initial vorticity amplitude $\varpi$ is of distinguished sign.}

\subsection{Discretization of  \eqref{BR-system-regular}}

Suppose that a single wavelength of the periodic interface $\Gamma(t)$ is parametrized by the function 
$z(\alpha,t)$, and that the parameter $\alpha$ is discretized with $N+1=2^r+1$ nodes,  
$$
\alpha_k = -L/2 + (k-1) {\scriptstyle\Delta} \alpha \,,
$$
with ${\scriptstyle\Delta} \alpha = L/N$. 

We spatially discretize the equations of motion, then use a standard third-order explicit Runge-Kutta
solver for time integration. A trapezoidal quadrature rule is used 
to approximate the right-hand side to the $z$-equation. 
Define for $k=1,\ldots,N+1$ the functions $G_k(\alpha,t) : \mathbb{T}_L \times [0,T] \to \mathbb{R}^2$ on the discretized domain as 
\begin{equation*}
G_k(\alpha_l,t) = \begin{cases}
				\mathcal{K}^{\delta}_{\mathbb{T}_L}(z_k(t)-z_l(t))\, \varpi_l(t)&, \text{if } l \neq k\,, \\
				0 &,\text{if } l=k\,,
				\end{cases}
\end{equation*}
where we have used the notation $f_k(t) = f(\alpha_k,t)$.  
The right-hand side to the $z$-equation \eqref{BR-eqn-regular} may then be approximated as
\begin{equation}\label{BR-eqn-regular-numerical}
\frac{\mathrm{d}}{\mathrm{d}t} z_k(t) = \frac{{\scriptstyle\Delta} \alpha}{2} G_k(\alpha_1,t) + {\scriptstyle\Delta} \alpha \sum_{l=2}^{N} G_k(\alpha_l,t) + \frac{{\scriptstyle\Delta} \alpha}{2} G_k(\alpha_{N+1},t)\,.
\end{equation}
\textcolor{black}{
The trapezoidal rule we employ, while in general 
only second order accurate, achieves spectral accuracy when the integrand is smooth and the mesh is 
uniform \cite{BaPh2006}. }

For the $\varpi$ equation, we follow \cite{GrSh2017} and convert the equation to Fourier space. Using 
the identities \eqref{ft-derivative} and \eqref{Hilbert}, we write the $\varpi$-equation 
\eqref{wbar-eqn-regular} in frequency space as 
\begin{align}
\partial_t \ft{\varpi}(\xi,t) &= -\frac{A}{2} \left( \frac{2 i \pi}{L} \xi \right) \mathcal{F} \left\{ \frac{1}{|\partial_\alpha z|^2} \mathcal{F}^{-1} \left\{ - i \sgn(\xi) \mathcal{F} \left\{ \varpi \mathcal{F}^{-1} \left\{ -i \sgn(\xi) \ft{\varpi} \right\} \right\} \right\} \right\}(\xi,t) \label{wbar-eqn-regular-freq} \\
& \qquad + 2Ag \left( \frac{2 i \pi}{L} \xi \right) \ft{z}_2(\xi,t) - \mu \left| \frac{2 \pi}{L} \xi \right|^2 \ft{\varpi}(\xi,t) \,, \nonumber
\end{align}
where $\mathcal{F}^{-1}\{ \cdot \}$ denotes the inverse Fourier transform operator.

The discretized version of \eqref{wbar-eqn-regular-freq} then becomes
\begin{align}
\frac{\mathrm{d}}{\mathrm{d}t}( \ft{\varpi}_k) &= -\frac{A}{2}  \left( D_k \tilde{\mathcal{F}} \left\{ \left( \frac{1}{|\partial_{\alpha} z_l|^2} \tilde{\mathcal{F}}^{-1} \left\{ \left( H_m \tilde{\mathcal{F}} \left\{ \left( \varpi_n \tilde{\mathcal{F}}^{-1} \left\{ \left( H_r \ft{\varpi}_r \right) \right\}_n \right)  \right\}_m \right)  \right\}_l  \right) \right\}_k \right)  \label{wbar-eqn-regular-freq-discrete} \\ 
& \qquad + 2Ag \left( D_k \ft{z}_2^k \right) -  {\scriptstyle\Delta} \alpha \cdot \tilde{\mu}  \left( D^2_k \ft{\varpi}_k  \right) \,, \nonumber 
\end{align}
where we have used \eqref{discrete-hilbert}-\eqref{discrete-partial-squared} to denote the discrete 
Hilbert and derivative operators in Fourier space. 
We remark that we are not using the usual summation convention in \eqref{wbar-eqn-regular-freq-discrete}, and 
instead use the notation $(f_k)$ to denote the vector with entries $f_k$ for $k=1,\ldots,N$, so that 
$(f_k g_k)$ denotes the vector with entries $f_k g_k $ for $k=1,\ldots,N$.

Let us note that we have  used the scaling $\mu = {\scriptstyle\Delta} \alpha \cdot \tilde{\mu}$ for the artificial 
viscosity parameter for the $\varpi$-equation. In general, we shall keep $\tilde{\mu}$ fixed as the resolution
${\scriptstyle\Delta} \alpha$ varies, but note that it is often necessary to vary $\tilde{\mu}$ with the resolution to 
stabilize small-scale noise that may occur in the variable $\varpi$. 

Equations \eqref{BR-eqn-regular-numerical} and \eqref{wbar-eqn-regular-freq-discrete} 
form a nonlinear system of coupled ordinary differential equations, to which we can apply a 
standard third-order explicit Runge-Kutta time integration scheme. We supplement the equations with 
initial data $z_k(0)$ and $\varpi_k(0)$, as well as periodic boundary conditions $z_1^{N+1}(t) = L + z_1^1(t)$, 
$z_2^{N+1}(t) = z_2^1(t)$, and $\varpi_{N+1}(t) = \varpi_1(t)$ for all $t \geq 0$. 

The direct summation method \eqref{BR-eqn-regular-numerical} employed for the integral 
calculation \eqref{BR-eqn-regular} is $\mathcal{O}(N^2)$, and is thus inefficient for large values of 
$N$. Other methods have been proposed to reduce the computational complexity of the velocity calculation. 
For instance, one technique is the so-called ``vortex-in-cell'' method 
\cite{Christiansen1973,Baker1979,Tryggvason1988}, in which the velocity of a point vortex is computed 
by solving a Poisson equation on an underlying mesh, and interpolation is used to compute 
values on the interface. This method reduces the computational cost by virtue of the use of fast 
Poisson solvers, and appears to accurately predict the large-scale behavior of the vortex sheet, but does not 
seem suitable for the study of small-scale behavior \cite{Tryggvason1989}.

Other fast summation methods 
include the Fast Multipole Method of \citet{GrRo1987} (see also \cite{CaGrRo1988}), 
the Barnes-Hut algorithm \cite{BaHu1986}, and various other so-called ``treecode'' algorithms 
\cite{Appel1985,vanDoRu1989,Anderson1992,HaMa1995,DrDr1995,SaOk1998,LiKr2001}. Such methods 
reduce the computational complexity of the summation to $\mathcal{O}(N)$ or $\mathcal{O}(N \log N)$ 
by combining large numbers of point vortices into single computational elements. However,
they are often complicated to implement since the computations must be organized in a manner that leads to 
an efficient and accurate algorithm. 
Moreover, such algorithms often have significant computational overhead that 
make them efficient only for large values of $N$. In the numerical simulations considered in the current 
paper, we restrict our attention to problems requiring only relatively small values of $N$; for such problems, 
the direct summation method we employ is likely comparable (in terms of efficiency and CPU time)
to the more sophisticated algorithms mentioned above. In future studies, we shall implement a fast summation 
method to study vortex sheet evolution for large values of $N$.

Following \citet{Krasny1987}, we reduce the computational expense of calculating 
\eqref{BR-eqn-regular-numerical} in the following two ways: 
first, we use the relation 
$\mathcal{K}^{\delta}_{\mathbb{T}_L}(z_k-z_l) = - \mathcal{K}^{\delta}_{\mathbb{T}_L}(z_l-z_k)$ so that 
the calculation \eqref{BR-eqn-regular-numerical} is required for only half the points; 
 second, for problems which are symmetric 
about $\alpha = 0$, we compute \eqref{BR-eqn-regular-numerical} for only half the points and use 
reflection to obtain the values for the rest.

\subsection{Numerical studies and discussion}\label{z-model-numerical-study}

We next conduct several numerical studies to validate our regularized $z$-model 
system, as well as its numerical implementation. 
\textcolor{black}{In particular, we shall  
compare the Krasny desingularization \eqref{smoothed-kernel} with the two other desingularizations 
\eqref{kernel-baker}. We provide numerical evidence to show that all three numerical methods produce 
similar solutions, and that the computed numerical solutions appear to converge.  
The situation is somewhat complicated by the fact that there is a large gap in the 
theory of vortex sheet evolution when the vorticity does not have distinguished sign. In particular, the 
question of existence of solutions to the incompressible Euler system 
when the vorticity is a measure but is not of distinguished sign is 
open; additionally, the question of uniqueness is open, even when the vorticity is of distinguished sign. 
Consequently, comparisons of the numerical methods as the mesh is refined is complicated due to the fact
that the convergence may be towards different solutions. Nonetheless, in agreement with prior numerical 
studies \cite{BaPh2006,Sohn2014}, we find that the computed 
numerical solutions agree, in a sense to be made precise below.}

\textcolor{black}{Specifically, the quantities that we shall be interested in with regards 
to our convergence studies are (1) the 
bubble and spike tip locations, (2) the radius of the spiral roll-up region, and (3)  the location of the center of the 
roll-up region. Quantity (1) provides some information about the convergence of solutions 
``at the large scales'', while the  quantities (2) and (3) provide information about the 
convergence of solutions ``at the small 
scales''. The bubble and spike tip locations are defined by $\max_i z_2(\alpha_i)$ and 
$\min_i z_2(\alpha_i)$, respectively, while the radius and center of spiral roll-up region are computed as 
follows. We first find the intersection points $\left\{ (x_1^*\,,x_2^*) \right\}$ of the computed curve $z$ with a 
fixed horizontal axis $x_2 = x_2^*$. These intersection points are computed by bilinear interpolation. 
The radius $r_{\delta}$ and location of the center $\sigma_{\delta}$ of the spiral region may then be 
approximated as $r_{\delta} \approx (\max x_1^* - \min x_1^*)/2$ and 
$\sigma_{\delta} \approx (\max x_1^* + \min x_1^*)/2$, respectively. The subscript $\delta$ indicates that 
these quantities depend on the regularization parameter $\delta$ (as well as the mesh resolution 
${\scriptstyle\Delta} \alpha$).}

\textcolor{black}{In the convergence studies presented, we will be interested in two different limits: the 
first is the limit $\delta \to 0$ with $N$ held fixed, and the second is the limit $\delta \to 0$ and 
$N^{-1} \to 0$. In the latter case, it is important exactly how the limits are taken \cite{Krasny1986b}. For 
the Krasny desingularization, we will use the scaling \eqref{delta-scaling}, and we will show that the resulting 
solutions are stable with increasing amounts of roll up as $N \to \infty$.}

\textcolor{black}{We are unaware of  scaling laws similar to  \eqref{delta-scaling} for the kernels of 
\eqref{kernel-baker}, and will instead use the following empirical (though tedious and computationally 
expensive)  procedure employed by 
\citet{Anderson1985} and \citet{Krasny1986b}. This empirical method amounts to fixing a value of ${\scriptstyle\Delta} \alpha$, say 
${\scriptstyle\Delta} \alpha = {\scriptstyle\Delta} \alpha_1$, then choosing the smallest $\delta=\delta_1$ such that the 
computed numerical solution is stable for every $\delta > \delta_1$. 
This procedure is then repeated for ${\scriptstyle\Delta} \alpha_2 < {\scriptstyle\Delta} \alpha_1$, yielding 
$\delta_2 < \delta_1$. 
In this way, a sequence $({\scriptstyle\Delta} \alpha_1\,,\delta_1), ({\scriptstyle\Delta} \alpha_2 \,,\delta_2), \ldots$ is constructed, with 
$\delta_i, {\scriptstyle\Delta} \alpha_i \to 0$, and we are able to discuss the limit $\delta \to 0$ and $N^{-1} \to 0$.}

\textcolor{black}{One of our goals in this section is to justify our use of the Krasny kernel in the numerical
implementation of  our multiscale algorithm in \Cref{sec:multi-scale-algorithm}. We shall show that 
the Krasny kernel produces solutions with similar asymptotic behavior (i.e. as $\delta, N^{-1} \to 0$) as 
those produced using the kernels \eqref{kernel-baker}. On the other hand, calculations using the 
Krasny kernel require less computational expense than the corresponding calculations using the 
kernels of \eqref{kernel-baker}. Consequently, as our goal in 
\Cref{sec:multi-scale-algorithm} is to produce a fast-running algorithm for compressible flow simulations, 
we will use the Krasny approximation rather than the kernels \eqref{kernel-baker}.}

\textcolor{black}{We remark that, in general, we will be restricted to using relatively large values of the 
regularization parameter $\delta$ for the 3rd-order kernel $\mathcal{K}_3^\delta$, compared to those 
for the lower order Krasny and $\mathcal{K}_1^\delta$ kernels. This is due to the fact that a large amount of 
nodes $N$ is required to fully resolve the small scale structure that is observed with the 3rd order kernel
\cite{BaBe2004}, which proves prohibitively computationally expensive for our purposes.}

\textcolor{black}{
Standard MATLAB plotting routines have been employed to present interface evolution;  in particular,  we follow \citet{Krasny1987} and use
trigonometric polynomials of degree $N/2$ to interpolate the discrete computed 
interface nodal positions $z_k(t)$.  
}

\subsubsection{KHI test on an ellipse: comparison with an exact solution}

\textcolor{black}{We begin with a numerical experiment  for which there is a known exact solution; namely, we consider  the KHI problem
 of \citet{BaBe2004}. For this test, we compute the velocity induced by a vorticity measure 
concentrated on an ellipse. More precisely, we set the Atwood number to be zero, $A=0$, so that the amplitude 
of vorticity $\varpi$ remains constant over time, and choose the initial data as 
\begin{subequations}
\begin{align*}
 z_1(\alpha,0) &= \lambda \cosh(\sigma) \cos(\alpha)  \,, \\
z_2(\alpha,0) &= \lambda \sinh(\sigma) \sin (\alpha) \,, \\
\varpi(\alpha,0) = \varpi(\alpha,t) &= \sin(\alpha) \,, 
\end{align*}
\end{subequations}
with $\alpha \in [0,2 \pi]$. The parameter $\lambda$ measures the eccentricity of the ellipse, with 
$\lambda \to 0$ yielding a circle, and $\lambda \to  1$ yielding a slit. The constant $\sigma$ is determined 
from $\lambda$ by the relation $\lambda \cosh(\sigma) = 1$.}

\textcolor{black}{The exact solution \cite{BaBe2004} is of the form
\begin{equation*}
 w_1(\alpha) = \frac{R}{4 \lambda D} \,,  \quad \text{and} \quad w_2(\alpha) = \frac{I}{4 \lambda D}  \,, 
\end{equation*}
with
\begin{gather*}
 R = 2 e^{-\sigma} \left(  2 \sinh^2 (\sigma) \cos^2 (\alpha) + e^{-\sigma} \cosh(\sigma) \sin^2(\alpha) \right)\,, \\
  I = \sinh(\sigma) \sin(2 \alpha) \,,  \qquad D = \cosh^2(\sigma) - \cos^2(\alpha)  \,.  
\end{gather*}
Since the curve $z$ is a closed curve, we use the regularized versions of the 
 $\mathbb{R}^2$ kernel \eqref{biot-savart-r2} i.e. \eqref{smoothed-kernel-r} and \eqref{kernel-baker}. We 
 set $\lambda = 0.01$, consider $N=16, \ldots,512$ in increasing powers of $2$, and consider two 
 different values of $\delta / {\scriptstyle\Delta} \alpha$.} 
 
 \textcolor{black}{We measure the $L^ \infty $ error $E_N = \max_{i = 1,\ldots,N} | w(\alpha_i) - w_i|$, where 
 $w(\alpha_i)$ is the exact  velocity and $w_i$ is the computed velocity.   Then, the quantity 
 $- \log_{10} E_N$ denotes the number of digits of accuracy of the computed solution (see \cite{BaBe2004}).   In
 \Cref{table:KH-errors}, we list   $- \log_{10} E_N$ for the three different methods of desingularizing the integral kernel.  
} 

\begin{table}[H]
\centering
\renewcommand{\arraystretch}{1.0}
\scalebox{0.8}{
\begin{tabular}{|c|ccc|ccc|}
\toprule
\multirow{2}{*}{$N$} &\multicolumn{3}{c|}{\textbf{$\delta = {\scriptstyle\Delta} \alpha$} } &\multicolumn{3}{c|}{\textbf{$\delta = {\scriptstyle\Delta} \alpha / 4$} }      \\[0.25em]

 & $\mathcal{K}^{\delta}_1$ & $\mathcal{K}^{\delta}_3$   & Krasny    & $\mathcal{K}^{\delta}_1$   & $\mathcal{K}^{\delta}_3$ & Krasny \\

\midrule

$16$ & $0.960$ & $2.440$ & $0.820$ & $1.204$ & $1.204$ &  $1.175$ \\[0.5em]

$32$ & $1.258$ & $3.320$ & $1.057$ & $1.505$ & $1.505$ &  $1.490$ \\[0.5em]

$64$ & $1.558$ & $4.031$ & $1.318$ & $1.806$ & $1.806$ &  $1.798$ \\[0.5em]

$128$ & $1.859$ & $4.484$ & $1.599$ & $2.107$ & $2.107$ &  $2.103$ \\[0.5em]

$256$ & $2.160$ & $4.833$ & $1.891$ & $2.408$ & $2.408$ &  $2.406$ \\[0.5em]

$512$ & $2.461$ & $5.147$ & $2.187$ & $2.709$ & $2.709$ &  $2.708$ \\[0.5em]

\bottomrule
\end{tabular}}
\caption{Error analysis and convergence results for the exact solution test of \citet{BaBe2004}. 
Shown are the number of digits of accuracy, $- \log_{10} E_N$, for the three different numerical schemes employed.}
\label{table:KH-errors}
\end{table} 

\textcolor{black}{For the larger value $\delta = {\scriptstyle\Delta} \alpha$, we see that the higher order kernel 
$\mathcal{K}^{\delta}_3$ is the most accurate of the three methods, with the computed 
velocity accurate to 5 digits when $N=512$; the lower order kernel $ \mathcal{K}^{\delta}_1$ and the Krasny 
kernel perform similarly, with the computed velocity accurate to 2 digits when $N=512$. For the smaller 
value  $\delta = {\scriptstyle\Delta} \alpha/4$, all three regularizations produce a velocity that is accurate to 2 digits. 
Thus, in the limit $\delta \to 0$ and $N \to \infty$, all three methods of regularization perform similarly; the main 
difference is then the fact that the kernels  $\mathcal{K}^{\delta}_i$ are more expensive to compute than the 
Krasny kernel \eqref{smoothed-kernel-r}. 
}

\subsubsection{KHI problem on a periodic curve: test of the Krasny $\delta$-regularization}\label{subsubsec-delta-tests}

\textcolor{black}{The purpose of the following test is to demonstrate how the 
regularization parameter $\delta$ scales with the mesh resolution ${\scriptstyle\Delta} \alpha$.
The procedure we employ for determining the appropriate value of $\tilde{\delta}$  for use 
in the Krasny desingularization method
is as follows: fix a relatively small value of $N$, say $N=32$ or 
$N=64$. Next, we find the smallest value of $\tilde{\delta}$ such that the computed interface demonstrates roll up, but 
without self-intersection. Finally, we fix this value of $\tilde{\delta}$, and use the scaling relation \eqref{delta-scaling} 
for larger values of $N$.}

\textcolor{black}{Below, we provide an example of the above procedure applied to a periodic 
KHI problem. In this case, the Atwood number vanishes $A=0$, and thus the amplitude of vorticity 
$\varpi$ remains constant over time. The initial data is \cite{BaBe2004} 
\begin{subequations}
\begin{align*}
 z_1(\alpha,0) &= \alpha  \,, \\
z_2(\alpha,0) &= 0 \,, \\
\varpi(\alpha,0) = \varpi(\alpha,t) &= 1 - 0.5 \cos(\alpha) \,, 
\end{align*}
\end{subequations}
with $\alpha \in [0,2 \pi]$. 
Numerical studies indicate that a curvature singularity forms at $\alpha = \pi$ and $t \approx 1.61$, 
after which time the sheet rolls up in a tightly wound spiral. 
}

\begin{figure}[h]
\centering
\subfigure[$N = 64$]{\label{fig:KH_BB_z_64}\includegraphics[width=27.38mm]{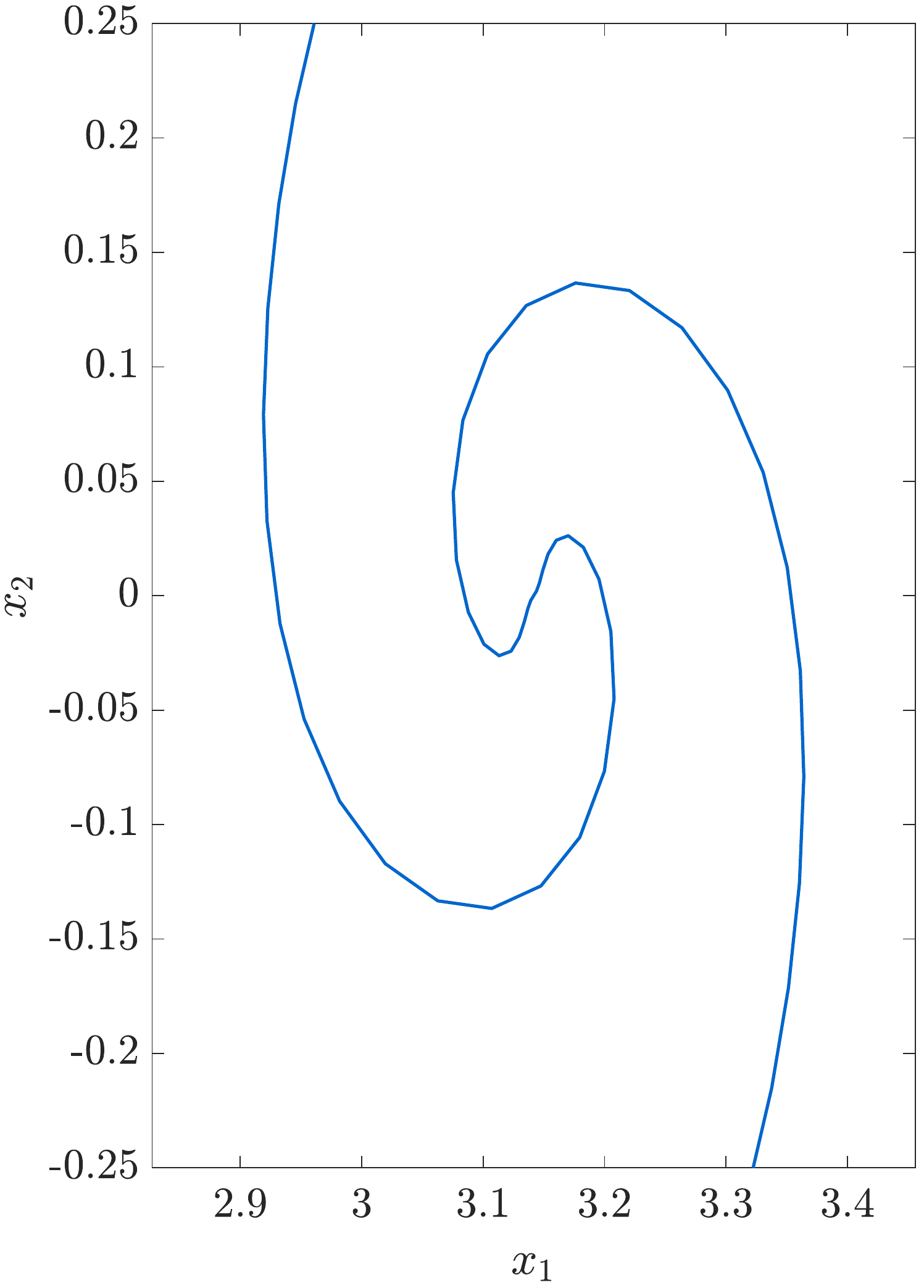}}
\hspace{0.5em}
\subfigure[$N = 128$]{\label{fig:KH_BB_z_128}\includegraphics[width=23mm]{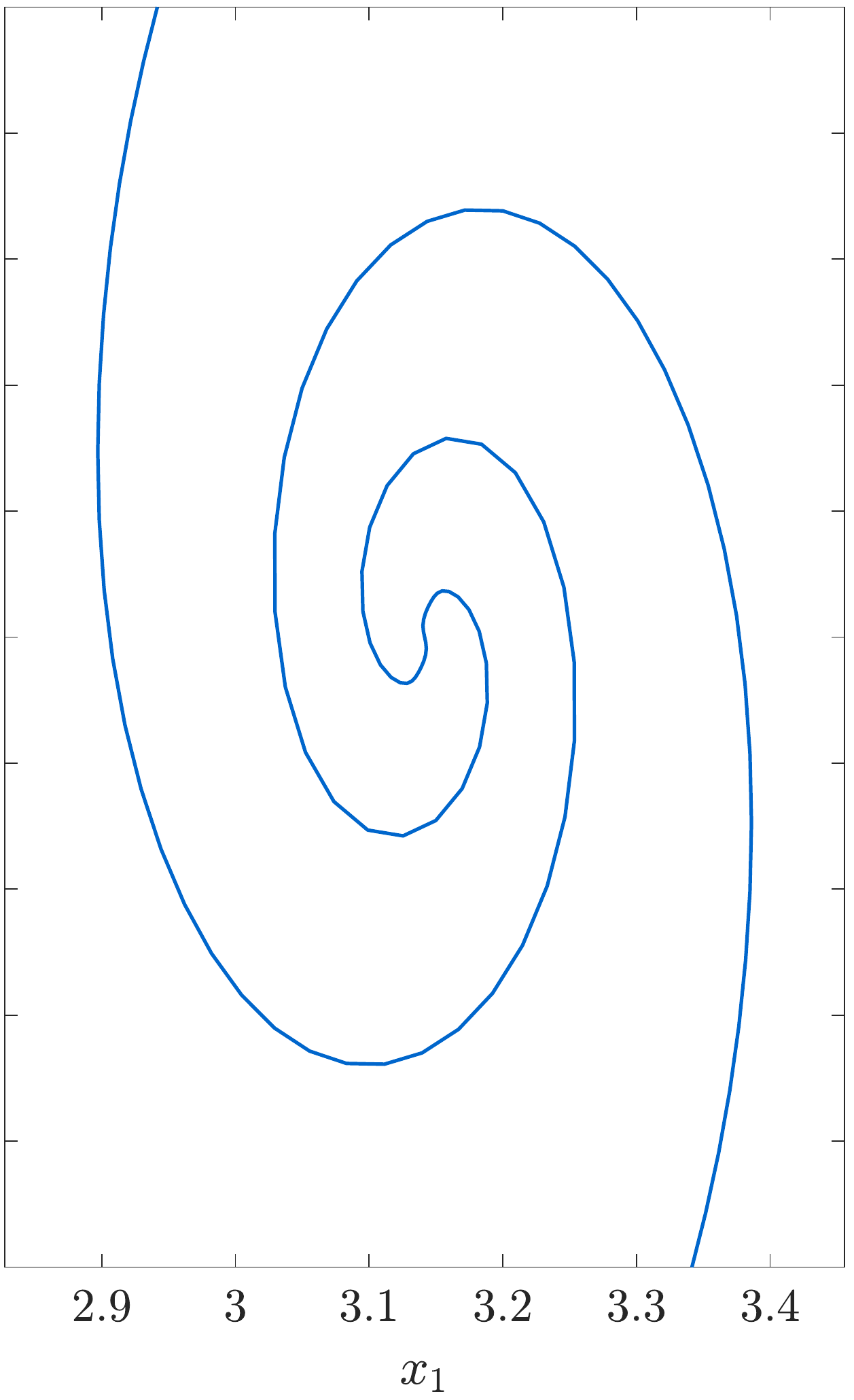}}
\hspace{0.5em}
\subfigure[$N = 256$]{\label{fig:KH_BB_z_256}\includegraphics[width=23mm]{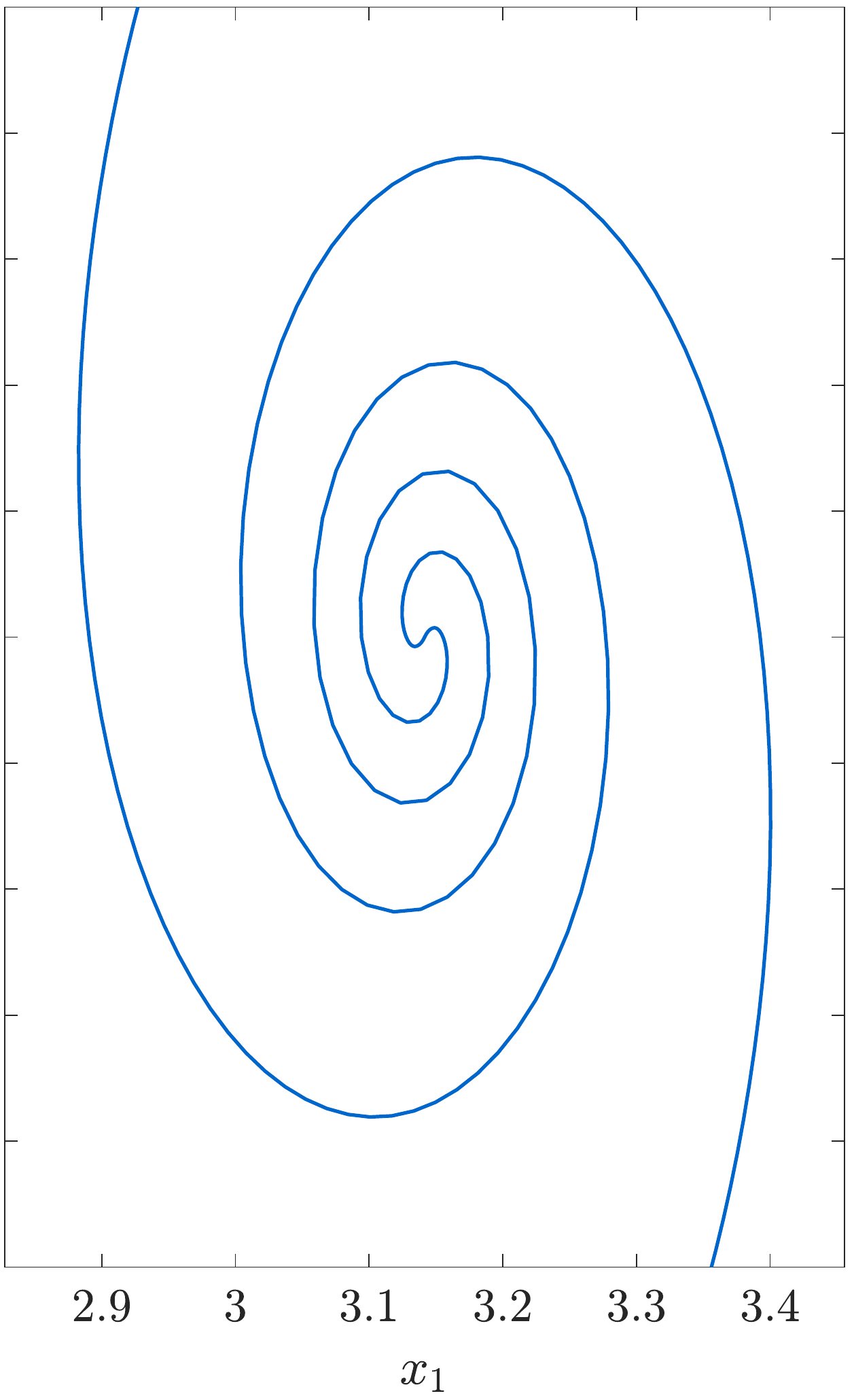}}
\hspace{0.5em}
\subfigure[$N = 512$]{\label{fig:KH_BB_z_512}\includegraphics[width=23mm]{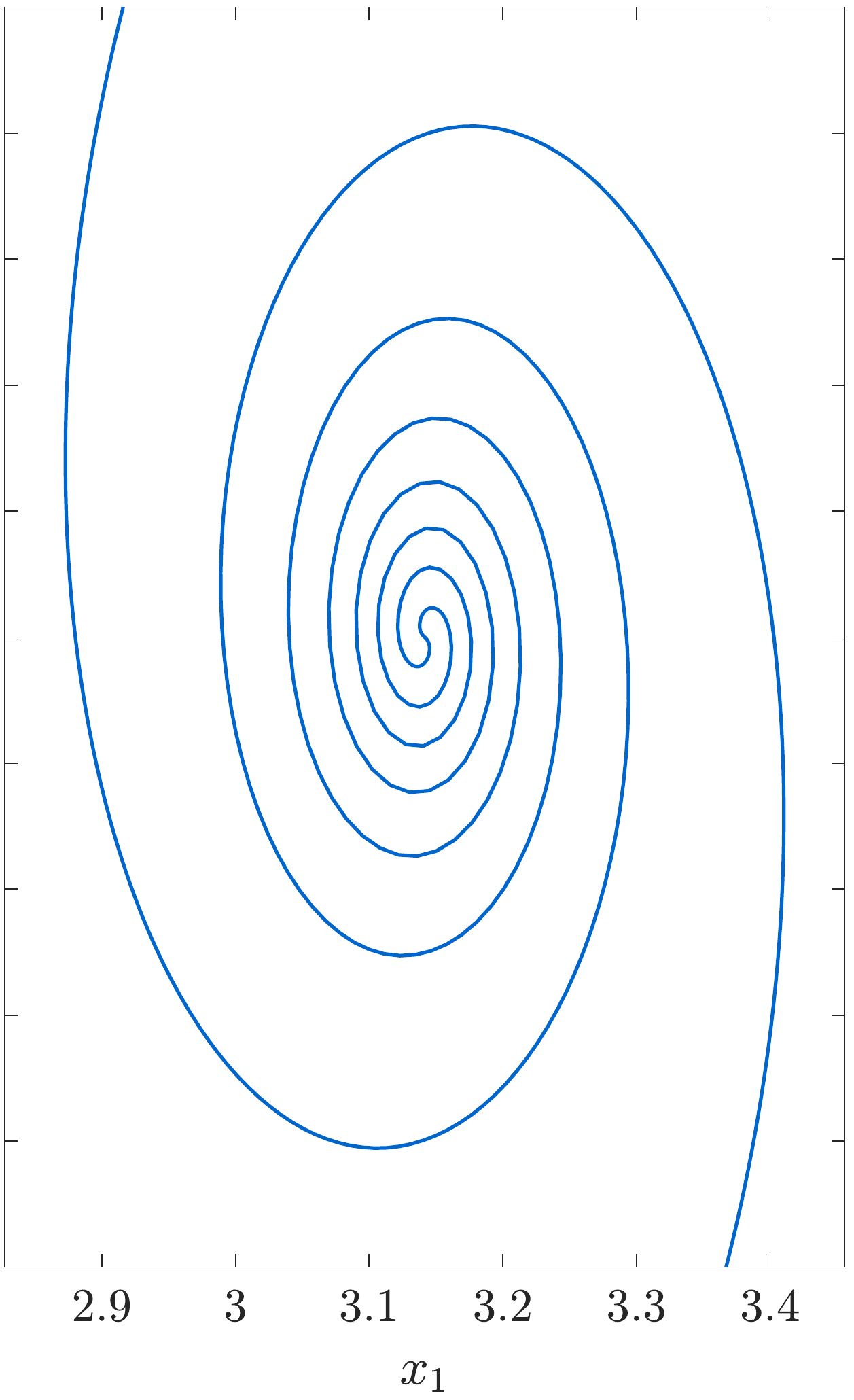}}
\hspace{0.5em}
\subfigure[$N = 1024$]{\label{fig:KH_BB_z_1024}\includegraphics[width=23mm]{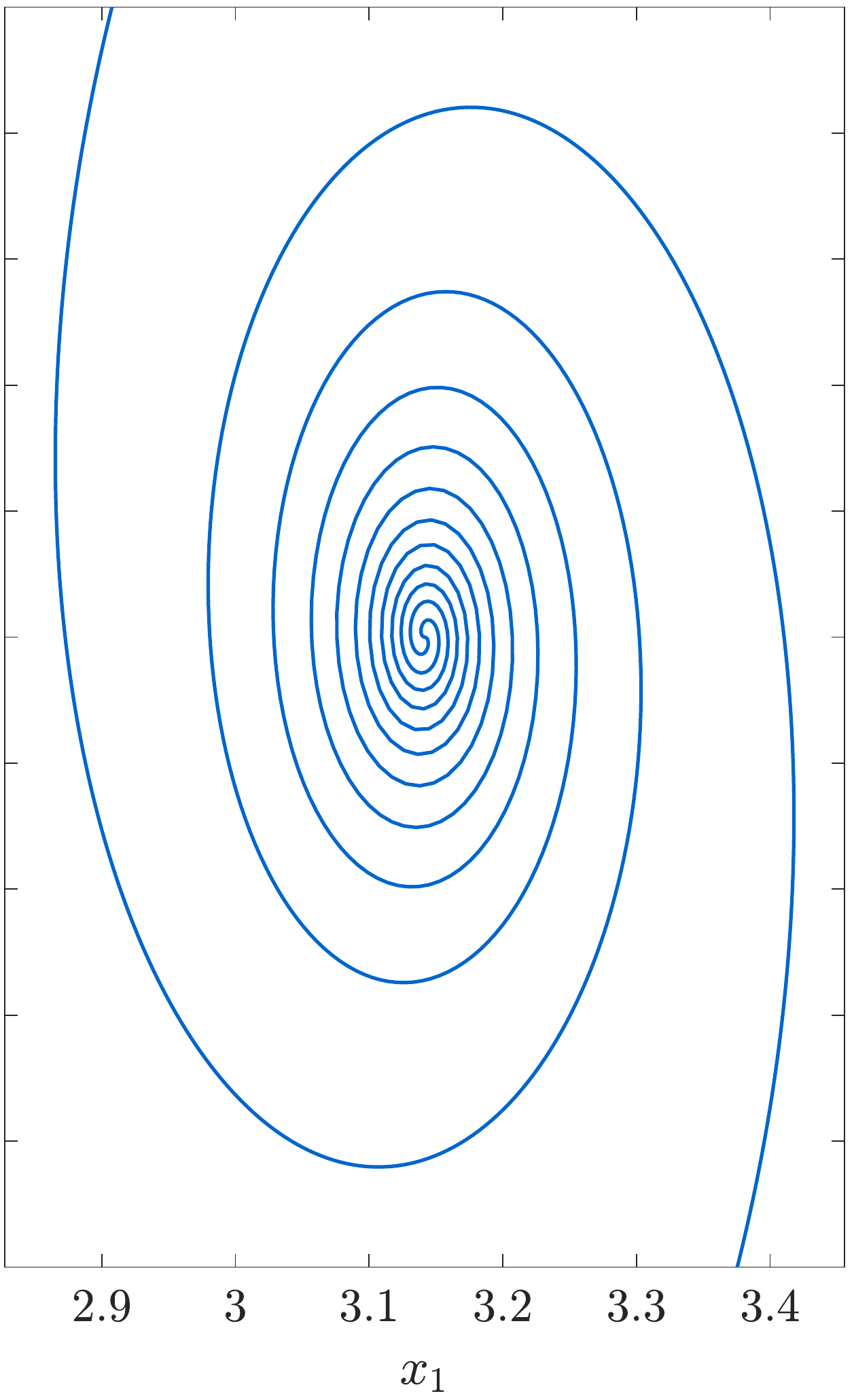}}
\hspace{0.5em}
\subfigure[$N = 2048$]{\label{fig:KH_BB_z_2048}\includegraphics[width=23mm]{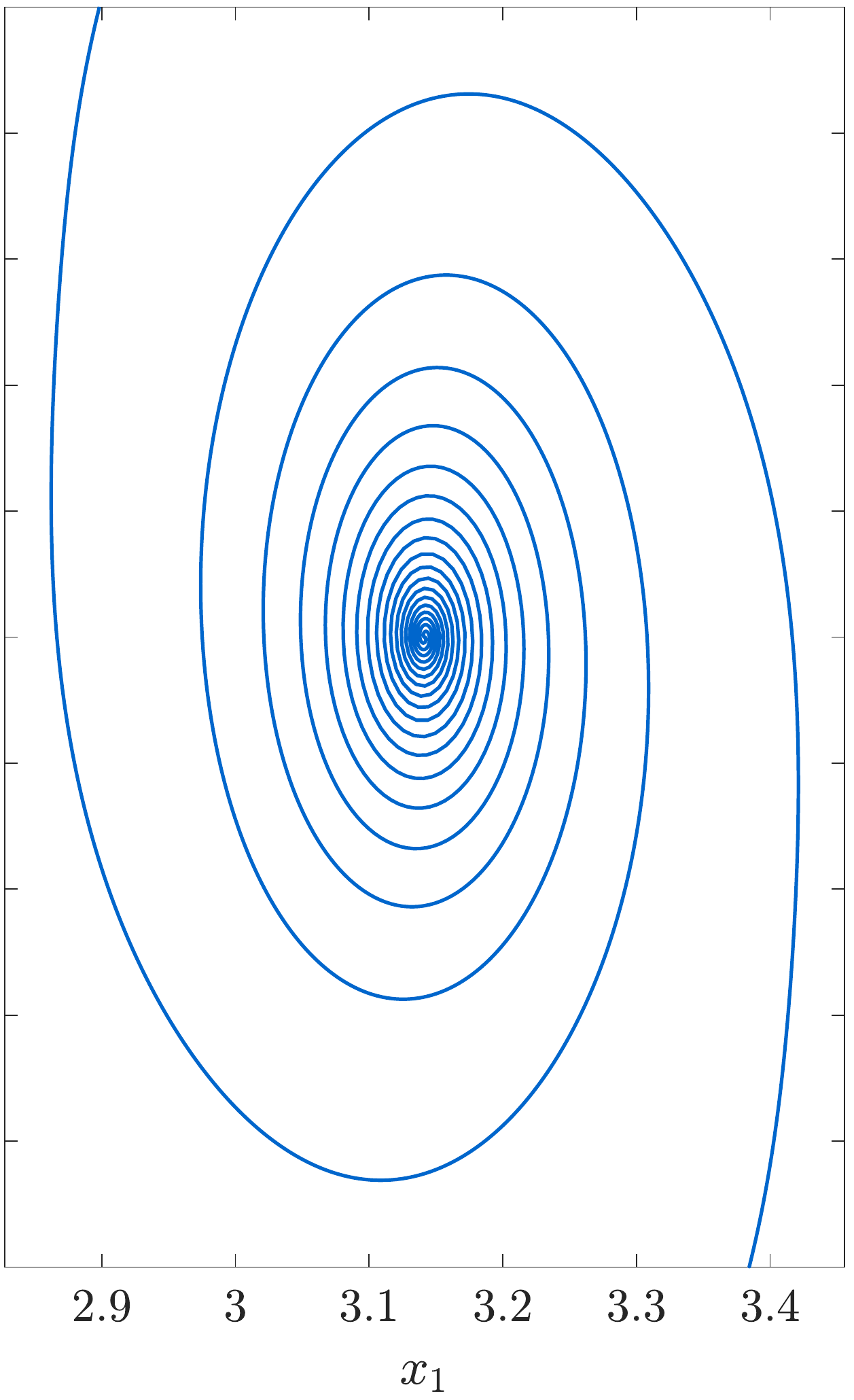}}
\hspace{0.5em}
\caption{Numerical simulation of a KHI test using the $z$-model with the Krasny desingularization. 
Shown  is the interface position $z(\alpha,t)$  at time $t=3.0$
for six simulations with 
resolution starting from $N=64$ and doubling until $N=2048$.}
\label{fig:KH_BB_z}
\end{figure}

\textcolor{black}{Experimentation with the value of $\tilde{\delta}$ with $N=64$ shows that choosing 
$\tilde{\delta} = 0.15$ produces an interface which demonstrates roll-up but for which 
self-intersection does not occur (see \Cref{fig:KH_BB_z_64}). A smaller value of $\tilde{\delta}$, 
say $\tilde{\delta} = 0.1$, produces an interface which self-intersects. The runtime for such a 
simulation is $ < 1$ s, so that experimentation with 
the precise value of $\tilde{\delta}$ is not computationally expensive. Once the value of $\tilde{\delta}$ is 
set for $N=64$, the same value is chosen for $N > 64$. The computed results are presented in 
\Cref{fig:KH_BB_z}. We observe that the scaling relation 
$\delta^2 = | {\scriptstyle\Delta} \alpha \log{{\scriptstyle\Delta} \alpha} | \cdot \tilde{\delta}^2$ results in more turns appearing in 
the core region as $N$ increases, but that the solution remains stable and the curve does 
not self-intersect or suffer from irregular vortex motion.}

\textcolor{black}{Next, we consider the convergence of the numerical solution as $\delta \,, N^{-1} \to 0$. 
The convergence of solutions as $\delta \to 0$ with $N$ fixed is considered in detail in
\cite{Krasny1986b,BaPh2006}, wherein it is shown that the computed 
numerical solution appears to converge.}

\begin{figure}[h]
\centering
\subfigure[Amplitude $\max z_2$.]{\label{fig:KH_BB_amplitude}\includegraphics[width=70mm]{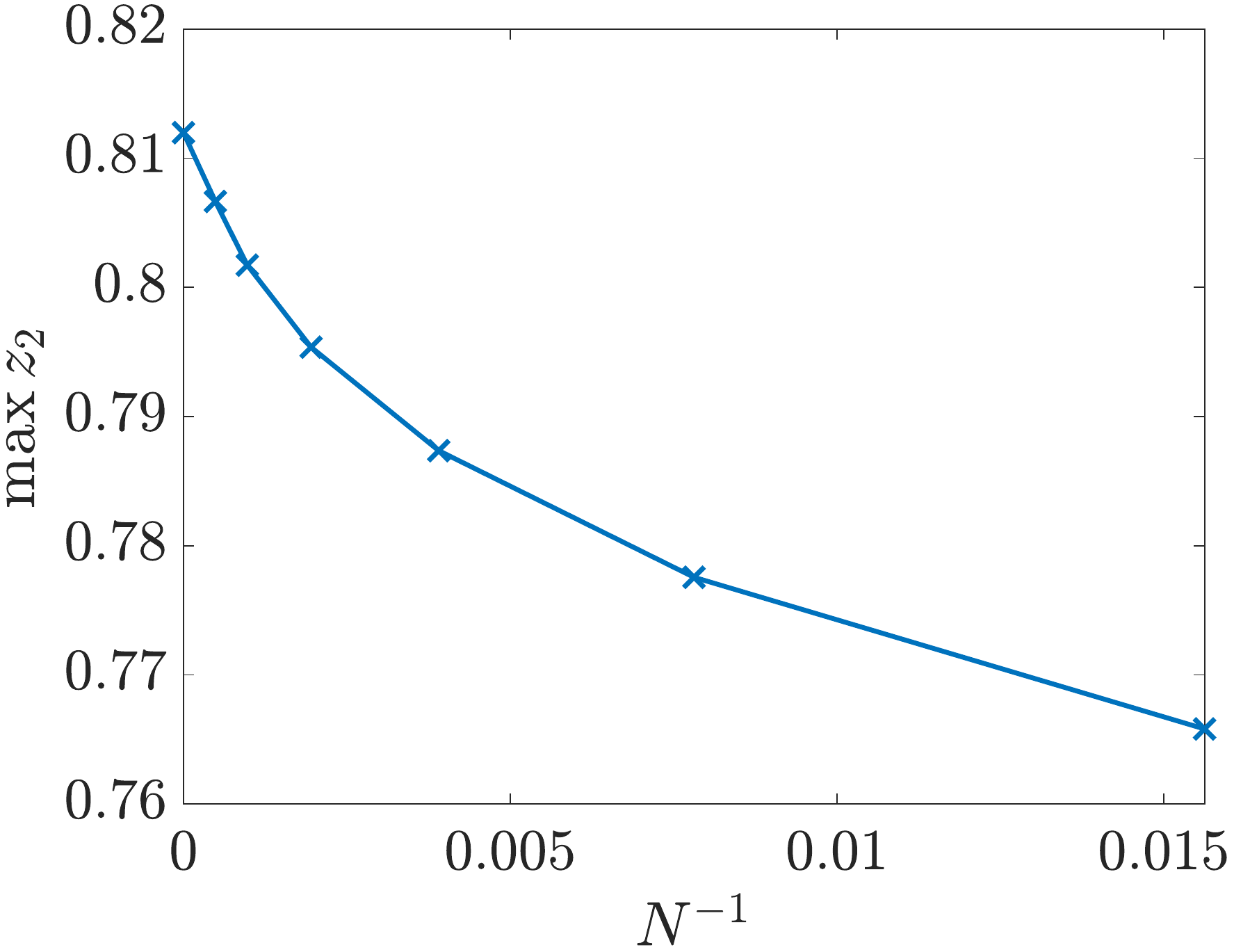}}
\hspace{2em}
\subfigure[Radius of the spiral region $r_\delta \approx \pi - \min x_1^*$.]{\label{fig:KH_BB_radius}\includegraphics[width=70.85mm]{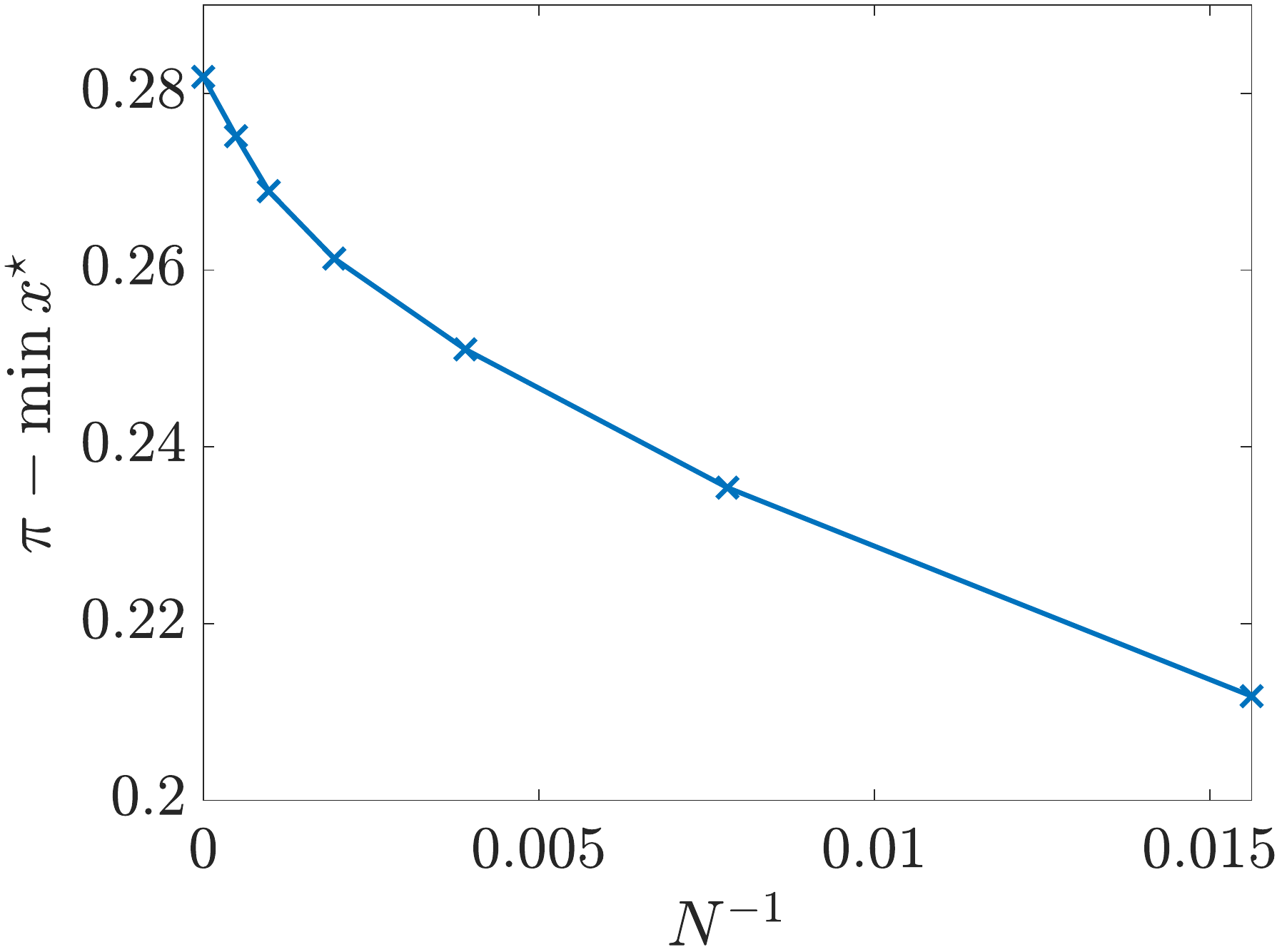}}
\caption{Convergence behavior for a KHI test using the $z$-model with the Krasny desingularization. 
Shown are (a) the amplitude of the curve $\max z_2$ and (b) the radius of the spiral region $r_\delta$.}
\label{fig:KH_BB_convergence}
\end{figure}

\textcolor{black}{In \Cref{fig:KH_BB_amplitude}, we show how the amplitude $\max z_2$ of the curve 
varies as $\delta, N^{-1} \to 0$. The value for $N^{-1}$ is obtained by cubic extrapolation. We see that 
the amplitude appears to converge to a finite value $\approx 0.81$. Similarly, in 
\Cref{fig:KH_BB_radius}, we estimate the radius of the spiral region by $r_{\delta} \approx \pi - \min x_1^*$, 
where $x^*$ are the intersection points with the axis $x_2 = x_2^* = 0$. 
By symmetry, the center of the spiral is located at $(\pi,0)$. Again, with the value for 
$N^{-1} = 0$ obtained by cubic extrapolation, we see that the $r_{\delta}$ appears to converge to a 
value close to 0.28. This demonstrates that the scaling 
\eqref{delta-scaling} appears to be appropriate 
for recovering a meaningful solution as $\delta, N^{-1} \to 0$.}

\subsubsection{Single-mode RTI: comparison with experiments}

We continue our numerical studies for the $z$-model by performing simulations for the low Atwood number
single-mode RTI experiments of \citet{WaNiJa2001}. The particular problem setup considered is 
a heavy fluid lying atop a lighter fluid, with the Atwood number given by $A=0.155$ and the two fluids 
subject to an approximately constant gravitational acceleration $g = 0.74 \times 9.8 \mathrm{ms}^{-2}$. 
The $z$-model is employed for this problem on the domain $\alpha \in [-0.027 , 0.027]$ with initial data 
\begin{subequations}
\begin{align*}
 z_1(\alpha,0) &= \alpha  \,, \\
z_2(\alpha,0) &= 0.0012 \cos(2 \pi \alpha / L) \,, \\
\varpi(\alpha,0) &= 0\,. 
\end{align*}
\end{subequations}

We perform six simulations with resolution starting from $N=32$ and doubling until $N=1024$. 
The strategy for parameter choice we adopt here is to keep the 
parameters $\tilde{\delta} =2$ and $\tilde{\mu} = 0.02$ fixed as $N$ varies, while allowing
the time-steps $\delta t$ to vary with $N$. Specifically, we first choose $\delta t$ for $N=64$ 
as the largest possible value that will allow the $N=64$ simulation to run until the final time $t=0.3795$. 
The values of $\delta t$ for larger $N$ are then determined by repeatedly halving this value 
until $\delta t$ is sufficiently small so as to allow the simulation to complete. 
%
%
%
%
%
%
%

\begin{figure}[h]
\centering
\subfigure[$N = 32$]{\label{fig:RT_WaNiJa2001_z_32}\includegraphics[width=28mm]{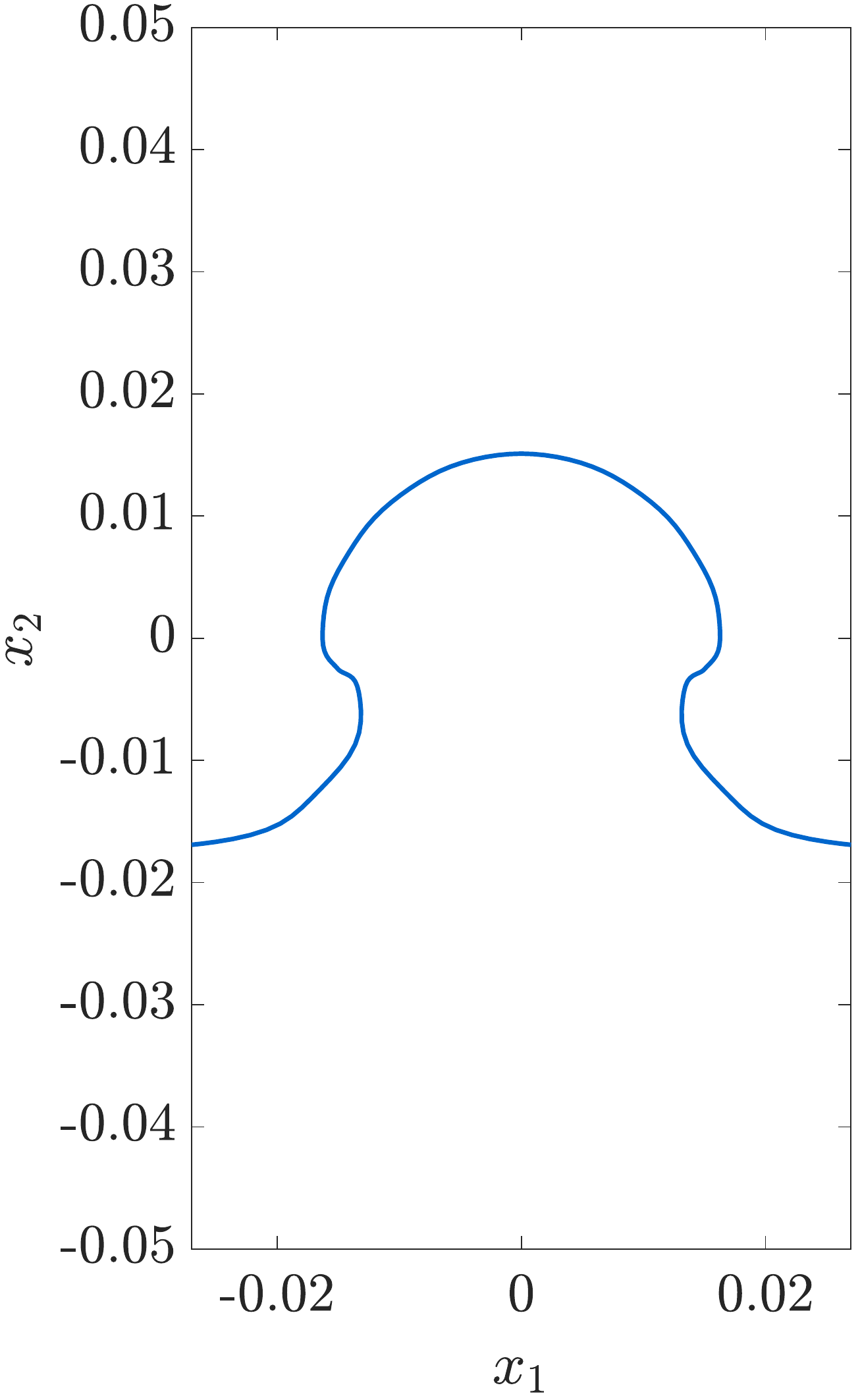}}
\hspace{0.5em}
\subfigure[$N = 64$]{\label{fig:RT_WaNiJa2001_z_64}\includegraphics[width=22mm]{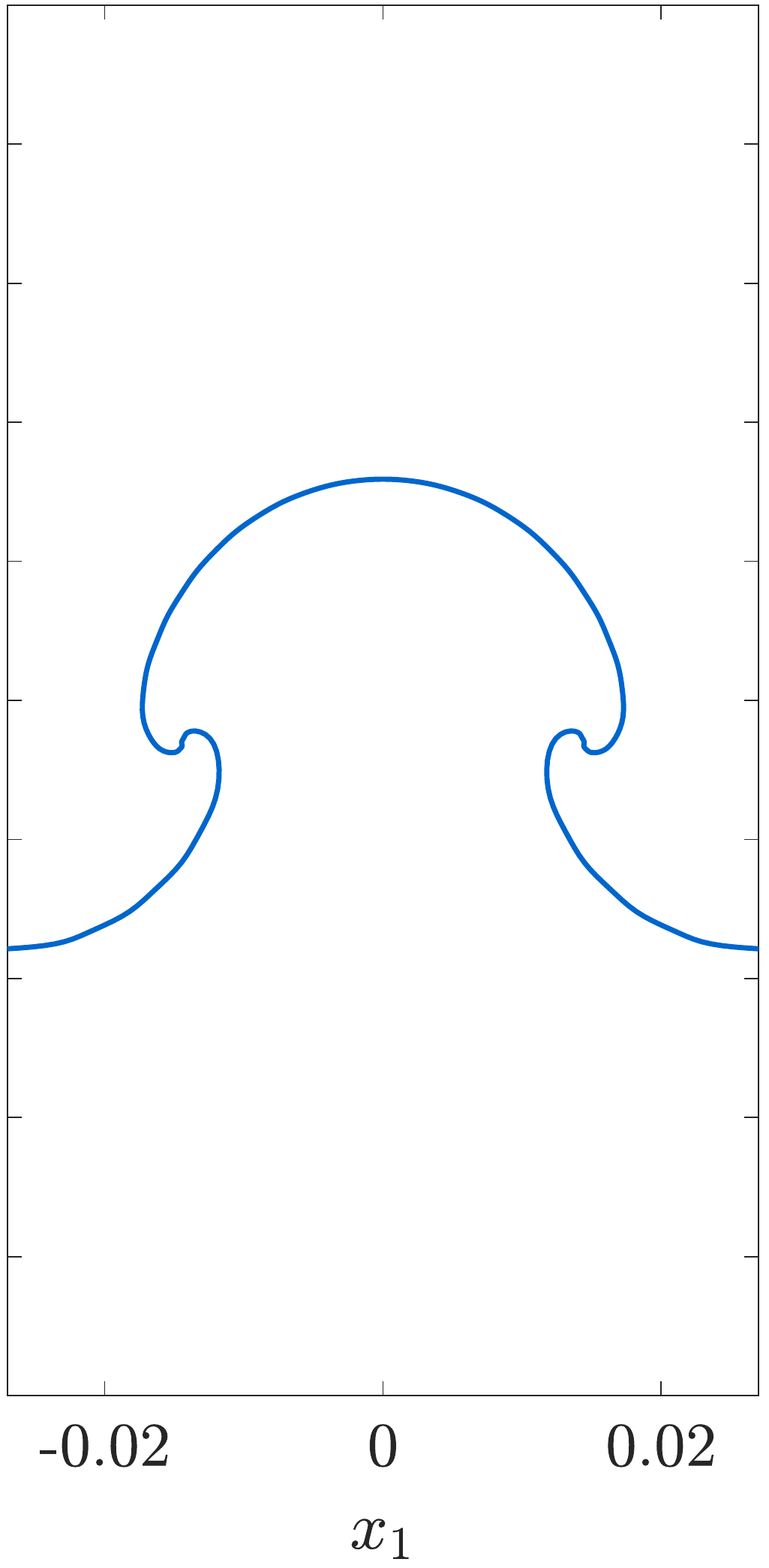}}
\hspace{0.5em}
\subfigure[$N = 128$]{\label{fig:RT_WaNiJa2001_z_128}\includegraphics[width=22mm]{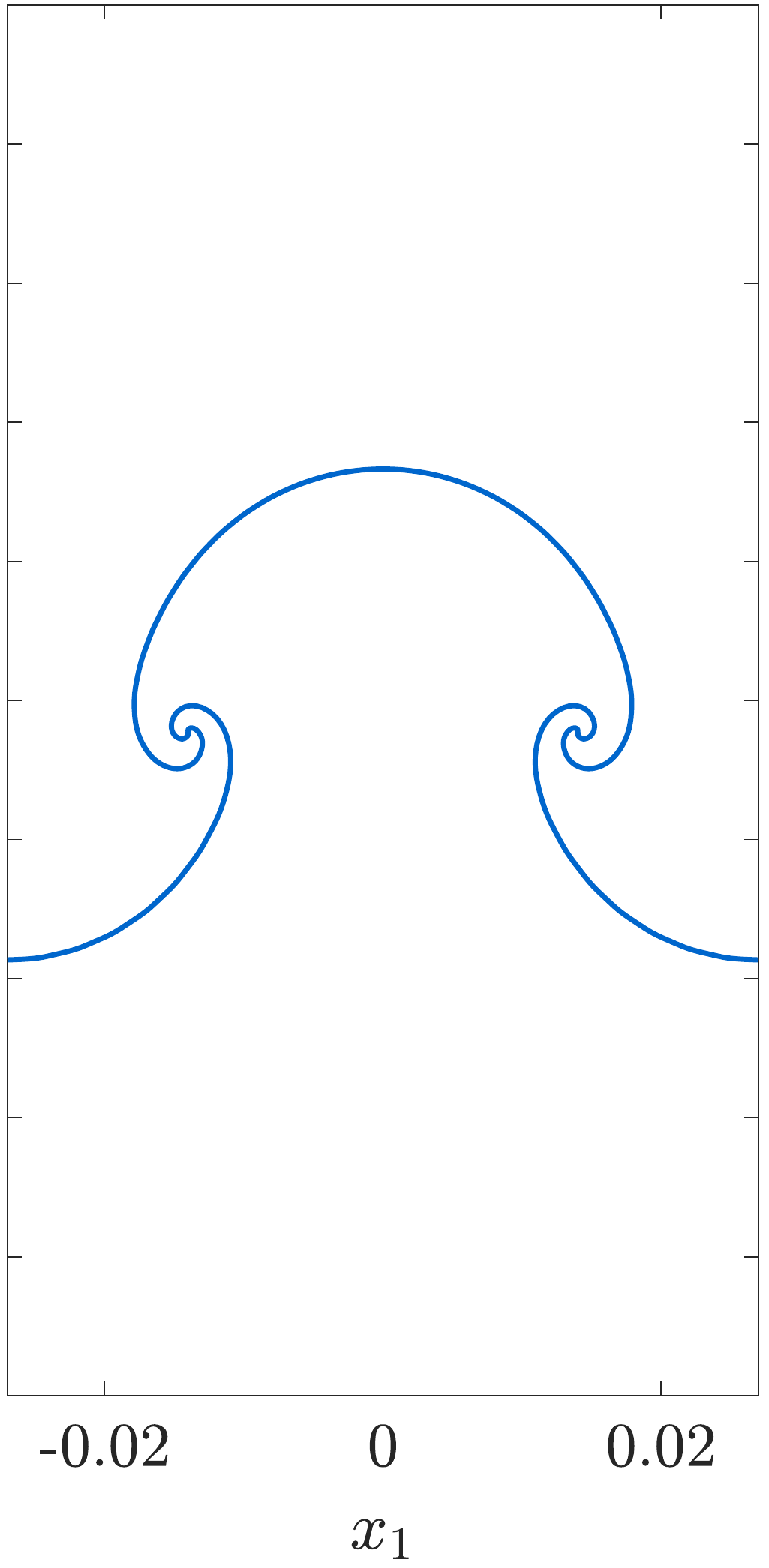}}
\hspace{0.5em}
\subfigure[$N = 256$]{\label{fig:RT_WaNiJa2001_z_256}\includegraphics[width=22mm]{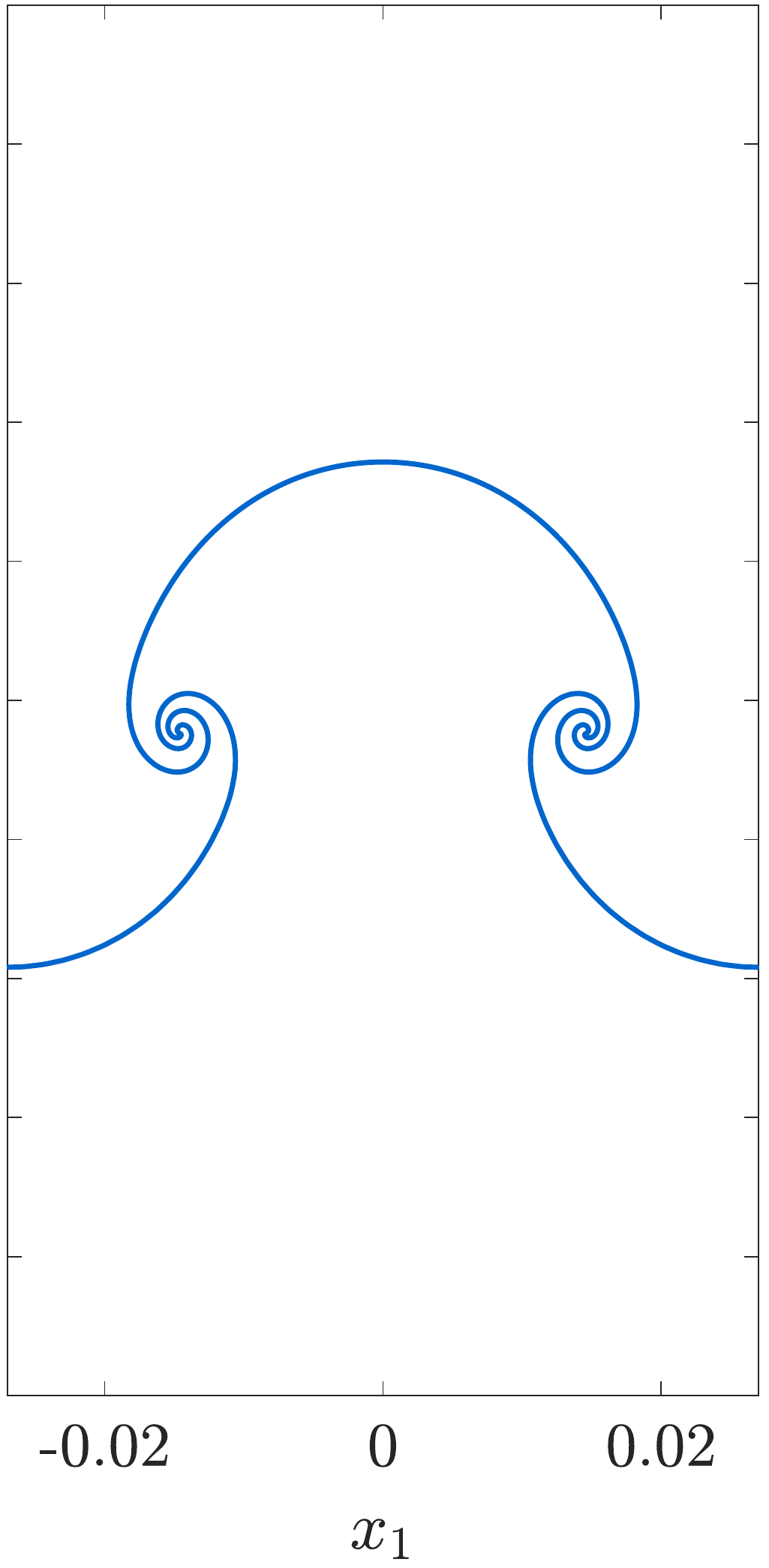}}
\hspace{0.5em}
\subfigure[$N = 512$]{\label{fig:RT_WaNiJa2001_z_512}\includegraphics[width=22mm]{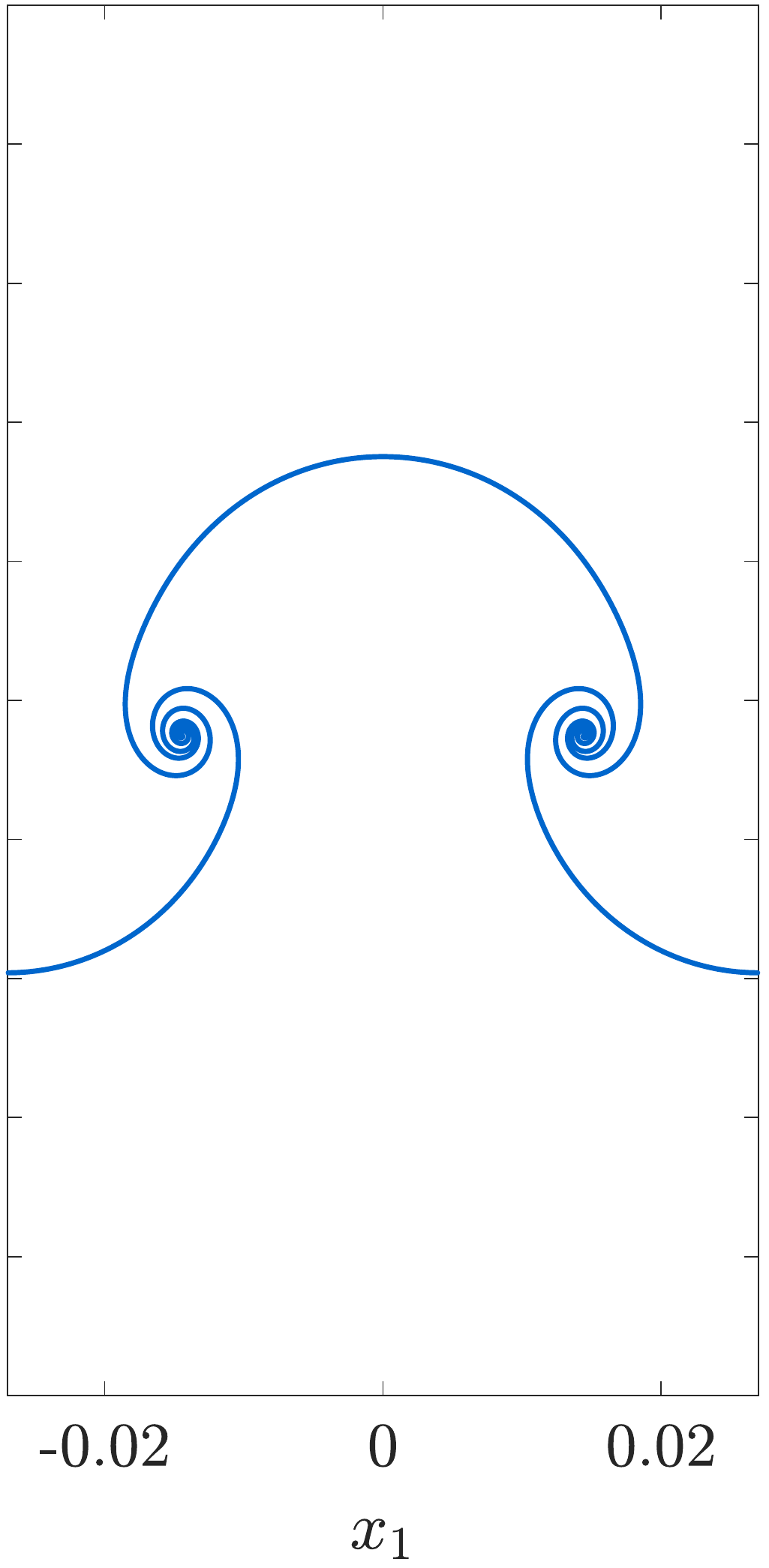}}
\hspace{0.5em}
\subfigure[$N = 1024$]{\label{fig:RT_WaNiJa2001_z_1024}\includegraphics[width=22mm]{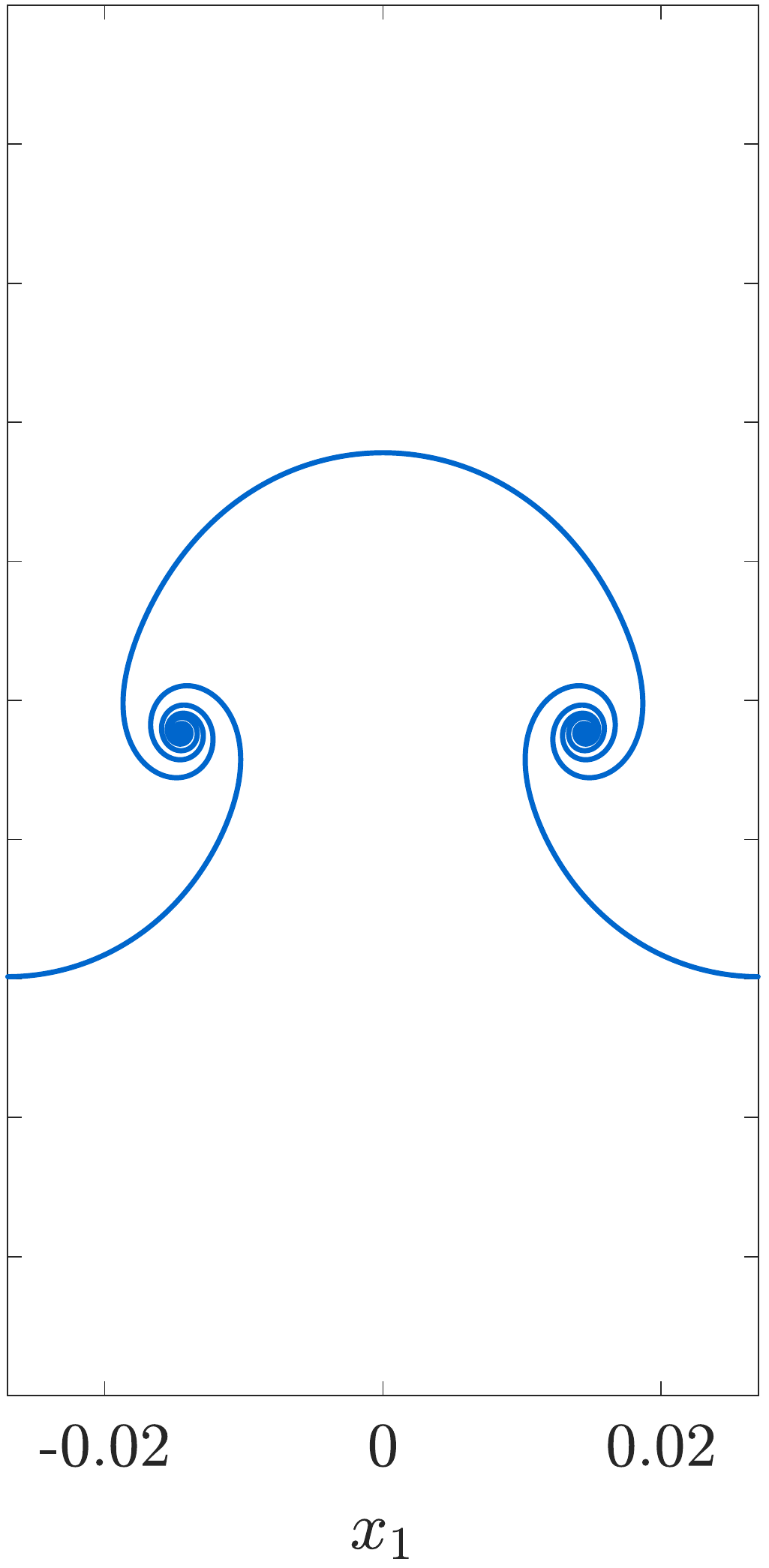}}
\hspace{0.5em}
\caption{Numerical simulation of the \citet{WaNiJa2001} RTI using the $z$-model. 
The interface parameterizations 
$z(\alpha,t)$ for six simulations with 
resolution starting from $N=32$ and doubling until $N=1024$ are shown at time $t=0.3795$.}
\label{fig:RT_WaNiJa2001_z}
\end{figure} 

 \begin{wrapfigure}{R}{0.24\textwidth}
\includegraphics[width=40mm]{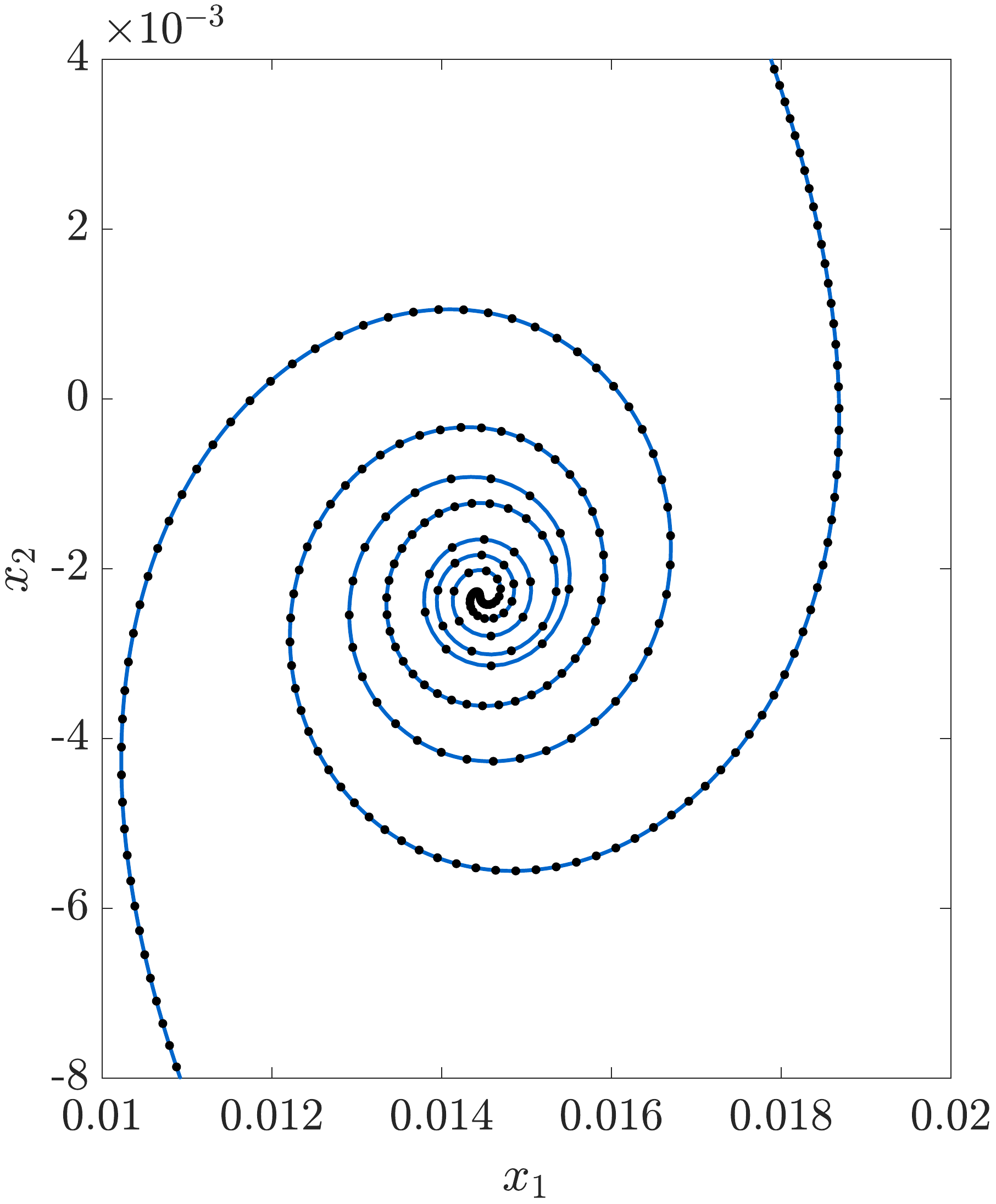}\
\caption{{Closeup view of the roll-up region for $N=1024$.}}
\label{fig:RT_WaNiJa2001_z_1024_zoom}
\end{wrapfigure}
The results of the six simulations are shown in \Cref{fig:RT_WaNiJa2001_z}, which should be 
compared with Figure 4(l) of \cite{WaNiJa2001}. The bubble and spike shapes are roughly symmetric, 
in agreement with the observations in \cite{WaNiJa2001}. The scaling 
for the regularization parameters $\tilde{\delta}$ and $\tilde{\mu}$ results in more roll-up of the 
vortex sheet as the resolution is increased. This is demonstrated in \Cref{fig:RT_WaNiJa2001_z_1024_zoom}, 
which shows that the interface for the $N=1024$ simulation has a tightly packed spiral region with several 
complete revolutions of each branch of the spiral.

\begin{figure}[h]
\centering
\subfigure[Comparison of the computed amplitude and linear theory predictions for the simulations in 
\Cref{fig:RT_WaNiJa2001_z}.]{\label{fig:RT_WaNiJa2001_amp_comparison1}\includegraphics[width=75mm]{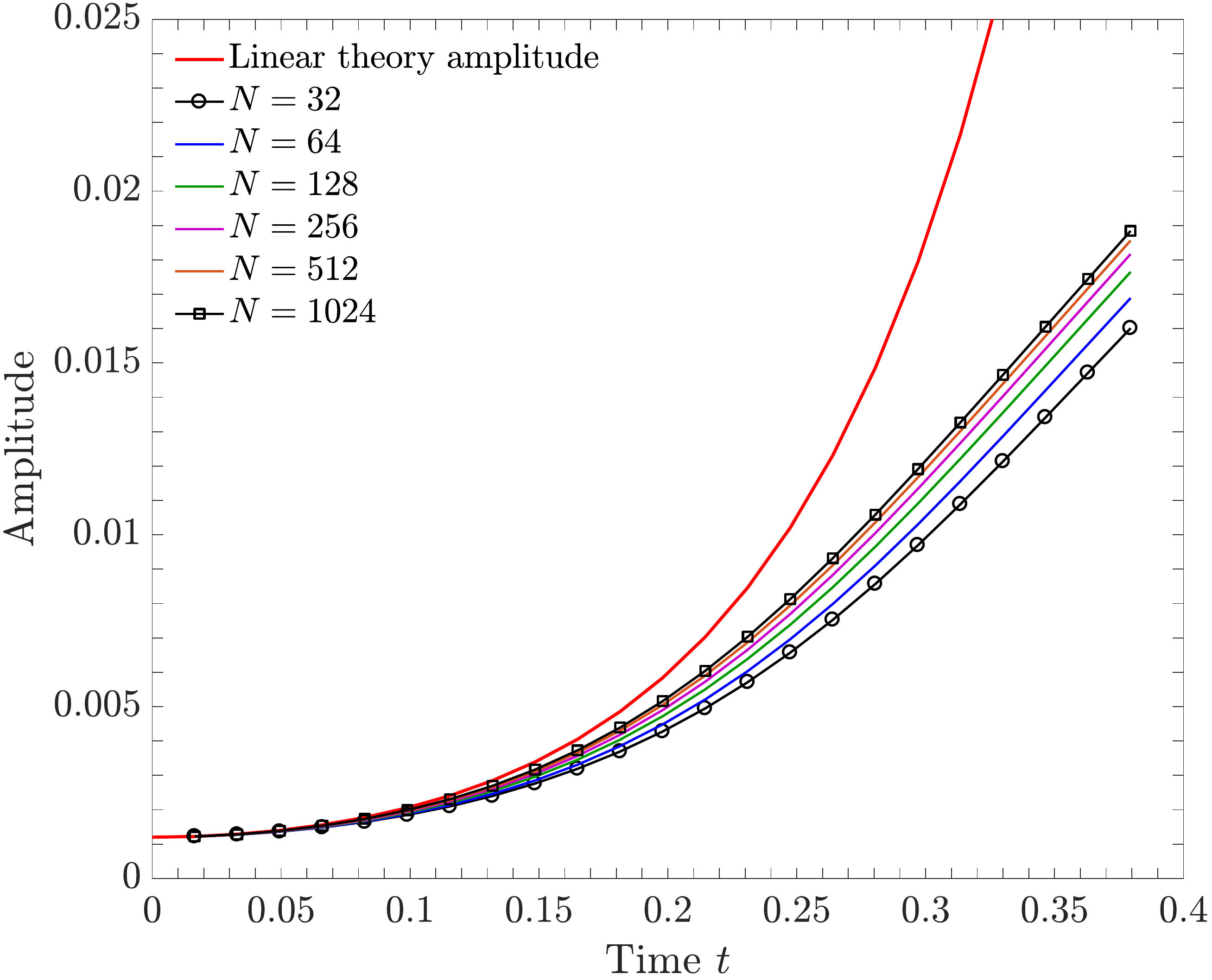}}
\hspace{2em}
\subfigure[Comparison of the computed amplitude for the $N=1024$ simulation with large time linear fits.]{\label{fig:RT_WaNiJa2001_amp_comparison2}\includegraphics[width=71.91mm]{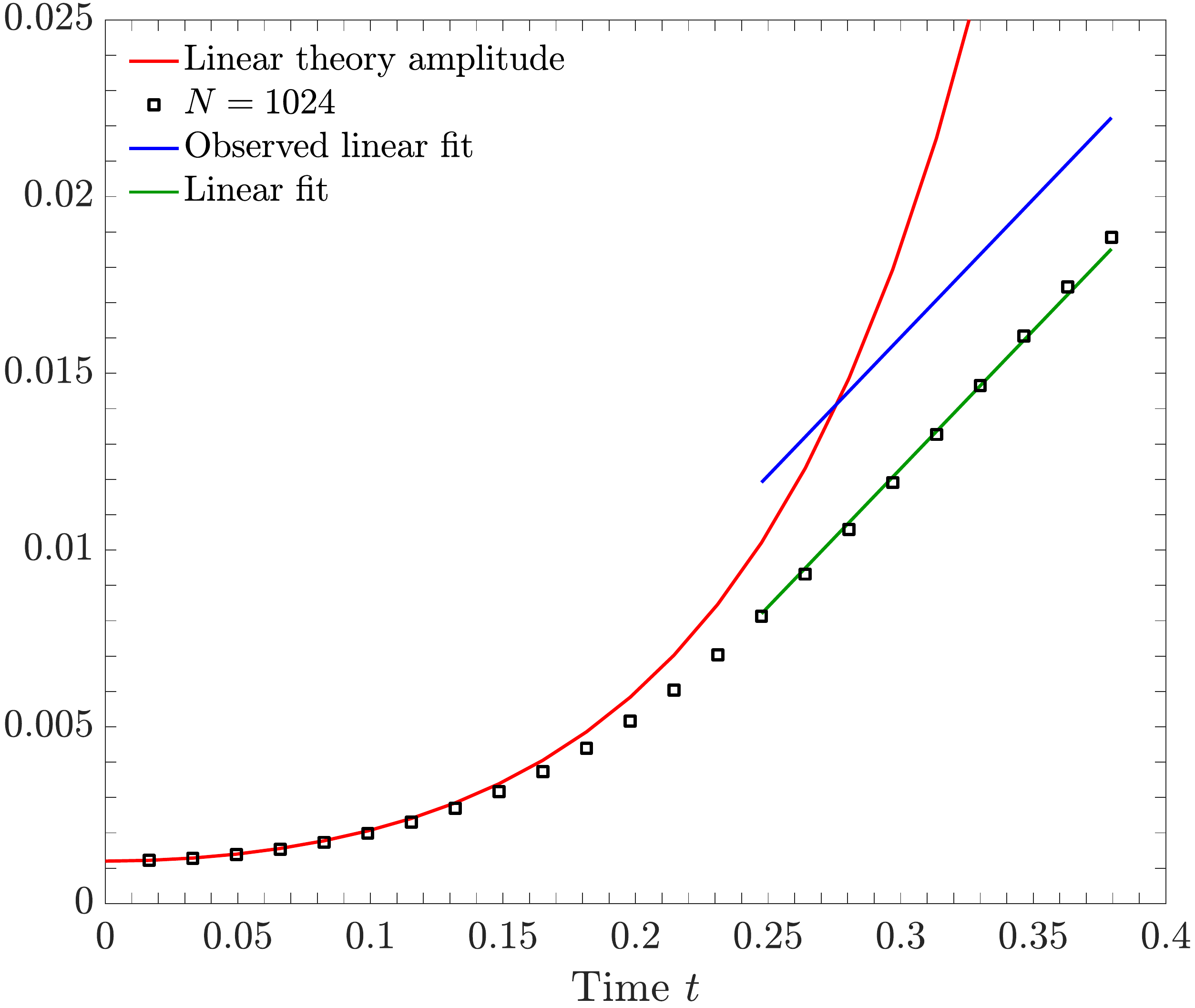}}
\caption{Plots of the amplitude $a(t)$ versus time 
$t$ for the \citet{WaNiJa2001} RTI problem.}
\label{fig:RT_WaNiJa2001_amp_comparison}
\end{figure}

We next compare the growth rate of the $z$-model solution interface with growth rates obtained from 
small-time linear theory predictions and late-time experimental observations. Define the 
amplitude of the interface as 
$a(t) = \frac{1}{2} \left[ \max_{\alpha} z_2(\alpha,t) - \min_{\alpha} z_2(\alpha,t) \right]$. Linear theory 
\cite{Rayleigh1882,Taylor1950} predicts that for early times $t$ before the non-linearity is activated, 
the amplitude satisfies $a(t) = a(0) \cosh (t \sqrt{2 \pi A g/L})$.  We plot in
 \Cref{fig:RT_WaNiJa2001_amp_comparison1} the linear theory amplitude and the computed amplitude $a(t)$ 
 versus time for the simulations shown in \Cref{fig:RT_WaNiJa2001_z}. It is clear from the graph that the 
 computed amplitude and linear prediction are in excellent agreement, as expected, 
 for small times $t \leq 0.15$. 

For large times, the nonlinearity is no longer negligible and the linear theory breaks down. Experimental 
observations indicate that the amplitude grows linearly at late times; a linear fit of the 
measured late time amplitude from experimental data is shown as the blue curve in 
\Cref{fig:RT_WaNiJa2001_amp_comparison2}. This curve is defined in \cite{WaNiJa2001} as 
$-0.007436 + 0.078177 t$. It is clear from the graph that the measured amplitude differs considerably from 
the amplitude computed using the $z$-model. This difference may be explained by the fact that the (effective) gravitational acceleration in the experiments is only approximately constant, 
and is in fact time-dependent, whereas we have used\footnote{We note that it is of course possible to have  time-dependent $g$ for $z$-model simulations. 
Our use of a constant value of $g$ for this experiment is due to the fact that only an approximate 
constant value of $g$ is provided in \cite{WaNiJa2001}.} 
a constant value of $g$. 

While the amplitudes themselves differ, we nonetheless observe that 
the growth \emph{rates} of the computed amplitude and measured amplitude are in excellent 
agreement. The green curve shown in \Cref{fig:RT_WaNiJa2001_amp_comparison2} has the same slope 
as the blue curve (i.e. the measured amplitude) and is given explicitly as 
$-0.01115+0.078177 t$. This curve matches almost exactly with the computed $a(t)$ for large times
$t \geq 0.25$. In fact, the true gravitational acceleration in the experimental setup is initially 
time-dependent, but eventually reaches a constant value; in our numerical simulation, we have used this final 
constant value for $g$, which is why the amplitude of the numerical solution at large times $t \geq 0.25$
grows at the same rate as that 
observed in the experiment. The numerical solution and the experimental data are thus in good agreement
(modulo the issue regarding the non-constant value of $g$), 
which consequently provides validating evidence for the $z$-model. 

\begin{figure}[h]
\centering
\subfigure[Bubble and spike tip locations.]{\label{fig:RT_convergence1_amplitude}\includegraphics[width=50mm]{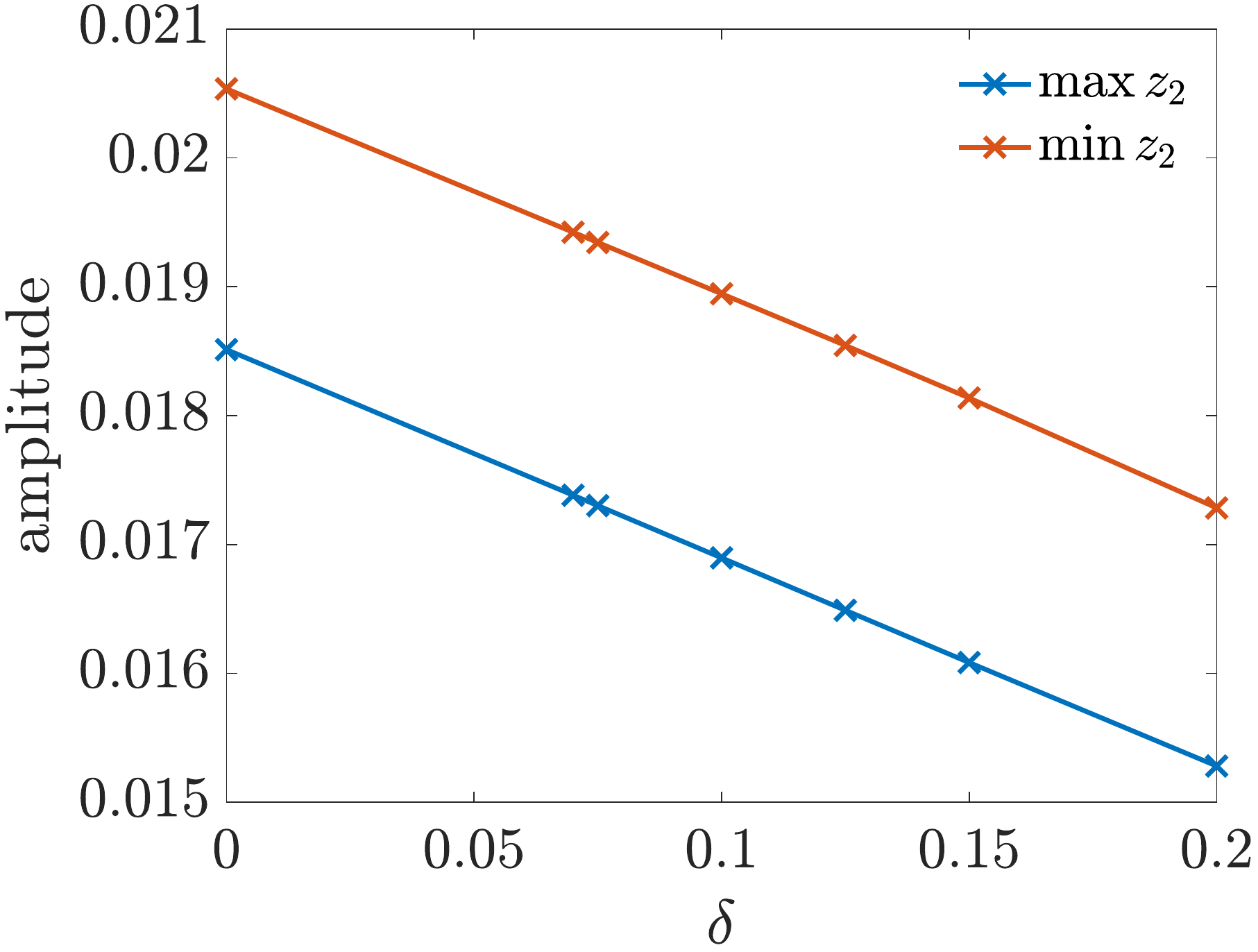}}
\hspace{0em}
\subfigure[Location of spiral center $\sigma_{\delta}$.]{\label{fig:RT_convergence1_sigma}\includegraphics[width=50mm]{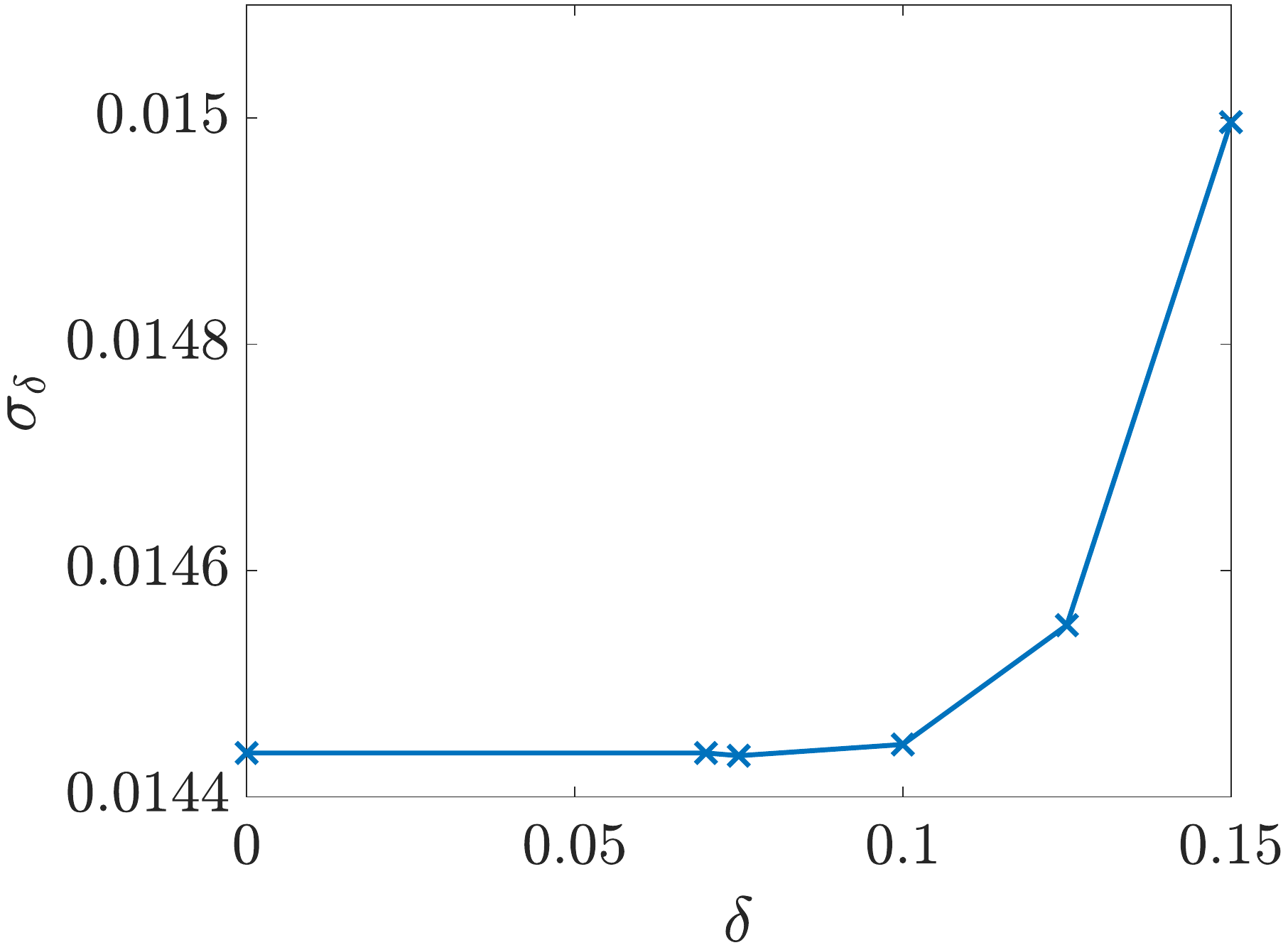}}
\hspace{0em}
\subfigure[Radius of spiral $r_{\delta}$.]{\label{fig:RT_convergence1_radius}\includegraphics[width=50mm]{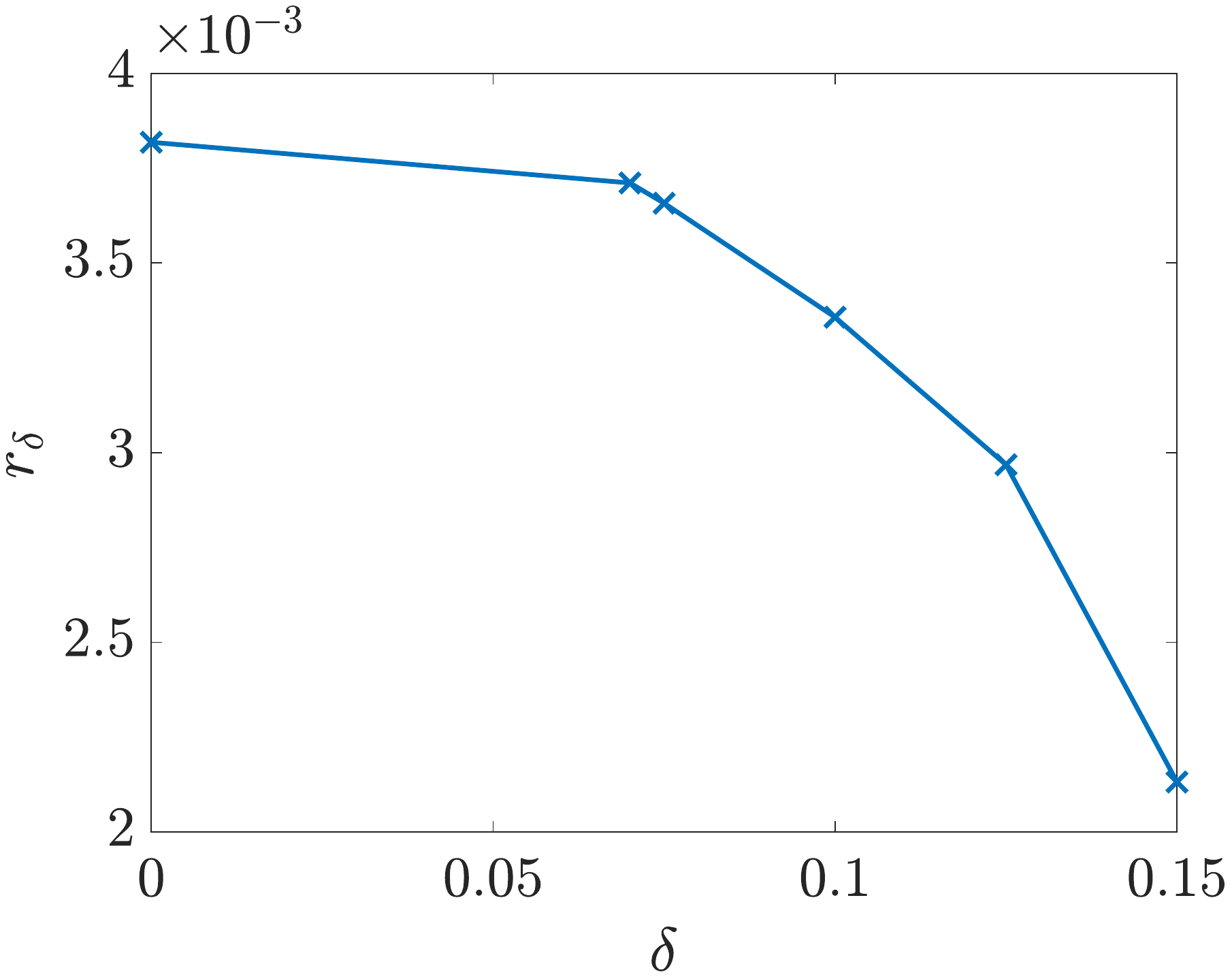}}
\caption{Convergence behavior for the RTI test of \citet{WaNiJa2001} 
using the $z$-model with the Krasny desingularization. 
Shown are (a) the bubble and spike tip locations, $| \max z_2 |$ and $| \min z_2 |$, respectively, 
(b) the location of the center of the spiral region $\sigma_\delta$, and (c) the radius of the 
spiral region $r_{\delta}$.}
\label{fig:RT_convergence1}
\end{figure}

\textcolor{black}{Next, we consider the convergence of the computed numerical solution in the 
limits $\delta \to 0$ with $N=256$ fixed. \Cref{fig:RT_convergence1_amplitude} shows that the 
bubble and spike tip locations appear to converge linearly, which agrees with the results of 
\cite{Krasny1986b} for the KHI test. Here, the value at $\delta = 0$ is obtained by linear extrapolation. 
In \Cref{fig:RT_convergence1_sigma} and \Cref{fig:RT_convergence1_radius}, we show the 
convergence of the computed spiral center $\sigma_\delta$ and radius $r_\delta$. 
We choose the axis $x_2 = x_2^* = -0.0023$ to compute the intersection points $x^*$. Again, the 
computed values appear to converge, though the precise nature of this convergence is less clear. In this case, 
we obtain the value at $\delta = 0$ via cubic interpolation.}

\begin{figure}[h]
\centering
\subfigure[Bubble (blue) and spike (red) tip locations.]{\label{fig:RT_convergence-comparison_amplitude}\includegraphics[width=50mm]{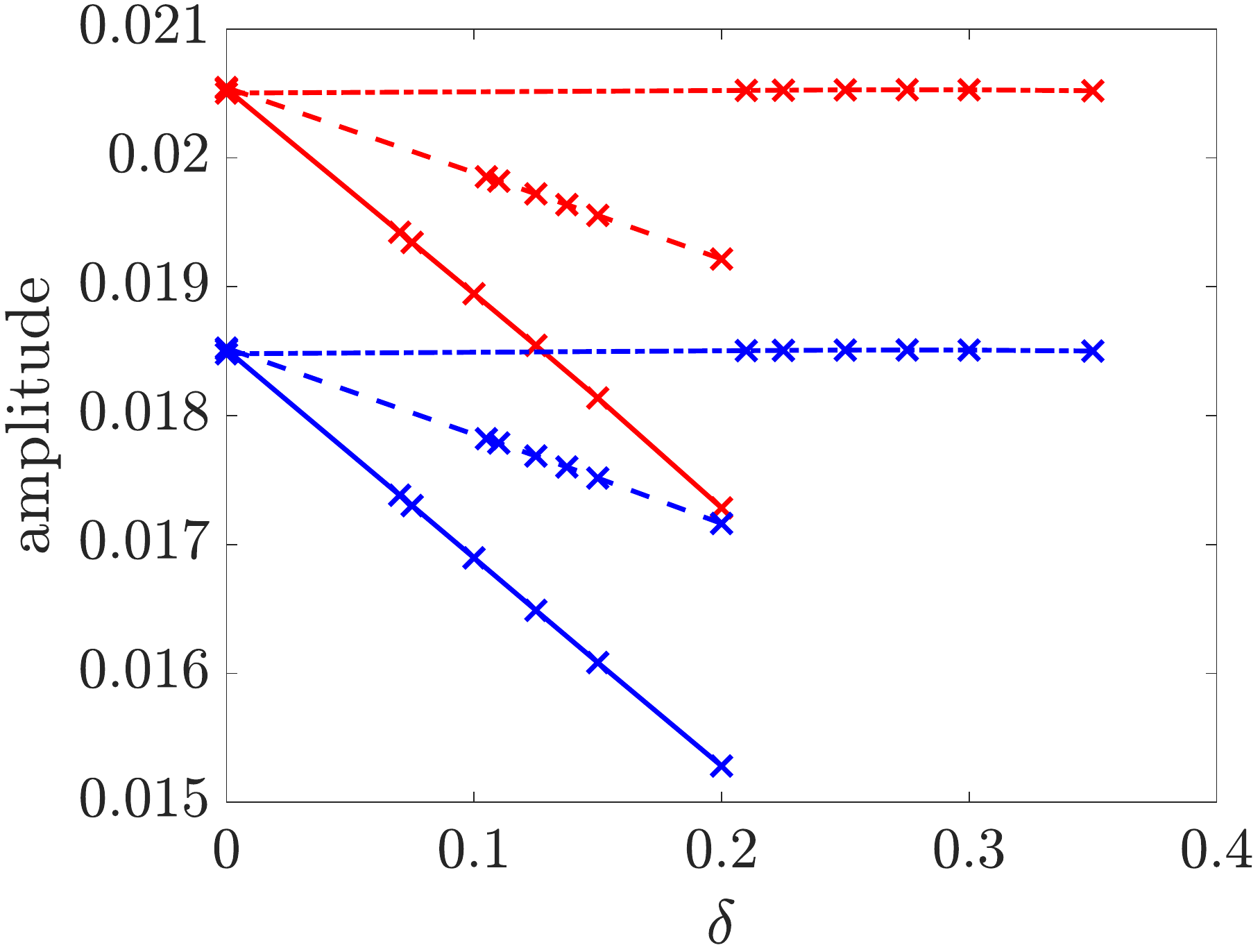}}
\hspace{0em}
\subfigure[Location of spiral center $\sigma_{\delta}$.]{\label{fig:RT_convergence-comparison_sigma}\includegraphics[width=50mm]{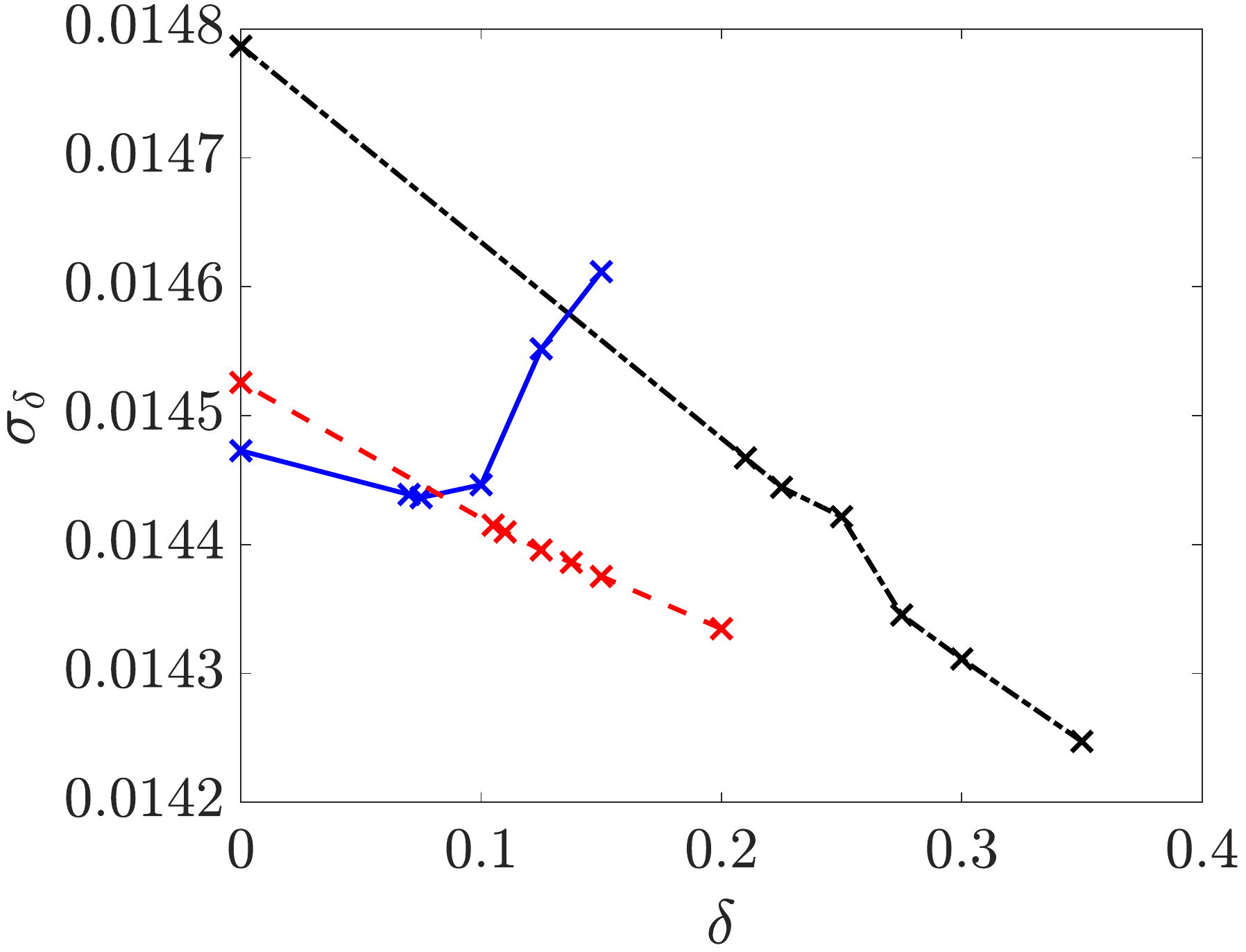}}
\hspace{0em}
\subfigure[Radius of spiral $r_{\delta}$.]{\label{fig:RT_convergence-comparison_radius}\includegraphics[width=50mm]{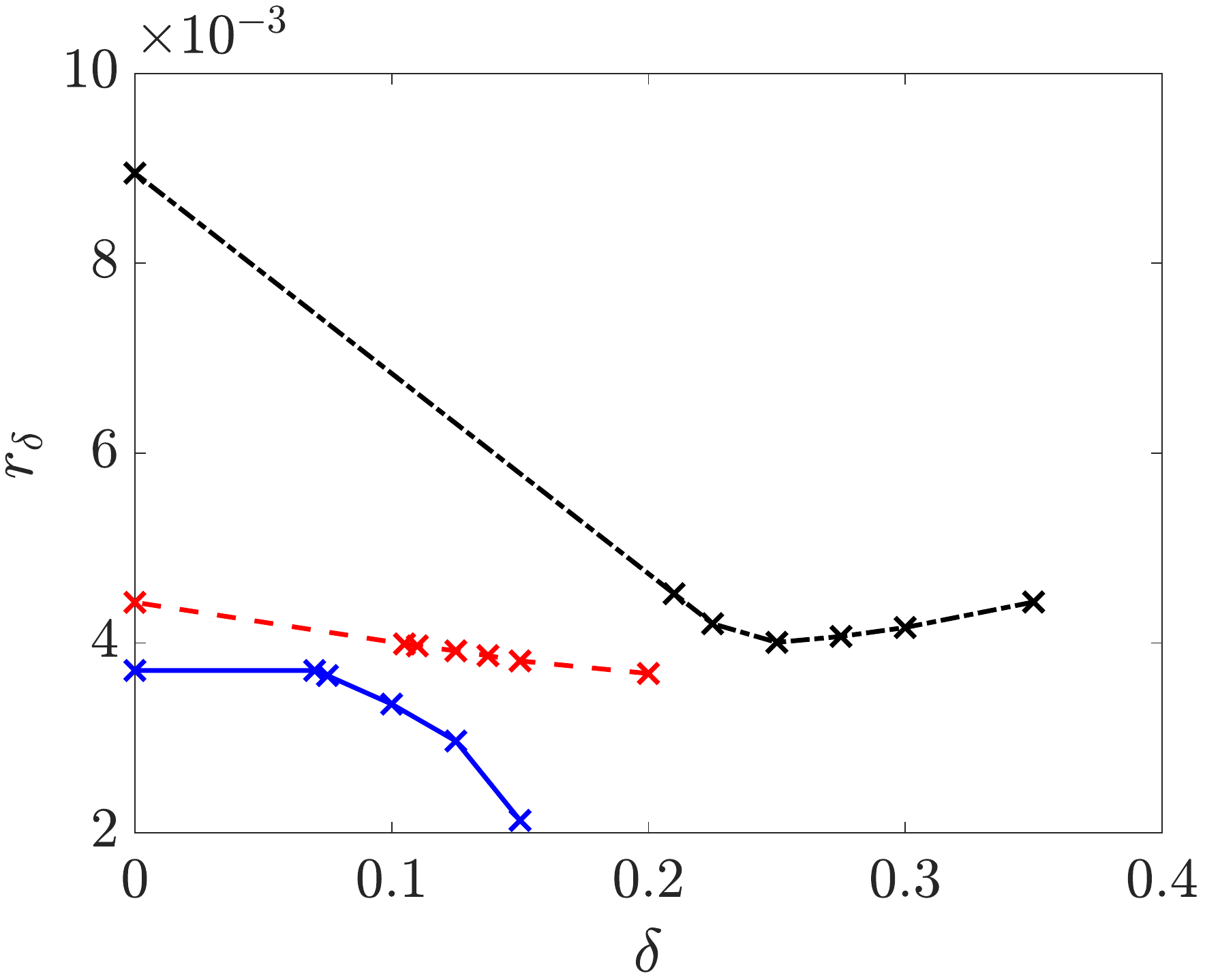}}
\caption{Convergence behavior as $\delta \to 0$ with $N=256$ fixed for the RTI test of \citet{WaNiJa2001} 
using the $z$-model. 
Shown are (a) the bubble and spike tip locations, $| \max z_2 |$ and $| \min z_2 |$, respectively, 
(b) the location of the center of the spiral region $\sigma_\delta$, and (c) the radius of the 
spiral region $r_{\delta}$. The solid, dashed, and dotted curves in (a) refer to the Krasny, 
$\mathcal{K}^\delta_1$, and $\mathcal{K}^\delta_3$ kernels, respectively.
The blue, red, and black curves in (b) and (c) refer to the Krasny, 
$\mathcal{K}^\delta_1$, and $\mathcal{K}^\delta_3$ kernels, respectively.}
\label{fig:RT_convergence-comparison}
\end{figure}

\textcolor{black}{We repeat our convergence tests for $N=256$ fixed and $\delta \to 0$ with the kernels 
\eqref{kernel-baker}, and compare the results with the Krasny kernel results in 
\Cref{fig:RT_convergence-comparison}. We find excellent agreement between all three methods in the 
limiting behavior of the bubble and spike tip locations, as shown in 
\Cref{fig:RT_convergence-comparison_amplitude}; this is in agreement with previous numerical 
studies \cite{BaPh2006,Sohn2014}. The small scale structure of the limiting solutions are slightly different, 
with the Krasny kernel and first order kernel $\mathcal{K}_1^\delta$ producing similar spiral 
center locations $\sigma_\delta$ and radii $r_\delta$. Here, the value at $\delta = 0$ is obtained by 
cubic extrapolation. Let us note that for $\delta > 0$, all three methods predict similar values of 
$\sigma_{\delta}$ and $r_\delta$.} 

\textcolor{black}{Next, we consider the limit $\delta, N^{-1} \to 0$, simultaneously; for the 
Krasny kernel, we use the scaling \eqref{delta-scaling}, while for the kernels 
$\mathcal{K}^\delta_i$ we use the empirical procedure discussed at the beginning of 
\Cref{z-model-numerical-study}. We consider six simulations with resolution starting from $N=32$ and 
doubling until $N=1024$, and again compute the bubble/spike tip locations and quantities 
$\sigma_\delta$ and $r_\delta$. We find excellent agreement between all three methods for 
the limiting values of each of the 
relevant quantities. Moreover, the Krasny scheme is the least computationally expensive method: for the 
$N=1024$ simulation, the runtime for the Krasny scheme is $T_{\mathrm{CPU}} \approx 333$ s, whereas 
the runtime for the 3rd order kernel $\mathcal{K}^\delta_3$ scheme is $T_{\mathrm{CPU}} \approx 469$ s, and thus the
simpler Krasny scheme is 40\% faster.
As such, we conclude that, as the simplest and least computationally expensive method, 
the Krasny desingularization is the most suitable method for our objective.}

\begin{figure}[h]
\centering
\subfigure[Bubble (blue) and spike (red) tip locations.]{\label{fig:RT_convergence2-comparison_amplitude}\includegraphics[width=50mm]{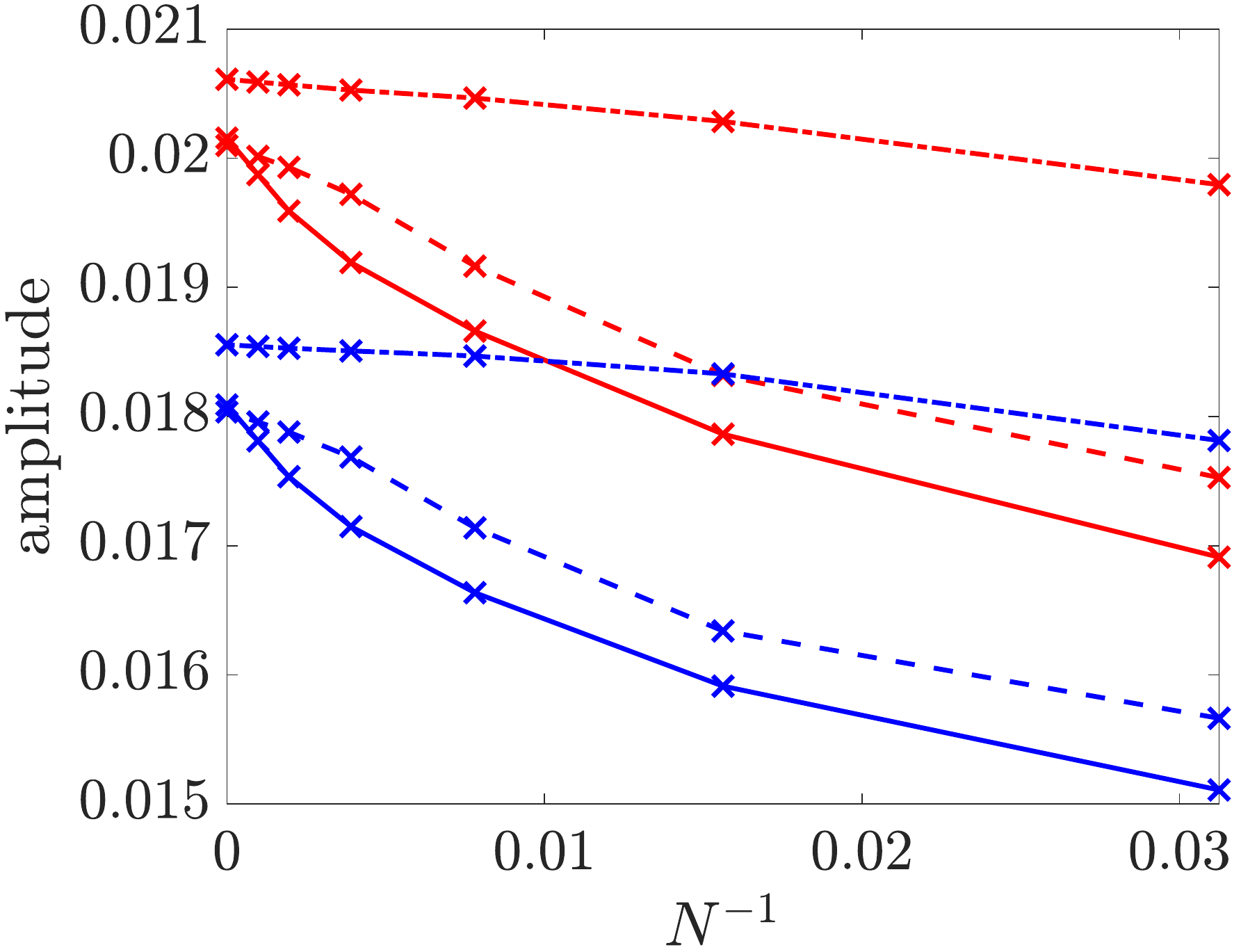}}
\hspace{0em}
\subfigure[Location of spiral center $\sigma_{\delta}$.]{\label{fig:RT_convergence2-comparison_sigma}\includegraphics[width=50mm]{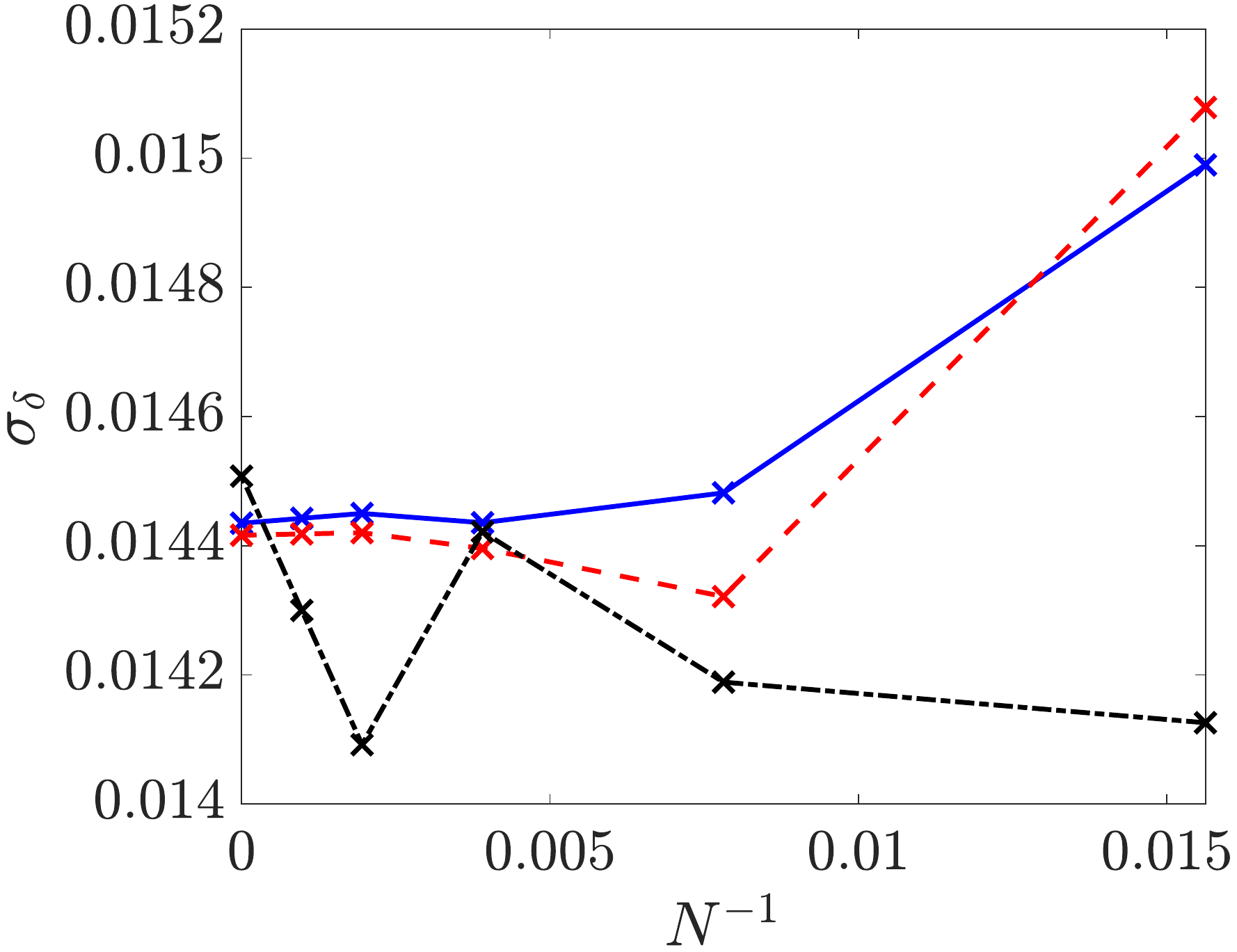}}
\hspace{0em}
\subfigure[Radius of spiral $r_{\delta}$.]{\label{fig:RT_convergence2-comparison_radius}\includegraphics[width=50mm]{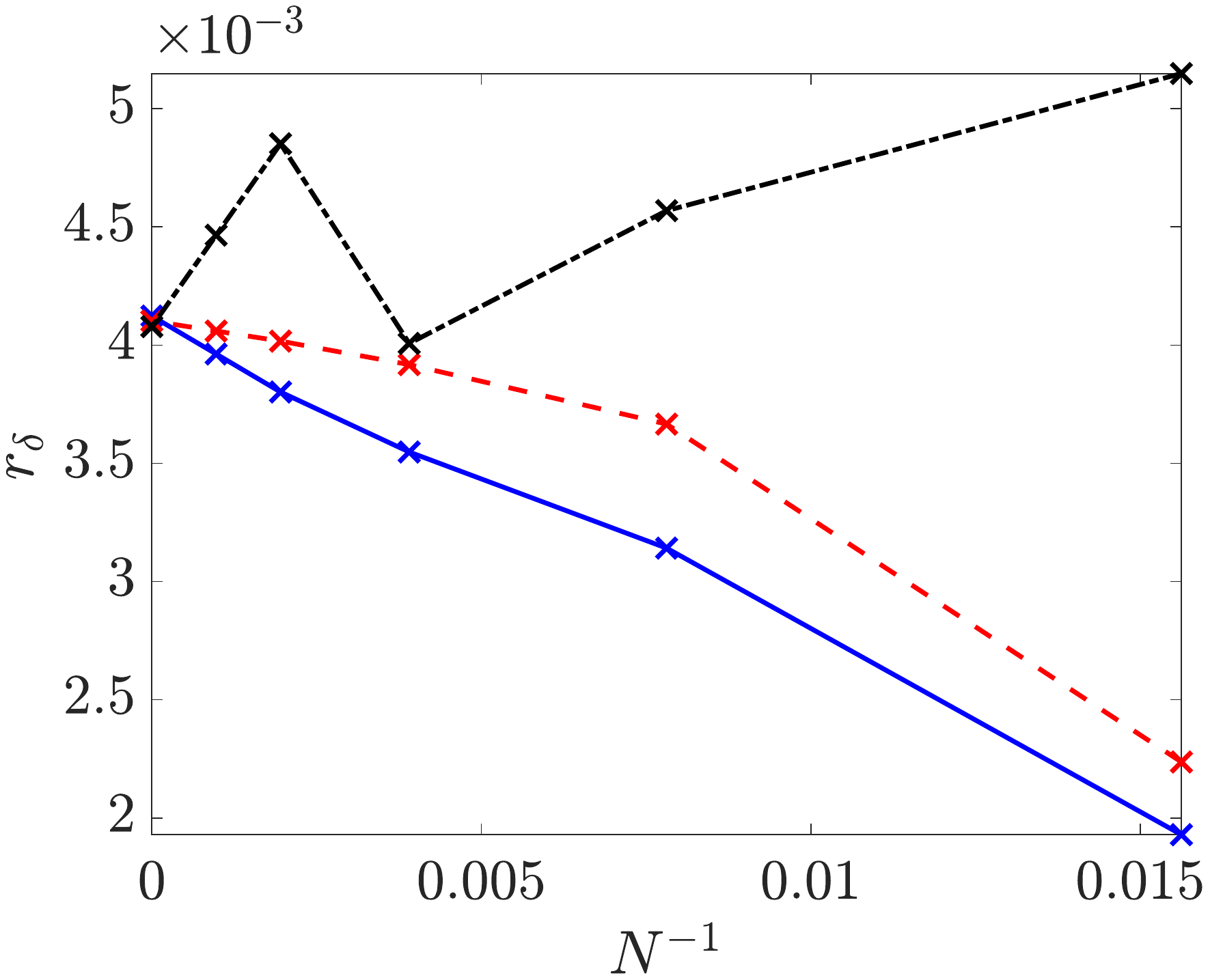}}
\caption{Convergence behavior as $\delta, N^{-1} \to 0$ for the RTI test of \citet{WaNiJa2001} 
using the $z$-model. 
Shown are (a) the bubble and spike tip locations, $| \max z_2 |$ and $| \min z_2 |$, respectively, 
(b) the location of the center of the spiral region $\sigma_\delta$, and (c) the radius of the 
spiral region $r_{\delta}$. The solid, dashed, and dotted curves in (a) refer to the Krasny, 
$\mathcal{K}^\delta_1$, and $\mathcal{K}^\delta_3$ kernels, respectively.
The blue, red, and black curves in (b) and (c) refer to the Krasny, 
$\mathcal{K}^\delta_1$, and $\mathcal{K}^\delta_3$ kernels, respectively.}
\label{fig:RT_convergence2-comparison}
\end{figure} 

\subsubsection{Single-mode RTI: comparison with numerical simulations}

Next, we compare the numerical simulations of \citet{Sohn2004a} that used the 
discretized equations \eqref{zeuler}, with our higher-order z-model solutions. A brief description of the 
numerical method in \cite{Sohn2004a} is as follows. The $z$-equation \eqref{zeuler-a} is treated in an 
identical fashion as in our numerical framework; in particular, the same Krasny $\delta$-desingularization and 
trapezoidal rule is used. The $\varpi$-equation \eqref{zeuler-b} is discretized in physical space, rather than in 
Fourier space as in our numerical method. The nonlinear term is treated using upwinding via the Godunov 
method, while an iterative procedure is used for the time-derivative term appearing on the right-hand side 
of \eqref{zeuler-b}. Let us remark that, on average, 6 iterations per time-step are required for this numerical 
method. 

Since we are interested in capturing vortex sheet roll-up, we consider the low Atwood number 
RTI test problem from \cite{Sohn2004a}. 
The domain is $\alpha \in [-\pi,\pi]$, the Atwood number is $A=0.05$, the gravitational constant is $g=1$, 
and the initial data is 
\begin{subequations}
\begin{align*}
 z_1(\alpha,0) &= \alpha  \,, \\
z_2(\alpha,0) &= 0.5 \cos(\alpha) \,, \\
\varpi(\alpha,0) &= 0\,. 
\end{align*}
\end{subequations}
The $z$-model is run for this problem with $N=400$, which is the same value employed in 
\cite{Sohn2004a}, but whereas the value $\delta t = 0.002$ is required in \cite{Sohn2004a}, we are able 
to use the much larger $\delta t = 0.025$. The regularizing parameters chosen as 
$\tilde{\delta} = 0.6$ and $\tilde{\mu} = 0.005$. This value of $\tilde{\delta}$ was chosen to agree with the 
parameter choices in \cite{Sohn2004a}. 

\begin{figure}[h]
\centering
\subfigure[$t=0$]{\label{fig:RT_Sohn2004_Atwd=s_t=0}\includegraphics[width=24mm]{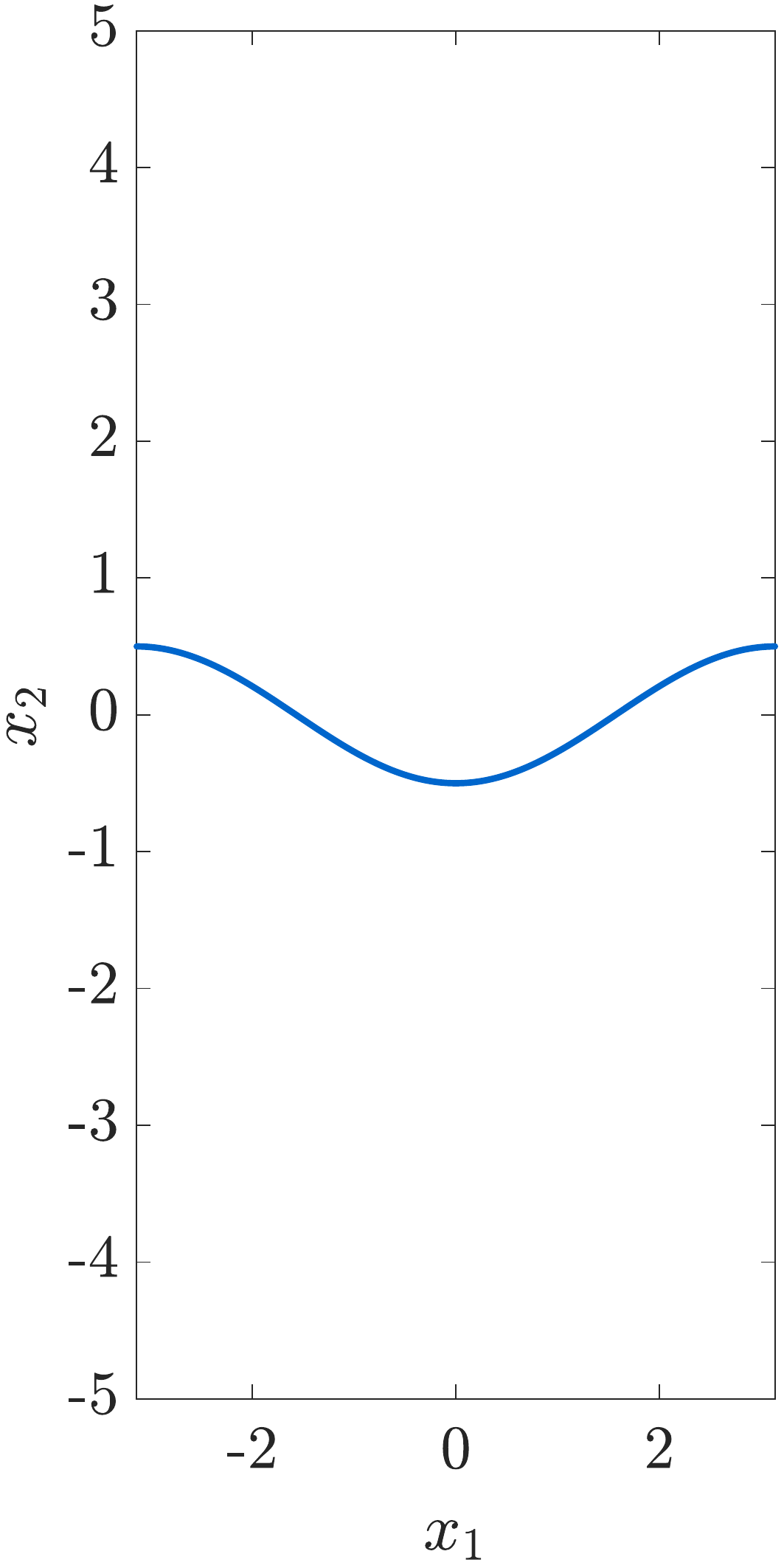}}
\hspace{1em}
\subfigure[$t=10$]{\label{fig:RT_Sohn2004_Atwd=s_t=10}\includegraphics[width=20mm]{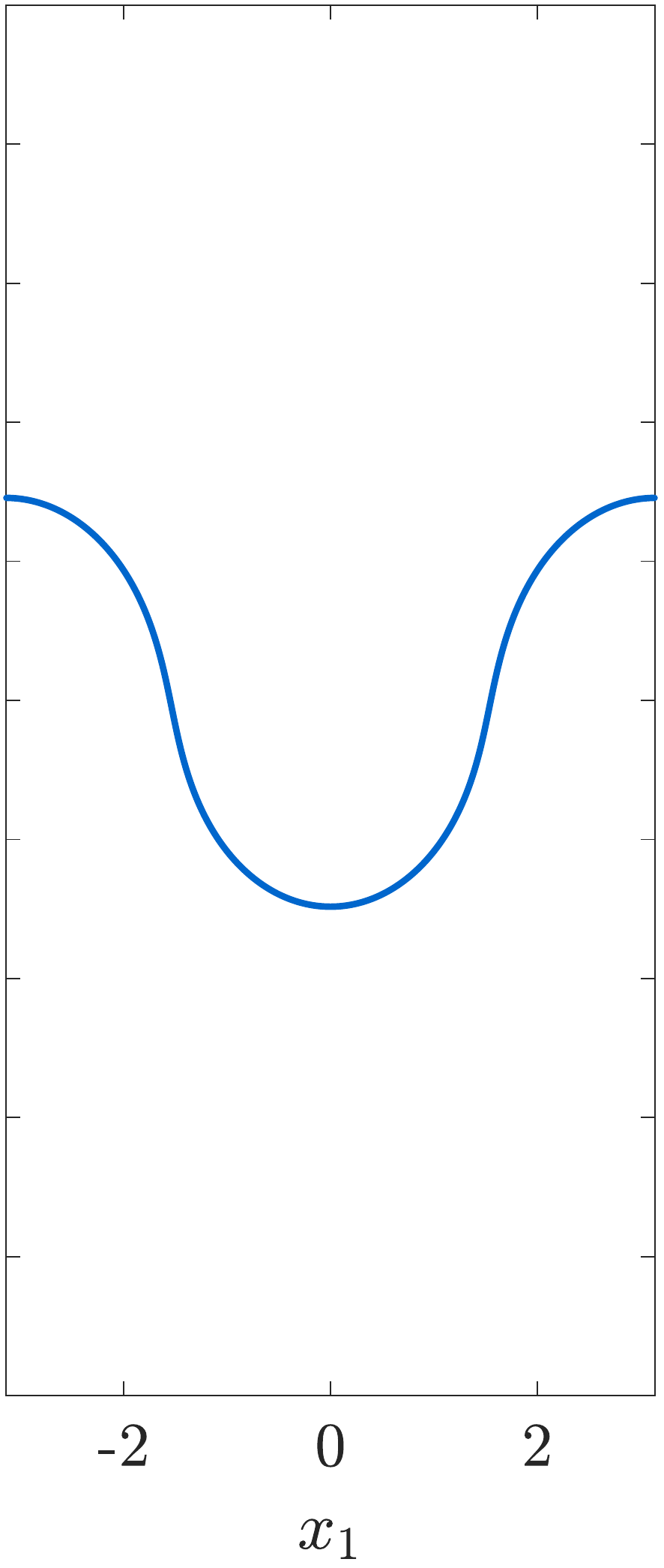}}
\hspace{1em}
\subfigure[$t=14$]{\label{fig:RT_Sohn2004_Atwd=s_t=14}\includegraphics[width=20mm]{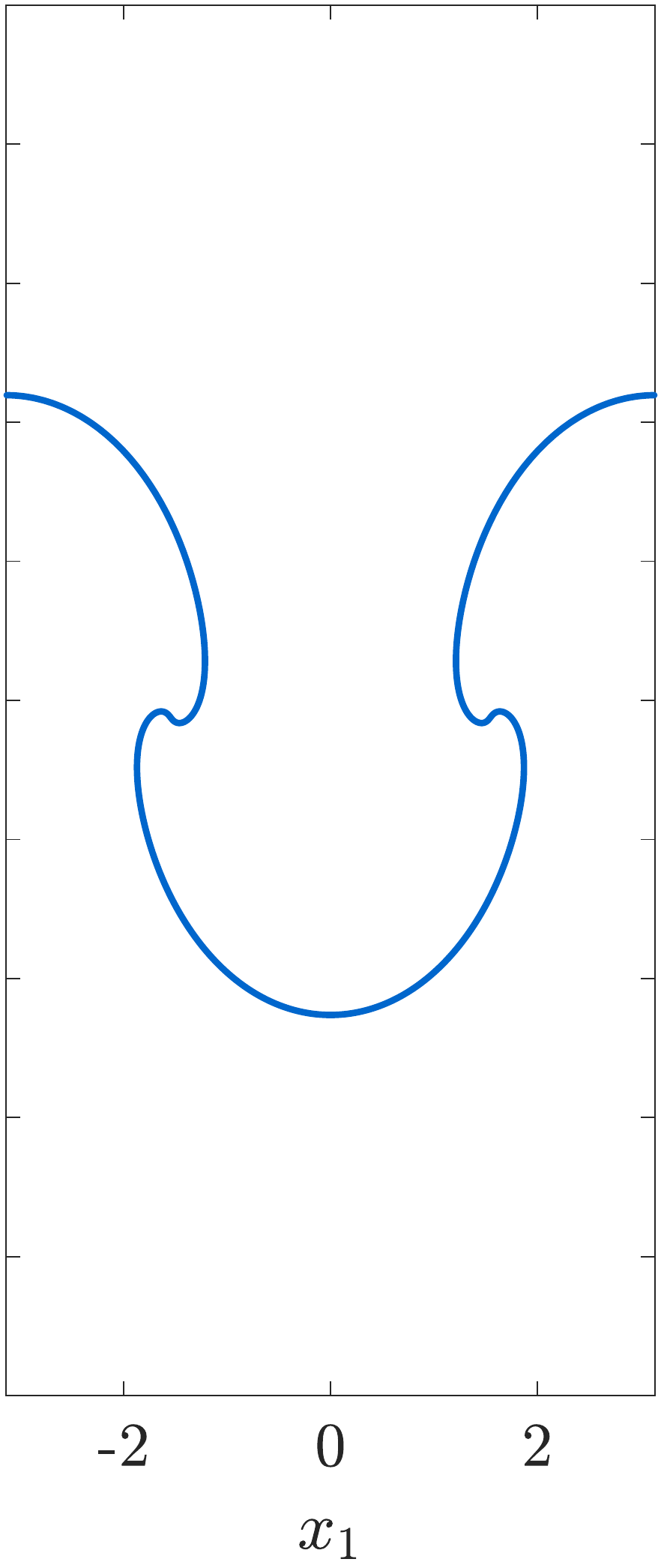}}
\hspace{1em}
\subfigure[$t=18$]{\label{fig:RT_Sohn2004_Atwd=s_t=18}\includegraphics[width=20mm]{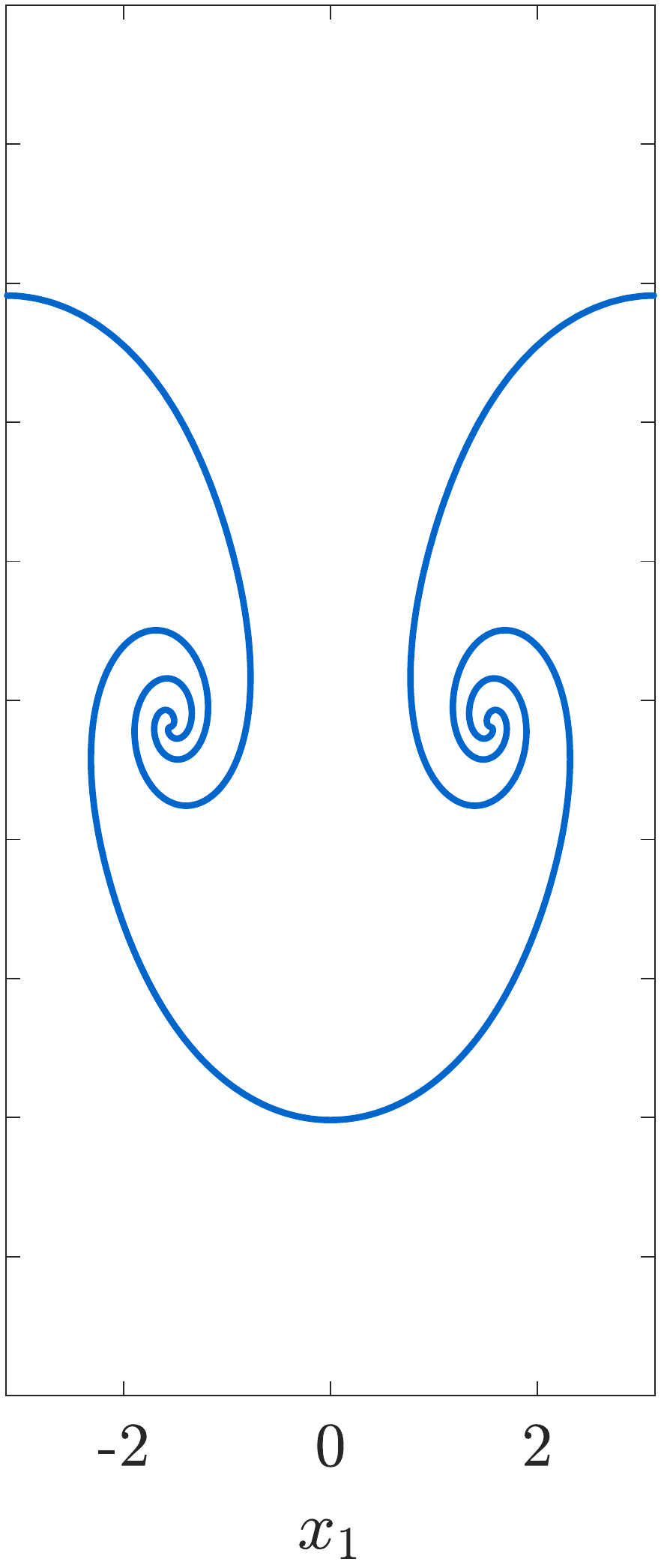}}
\hspace{1em}
\subfigure[$t=22$]{\label{fig:RT_Sohn2004_Atwd=s_t=22}\includegraphics[width=20mm]{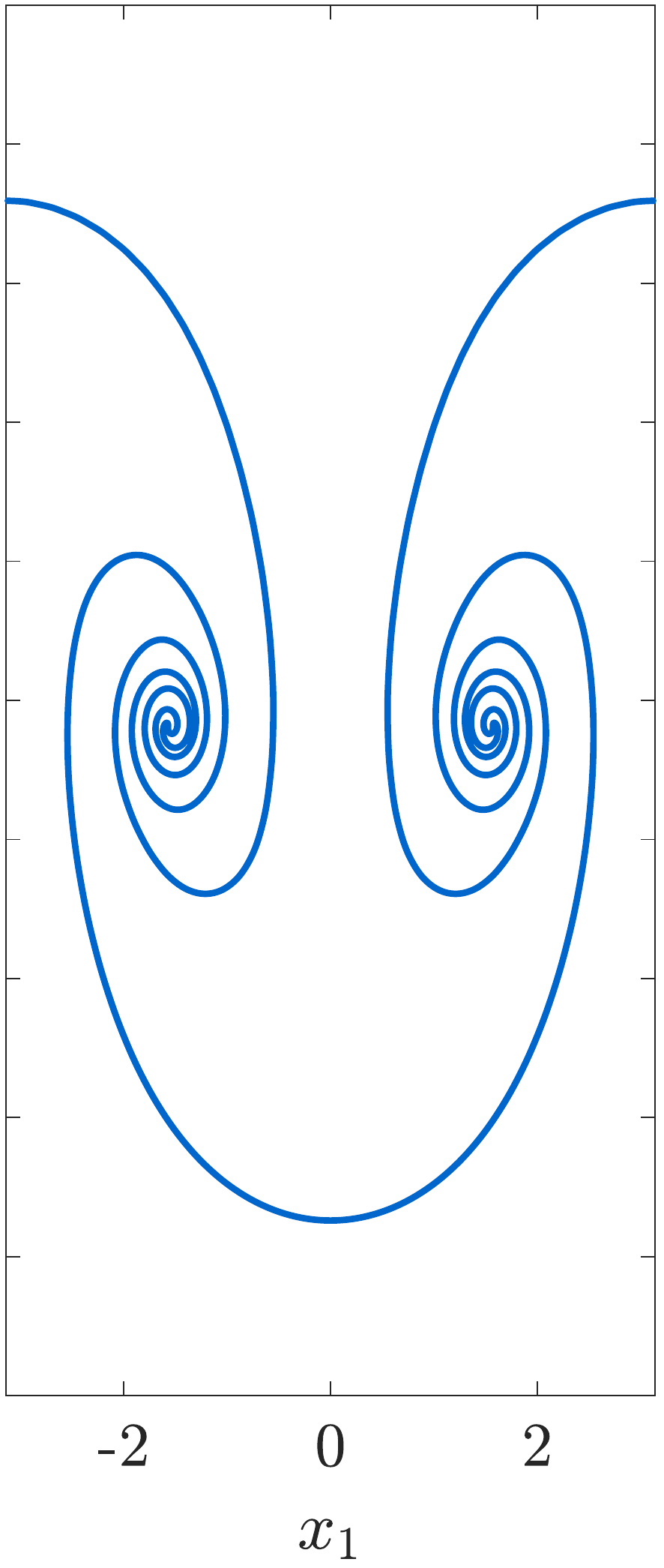}}
\hspace{1em}
\caption{Numerical simulation of the RTI using the $z$-model with the setup as in the 
numerical studies of \citet{Sohn2004a}. Plots of the interface $z$ 
for a simulation with Atwood number $A = 0.05$ are 
shown at various times $t$.}
\label{fig:RT_Sohn2004_Atwd=s}
\end{figure}

The computed $z$ for the above numerical experiment is shown in \Cref{fig:RT_Sohn2004_Atwd=s} at 
various time $t$. This figure should be compared with Figure 1(a) in \cite{Sohn2004a}, upon which it is clear 
that the two are essentially indistinguishable. In particular, we note that the two solutions are in excellent 
agreement in the roll-up region, with both branches of the spirals having almost four full rotations at the 
final time $t=22$. 

\begin{figure}[h]
\centering
\subfigure[Bubble and spike speeds]{\label{fig:RT_Sohn2004_Atwd=s_speed}{\includegraphics[width=55mm]{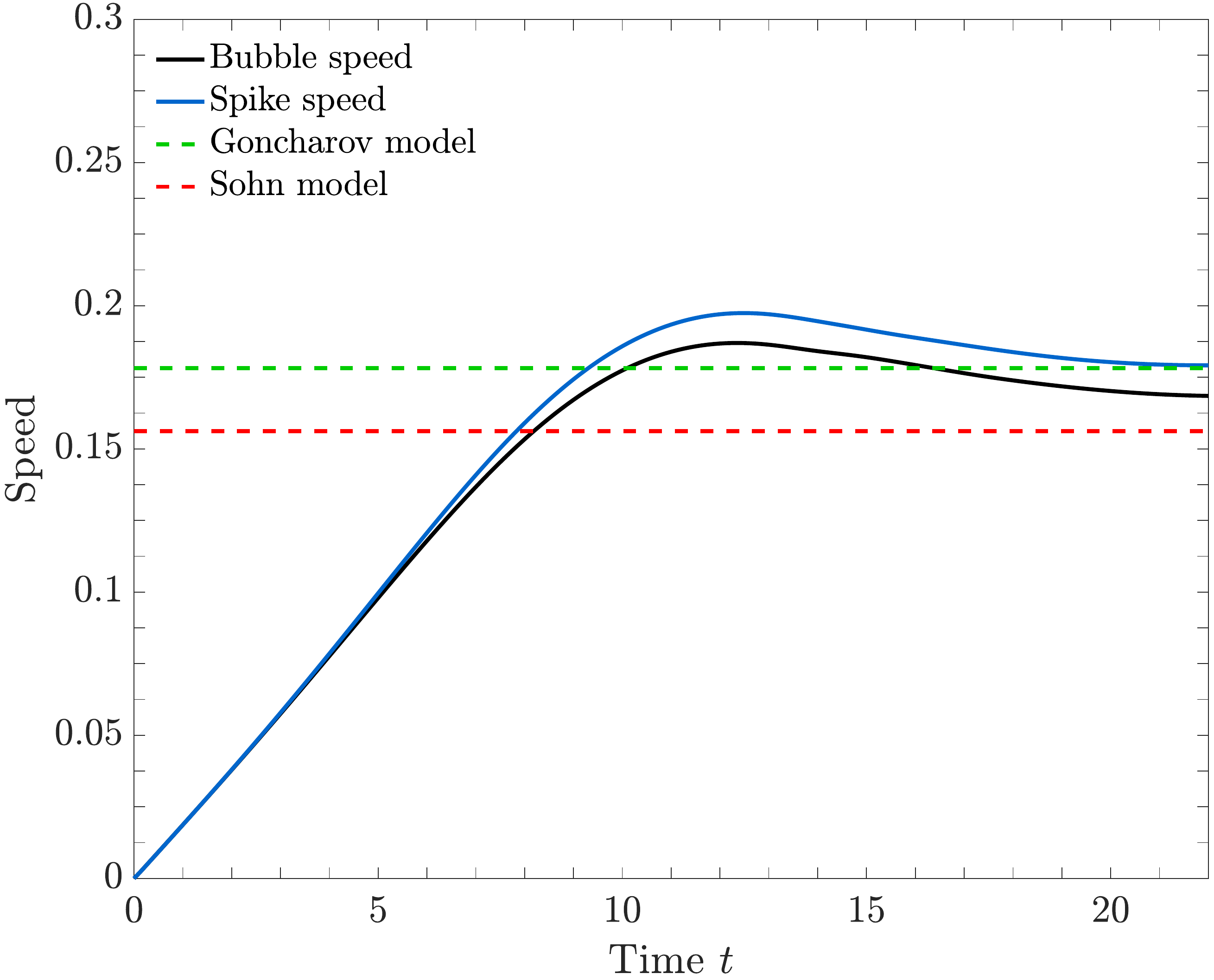}}}
\hspace{4em}
\subfigure[Vortex strength]{\label{RT_Sohn2004_Atwd=s_gamma}\includegraphics[width=54.5mm]{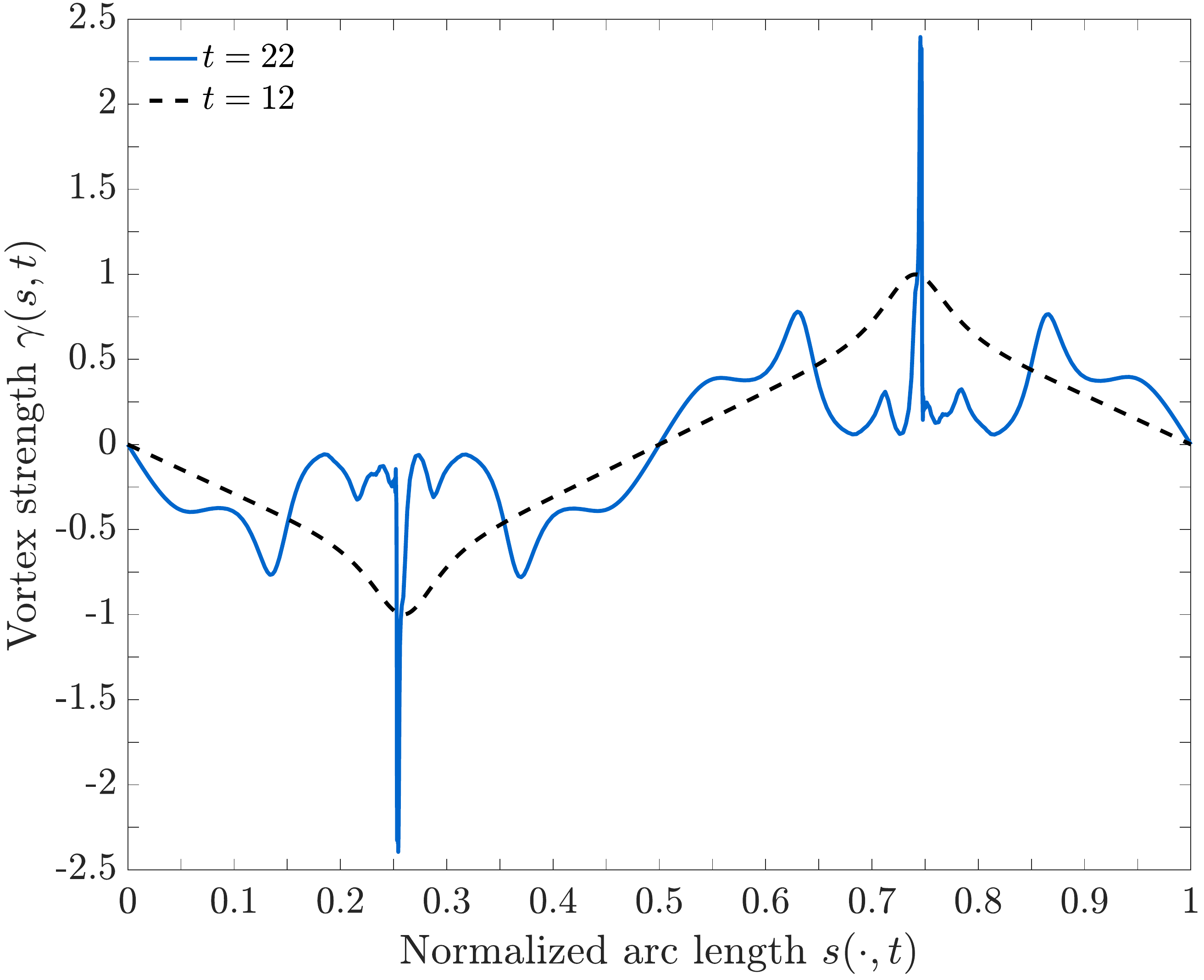}}
\hspace{0em}
\subfigure[Tip curvature]{\label{RT_Sohn2004_Atwd=s_kappa}\includegraphics[width=54.5mm]{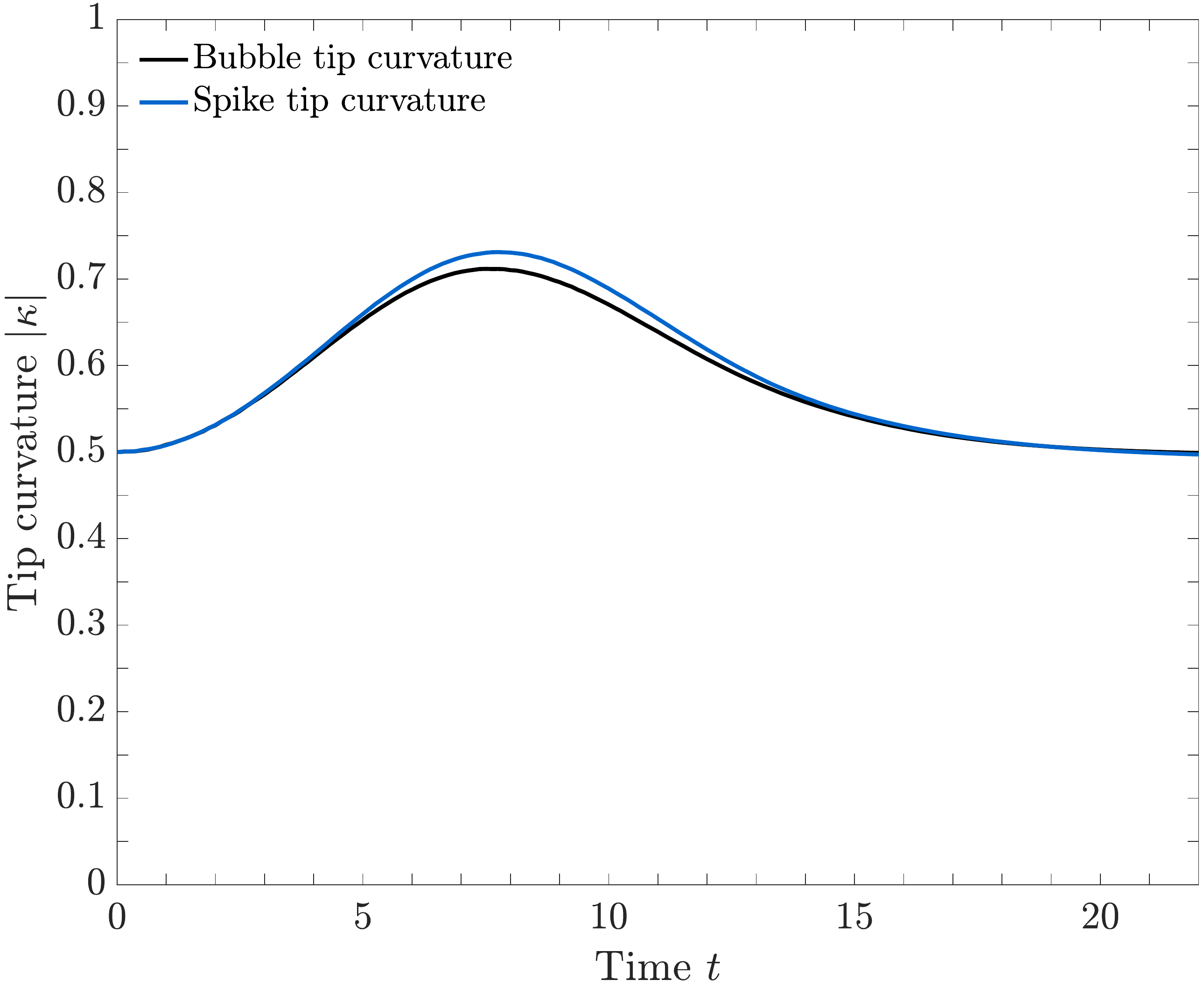}}
\caption{Plots of various computed quantities for the \citet{Sohn2004a} RTI problem.}
\label{fig:RT_Sohn2004_speed}
\end{figure}

To quantify the agreement between our $z$-model solution and the reference solution from \cite{Sohn2004a}, 
we plot various computed quantities in \Cref{fig:RT_Sohn2004_speed}. The bubble tip and spike tip 
speeds versus time are shown in \Cref{fig:RT_Sohn2004_Atwd=s_speed} (compare with 
Figure 3(a) in \cite{Sohn2004a}). Shown also are theoretical predictions for the bubble speed from 
asymptotic potential flow models. Such models aim to describe analytically the evolution of the amplitude of 
the interface $z$, {after} the transition from exponential growth $z \sim z_0 e^{t\sqrt{Ag}}$ at 
small times (as predicted by the linear theory) to linear-in-time growth $z \sim w_{\infty} t$  
(as observed in experiments, for instance), where $w_{\infty}$ is the asymptotic bubble velocity. The 
Goncharov \cite{Goncharov2002} and Sohn \cite{Sohn2003} models estimate $w_\infty$ as 
$$
w_{\infty} = \sqrt{\frac{2 A}{1+ A} \frac{g}{3}} \qquad \text{ and } \qquad w_{\infty} = \sqrt{\frac{Ag}{2+ A}} \,, 
$$
respectively. As shown in \Cref{fig:RT_Sohn2004_Atwd=s_speed}, the computed speed and asymptotic 
predictions are in good agreement; let us note that the bubble and spike speeds computed using the 
$z$-model appear identical to those in \cite{Sohn2004a}. 

Next, we compute the \emph{vortex sheet strength} 
$\gamma(\cdot,t) = \varpi(\cdot,t) / |\partial_{\alpha} z(\cdot,t)|$. We plot $\gamma(s,t)$ versus $s$ in 
\Cref{RT_Sohn2004_Atwd=s_gamma}, where $s(t)$ is the normalized arclength. Comparing 
\Cref{RT_Sohn2004_Atwd=s_gamma} with Figure 7(a) in \cite{Sohn2004a}, we see that $\gamma(s,t)$
 computed using the $z$-model is slightly larger in magnitude at the shock, but is otherwise identical to the 
 solution in \cite{Sohn2004a}. Finally, we compute the magnitude of the bubble tip and spike tip
 curvatures $| \kappa |$ in \Cref{RT_Sohn2004_Atwd=s_kappa}, which is in excellent agreement with 
 Figure 6(a) in \cite{Sohn2004a}. 
 
 The above observations indicate that the $z$-model is able to accurately simulate interface turnover and 
 roll-up, while minimizing the cost of the numerical computations. In particular, since six iterations are 
 required, on average, for the algorithm in \cite{Sohn2004a}, the $z$-model computation is at least 
 6 times faster if the same time-step $\delta t$ used. For the above simulation, we were able to use a much 
 larger time-step, which shows that that the $z$-model computation is a factor of at least 
 $6 \times 0.025/0.002 = 75$ times faster. In fact, since the Fast Fourier Transform (FFT) is used
 to efficiently compute the $\varpi$-equation \eqref{wbar-eqn-regular-freq-discrete}, 
 whereas costly upwinding is required for the 
 algorithm in \cite{Sohn2004a}, it is highly likely that the $z$-model computation is much greater than 75 times 
 faster than the algorithm in \cite{Sohn2004a}.

\subsubsection{Multi-mode RTI: the rocket rig experiment of Read and Youngs}

We next consider the rocket rig experiment of  \citet{Read1984} and \citet{Youngs1984}, in which the
initial interface separating the two fluids of densities $\rho^{+}$ and $\rho^{-}$ is given by a small and 
random perturbation of the flat interface. Our aim is to compare the growth rates of the mixing layer 
computed using our $z$-model with the growth rates observed in experiments \cite{Read1984} and 
2-$D$ DNS simulations \cite{Youngs1984}. The experimental and numerical evidence indicate that the 
width of the mixing layer grows like 
\begin{equation}\label{rocket-rig-quadratic}
\Theta A g t^2 \,, 
\end{equation}
where $\Theta$ is some constant. The experimental 
evidence suggests that the constant $\Theta$ lies in the range $\Theta \in (0.05,0.775)$, while the 
numerical simulations give $\Theta \in (0.04,0.05)$. We shall follow \citet{Youngs1989}, and employ the 
value $\Theta = 0.06$ for comparison purposes. 

We set up the problem as follows: the domain is $\alpha \in [-\pi,+\pi]$, the densities of the two fluids are 
$\rho^{+} = 0.66$ and $\rho^{-}= 1.89$, giving an Atwood number of $A \approx 0.482$, 
the gravitational acceleration is $g = - 9.8 \times \frac{2 \pi}{0.3}$, and  the initial data is given by 
\begin{subequations}
\label{rocket-rig-initial-data}
\begin{align*}
 z_1(\alpha,0) &= \alpha  \,, \\
z_2(\alpha,0) &=  \frac{1}{\sigma} \sum_{r = 1}^{s} a_r \cos(r \alpha) + b_r \sin(r \alpha) \,, \\
\varpi(\alpha,0) &= 0\,,
\end{align*}
\end{subequations}
where $a_r$ and $b_r$ are random number chosen from a standard Gaussian distribution, $s=32$, and 
$\sigma$ is a constant chosen such that $|| z_2(\cdot,0) ||_{L^2} = 0.01$. 

\begin{table}[H]
\centering
\renewcommand{\arraystretch}{1.0}
\scalebox{0.8}{
\begin{tabular}{|c|ccccc|}
\toprule
\multirow{2}{*}{} & \multicolumn{5}{c|}{\textbf{$N$}}\\

& 128   & 256    & 512   & 1024 & 2048 \\

\midrule

$\delta t$ & $1 \times 10^{-3}$ & $1 \times 10^{-3}$ & $5 \times 10^{-4}$ & $1.25 \times 10^{-4}$ &  $3.125 \times 10^{-5}$\\

\bottomrule
\end{tabular}}
\caption{Time-step $\delta t$ choices for the rocket rig test.}
\label{table:rocket-rig-parameters}
\end{table}

\begin{figure}[h]
\centering
\includegraphics[width=120mm]{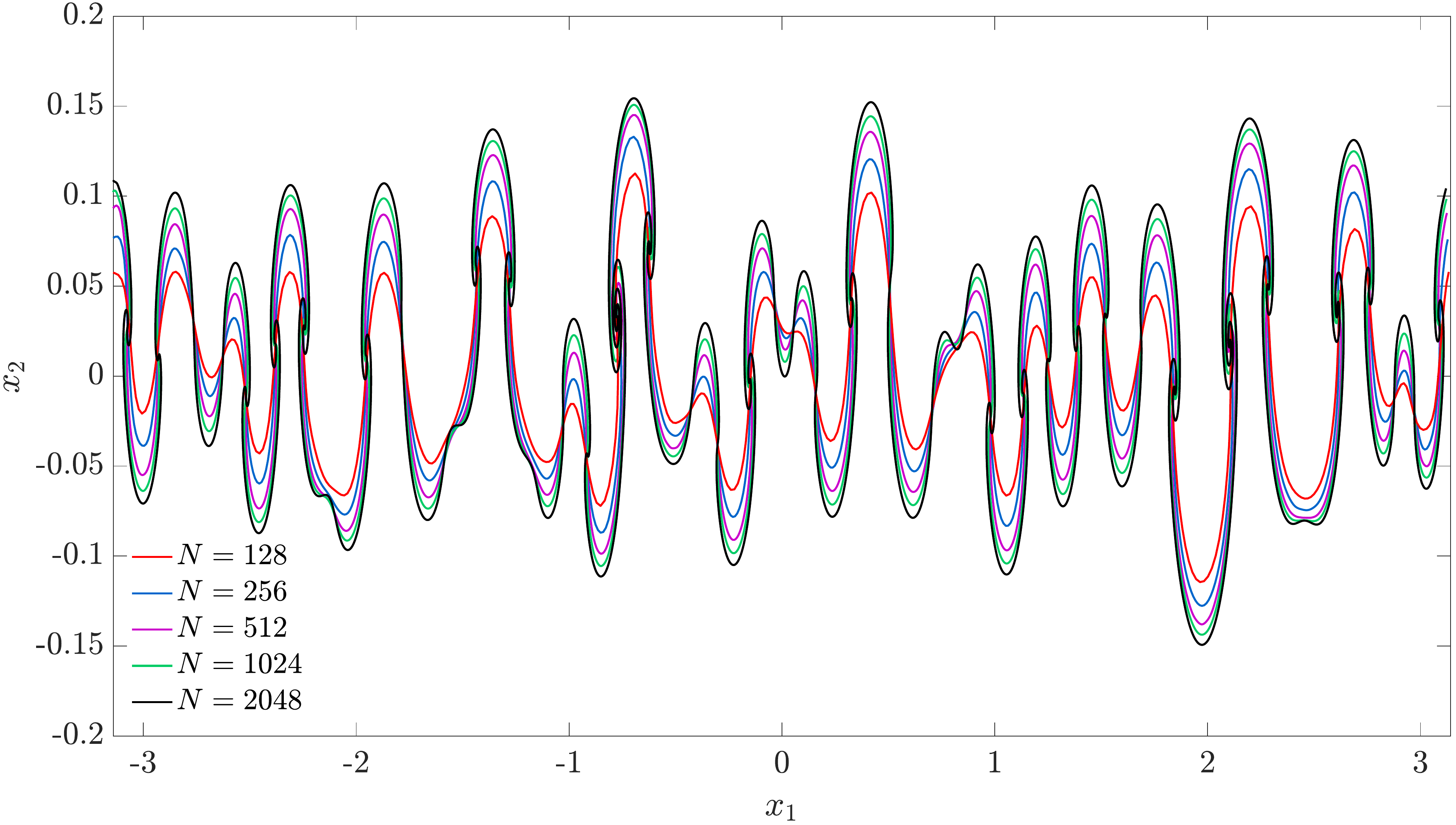}
\caption{Numerical simulation of the rocket rig test using the $z$-model. The results for five simulations with 
resolution starting from $N=128$ and doubling until $N=2048$ are shown at time $t=0.15$.}
\label{fig:rocket_rig_comparison}
\end{figure}

\begin{figure}[h]
\centering
\includegraphics[width=70mm]{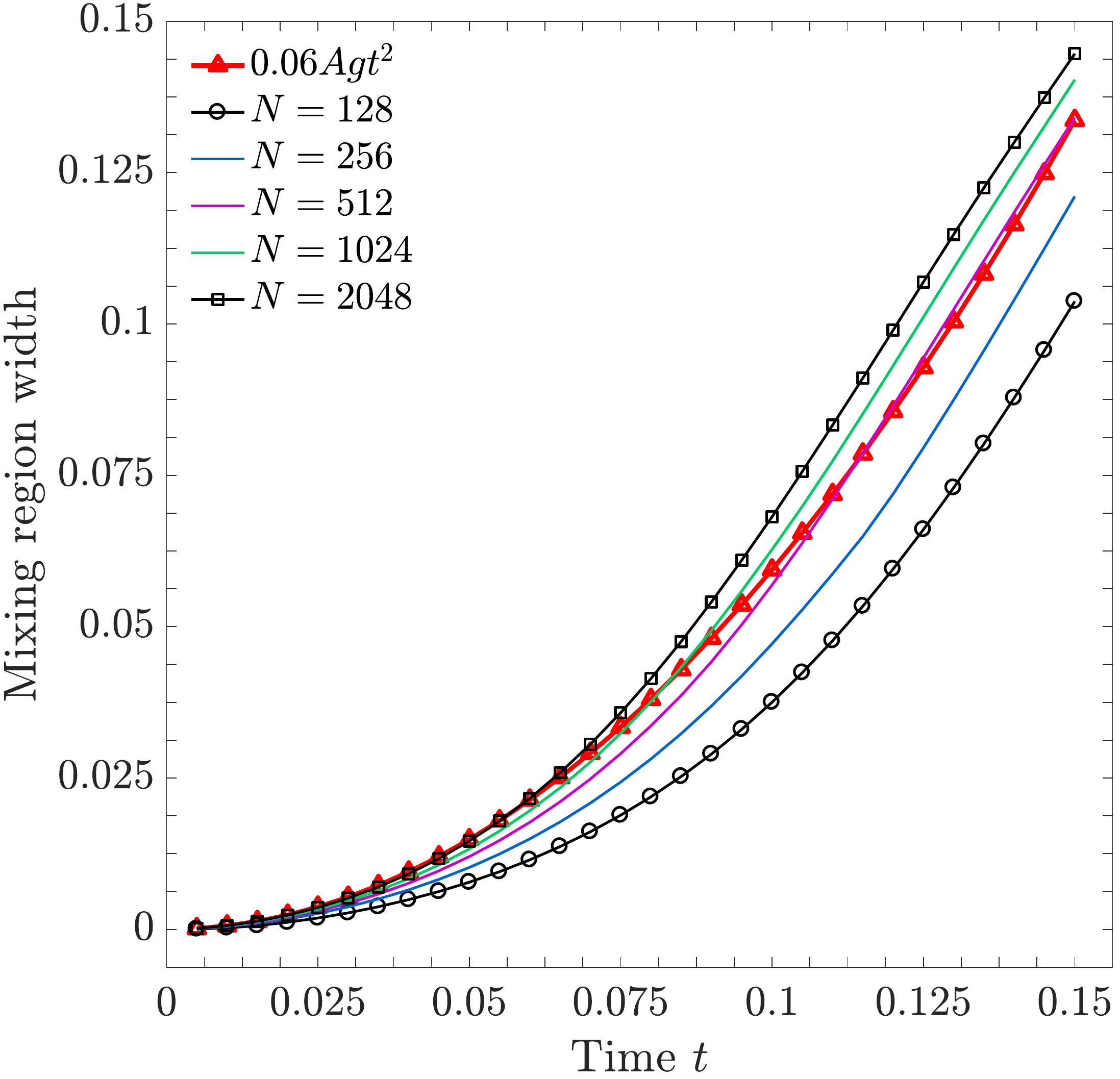}
\caption{Plot of the mixing region width $\max z_2(\cdot,t) - \min z_2(\cdot,0)$ versus time 
$t$ for the rocket rig test. Results are shown for five simulations using the $z$-model
with increasing resolution $N$. The red curve is the predicted quadratic growth rate 
\eqref{rocket-rig-quadratic} with $\Theta=0.06$.}
\label{fig:rocket_rig_mixing_width}
\end{figure}

We perform 5 simulations using the $z$-model and  with resolution starting from $N=128$ and 
doubling until $N=2048$. The regularization parameters are fixed as $\tilde{\delta} = 0.065$ and 
$\mu = 0.06$ for all of the simulations. 
The time-step $\delta t$ varies
with $N$, and is listed in \Cref{table:rocket-rig-parameters}. We show plots of the computed 
interface parametrization $z(\alpha,t)$ at the final time $t=0.15$ 
in \Cref{fig:rocket_rig_comparison}. As expected, there is more-roll up of the interface as the 
resolution is increased. To quantify the amount of mixing, 
 the width of the mixing region is approximated as $\max_{\alpha} z_2(\alpha,t) - \max_{\alpha} z_2(\alpha,0)$. 
A comparison of the computed mixing region width and the predicted quadratic growth rate 
\eqref{rocket-rig-quadratic} with $\Theta = 0.06$ is shown in \Cref{fig:rocket_rig_mixing_width}, from which 
it is clear that the two are in very good agreement. 

\begin{figure}[h]
\centering
\subfigure[Interface position $z$]{\label{fig:rocket_rig_ensemble}\includegraphics[width=80mm]{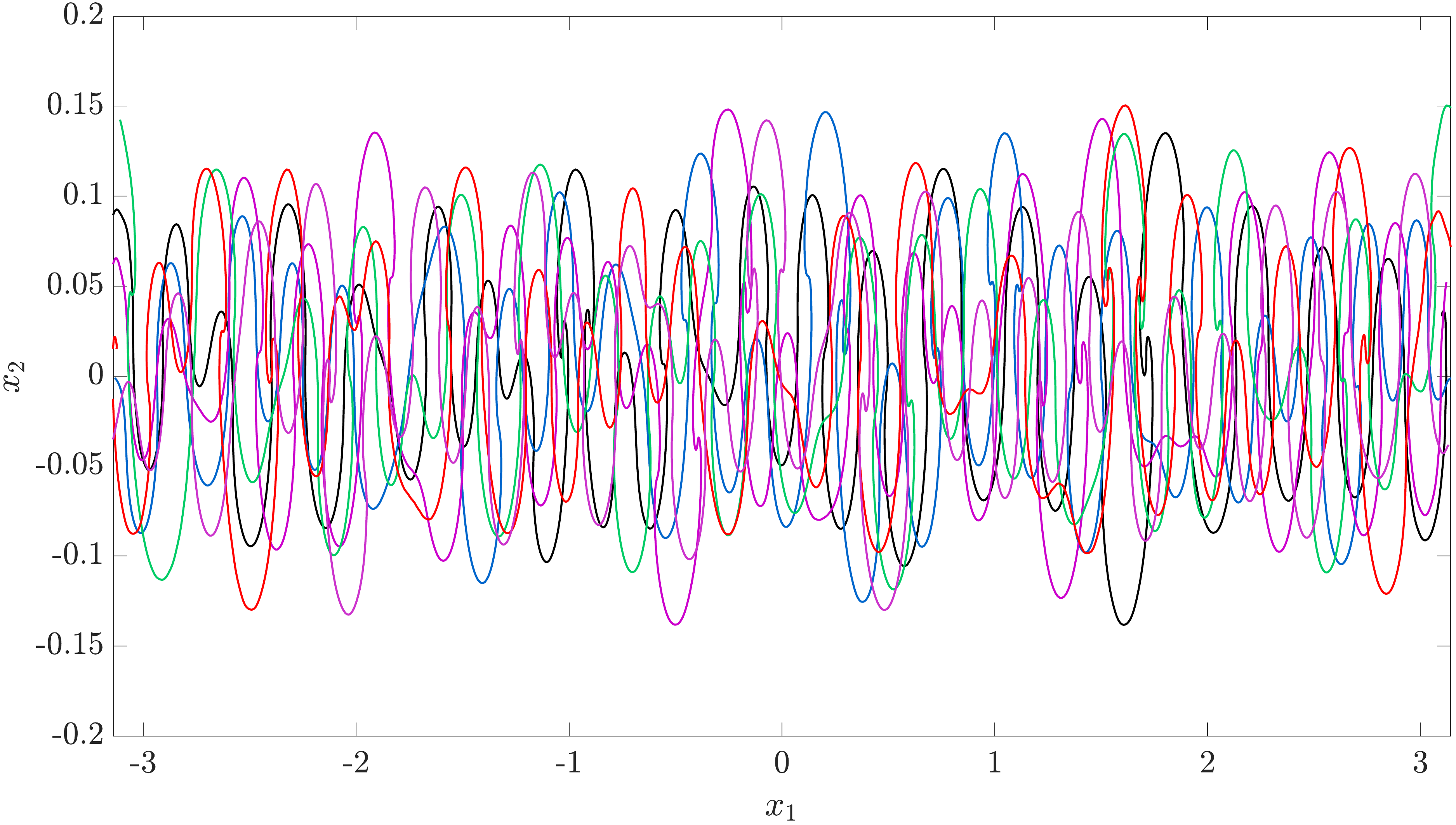}}
\hspace{2em}
\subfigure[Mean mixing region width]{\label{fig:rocket_rig_ensemble_mixing_width}\includegraphics[width=50mm]{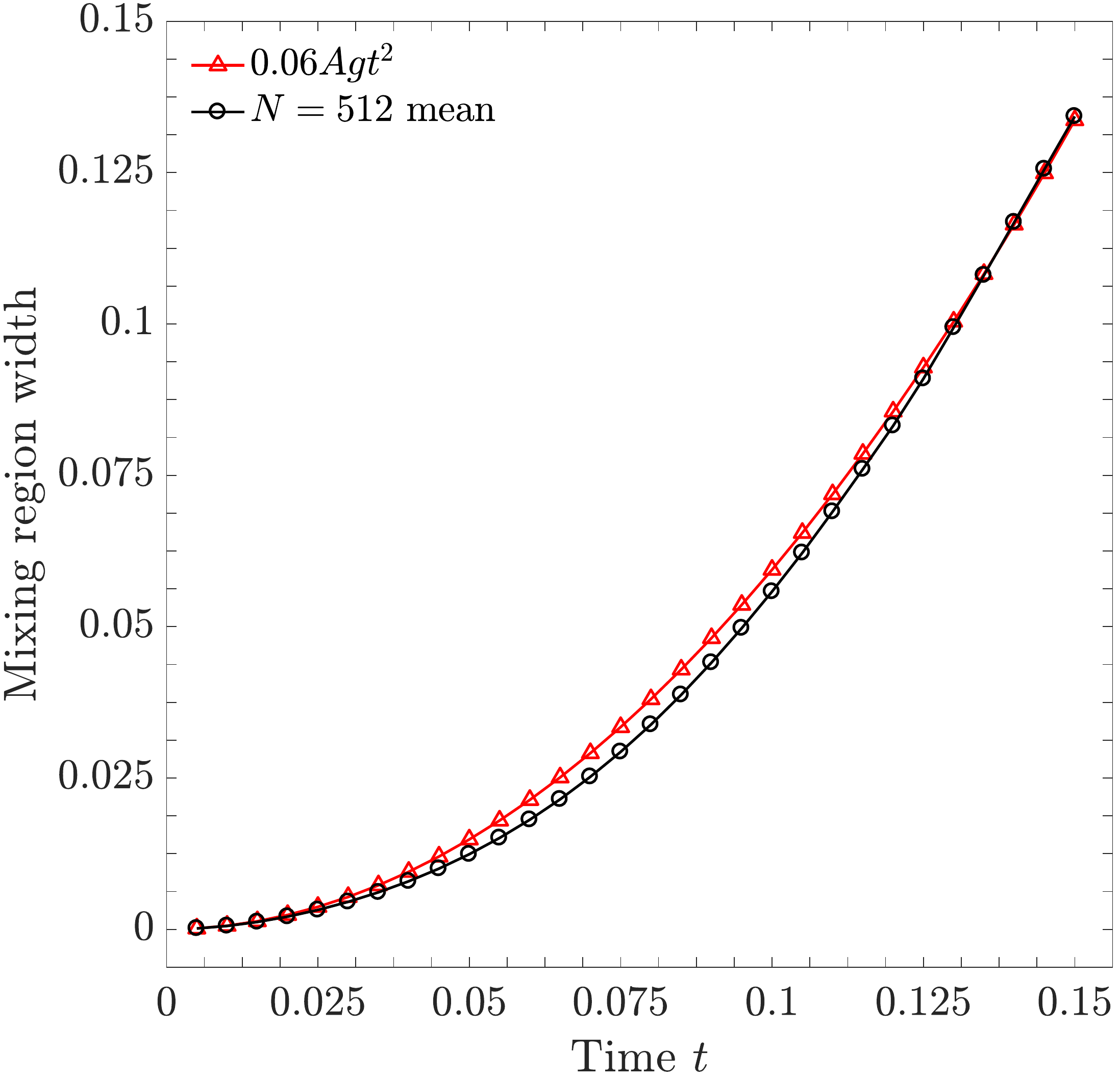}}
\hspace{2em}
\caption{Ensemble of simulations for the rocket rig test with $N=512$. Shown are 
(a) the interface position $z$ at time $t=0.15$, and (b) the mean mixing region width versus time.}
\label{fig:rocket_rig_512_ensemble}
\end{figure}

Since the initial conditions for the rocket rig test are random, 
it may be difficult to obtain a clear qualitative picture of the mixing region from a single 
simulation alone. Consequently, we repeat the test for $N=512$ with six different random initial 
conditions. The interface position for the ensemble of runs is shown in 
\Cref{fig:rocket_rig_ensemble}, while the mean mixing region width is compared with the quadratic growth 
rate in \Cref{fig:rocket_rig_ensemble_mixing_width}. It is clear from 
\Cref{fig:rocket_rig_ensemble_mixing_width} that the computed mixing region width is in excellent 
agreement with the predicted quadratic growth rate, thus providing strong evidence for the validity of the 
$z$-model. The average runtime for the $N=512$ simulations is only 
$T_{\mathrm{CPU}} \approx 20\mathrm{s}$; thus, the 
use of the $z$-model permits the inference of large-scale qualitative and quantitive information 
with minimal computational expense. 

\section{A mutliscale  model for interface evolution in compressible flow}\label{sec:multi-scale-algorithm}
We now derive a multiscale interface model, founded upon our incompressible-compressible decomposition of the Euler equations.   
The discontinuous incompressible velocity $w$ solving \eqref{w-eqn} (and, in particular, obtained from  \eqref{zmodel}) can be used to compute small scale structures on the
interface $\Gamma(t)$ and the Kelvin-Helmholtz instability (KHI).   It is often the case that vortex sheet roll-up, caused by the KHI,  occurs at spatial scales
which are smaller than the scales along which bulk vorticity is transported and for which sound waves propagate.    When this occurs, the continuous velocity $v$ solving \eqref{v-eqn},
which is only forced by bulk compression and vorticity,  may be computed at larger spatial scales than the velocity $w$.    

\subsection{A multiscale model for the compressible RTI}   In order to produce a fast-running model of the compressible RTI,
we begin by using the  higher-order $z$-model \eqref{zmodel} to generate the velocity $w$.   Our model is generated by coupling the equation for $v$ \eqref{v-eqn} together with
\begin{subequations}
\label{zmodelRT}
\begin{align}
\partial_t z(\alpha,t) &= \pvint_{\mathbb{T}_L} \mathcal{K}_{\mathbb{T}_L}(z(\alpha,t)-z(\beta,t)) \varpi(\beta,t) \,\mathrm{d}\beta  + v(z( \alpha ,t),t) \,, \\
\partial_t \varpi(\alpha,t) &= - \partial_{\alpha} \left[ \frac{A}{2|\partial_{\alpha} z (\alpha,t)|^2}  \mathcal{H} \left( \varpi(\alpha,t) \mathcal{H} \varpi(\alpha,t) \right) - 2Ag z_2(\alpha,t) \right]\,,
\end{align}
\end{subequations}
and computing $w$ using \eqref{BS-velocity-R}.   As we explain below, the compressible equations \eqref{v-eqn} will be solved on a coarse grid, while \eqref{zmodelRT} will be solved
on a fine, but one-dimensional, grid.

\subsection{A multiscale model for the compressible RMI}
\subsubsection{Vorticity production}
For flows in which shock waves collide with contact discontinuities and initiate the RMI, we shall
derive a modified form of the $z$-model which accounts for the vorticity production that is caused by the 
misalignment of the pressure gradient 
at the shock wave and the density gradient at the interface. 

Computing the curl of (\ref{ceuler}b), we obtain the two-dimensional 
vorticity equation for compressible flow as
\begin{equation}\label{vorticity-equation}
\frac{\mathrm{D} \omega}{\mathrm{D}t}   + \omega  \operatorname{div} u = -\frac{\nabla \rho \cdot \nabla^{\perp} p}{\rho^2} \,.
\end{equation}
The  term on the right-hand side of \eqref{vorticity-equation} is the \emph{baroclinic} term, responsible 
for vorticity production on the interface when a shock-wave collides with a vortex sheet. The  amplitude of vorticity 
$\varpi$ is the weak (or distributional) form of the vorticity, and consequently it is important to include a weak form of the baroclinic
term in the  dynamics of $\varpi$.  We thus introduce the following modification of the $z$-model
\begin{subequations}
\label{zmodelRM}
\begin{align}
\partial_t z(\alpha,t) &= \pvint_{\mathbb{T}_L} \mathcal{K}_{\mathbb{T}_L}(z(\alpha,t)-z(\beta,t)) \varpi(\beta,t) \,\mathrm{d}\beta  + v(z( \alpha ,t),t) \,, \\
\partial_t \varpi(\alpha,t) &= - \partial_{\alpha} \left[ \frac{A}{2|\partial_{\alpha} z (\alpha,t)|^2}  \mathcal{H} \left( \varpi(\alpha,t) \mathcal{H} \varpi(\alpha,t) \right) - 2Ag z_2(\alpha,t) \right] \nonumber  \\
& \qquad  \qquad - \llbracket \nabla p \cdot \partial_{\alpha} z(\alpha,t) / \rho \rrbracket + \mu \partial^2_{\alpha} \varpi(\alpha,t) \,, \label{zmodelRM-b}
\end{align}
\end{subequations}
which will be used for flows in which shocks collide with contact discontinuities. 

\textcolor{black}{ 
\begin{remark} \label{rem:TH}
Let us mention that for Richtmyer-Meshkov problems, at the time at which the  planar shock collides with the perturbed contact, the 
pressure is discontinuous along points in the intersection of the shock and contact.  As such the numerator in the baroclinic term,
$  \llbracket \nabla p \cdot \partial_{\alpha} z(\alpha,t) \rrbracket$ can be large at such points.   Such a pressure profile does not occur
in
the Rayleigh-Taylor problems, for which  the numerator $  \llbracket \nabla p \cdot \partial_{\alpha} z(\alpha,t) \rrbracket$ vanishes along
the contact.   Thus, the use of this baroclinic term in the $\varpi$-equation is imperative for the simulation of the RMI problems.
\end{remark} 
}

\subsubsection{Taylor's frozen turbulence hypothesis}
Taylor's ``frozen turbulence'' hypothesis  \cite{Taylor1938}, roughly speaking, 
states that if the mean flow velocity is much larger than the velocity of the turbulent eddies, then the 
advection of the turbulent flow past a fixed point can be taken to be due entirely to the mean flow, or in other words, that the turbulent fluctuations are
transported by the mean flow.  For shock-contact collisions that initiate the RMI, 
the velocity $v$ at the shock front is much larger in 
magnitude than the velocity $w$. That is to say, the relation $\max_x |v| \gg \max_x |w|$ holds true. 
\textcolor{black}{
For instance, for the RMI problem considered in \Cref{subsec:RM_TeTr2009}, the quantity 
$\max_x |w|$ is two orders of magnitude smaller than the quantity $\max_x |v|$. 
} 
For such flows,
we view the velocity $v$ as the mean velocity and $w$ as the fluctuation velocity, and impose the Taylor hypothesis that for very short time-intervals (i.e.,  about one time-step
in an explicit numerical simulation), 
$w$ is transported by $v$, so that
\begin{equation}\label{w-advection-by-v}
\partial_t w + (v \cdot \nabla) w =0 \,.
\end{equation}
\textcolor{black}{ As we described in Remark \ref{rem:TH}, at points along the interface at which the shock front intersects the contact
discontinuity, there exists a large increase in the baroclinic term, which in turn, produces a large increase in the amplitude of vorticity and this 
then leads to a localized increase in the small-scale velocity field $w$ via equation \eqref{zmodelRM-b} at each such intersection point at each
time-step.   As the shock passes through the contact discontinuity these points of intersection evolve, and this evolution causes large (and localized)
space gradients $ \nabla w $ and temporal gradients $\p_t w$.      The Taylor hypothesis \eqref{w-advection-by-v} ensures that a proper
balance is retained between the space and time gradient of $w$ in a numerical implementation which approximates these two different 
types of derivatives in a very different manner.  
}

Using \eqref{w-advection-by-v} together with
\eqref{v-eqn-a}, we find that
$$
\partial_t (\rho w) + \operatorname{div} (\rho w \otimes v) + \operatorname{div}  (\rho w \otimes w) = \rho (w \cdot \nabla)w \,, 
$$
and hence \eqref{v-eqn-b} must be replaced by
\begin{equation} 
\partial_t (\rho v) + \nabla \cdot (\rho v \otimes v) + \nabla p + \rho g e_2 =   - \nabla \cdot (\rho v \otimes w) - \rho (w \cdot \nabla) w\,.
\tag{\ref*{v-eqn}b'} \label{v-eqn-b'}
\end{equation} 

Therefore, our model for the RMI problem couples \eqref{zmodelRM} with \eqref{v-eqn}, but with  
\eqref{v-eqn-b'} replacing \eqref{v-eqn-b}. \textcolor{black}{Using a   Richtmyer-Meshkov  test problem, we explain in 
detail in \Cref{subsec:RM_TeTr2009} the reasons for  using  the Taylor hypothesis and 
\eqref{v-eqn-b'} in our multiscale algorithm for RMI flows.}

\subsection{The multiscale algorithms for the RTI and RMI problems}\label{subsec:algorithms}
We are now ready to give a precise description of the multiscale algorithm. Denote by 
${\bf V}(x,t) = (\rho,\rho v_1,\rho v_2,E)^{\mathcal{T}}$,  the solutions to the compressible equations 
\eqref{v-eqn}. We shall use a standard 3rd order Runge-Kutta procedure for time-integration; in the following 
algorithms, we use the superscript notation to denote the Runge-Kutta stage.

We shall use two slightly different algorithms; the first is for our multiscale model for RTI problems, while 
the second is for the multiscale model for RMI problems.

We note that the ordering of Steps \ref{RT-step4b}, \ref{RT-step4c}, and \ref{RT-step4d} in the RTI 
multiscale algorithm is important; our numerical 
experiments have shown that it is important that the velocity $w'(x)$ is computed \emph{before} the 
auxiliary interface $\tilde{z}$ is updated. 

For Richtmyer-Meshkov problems, we use a slightly different algorithm. In particular, we are 
no longer required to compute the time-derivative $\partial_t(\rho w)$ and so shall omit those steps from 
the algorithm. Notice that this means that we can omit in particular Step \ref{RT-step4b} of the RTI algorithm, 
which removes at each time-step one of the costly integral computations. 
On the other hand, we must compute in Step \ref{RM-step4d} of the RMI algorithm 
the baroclinic term in the $\varpi$-equation \eqref{zmodelRM}. 

\begin{remark}\label{remark:wbar-odd}
In this work, we apply our multiscale algorithm to 2-$D$ flows that 
are symmetric across the line $x_1 = 0$. In particular, this means that the velocities $w_1$ and $v_1$ are 
odd functions of $x_1$, and the vorticity amplitude $\varpi$ is an odd function of $\alpha$. 
\end{remark}

\begin{algorithm}[h]
\caption*{\textbf{\textsc{RTI Multiscale Algorithm}}}\label{RTI-multiscale-algorithm-alg}
\begin{enumerate}[itemsep=0.0em,leftmargin=2.0cm,label=\textbf{Step \arabic*}]\vspace{0em}\setcounter{enumi}{-1}
\item Suppose that we are given the solution $z(\alpha,t)$, $\varpi(\alpha,t)$, and $\mathbf{V}(x,t)$
 at time-step $t$, as well as a velocity ${w'}(x)$, 
 and we wish to compute the solution at $t+\delta t$. 
 Define $z^0(\alpha) \coloneqq z(\alpha,t)$, $\varpi^0(\alpha) \coloneqq \varpi(\alpha,t)$, and 
 $ \mathbf{V}^0(x) \coloneqq \mathbf{V}(x,t)$, and let $\rho'(x) \coloneqq \rho^0(x) = \rho(x,t)$.\label{RT-step0}
 
 \item \hfill \vspace{-0.5em}
 	\begin{enumerate}[itemsep=0.0em,leftmargin=0.5cm,label=\textbf{1(\alph*)}]\vspace{0em}
	\item Compute a velocity $\tilde{w}(x)$ from $z^0(\alpha)$ and $\varpi^0(\alpha)$  
		using the Biot-Savart law.
	\item Approximate $\partial_t (\rho \tilde{w})(x) = \left[ \rho^0(x) \tilde{w}(x) - \rho'(x) w'(x) \right]/\delta t$.  
	\item Solve the compressible equations \eqref{v-eqn} (with $w=\tilde{w}$) to obtain 
		$\mathbf{V}^1(x)$. 
	\item Compute an auxiliary $\tilde{z}(\alpha)$ and $\varpi(\alpha)$ by solving the system 
		\eqref{zmodelRT}. 
	\item Calculate the interfacial velocity 
		$v(\tilde{z}(\alpha)) = v^1(x) |_{\tilde{z}(\alpha)}$ using interpolation. 
	\item Update $z^1(\alpha)=\tilde{z}(\alpha)+\delta t \cdot v(\tilde{z}(\alpha))$, and set 
		$\varpi^1(\alpha) = \tilde{\varpi}(\alpha)$. 
	\end{enumerate} \label{RT-step1}
\item Repeat  \ref{RT-step1} but with quantities evaluated at the next Runge-Kutta stage.  \label{RT-step2}
\item Repeat  \ref{RT-step2} but with quantities evaluated at the next Runge-Kutta stage.  \label{RT-step3}
\item \hfill \vspace{-0.5em}
	\begin{enumerate}[itemsep=0.0em,leftmargin=0.5cm,label=\textbf{4(\alph*)}]\vspace{0em}
	\item Use the standard 3rd order Runge-Kutta formula to produce $\mathbf{V}(x,t+\delta t)$, and an 
		auxiliary interface $\tilde{z}(\alpha)$ and vorticity amplitude $\varpi(\alpha)$. 
	\item Compute a velocity $w'(x)$ from $\tilde{z}(\alpha)$ and $\tilde{\varpi}(\alpha)$ using the 
		Biot-Savart law. \label{RT-step4b}
	\item Calculate the interfacial velocity 
		$v(\tilde{z}(\alpha)) = v(x,t+\delta t) |_{\tilde{z}(\alpha)}$ using interpolation. \label{RT-step4c}
	\item Update $z(\alpha,t+\delta t)=\tilde{z}(\alpha)+\delta t \cdot v(\tilde{z}(\alpha))$, and set 
		$\varpi(\alpha,t+\delta t) = \tilde{\varpi}(\alpha)$, then return to \ref{RT-step0}.  \label{RT-step4d}
	\end{enumerate} \label{RT-step4} 
\end{enumerate}
\end{algorithm} 

\begin{algorithm}[H]
\caption*{\textbf{\textsc{RMI Multiscale Algorithm}}}\label{RMI-multiscale-algorithm}
\begin{enumerate}[itemsep=0.0em,leftmargin=2.0cm,label=\textbf{Step \arabic*}]\vspace{0em}\setcounter{enumi}{-1}
\item Suppose that we are given the solution $z(\alpha,t)$, $\varpi(\alpha,t)$, and $\mathbf{V}(x,t)$
 at time-step $t$, as well as the baroclinic term 
 $\phi(\alpha,t) = \llbracket \nabla p \cdot \partial_{\alpha} z / \rho \rrbracket$ and we wish to compute the solution at $t+\delta t$. 
 Define $z^0(\alpha) \coloneqq z(\alpha,t)$, $\varpi^0(\alpha) \coloneqq \varpi(\alpha,t)$, and 
 $ \mathbf{V}^0(x) \coloneqq \mathbf{V}(x,t)$. \label{RM-step0}
 
 \item \hfill \vspace{-0.5em}
 	\begin{enumerate}[itemsep=0.0em,leftmargin=0.5cm,label=\textbf{1(\alph*)}]\vspace{0em}
	\item Compute a velocity $\tilde{w}(x)$ from $z^0(\alpha)$ and $\varpi^0(\alpha)$  
		using the Biot-Savart law.
	\item Solve the modified compressible equations \eqref{v-eqn} (with $w=\tilde{w}$), using 
		\eqref{v-eqn-b'} in place of \eqref{v-eqn-b}, to obtain $\mathbf{V}^1(x)$. 
	\item Compute an auxiliary $\tilde{z}(\alpha)$ and $\varpi(\alpha)$ by solving the system 
		\eqref{zmodelRM}. 
	\item Calculate an auxiliary baroclinic term $\tilde{\phi}(\alpha)$ using $\tilde{z}(\alpha)$, $p^1(x)$, and $\rho^1(x)$. 
	\item Calculate the interfacial velocity 
		$v(\tilde{z}(\alpha)) = v^1(x) |_{\tilde{z}(\alpha)}$ using interpolation. \label{RM-step1e}
	\item Update $z^1(\alpha)=\tilde{z}(\alpha)+\delta t \cdot v(\tilde{z}(\alpha))$, and set 
		$\varpi^1(\alpha) = \tilde{\varpi}(\alpha)$. \label{RM-step1f}
	\end{enumerate} \label{RM-step1}
\item Repeat  \ref{RM-step1} but with quantities evaluated at the next Runge-Kutta stage.  \label{RM-step2}
\item Repeat  \ref{RM-step2} but with quantities evaluated at the next Runge-Kutta stage.  \label{RM-step3}
\item \hfill \vspace{-0.5em}
	\begin{enumerate}[itemsep=0.0em,leftmargin=0.5cm,label=\textbf{4(\alph*)}]\vspace{0em}
	\item Use the standard 3rd order Runge-Kutta formula to produce $\mathbf{V}(x,t+\delta t)$, and an 
		auxiliary interface $\tilde{z}(\alpha)$ and vorticity amplitude $\varpi(\alpha)$. 
	\item Calculate the interfacial velocity 
		$v(\tilde{z}(\alpha)) = v(x,t+\delta t) |_{\tilde{z}(\alpha)}$ using interpolation.  \label{RM-step4b}
	\item Update $z(\alpha,t+\delta t)=\tilde{z}(\alpha)+\delta t \cdot v(\tilde{z}(\alpha))$, and set 
		$\varpi(\alpha,t+\delta t) = \tilde{\varpi}(\alpha)$.  \label{RM-step4c}
	\item Compute the baroclinic term $\phi(\alpha,t+\delta t)$ using $z(\alpha,t+\delta t)$, $p(x,t+\delta t)$, 			and $\rho(x,t+\delta t)$, then return to \ref{RM-step0}.  \label{RM-step4d}
	\end{enumerate} \label{RM-step4} 
\end{enumerate}
\end{algorithm}

\subsection{Numerical implementation of the multiscale algorithm}

We now describe how we numerically implement the multiscale algorithm described in \Cref{sec:multi-scale-algorithm}. 
Suppose that the conservative variables ${\bf V}$ are computed in the bounded domain 
$\Omega = \mathbb{T}_L \times [x^1_2\,,x^n_2]$, and that the flow is periodic in the horizontal variable $x_1$. 
Suppose also that a single wavelength of the periodic interface $\Gamma(t)$ is parametrized by the function 
$z(\alpha,t)$. 

Discretize the domain $\Omega$ with $(2m-1) \times n$ cells with cell centers at
\begin{align*}
x^i_1 &= -L/2 + (i-1) \delta x_1\,, \\
x^j_2 &= x^1_2 + (j-1) \delta x_2\,,
\end{align*}
with $\delta x_1 = L/(2m-2)$ and $\delta x_2 = (x^n_2-x^1_2)/(n-1)$, and suppose that the parameter 
$\alpha$ is discretized with $N=2^r+1$ nodes,  
$$
\alpha_k = -L/2 + (k-1) {\scriptstyle\Delta} \alpha \,,
$$
with ${\scriptstyle\Delta} \alpha = L/(N-1)$. 
We spatially discretize the equations of motion, then use a standard third-order explicit Runge-Kutta
solver for time integration. 
We shall use a space-time smooth artificial viscosity method, which we call the $C$-method
 \cite{RaReSh2019b}, for the compressible
$v$-equations \eqref{v-eqn} to stabilize shock fronts and contact discontinuities, and thereby 
prevent the onset of Gibbs oscillations. 

It remains to describe the following: 
first, the numerical implementation of the space-time smooth artificial viscosity $C$-method; 
second, the computation of the velocity $w$ on the plane; third, the bilinear 
interpolation scheme; and finally, the calculation of the weak baroclinic term 
$\llbracket \nabla p \cdot \partial_{\alpha} z / \rho \rrbracket$. 

\subsubsection{Numerical implementation of the $C$-method} \label{C-method-numerical-implementation}
We implement a simple finite difference WENO-based scheme to spatially discretize the
system \eqref{v-eqn}. Our simplified WENO scheme is 
devoid of any exact or approximate Riemann solvers, and instead relies on the sign of the velocity to perform 
upwinding. A fourth-order central difference approximation is used to compute the pressure gradient $\nabla p$, 
while a second-order central difference approximation is employed to compute the diffusion terms 
\eqref{artificial-visc-Euler-C-Chat}. For brevity, we omit further details of the numerical 
implementation of the $C$-method and refer the reader to Appendix \ref{sec-appendices-a} and 
 \cite{RaReSh2019b} for further details.

\subsubsection{Computing the $w$ velocity}\label{multiscale-w-computation}
We first describe our method for calculating the discrete velocity $w_{i,j} = w(x_1^i,x_2^j)$ from a given
discretized interface parametrization $z_k$ and vorticity amplitude $\varpi_k$.
 
Since integral-kernel calculations can be computationally very expensive, we begin by proposing a 
simplification to speed up such calculations. 
Suppose first that we wish to compute the velocity $w$ at a point $x_{i,j}$ such that 
$|x_2^j - z_2^k| \gg 1$ for every $k$. Then the following approximations are valid:
\begin{gather*}
-\frac{\sinh(2 \pi (x^j_2 - z_2^k)/L)}{\cosh(2 \pi (x^j_2 - z_2^k)/L) - \cos(2 \pi (x^i_1 - z_1^k)/L)} \approx \pm 1 \,, \\ 
\frac{\sin(2 \pi (x^i_1 - z_1^k)/L)}{\cosh(2 \pi (x^j_2 - z_2^k)/L) - \cos(2 \pi (x^i_1 - z_1^k)/L)} \approx 0 \,.
\end{gather*}
Then, using the fact that $\varpi$ is an odd function (c.f. \Cref{remark:wbar-odd}),
it follows that $w_{i,j} \approx 0$. Consequently, it is
sufficient to compute $w$ only for those $x_{i,j}$ that lie in the horizontal strip 
\begin{equation*}\label{strip}
\Omega_z = \left\{x \in \mathbb{T}_L \times [x_2^1,x_2^n] \text{ s.t. } \min_k z^k_2 - \lambda \leq x_2 \leq \max_k z^k_2 + \lambda \right\}\,.
\end{equation*}
For the simulations considered here, we set $\lambda=0.075$, but note that this parameter is
problem dependent. Computing the velocity only in the strip 
$\Omega_z$ speeds up an otherwise time-consuming calculation.

We now define the scalar function $\mathcal{J}^{\delta}(x)$ by 
\begin{equation*}
\mathcal{J}^{\delta}(x) = \frac{1}{4 \pi} \log \left\{\delta^2 + \cosh(2 \pi x_2/L) - \cos(2 \pi x_1 /L) \right\}\,.
\end{equation*}
The function $\mathcal{J}^{\delta}$ is a smoothed version of the singular kernel used in the integral 
representation of $\Delta^{-1}$, so that 
$\mathcal{K}^{\delta}_{\mathbb{T}_L}(x) = \nabla^{\perp} \mathcal{J}^{\delta}(x)$.

The velocity $w$ is determined by first calculating the \emph{stream function} $\psi(x,t)$ using the formula
\begin{equation*}
\psi(x,t) = \int_{\mathbb{T}_L} \mathcal{J}^{\delta}(x-z(\beta,t))\varpi(\beta,t)\,\mathrm{d}\beta\,,
\end{equation*} 
then using the 
relation $w = \nabla^{\perp} \psi$. Since
integral-kernel calculations are computationally expensive, it is advantageous to perform a single 
such computation then take derivatives, rather than perform two such computations. Moreover, the 
function $\mathcal{J}^{\delta}(x)$ is (roughly speaking) one derivative smoother than the kernel 
$\mathcal{K}^{\delta}_{\mathbb{T}_L}(x)$, so that we may hope that the integral calculation is numerically 
more stable. Using the stream-function formulation also guarantees that the velocity field $w$ produced is 
divergence free, whereas a direct singular integral calculation of $w$ can produce inaccuracies so that the 
resultant velocity field does not satisfy $\nabla \cdot w = 0$. 

For $x_{i,j} \in \Omega_z$, we can compute the stream function $\psi_{i,j}$ using the same trapezoidal 
method as described in \Cref{z-model-numerical-implementation}. We define $\psi_{i,j} = 0$ for 
$x_{i,j} \notin \Omega_z$. We then use a standard second-order central difference approximation to 
determine the velocity $w_{i,j}$ from $\psi_{i,j}$. 

\subsubsection{Bilinear interpolation scheme}\label{multiscale-bilinear}
We shall employ a simple bilinear interpolation scheme as follows. Let $z_k$ be the discretized interface 
parametrization, and suppose that we are given a scalar function $f_{i,j}$ defined at the cell centers 
$x_{i,j} \in \Omega$. We wish to determine an approximation $f^k$ to the value of $f_{i,j}$ at 
the points $\alpha_k$. For fixed $k$, we determine for which $i$ and $j$ the point $z_k$ lies in the rectangle 
$[x_1^i,x_1^{i+1}] \times [x_2^j,x_2^{j+1}]$ by requiring that $(z^1_k-x_1^i)(z^1_k-x_1^{i+1}) \leq 0$ and 
$(z^2_k-x_2^j)(z^2_k-x_2^{j+1}) \leq 0$. The interpolated quantity $f_k$ is then defined as 
\begin{multline*}
f^k = f_{i,j} \left(1 - \frac{(z_1^k -x_1^i)}{\delta x_1} \right) \left(1 - \frac{(z_2^k -x_2^j)}{\delta x_2} \right) \\
+ f_{i+1,j} \left(1 - \frac{(x_1^{i+1}-z_1^k)}{\delta x_1} \right) \left(1 - \frac{(z_2^k -x_2^j)}{\delta x_2} \right) \\ 
+ f_{i,j+1} \left(1 - \frac{(z_1^k -x_1^i)}{\delta x_1} \right) \left(1 - \frac{(x_2^{j+1}-z_2^k)}{\delta x_2} \right) \\
+ f_{i+1,j+1} \left(1 - \frac{(x_1^{i+1}-z_1^k)}{\delta x_1} \right) \left(1 - \frac{(x_2^{j+1}-z_2^k)}{\delta x_2} \right)\,.
\end{multline*}

\subsubsection{Calculation of the weak baroclinic term}
Next, we discuss the calculation of the weak baroclinic term 
$\llbracket \nabla p \cdot \partial_{\alpha} z /\rho \rrbracket$ on the interface $\Gamma(t)$. Suppose that 
we are given the pressure $p_{i,j}$ and density $\rho_{i,j}$ defined on the plane, and the discretized interface 
parametrization $z_k$. 

We begin by computing the unit normal $n$ to the interface as 
$$
n = \frac{\partial_{\alpha} z^{\perp}}{|\partial_{\alpha} z^{\perp}|} = \frac{(-\partial_{ \alpha} z_2 \,,  \partial_{ \alpha} z_1 )}{ | ( -\partial_{ \alpha} z_2 \,,\partial_{ \alpha} z_1 )|  }\,.
$$
The jump across the interface in a quantity $f$ defined at grid points is approximated as
\begin{equation}\label{jump-approx}
\llbracket f \rrbracket \approx - \delta n \cdot \nabla f \approx - \frac{ |\delta x|}{2} \, n \cdot \nabla f |_{z(\alpha,t)}\,,
\end{equation}
where $\nabla f |_{z(\alpha,t)}$ denotes the evaluation of the quantity $\nabla f(x_1,x_2)$ at the 
interface parametrization 
$z(\alpha,t)$; this is accomplished using the bilinear interpolation scheme described  above. 
All derivatives are approximated using second-order accurate central difference approximations. 

Thus, to evaluate the baroclinic term $\llbracket \nabla p \cdot \partial_{\alpha} z /\rho \rrbracket$, we compute 
$\nabla (\partial_{i} p /\rho)$ for $i=1,2$ on the fixed Eulerian grid, 
interpolate onto the interface, and use formula 
\eqref{jump-approx}. More explicitly, 
\begin{equation}\label{baroclinic-term-calculation}
\llbracket \nabla p / \rho \rrbracket \cdot \partial_{\alpha} z \approx  - \frac{|\delta x |}{2 | \partial_{\alpha} z |} \partial_{\alpha} z^{\perp} \mathbf{M} \partial_{\alpha} z \,, 
\end{equation}
where $\mathbf{M}$ is the $2 \times 2$ matrix defined as
$$
\mathbf{M} = \left. \begin{bmatrix}
    			\partial_1(\partial_1 p/\rho) & \partial_1(\partial_2 p/\rho) \\
    			\partial_2(\partial_1 p/\rho) & \partial_2(\partial_2 p/\rho)
 		     \end{bmatrix} \,\,  \right\vert_{z(\alpha,t)} \,,
$$
and $|_{z(\alpha,t)}$ once again denotes evaluation at the interface parametrization (using bilinear 
interpolation). 

\section{Numerical simulations of the RTI and RMI using the multiscale model}\label{sec:multi-scale_numerical_simulations}

We next present results from four numerical experiments to demonstrate the efficacy of our 
multiscale model and its numerical implementation. 
The objective of this section is (1) to show
that the multiscale model  produces solutions with accurate interface motion  that correctly captures the KH structures during RTI and RMI, and (2)
to demonstrate that the multiscale algorithm is roughly two orders of magnitude times 
faster to run that a standard gas dynamics simulation.

\subsection{The compressible RTI test of Almgren et al.}\label{subsec:RT_CASTRO}
We first consider the compressible single-mode RTI test from the paper of \citet{AlmgrenEtAl2010}. The domain 
is $(x_1,x_2) \in [-0.25,0.25] \times [0,1]$ and the gravitational constant is $g=1$. 
Periodic conditions are applied in the horizontal $x_1$ direction and free-flow conditions are 
imposed at the boundaries in the $x_2$ direction. 
In particular, the pressure is extended linearly to satisfy the hydrostatic assumption at the 
top and bottom boundaries. 
The initial data is given as follows:
the initial velocity is identically zero $u_0 = 0$, and the pressure $p_0$ is defined as 
\begin{equation}\label{initial-pressure_CASTRO-hydrostatic}
 p_0 = \begin{cases}
 	   5 - \rho^{-}gx_2 \, &, \text{ if } x_2 < 0.5 \\
 	   5 - 0.5\rho^{-}g - \rho^{+}g(x_2-0.5) &, \text{ if } x_2 \geq 0.5 \\	
 	   \end{cases} \,,
\end{equation}
where $\rho^{+}=2$ and $\rho^{-}=1$. The initial density $\rho_0$ is defined as 
\begin{equation}\label{RT_CASTRO_initial-rho}
\rho_0(x_1,x_2) = \rho^{-} + \frac{\rho^{+}-\rho^{-}}{2} \left[  1+ \tanh \left(\frac{x_2 - \eta_0(x_1)}{h} \right) \right]\,,
\end{equation}
with $\eta_0(x_1) = 0.5-0.01 \cos (4 \pi x_1)$. The $\tanh$ profile introduces a small length scale $h$ over which 
the initial density is smeared.

We begin by computing a {\it benchmark} or {\it high-resolution reference} solution using the anisotropic 
$C$-method\footnote{This is a spacetime smooth artificial viscosity method employed with
a highly simplified WENO discretization of the compressible Euler equations 
(see \cite{ReSeSh2012,RaReSh2019a,RaReSh2019b}).} on a fine mesh consisting of $128 \times 512$ cells,
a CFL number of $\mathrm{CFL} \approx 0.4$, and $h=0.005$ as in \cite{AlmgrenEtAl2010}. 

\begin{figure}[h]
\centering
\includegraphics[width=30mm]{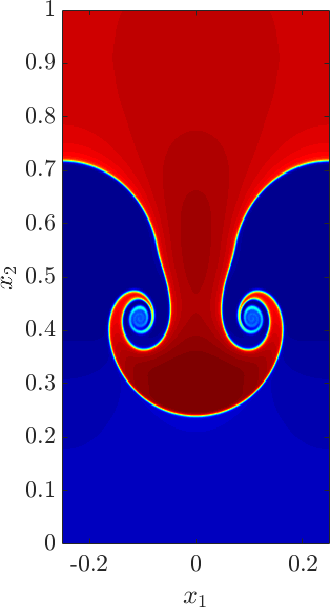}
\caption{The density profile at time $t=2.5$ for the RTI test of \citet{AlmgrenEtAl2010}. The solution is 
computed using the anisotropic $C$-method on a mesh with $128 \times 512$ cells.}
\label{fig:RT_CASTRO_rho_highres}
\end{figure}

 The computed density is shown in 
\Cref{fig:RT_CASTRO_rho_highres} at the final time $t=2.5$. 
In \cite{AlmgrenEtAl2010}, the authors compare solutions computed 
using the piecewise-parabolic method (PPM) and the piecewise-linear method. It is shown that the (dimensionally)
operator-split versions of the  methods result in spurious secondary instabilities, whereas the unsplit 
versions of the methods suppress these instabilities while keeping a sharp interface and well-defined roll-up
regions (see Figure 9 in \cite{AlmgrenEtAl2010}). 
While the $C$-method is dimensionally split, the use of the anisotropic artificial viscosity 
\cite{RaReSh2019b}  prevents  the onset of the spurious secondary instabilities,  while ensuring that the KHI roll-up 
region and mixing zones are not smeared\footnote{We note that both the CPU time and memory usage 
are approximately a factor of 2 larger for unsplit methods than for split methods \cite{AlmgrenEtAl2010}. 
The $C$-method uses a highly simplified dimensionally-split WENO-type 
scheme and  is thus  relatively fast.}.


\subsubsection{The multiscale algorithm applied to the RTI}
We next employ the RTI multiscale algorithm described in \Cref{RTI-multiscale-algorithm-alg} with the following 
parameter values: the underlying coarse mesh used to solve for $v$ contains $8 \times 32$ cells, while the interface is discretized with 
a fine mesh consisting of
$N=128$ nodes;  the time-step is $\delta t = 2.5 \times 10^{-3}$, giving $\mathrm{CFL} \approx 0.4$.    
The Atwood number is $A = 1/3$, the initial velocity is $v_0=0$, the initial pressure is given by 
\eqref{initial-pressure_CASTRO-hydrostatic}, the initial density is given by \eqref{RT_CASTRO_initial-rho} with 
$h=0.02$, and
\begin{align*}\label{RT_CASTRO_initialdata}
 z_1(\alpha,0) &= \alpha  \,, \\
z_2(\alpha,0) &= 0.5 -  0.01\cos(4 \pi \alpha) \,, \\
\varpi(\alpha,0) &= 0\,. 
\end{align*}
The artificial viscosity parameters are chosen as $\beta=50$, $\tilde{\delta}=1.0$, and 
$\mu=1.5 \times 10^{-3}$.

\begin{figure}[h]
\centering
\subfigure[Multiscale $z$]{\label{fig:RT_CASTRO_z_8x32_128}\includegraphics[width=30mm]{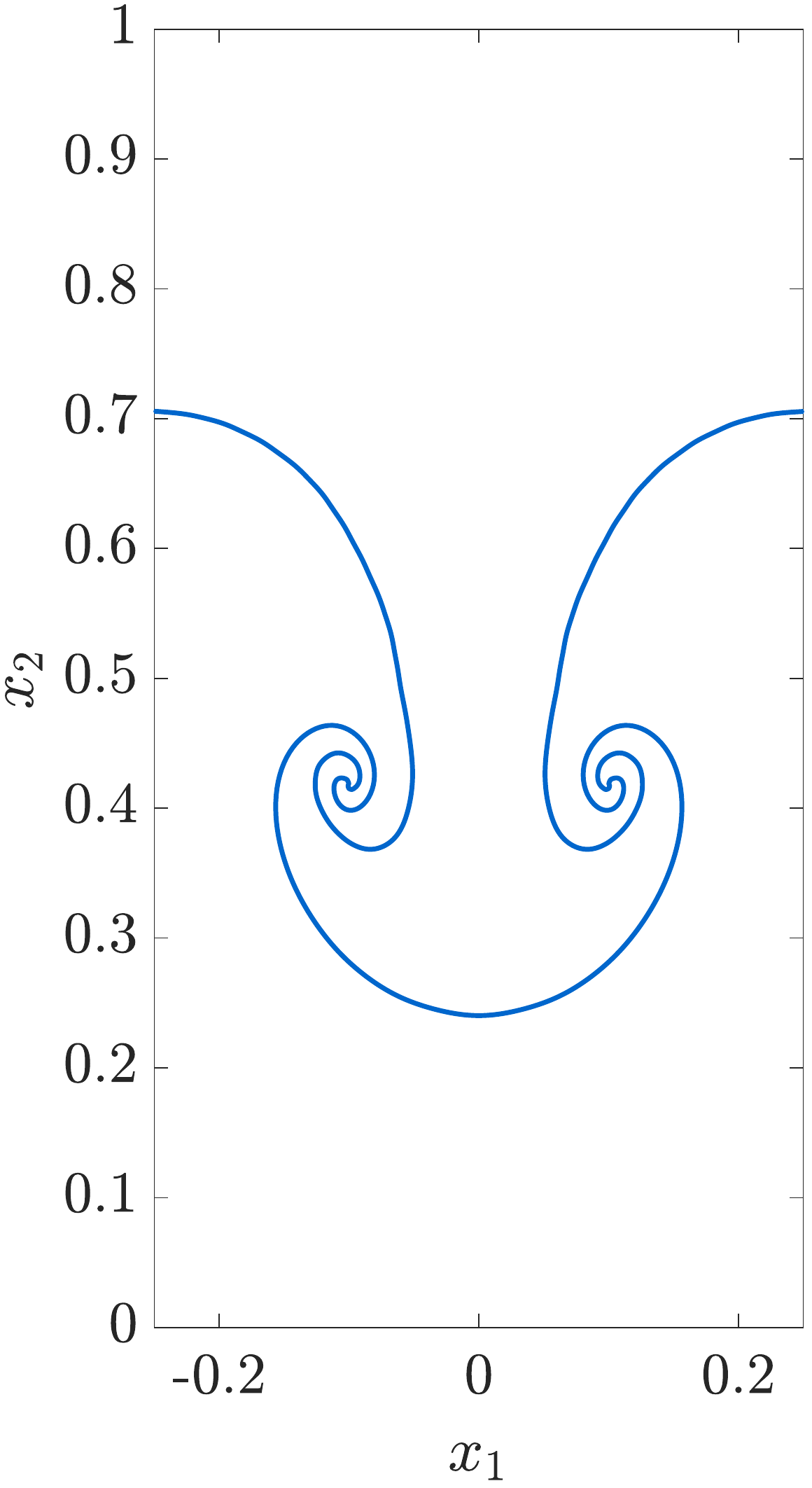}}
\hspace{2em}
\subfigure[High-res.]{\label{fig:RT_CASTRO_z_highres}\includegraphics[width=24.69mm]{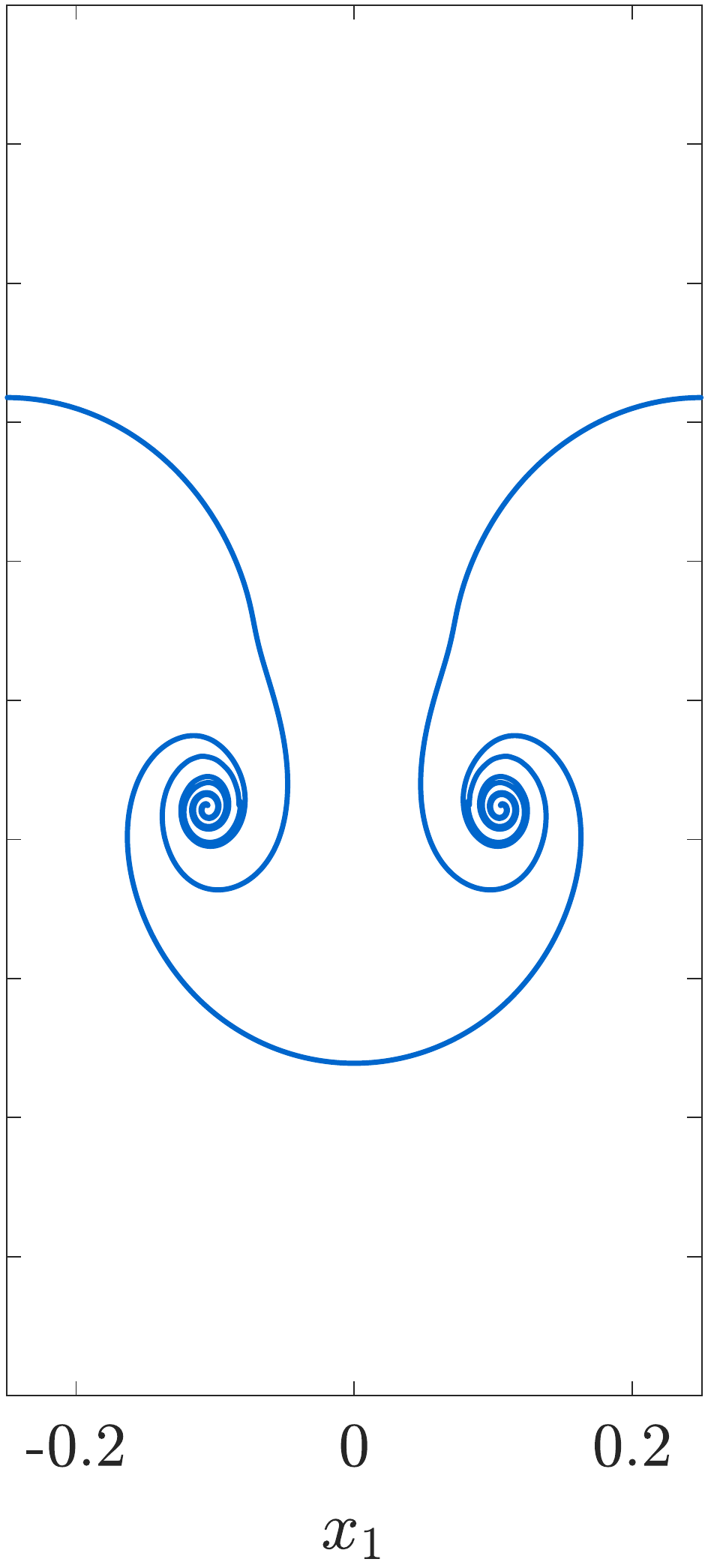}}
\hspace{2em}
\subfigure[Comparison]{\label{fig:RT_CASTRO_z_compare_8x32_128}\includegraphics[width=24.69mm]{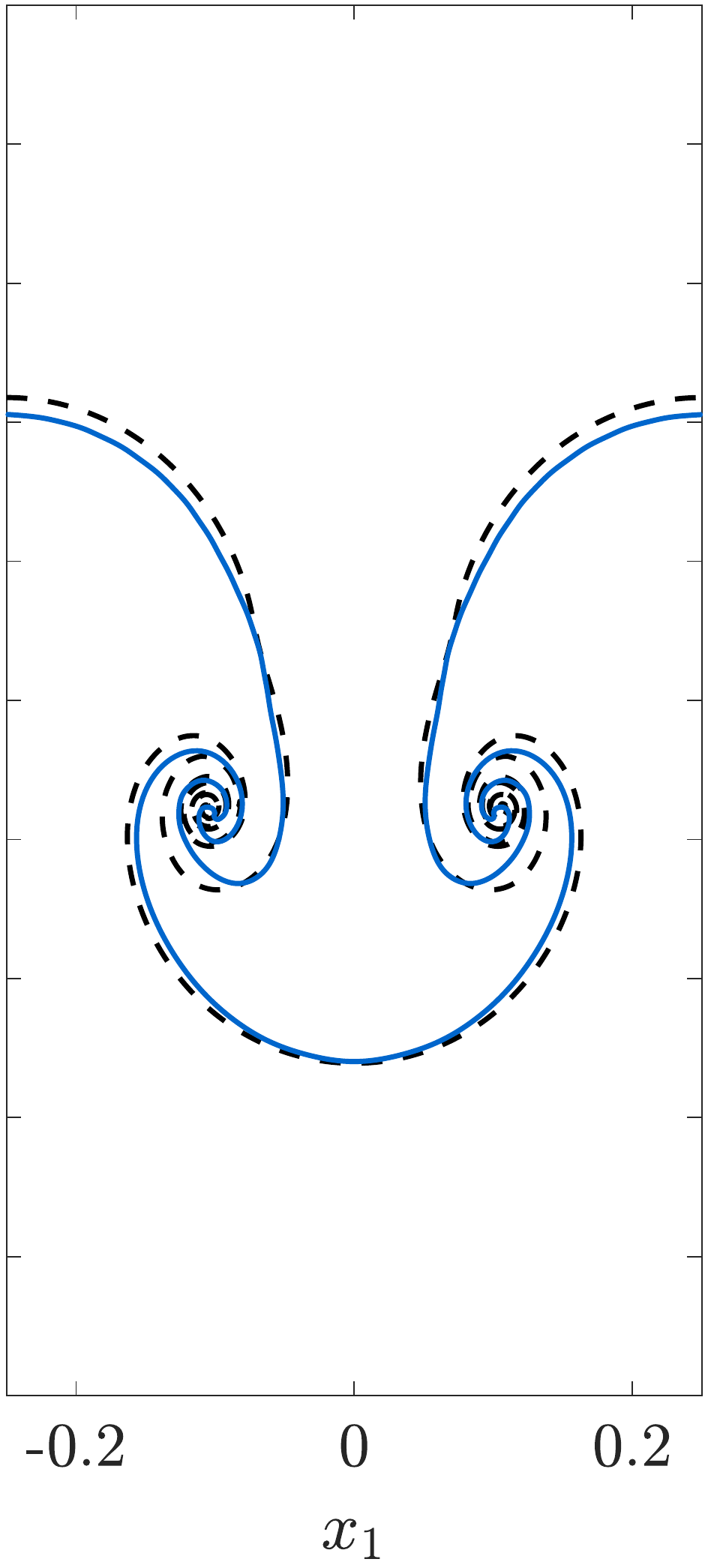}}
\caption{Results for the multiscale algorithm applied to the compressible single-mode RTI test 
of \citet{AlmgrenEtAl2010}, with the 
underlying grid containing $8 \times 32$ cells and the interface discretized with $N=128$. Shown are (a)  
computed interface parametrization $z$, (b) benchmark interface position, and (c) 
the computed interface $z$ (blue) overlaying the benchmark solution (dashed black)
at the final time $t=2.5$.}
\label{fig:RT_CASTRO_8x32_128}
\end{figure}

We provide plots  of the resulting solutions using our multiscale algorithm. The  interface position $z$, at the final time $t=2.5$, is shown in
\Cref{fig:RT_CASTRO_z_8x32_128}, and the high-resolution reference solution is shown in 
\Cref{fig:RT_CASTRO_z_highres}.   In \Cref{fig:RT_CASTRO_z_compare_8x32_128}, we compare the interface positions computed using
the multiscale algorithm and the high-resolution run;  we see that the two solutions are in excellent agreement, with nearly 
identical spike tip and bubble tip positions; moreover, the multiscale algorithm successfully simulates 
the roll up of the vortex sheet.
The reference solution computation had a runtime of $T_{\mathrm{CPU}} \approx 5015$ s, whereas the 
multiscale algorithm runtime was only $T_{\mathrm{CPU}} \approx 7$ s, giving a speed-up of 
approximately 683 times. 

\begin{figure}[h]
\centering
\subfigure[Multiscale $z$]{\label{fig:RT_CASTRO_z_16x64_256}\includegraphics[width=30.mm]{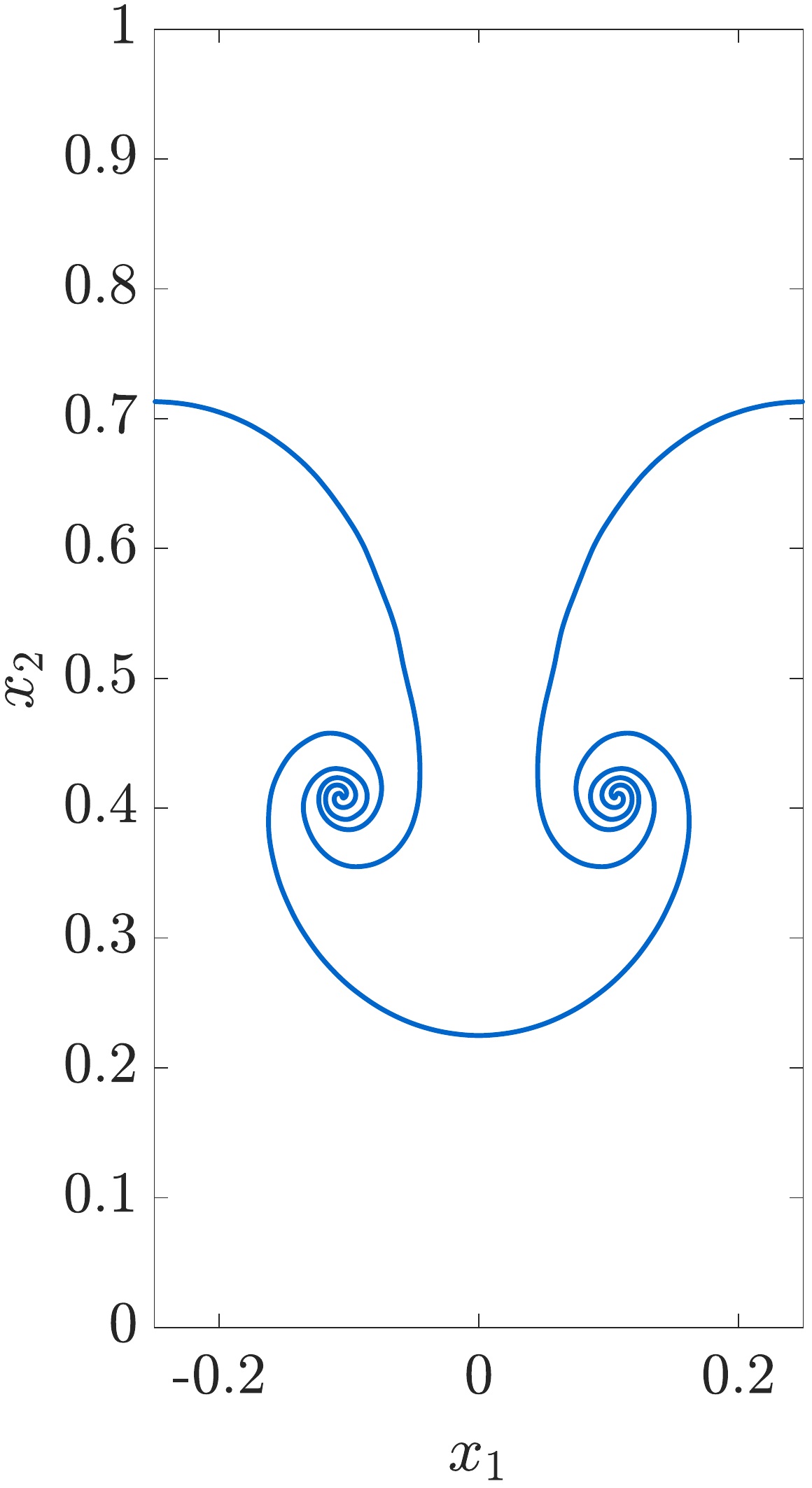}}
\hspace{2em}
\subfigure[High-res.]{\label{fig:RT_CASTRO_z_highres_2}\includegraphics[width=24.69mm]{RT_CASTRO_z_highres.pdf}}
\hspace{2em}
\subfigure[Comparison]{\label{fig:RT_CASTRO_z_compare_16x64_256}\includegraphics[width=24.69mm]{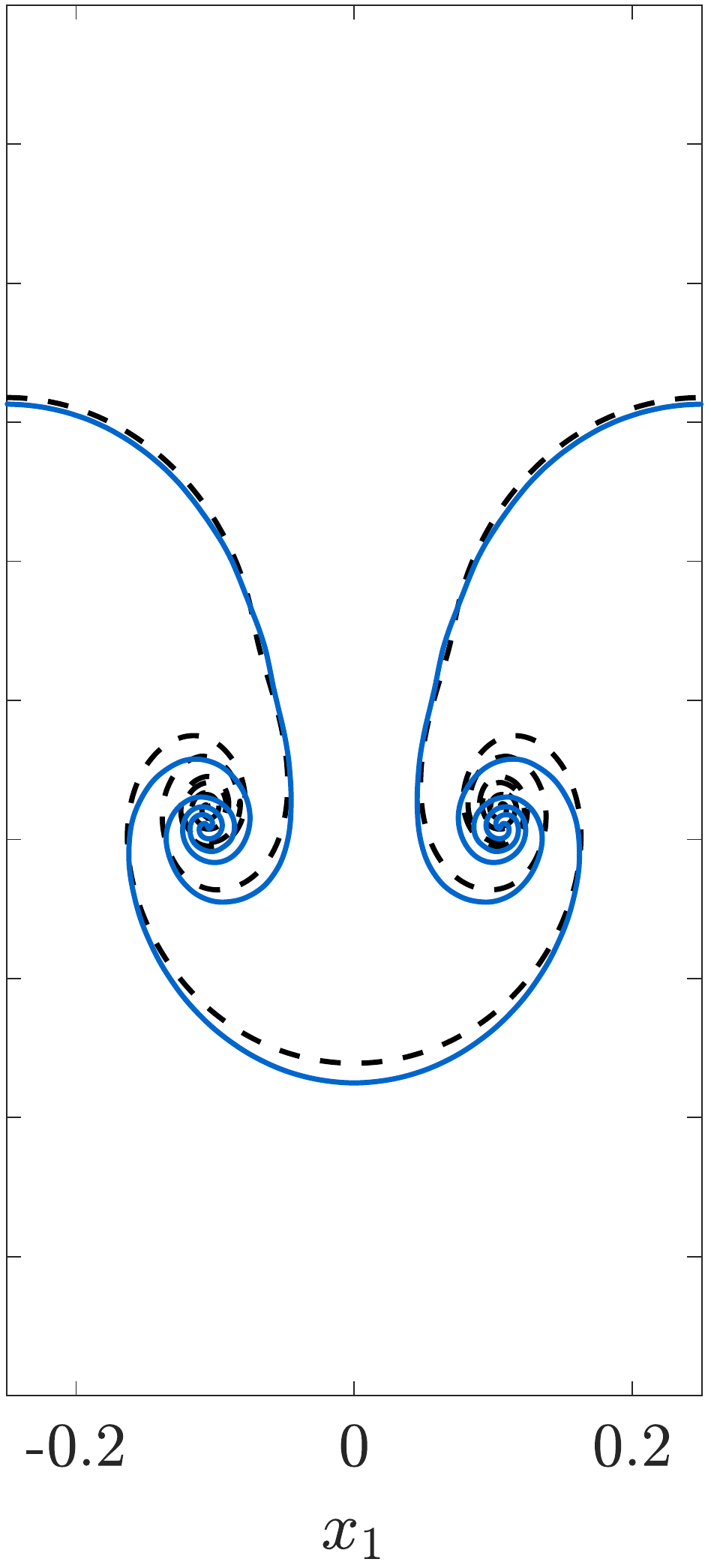}}
\caption{Results for the multiscale algorithm applied to the compressible single-mode RTI test 
of \citet{AlmgrenEtAl2010}, with the 
underlying grid containing $16 \times 64$ cells and the interface discretized with $N=256$. Shown are (a)  
computed interface parametrization $z$, (b) benchmark interface position, and (c) 
the computed interface $z$ (blue) overlaying the benchmark solution (dashed black)
at the final time $t=2.5$.}
\label{fig:RT_CASTRO_16x64_256}
\end{figure}

Increasing the resolution of both the coarse grid for $v$ and the fine grid for $w$   results in a solution 
with even more roll-up and accuracy.  We show in \Cref{fig:RT_CASTRO_z_16x64_256} the multiscale solution computed
using  $16 \times 64$ cells for the coarse mesh and
$N=256$ nodes for the fine mesh.   The time-step is $\delta t \approx 8.333 \times 10^{-4}$, giving 
$\mathrm{CFL} \approx 0.27$, and the artificial viscosity parameters are unchanged from the previous run. 
We observe that the computed interface shows more roll up, and is in even better agreement with the 
reference solution. The runtime of this simulation was $T_{\mathrm{CPU}} \approx 105$ s, resulting in 
 a speed-up factor of approximately 48.  
 
\subsubsection{Comparison of the low resolution density with the multiscale density}
The multiscale algorithm solves for the velocity $v$ on a coarse mesh, using fine-scale information from the velocity $w$.
In order to show how small-scale information improves the coarse-mesh simulation, we shall compare the solution
obtained with our multiscale algorithm with the solution obtained by  solving the Euler equations (using the same gas dynamics code as used for the
high-resolution reference solution)  on the same coarse meshes used to solve for $v$ in the multiscale algorithm.    We shall refer to this gas dynamics
simulation as the {\it low resolution solution}.

\begin{figure}[h]
\centering
\subfigure[Multiscale $\rho$]{\label{fig:RT_CASTRO_rho_16x64_256_2}\includegraphics[width=28mm]{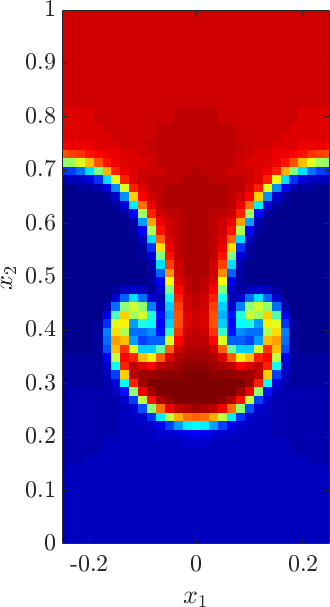}}
\hspace{2em}
\subfigure[Low-res. $\rho$]{\label{fig:RT_CASTRO_rho_stand-alone0}\includegraphics[width=22.74mm]{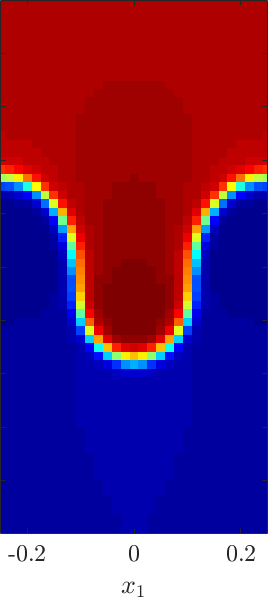}}
\caption{Comparison of the low resolution density with the multiscale density for the RTI test of 
\citet{AlmgrenEtAl2010}. \Cref{fig:RT_CASTRO_rho_16x64_256_2} is a plot of the density computed using the 
multiscale algorithm, and \Cref{fig:RT_CASTRO_rho_stand-alone0} is a plot of the low resolution Euler density,  
computed using the $C$-method on a coarse grid. Both solutions are computed using identical parameters.}
\label{fig:RT_CASTRO_stand-alone}
\end{figure}

To be more precise, we obtain the  low resolution density profile (using the $C$-method) 
on a grid with $16 \times 64$ cells and with a time-step of 
$\delta t \approx 8.333 \times 10^{-4}$. In \Cref{fig:RT_CASTRO_stand-alone},  
this solution is compared against the multiscale density function obtained from the simulation 
shown in \Cref{fig:RT_CASTRO_16x64_256}.   
 It is clear  that the  low resolution density profile does not have any of the basic KHI structure of the actual solution, and
 is very
different from the multiscale density which (although solved for on the same coarse mesh) shows the KH roll-up structure.
The multiscale algorithm
allows for the recovery of small-scale information on the coarse grid via the computation of the velocity 
$w$ using a fine mesh for the  interface. This small-scale information subsequently results in more structure in the 
roll-up region, and, therefore, a solution that is qualitatively more similar to high resolution simulations. 

\subsubsection{ \textcolor{black}{Comparison with other schemes and convergence studies}}

\textcolor{black}{
We next compare our multiscale algorithm with a modified version of the multiscale algorithm, in which 
the incompressible, irrotational velocity $w$ is computed using the kernels \eqref{kernel-baker}, rather than 
the Krasny kernel \eqref{smoothed-kernel}. As in \Cref{z-model-numerical-study}, we are interested in 
the convergence properties of 
(1)  the bubble and spike tip locations,  and (2) the 
radius $r_\delta$ and location $\sigma_\delta$ of the spiral roll up region. 
We choose the axis $x_2 = x_2^* = 0.4125$ to compute the intersection points $x^*$. 
}

\begin{figure}[h]
\centering
\subfigure[Bubble (blue) and spike (red) tip locations.]{\label{fig:RT_CASTRO_convergence1-comparison_amplitude}\includegraphics[width=50mm]{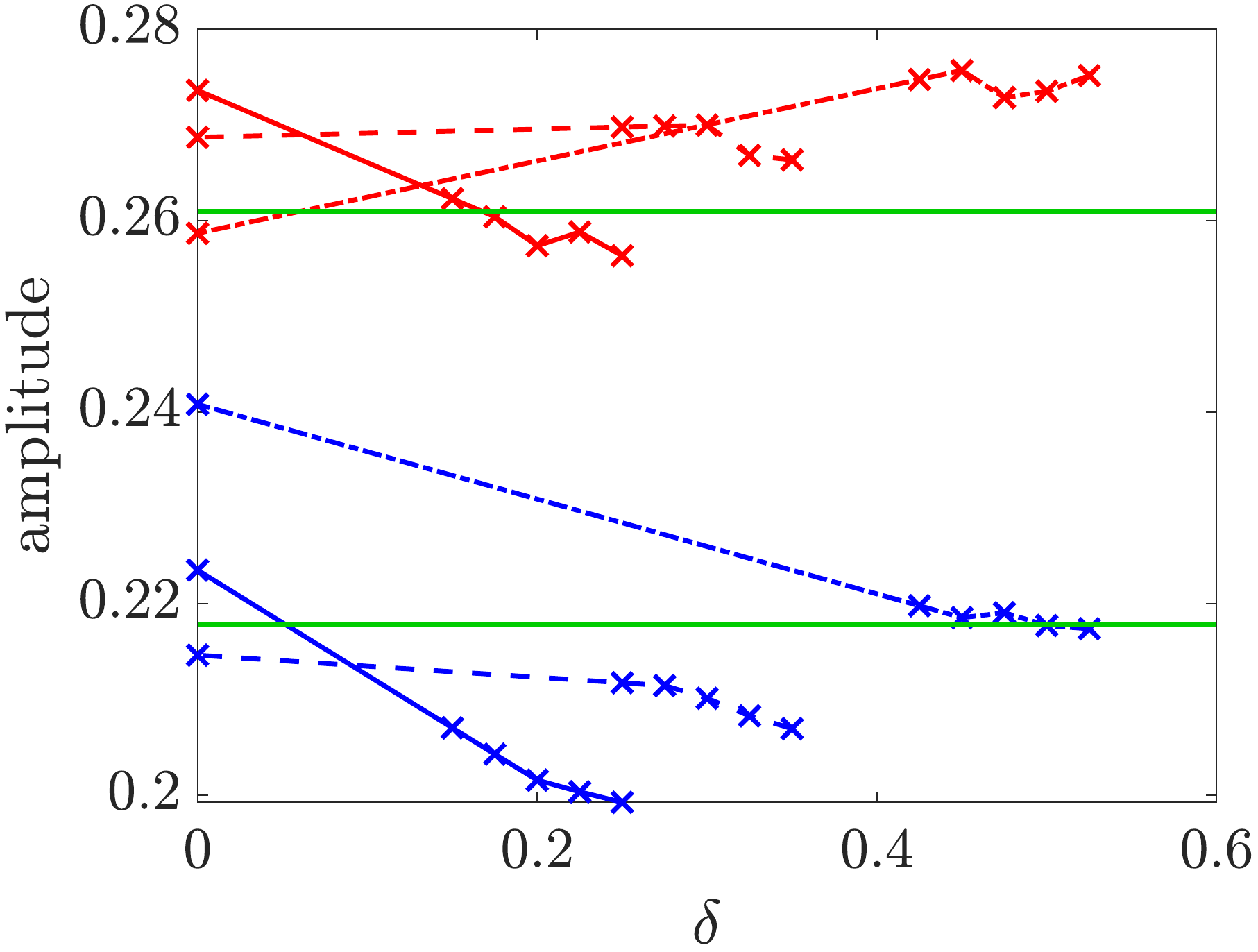}}
\hspace{0em}
\subfigure[Location of spiral center $\sigma_{\delta}$.]{\label{fig:RT_CASTRO_convergence1-comparison_sigma}\includegraphics[width=50mm]{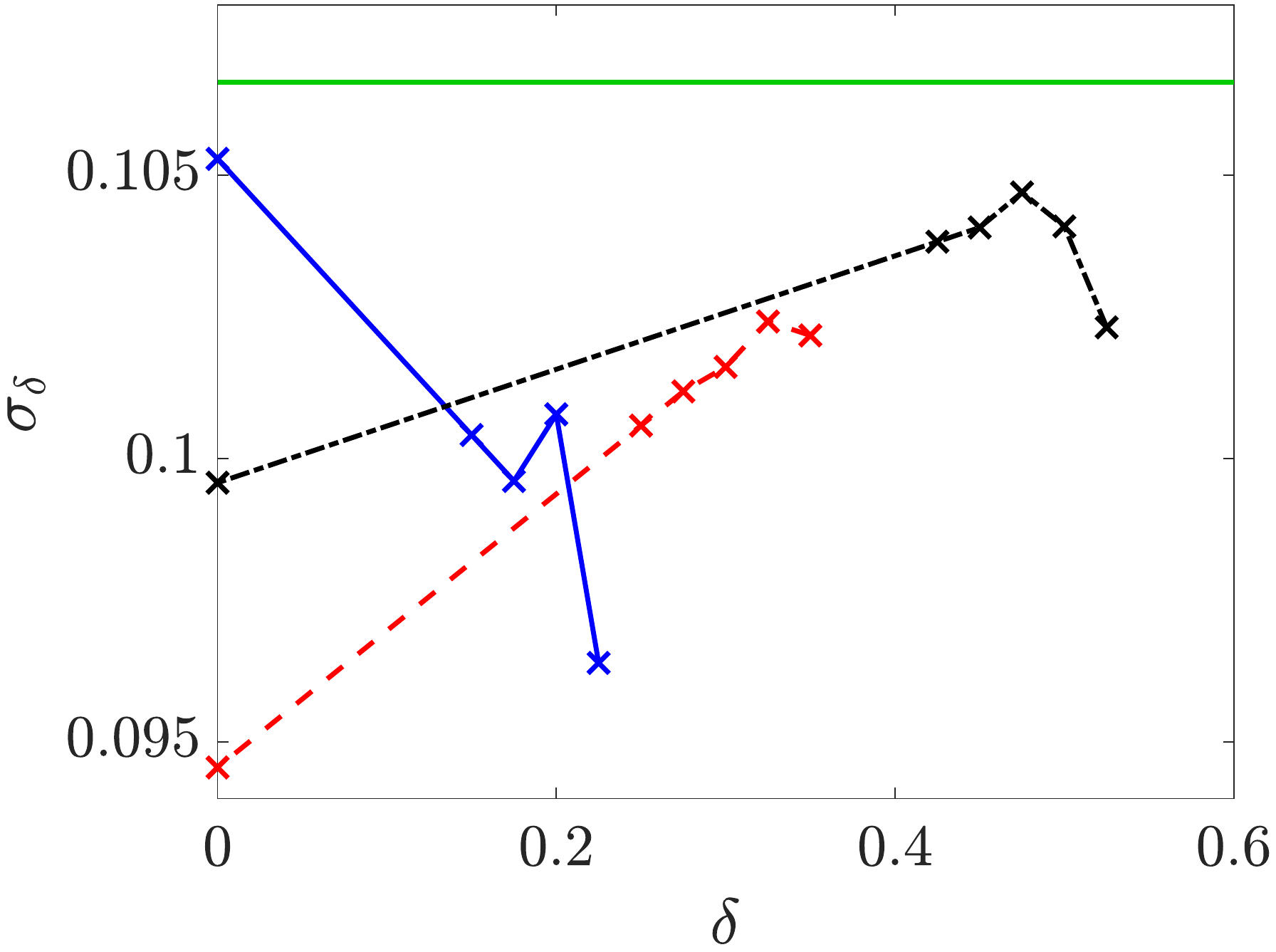}}
\hspace{0em}
\subfigure[Radius of spiral $r_{\delta}$.]{\label{fig:RT_CASTRO_convergence1-comparison_radius}\includegraphics[width=50mm]{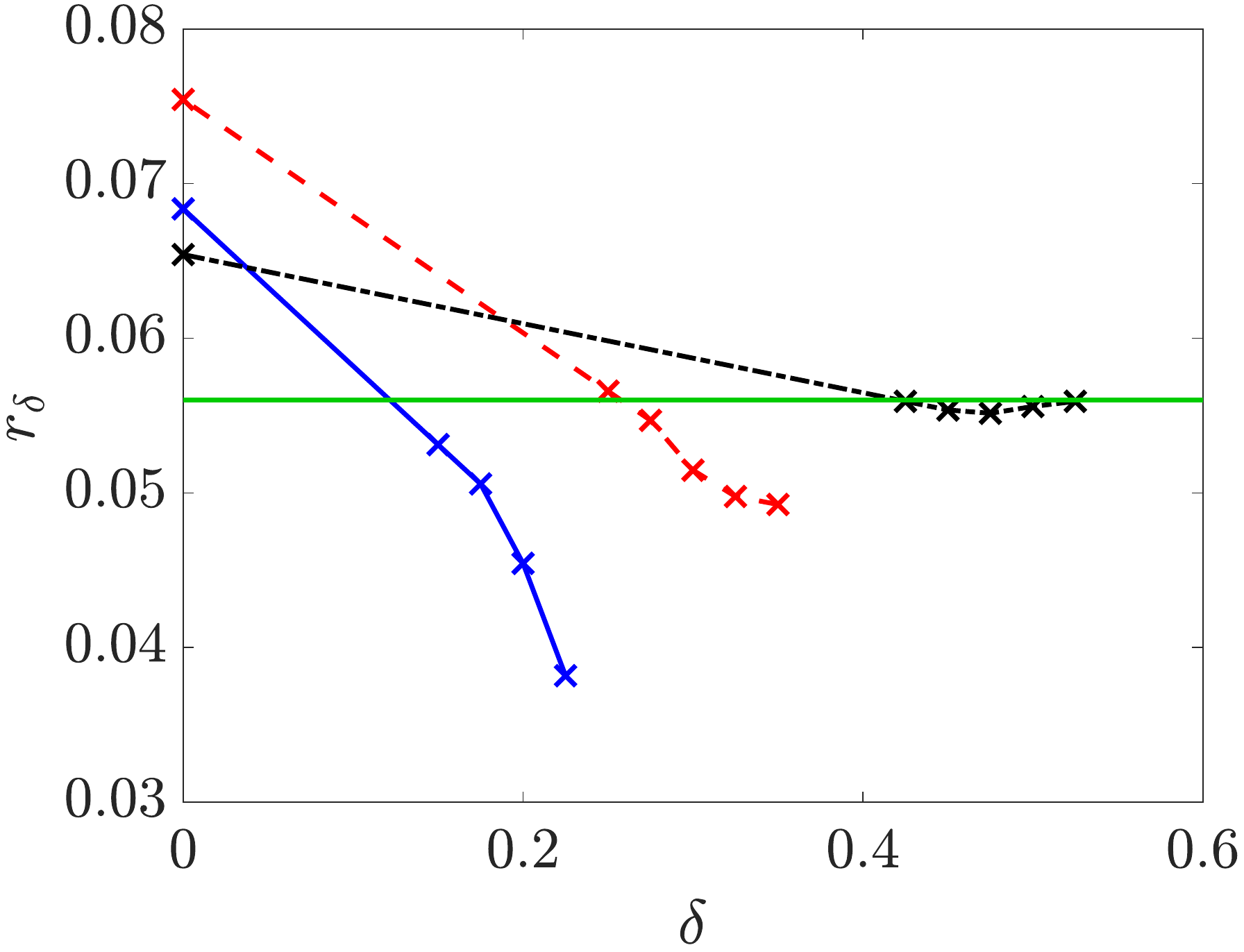}}
\caption{Convergence behavior as $\delta \to 0$ with $N=128$ fixed for the 
compressible RTI test of \citet{AlmgrenEtAl2010} using the $z$-model. 
Shown are (a) the bubble and spike tip locations, $| \max z_2 - 0.5 |$ and $| \min z_2 - 0.5 |$, respectively, 
(b) the location of the center of the spiral region $\sigma_\delta$, and (c) the radius of the 
spiral region $r_{\delta}$. The solid, dashed, and dotted curves in (a) refer to the Krasny, 
$\mathcal{K}^\delta_1$, and $\mathcal{K}^\delta_3$ kernels, respectively.
The blue, red, and black curves in (b) and (c) refer to the Krasny, 
$\mathcal{K}^\delta_1$, and $\mathcal{K}^\delta_3$ kernels, respectively.
The green curves indicate the corresponding quantities for the exact solution.}
\label{fig:RT_CASTRO_convergence1-comparison}
\end{figure}

\textcolor{black}{
We first take $N=128$ fixed and consider the limit $\delta \to 0$. The results are shown 
\Cref{fig:RT_CASTRO_convergence1-comparison}. We observe that all three methods are in reasonable 
agreement with regards to the computed bubble and spike tip locations. The 3rd order kernel is in better 
agreement with the exact solution for the bubble position, whereas the Krasny kernel is in better agreement
with the exact solution for the spike position. All three methods produce similar spiral radii $r_\delta$, and 
the computed values are in good agreement with the exact solution for $\delta > 0$. 
The 3rd order kernel, in particular, 
is in excellent agreement with the exact solution. The spiral center locations are similarly 
reasonably accurate for $\delta > 0$. 
}

\begin{figure}[h]
\centering
\subfigure[Bubble (blue) and spike (red) tip locations.]{\label{fig:RT_CASTRO_convergence2-comparison_amplitude}\includegraphics[width=50mm]{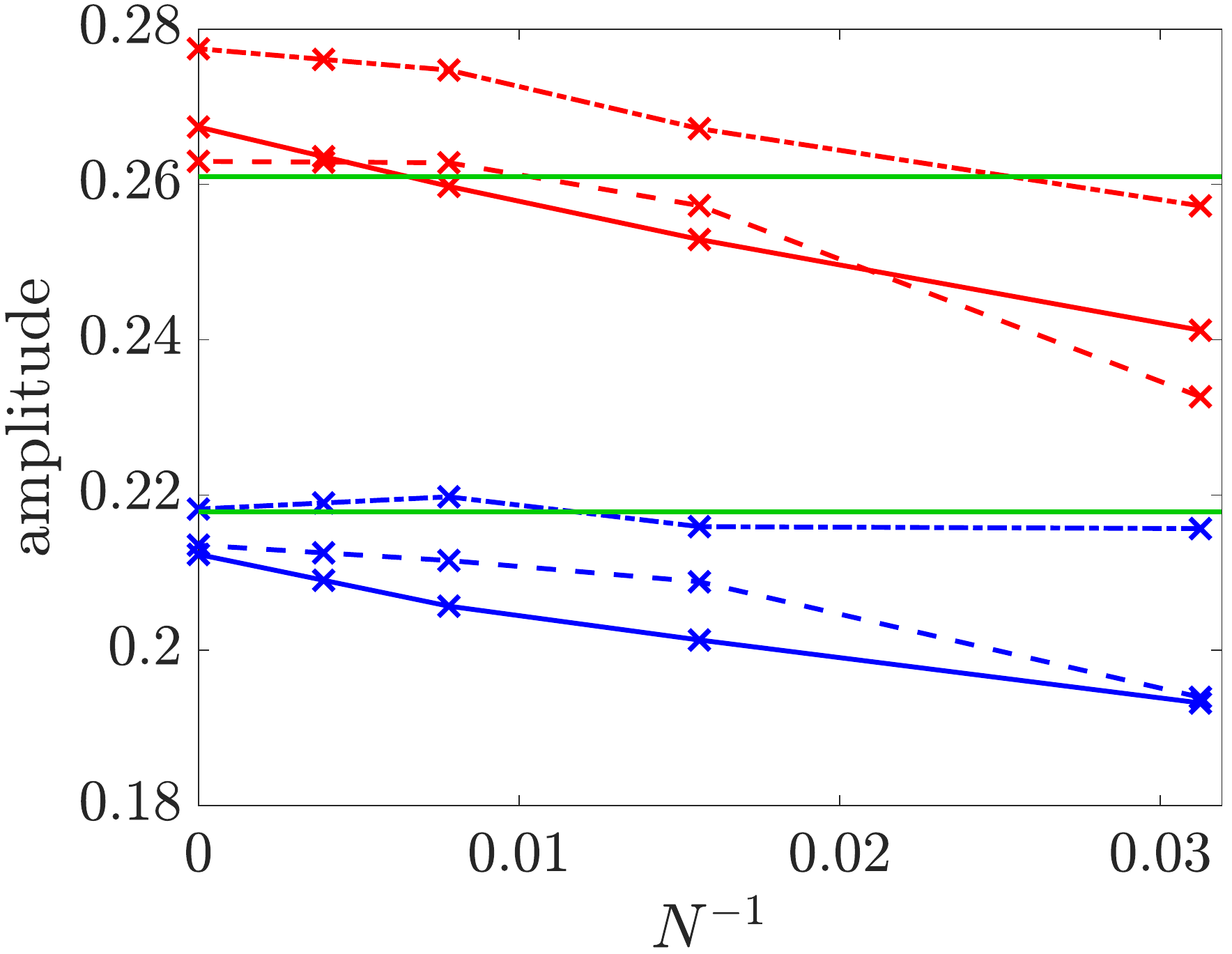}}
\hspace{0em}
\subfigure[Location of spiral center $\sigma_{\delta}$.]{\label{fig:RT_CASTRO_convergence2-comparison_sigma}\includegraphics[width=50mm]{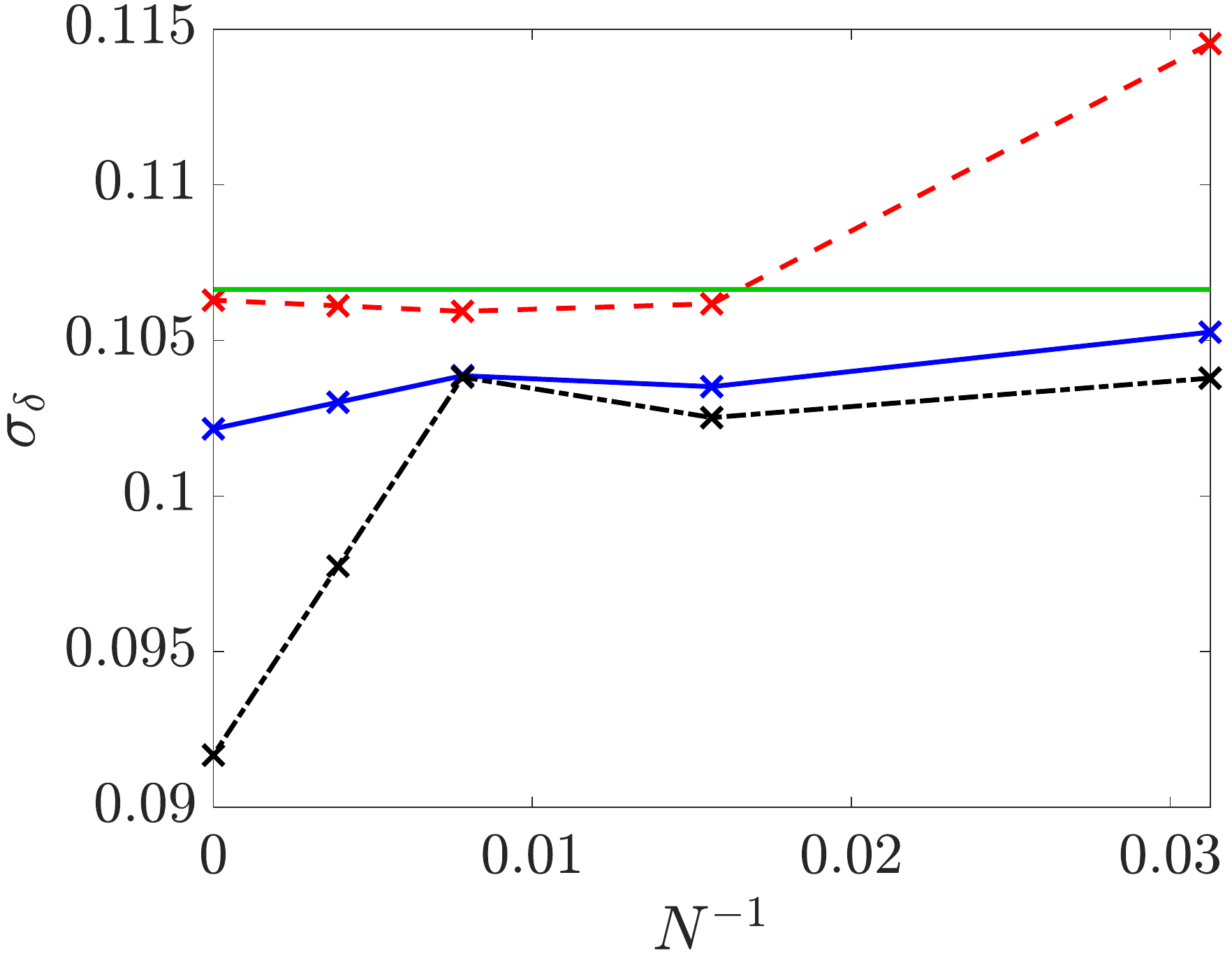}}
\hspace{0em}
\subfigure[Radius of spiral $r_{\delta}$.]{\label{fig:RT_CASTRO_convergence2-comparison_radius}\includegraphics[width=50mm]{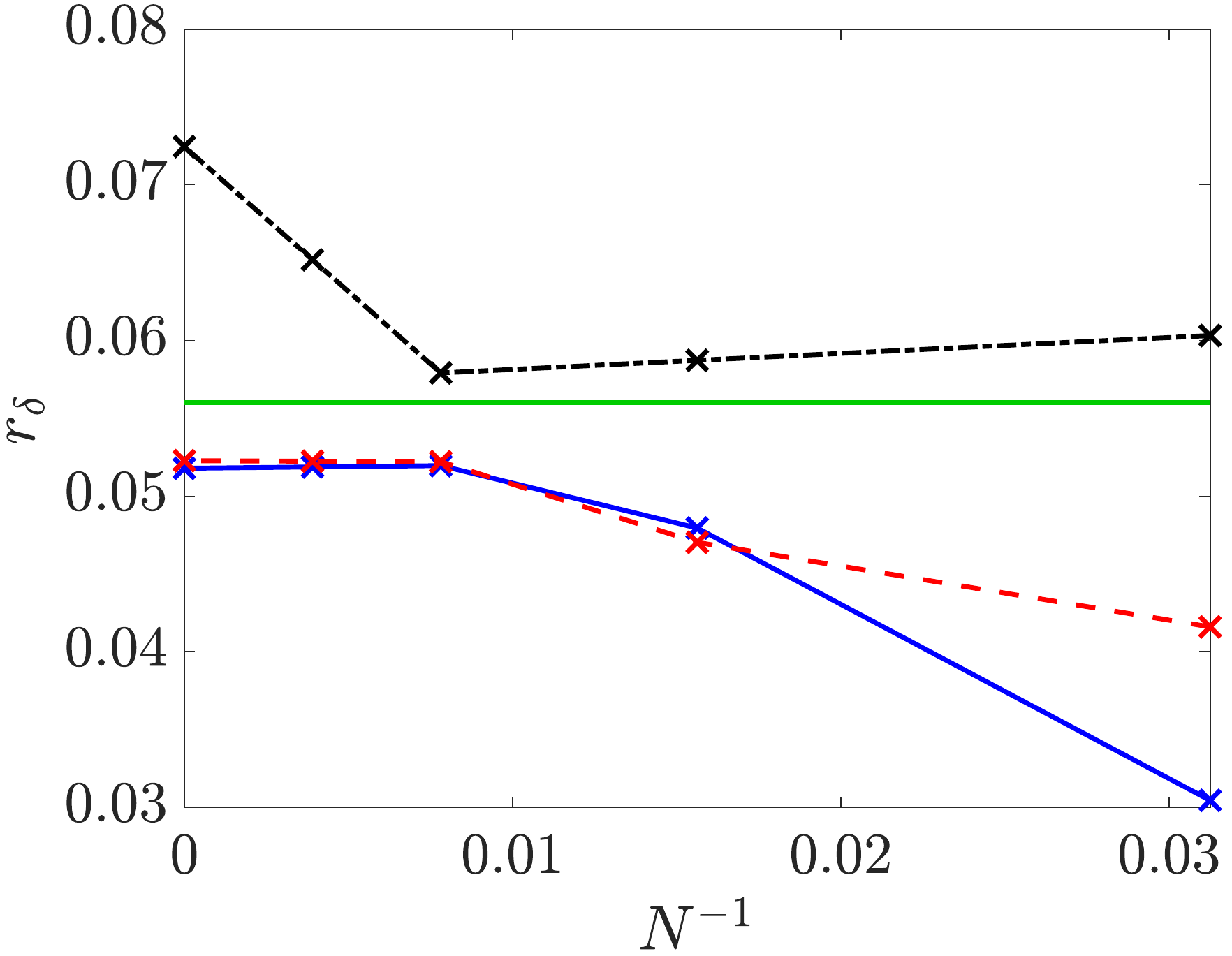}}
\caption{Convergence behavior as $\delta, N^{-1} \to 0$ for the 
compressible RTI test of \citet{AlmgrenEtAl2010} using the $z$-model. 
Shown are (a) the bubble and spike tip locations, $| \max z_2 - 0.5 |$ and $| \min z_2 - 0.5 |$, respectively, 
(b) the location of the center of the spiral region $\sigma_\delta$, and (c) the radius of the 
spiral region $r_{\delta}$. The solid, dashed, and dotted curves in (a) refer to the Krasny, 
$\mathcal{K}^\delta_1$, and $\mathcal{K}^\delta_3$ kernels, respectively.
The blue, red, and black curves in (b) and (c) refer to the Krasny, 
$\mathcal{K}^\delta_1$, and $\mathcal{K}^\delta_3$ kernels, respectively.
The green curves indicate the corresponding quantities for the exact solution.}
\label{fig:RT_CASTRO_convergence2-comparison}
\end{figure}

\textcolor{black}{
Next, we consider the limit $\delta, N^{-1} \to 0$. In particular, we consider $N=32,64,128,256$. 
We use the scaling \eqref{delta-scaling} for the Krasny scheme, and the empirical procedure described 
at the beginning of \Cref{z-model-numerical-study} for the $\mathcal{K}^\delta_i$ kernel schemes. 
The results are shown in \Cref{fig:RT_CASTRO_convergence2-comparison}. All of the schemes 
perform similarly, with the Krasny and $\mathcal{K}^{\delta}_1$ schemes in better agreement 
with the exact solution with regards to 
spike tip position, spiral center location $\sigma_\delta$, and spiral radius $r_\delta$, while the  
3rd order kernel more accurately predicts the bubble tip location.  We note that 
the convergence behavior of the 
3rd order kernel scheme with regards to the quantities $\sigma_\delta$ and $r_\delta$ is rather erratic; this 
is a result of the sensitivity of the scheme to the choice of parameter $\delta$. The 
asymptotic behavior of $\sigma_\delta$ and $r_\delta$ in the Krasny scheme with the 
scaling \eqref{delta-scaling},  on the other hand, is more uniform. 
}

\subsection{The compressible RTI test of Liska \& Wendroff}
For our next numerical experiment, we 
consider the compressible single-mode RTI from the review paper of \citet{LiWe2003}. The domain 
is $(x_1,x_2) \in [-1/6, +1/6] \times [0,1]$, the gravitational constant is $g=0.1$, and the initial data is 
\begin{equation}\label{RT-initialdata}
\begin{bmatrix}
\rho_0 \\ (\rho u)_0 \\ E_0 
\end{bmatrix}
=
\begin{bmatrix}
1 \\ 0 \\ p_0/(\gamma-1) 
\end{bmatrix}
\mathbbm{1}_{[0,\eta_0)}(x_2)
+
\begin{bmatrix}
2 \\ 0 \\ p_0/(\gamma-1)
\end{bmatrix}
\mathbbm{1}_{[\eta_0,1]}(x_2)  \,, 
\end{equation}
where the initial interface $\Gamma_0$ is parameterized by $(x_1,\eta_0(x_1))$ with $\eta_0(x_1)=0.5 + 0.01\cos(6 \pi x_1)$, and $p_0$ is 
the initial pressure, defined as
\begin{equation*}\label{initial-pressure-hydrostatic}
 p_0 = \begin{cases}
 	   2.4 + g \, (\eta_0(x_1) - x_2) + 2g\,(1-\eta_0(x_1)) &, \text{ if } x_2 < \eta_0(x_1) \\
 	   2.4 +2g \, (1 - x_2) &, \text{ if } x_2 \geq \eta_0(x_1) \\	
 	   \end{cases} \,,
\end{equation*}
Periodic conditions are applied in the horizontal direction $x_1$, while free-flow conditions are 
applied at the boundaries in the $x_2$ direction. 

In \cite{RaReSh2019b}, we used the anisotropic $C$-method to solve this problem; the resulting solutions 
have the classic mushroom-shaped interface profile without overly diffused KHI roll-up 
regions and mixing zones. We shall use this solution, computed on a fine mesh of $50 \times 200$ cells 
with $\mathrm{CFL} \approx 0.1$ as our {\it high resolution reference} solution\footnote{We have found that this 
smaller CFL number is required for this low Mach number flow calculation to prevent high frequency noise from 
corrupting the solution.}.

\subsubsection{The multiscale algorithm applied to the RTI}
We begin by employing the multiscale algorithm to the RTI problem \eqref{RT-initialdata} with the following parameter values: the coarse mesh for $v$
contains $8 \times 32$ cells, the  fine mesh for $w$ uses $N=64$, and the time-step is 
$\delta t = 2.5 \times 10^{-3}$, giving $\mathrm{CFL} \approx 0.35$.  
The Atwood number is set as 
$A = 1/3$, and the initial data is given by \eqref{RT-initialdata} (with $u$ replaced by $v$) with the 
following modification: as in the numerical experiment performed in \Cref{subsec:RT_CASTRO}, we 
smooth the initial density field $\rho_0$ over a length scale. The initial density is thus given by 
formula \eqref{RT_CASTRO_initial-rho} with $\eta_0(x_1)=0.5+0.01 \cos (6 \pi x_1)$, and the smoothing 
length scale is chosen as $h=0.02$. We also set 
\begin{align*}\label{RT_LiWe2003_initialdata}
 z_1(\alpha,0) &= \alpha  \,, \\
z_2(\alpha,0) &= 0.5 +  0.01\cos(6 \pi \alpha) \,, \\
\varpi(\alpha,0) &= 0\,. 
\end{align*}
The artificial viscosity parameters are chosen as $\beta=5$, $\tilde{\delta} = 1.5$, and 
${\mu} = 0.001$.

\begin{figure}[h]
\centering
\subfigure[Multiscale $z$]{\label{fig:RT_LiWe2003_z_8x32_64}\includegraphics[width=32mm]{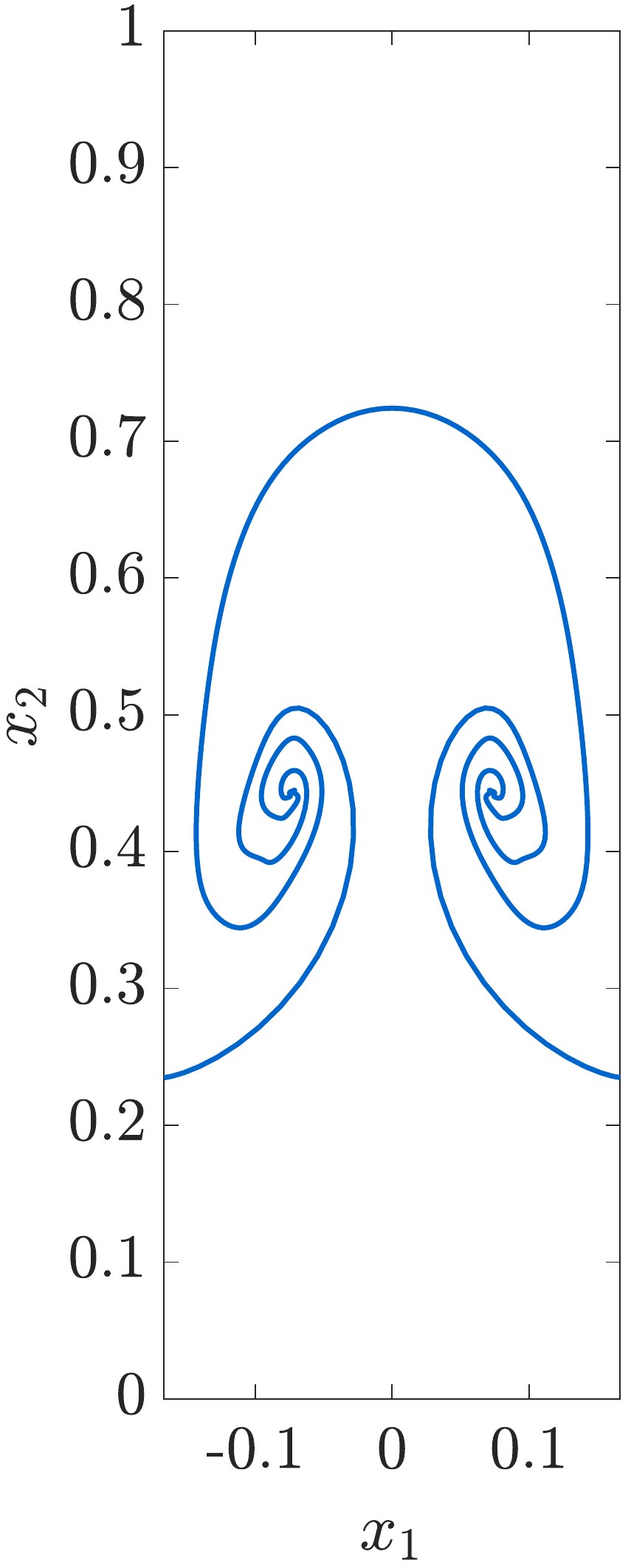}}
\hspace{2em}
\subfigure[High-res.]{\label{fig:RT_LiWe2003_z_highres}\includegraphics[width=24.065mm]{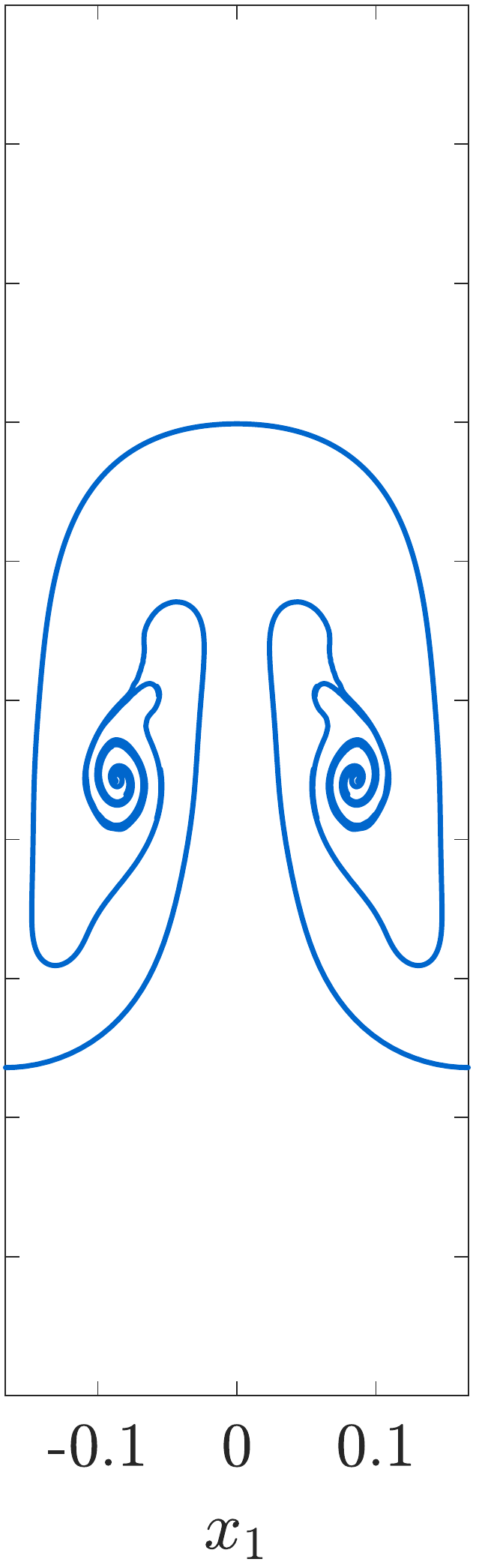}}
\hspace{2em}
\subfigure[Comparison]{\label{fig:RT_LiWe2003_z_compare_8x32_64}\includegraphics[width=24.065mm]{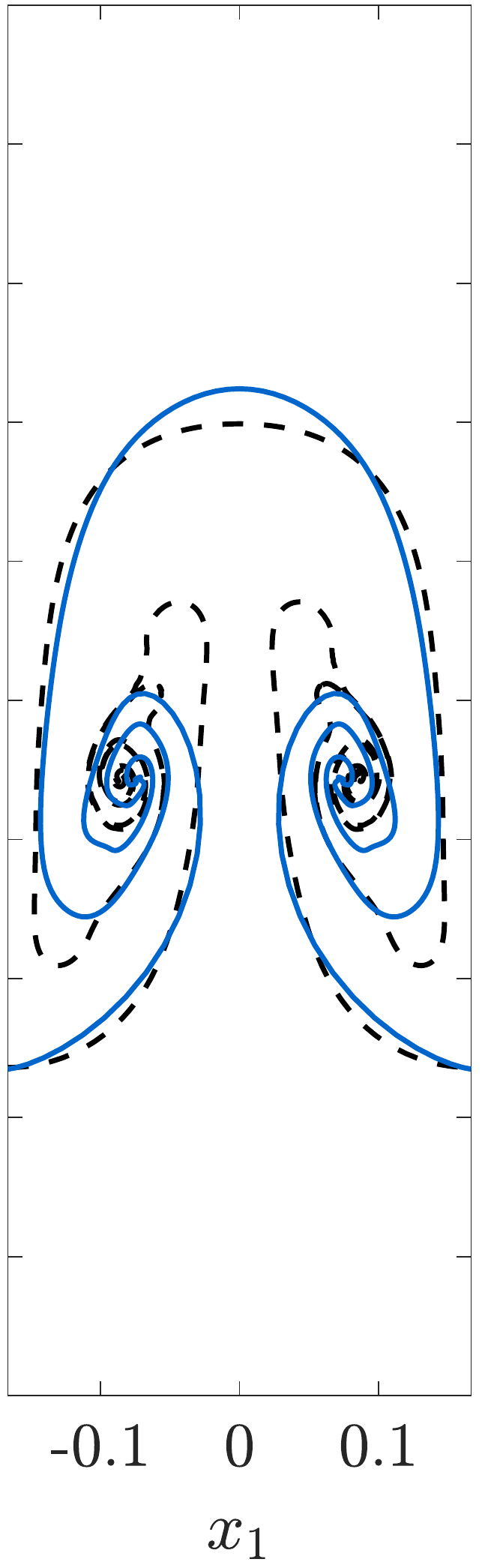}}
\caption{Results for the multiscale algorithm applied to the compressible single-mode RTI test 
of \citet{LiWe2003}, with the 
underlying grid containing $8 \times 32$ cells and the interface discretized with $N=64$. Shown are (a)  
computed interface parametrization $z$, (b) benchmark interface position, and (c) 
the computed interface $z$ (blue) overlaying the benchmark solution (dashed black)
at the final time $t=8.5$.}
\label{fig:RT_LiWe2003_8x32_64}
\end{figure}

The multiscale interface position $z$ at time $t=8.5$ is shown in 
\Cref{fig:RT_LiWe2003_8x32_64}; the  high resolution reference solution computation is presented in 
 \Cref{fig:RT_LiWe2003_z_highres}, and the comparison of the two solution is made in \Cref{fig:RT_LiWe2003_z_compare_8x32_64}.  We see that the spike tip position of the 
 computed solution matches almost exactly with the spike tip position of the reference solution, and the 
 bubble tip positions are also in good agreement. 
 Moreover, the multiscale solution successfully 
simulates the roll-up of the contact discontinuity and, by comparing with \Cref{fig:RT_LiWe2003_z_highres}, we
observe that this roll-up  occurs in the correct region of the flow and matches well with  the {\it scale} of the high resolution solution.

\begin{figure}[h]
\centering
\subfigure[$t=2.0$]{\label{fig:RT_LiWe2003_z_8x32_64_t=2}\includegraphics[width=30.03mm]{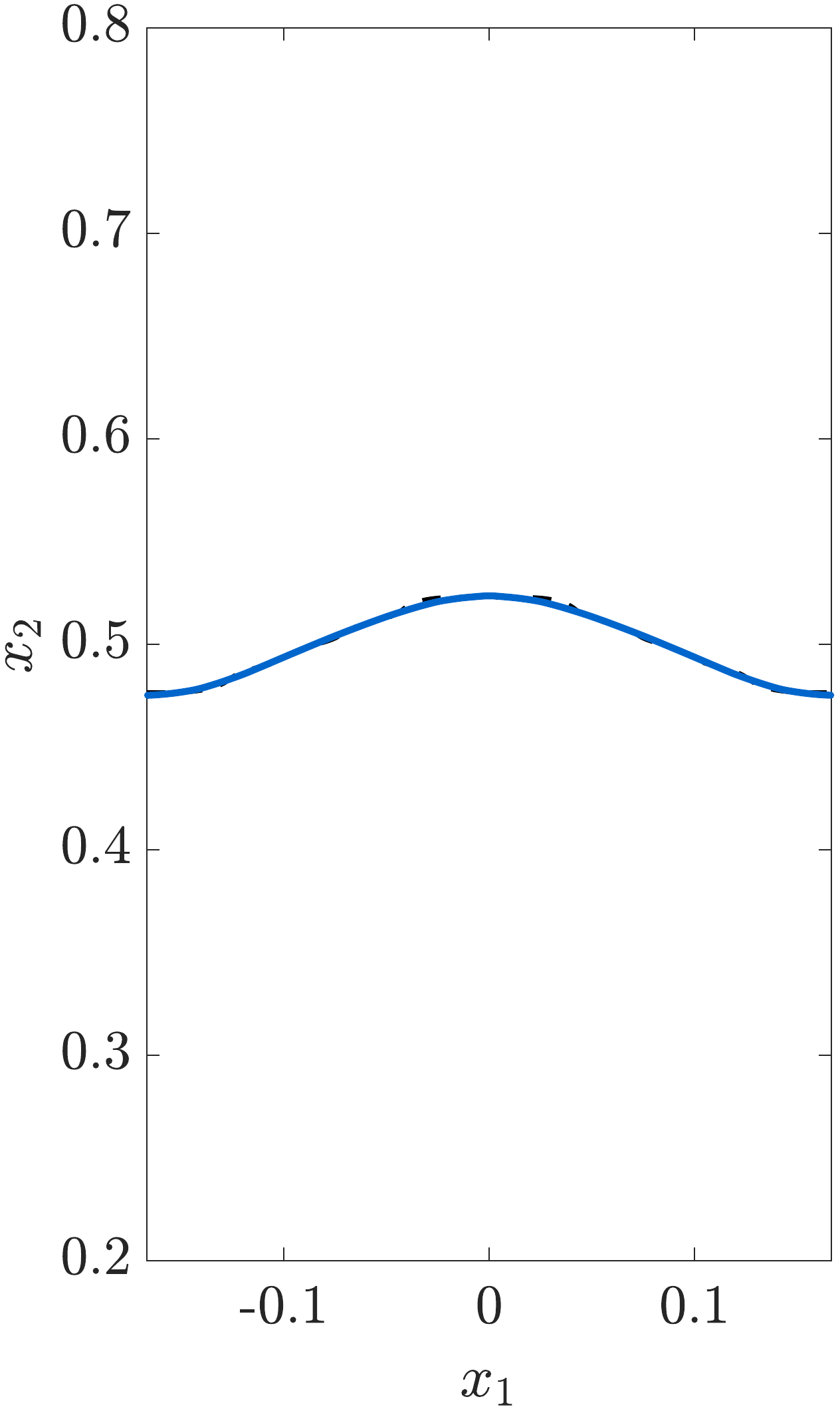}}
\hspace{1em}
\subfigure[$t=5.0$]{\label{fig:RT_LiWe2003_z_8x32_64_t=5}\includegraphics[width=25.0mm]{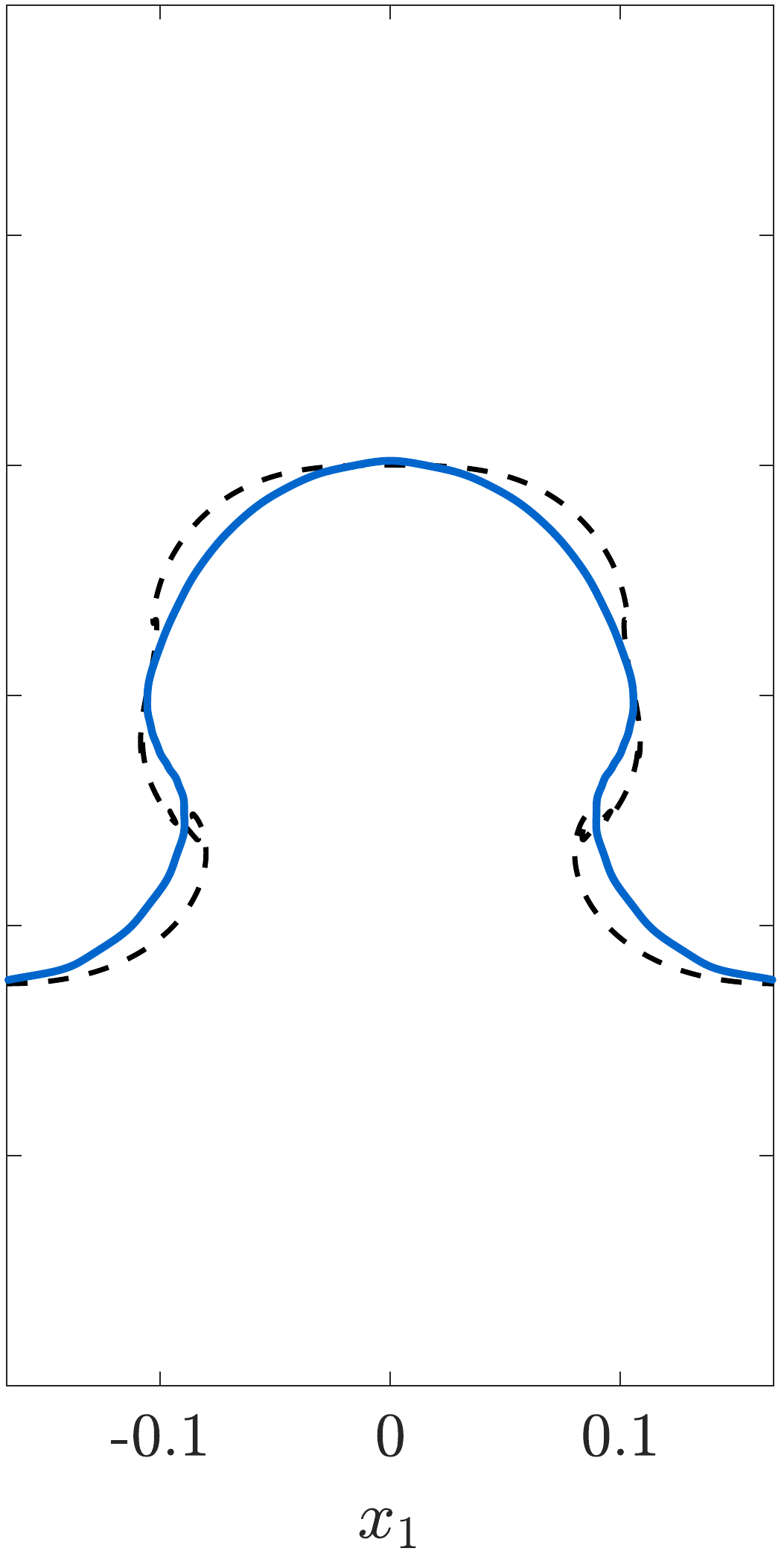}}
\hspace{1em}
\subfigure[$t=6.0$]{\label{fig:RT_LiWe2003_z_8x32_64_t=6}\includegraphics[width=25.0mm]{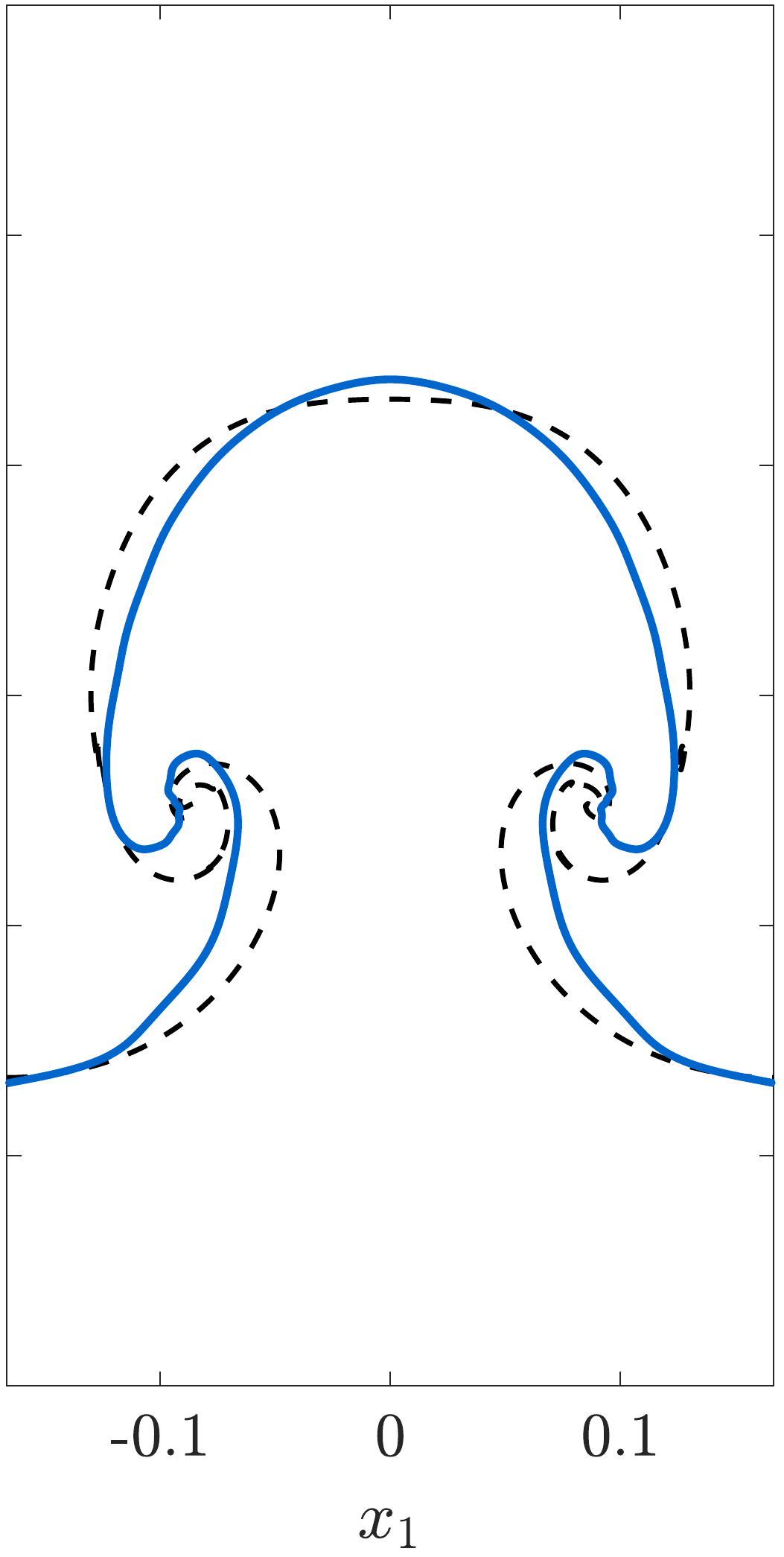}}
\hspace{1em}
\subfigure[$t=7.0$]{\label{fig:RT_LiWe2003_z_8x32_64_t=7}\includegraphics[width=25.0mm]{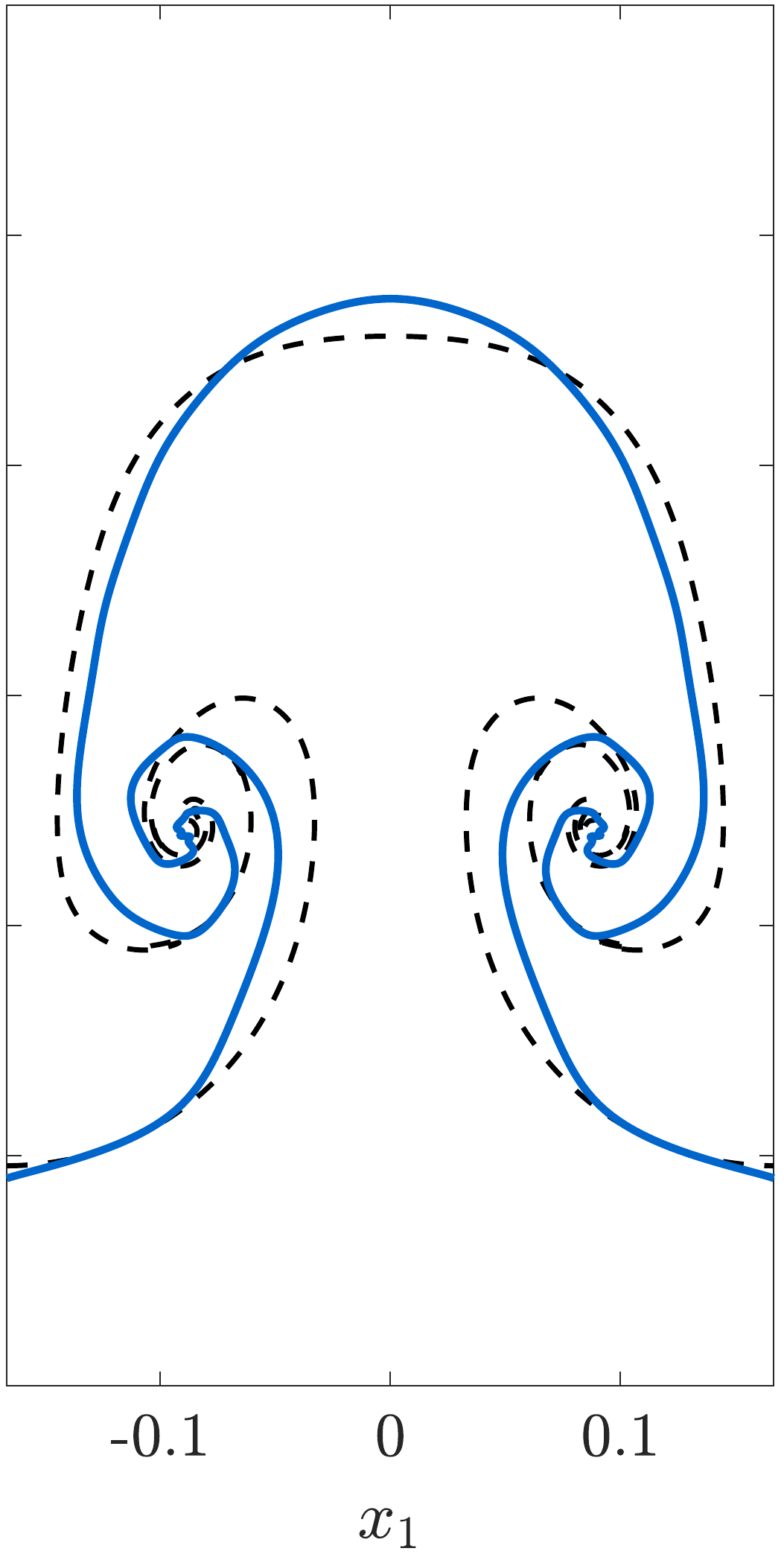}}
\hspace{1em}
\subfigure[$t=8.5$]{\label{fig:RT_LiWe2003_z_8x32_64_t=final}\includegraphics[width=25.0mm]{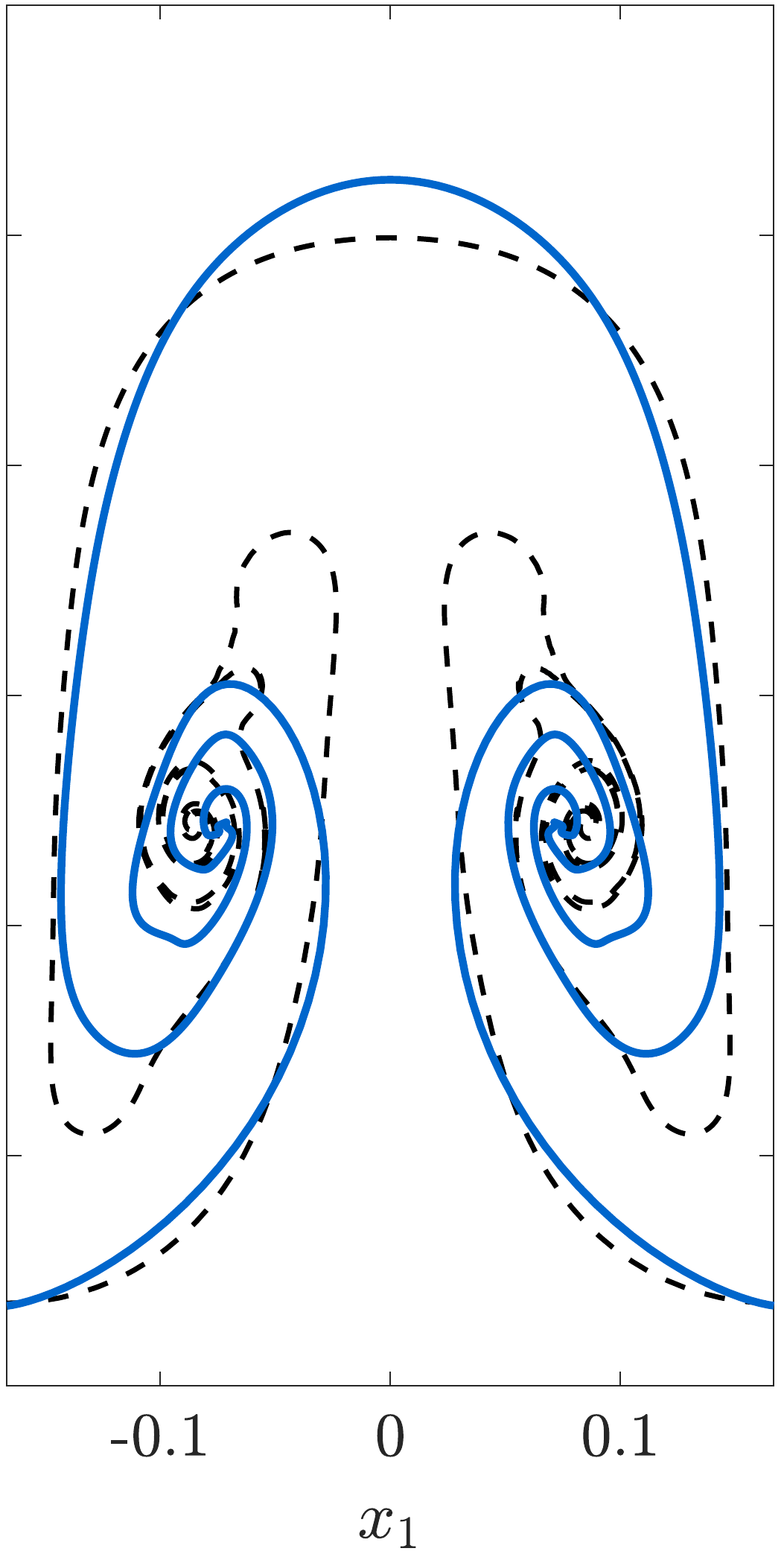}}
\hspace{1em}
\caption{Evolution over time $t$ of the interface for the compressible single-mode RTI. 
Here $z$ is computed using the multiscale algorithm on a mesh with $8 \times 32$ cells and an 
interface discretized with $N=64$. The blue curve is the computed $z(\alpha,t)$, and the dashed black 
curve is the reference solution.}
\label{fig:RT_LiWe2003_8x32_64_time_evolution}
\end{figure}

In \Cref{fig:RT_LiWe2003_8x32_64_time_evolution}, we compare the interface $z(\alpha,t)$ from the multiscale run and the high-resolution run
at various times $t$.  As can be seen, the  two solutions are in very good agreement throughout 
 the time interval of the simulation. 

\begin{figure}[h]
\centering
\includegraphics[width=50mm]{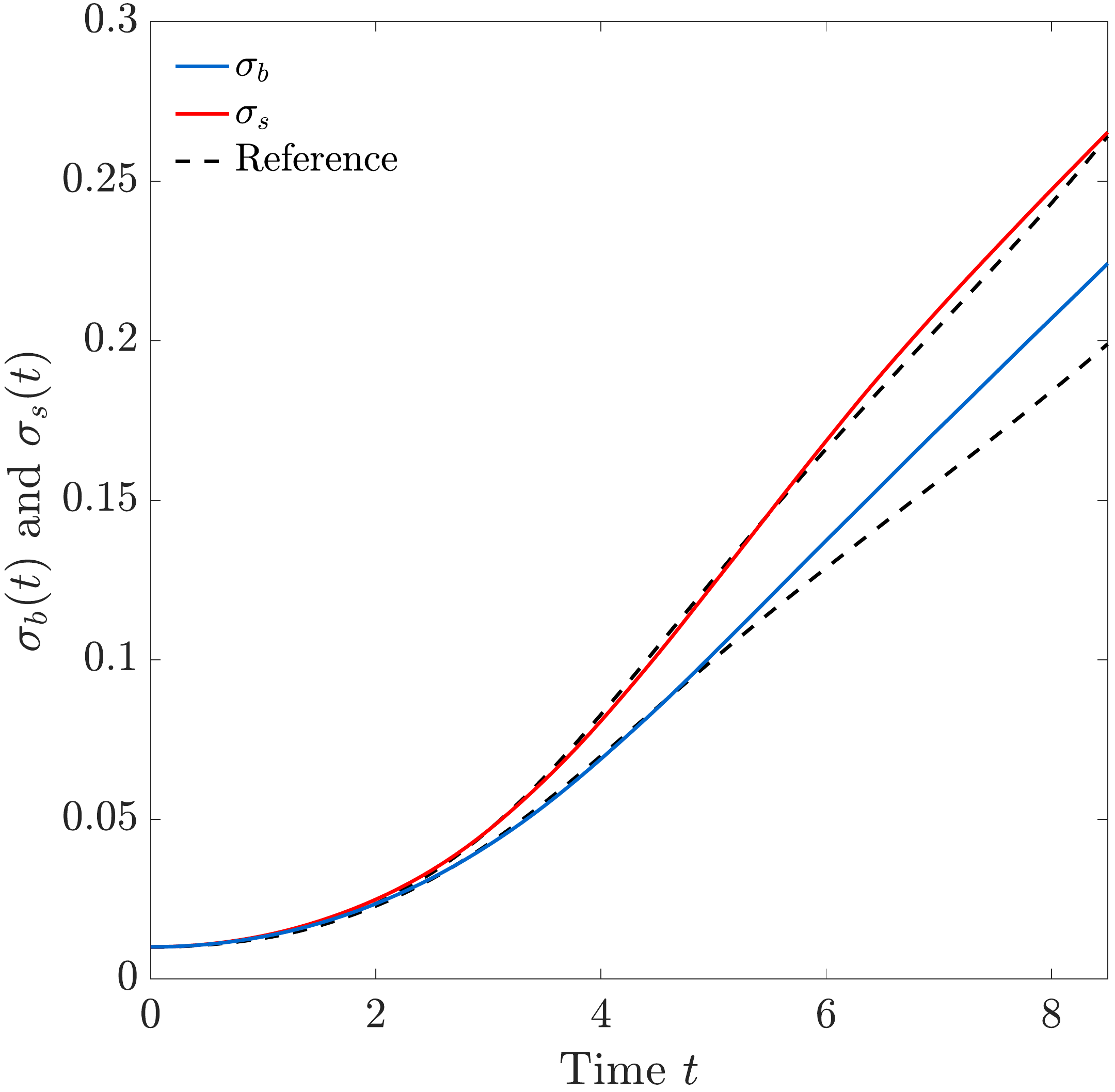}
\caption{Plots of the quantities $\sigma_b(t)$ and $\sigma_s(t)$ for the solution computed using the 
multiscale algorithm applied to the compressible single-mode RTI on a mesh with 
$8 \times 32$ cells and an interface with $N=64$.}
\label{RT_LiWe2003_z_8x32_64_sigma.pdf}
\end{figure}

To compare the bubble and spike tip positions, we plot in \Cref{RT_LiWe2003_z_8x32_64_sigma.pdf}
 the quantities  $\sigma_b(t) = |0.5-\max_{\alpha} z(\alpha,t)|$ and 
 $\sigma_s(t) = |0.5-\min_{\alpha} z(\alpha,t)|$. 
The spike tip position is in excellent agreement with the reference solution for all times $t$, while the bubble 
tip position is also in excellent agreement for times $t < 6$. For $t > 6$, the bubble tip position diverges slightly 
from the reference solution, but the error is still relatively small; at the final time $t=8.5$, the error is 
$\approx 2.5 \%$ of the height of the computational domain. 

Finally, we turn to the issue of the runtimes of the computations. 
The reference solution computation had a runtime of $T_{\mathrm{CPU}} \approx 2906$ s, whereas 
the multiscale algorithm runtime was only $T_{\mathrm{CPU}} \approx 14$ s, which gives a 
speed-up of approximately 203 times. We are thus able to infer both large-scale (amplitude growth rates) as 
well as small-scale (roll-up region structure) information by use of the multiscale model and algorithm, while 
drastically reducing the computational burden and runtime when compared with the reference solution 
computation.  

\subsubsection{Comparing the mutliscale solution with both the low resolution simulation and the  incompressible $z$-model}
To demonstrate the efficacy of our multiscale model, we compare the multiscale solution  with
the low resolution solution as well as with the incompressible and irrotational $z$-model.

\begin{figure}[h]
\centering
\subfigure[Multiscale $z$]{\label{fig:RT_LiWe2003_z_compare_8x32_64_2}\includegraphics[width=32mm]{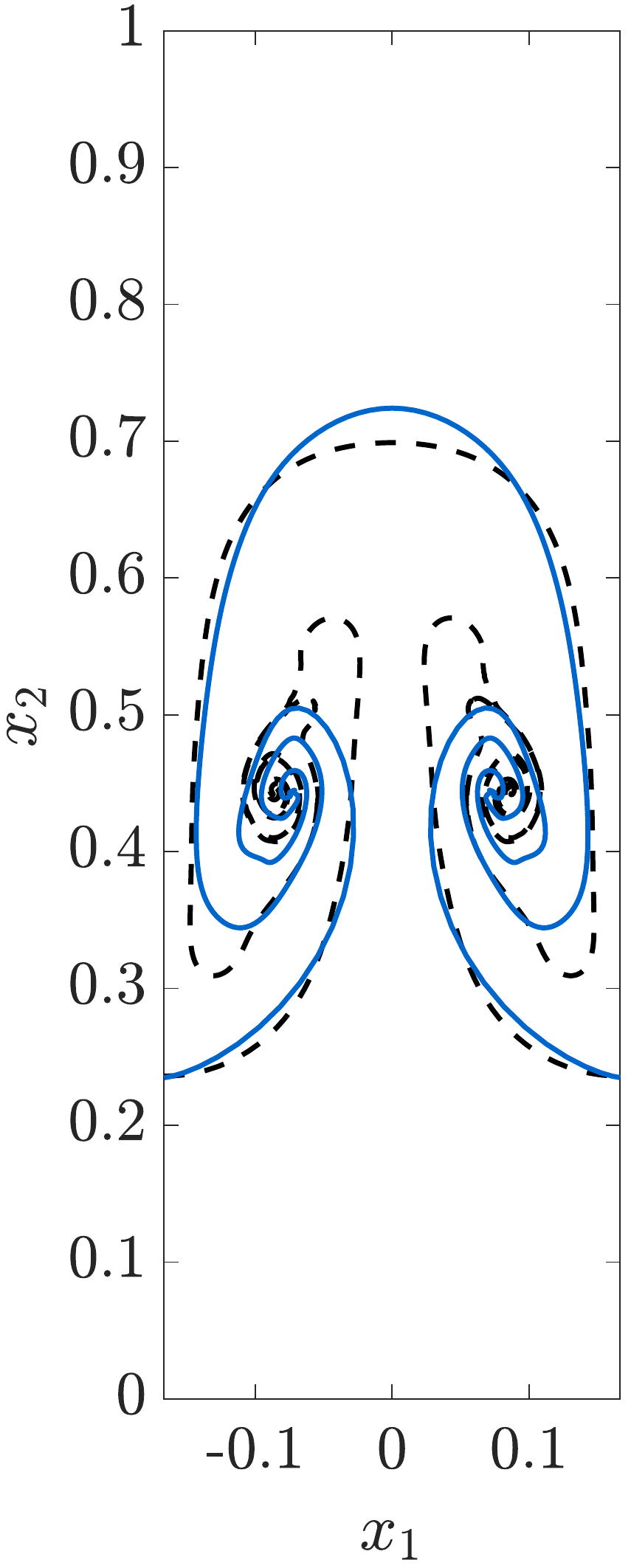}}
\hspace{2em}
\subfigure[$z$-model]{\label{fig:RT_LiWe2003_z_64}\includegraphics[width=24.065mm]{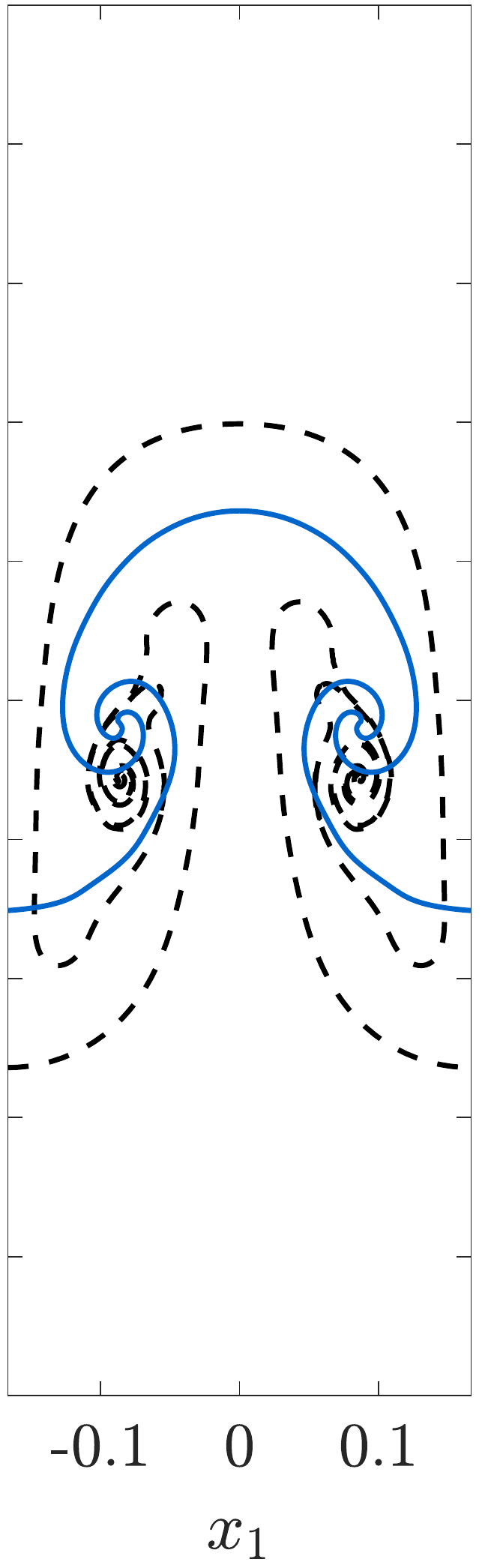}}
\hspace{2em}
\subfigure[Multiscale $\rho$]{\label{fig:RT_LiWe2003_rho_8x32_64}\includegraphics[width=24.065mm]{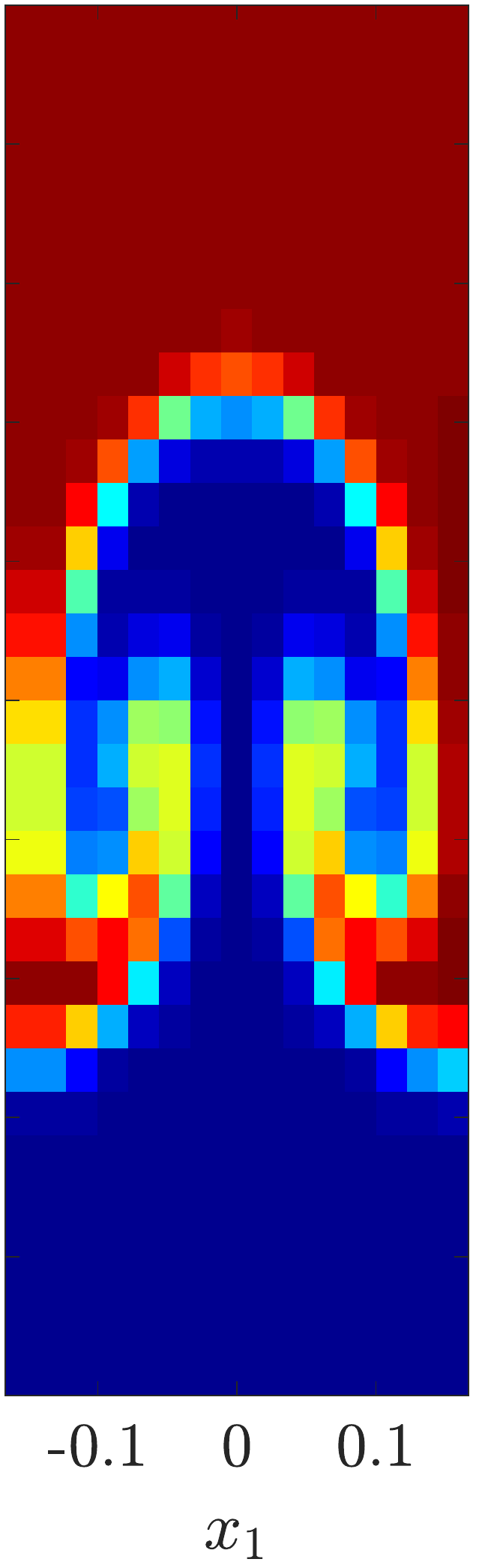}}
\hspace{2em}
\subfigure[Low res. $\rho$]{\label{fig:RT_LiWe2003_rho_8x32}\includegraphics[width=24.065mm]{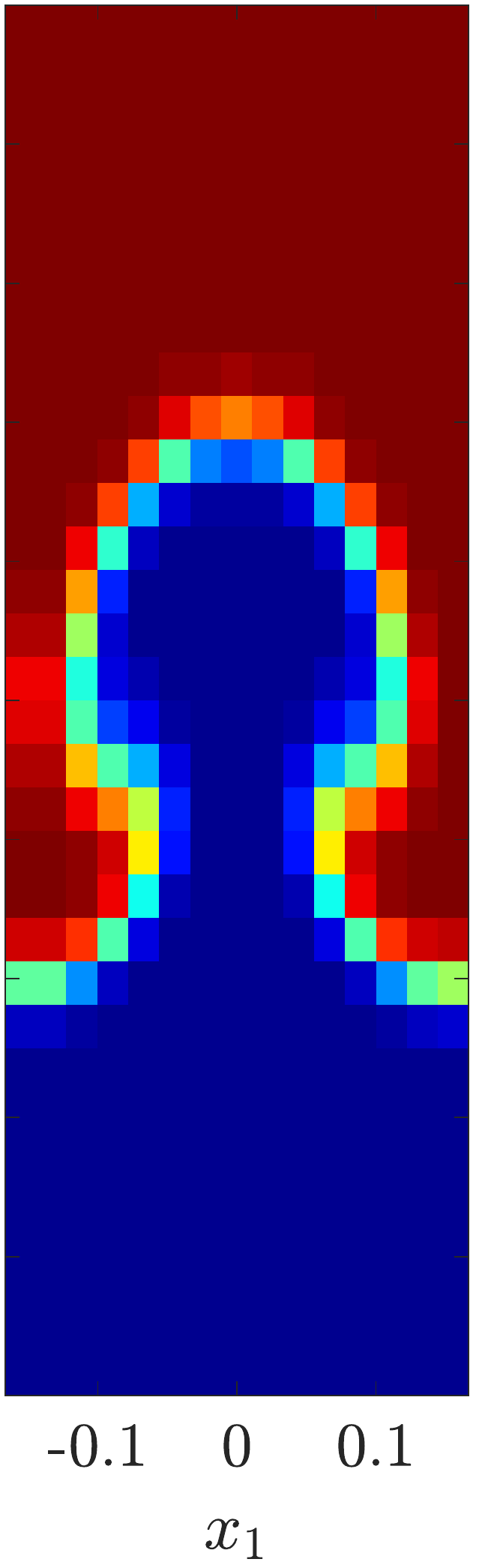}}
\caption{Comparison of the solutions  to the compressible single-mode RTI test of 
\citet{LiWe2003}. Figures \ref{fig:RT_LiWe2003_z_compare_8x32_64_2} and 
\ref{fig:RT_LiWe2003_rho_8x32_64} are plots of the interface $z$ and density $\rho$, respectively, 
computed using the multiscale algorithm. 
\Cref{fig:RT_LiWe2003_z_64} is the interface $z$ computed using the incompressible and $z$-model. 
\Cref{fig:RT_LiWe2003_rho_8x32} is the low resolution density $\rho$ computed using the $C$-method. 
All the relevant parameters are fixed across the simulations.}
\label{fig:RT_LiWe2003_stand-alone_8x32_64}
\end{figure}

In \Cref{fig:RT_LiWe2003_stand-alone_8x32_64}, the results from these simulations are 
compared with the simulations performed using the multiscale algorithm. 
We see that the solutions are drastically different; the low resolution density $\rho$ 
has no roll-up of the contact 
 discontinuity, while the interface $z$ computed using the incompressible $z$-model 
 has completely incorrect bubble and spike positions. 
 On the other hand, the use of the multiscale model allows the coarse grid calculation to 
 ``correct'' the interface position $z$, while the fine resolution computation on the interface $z$ similarly 
 ``corrects'' for the lack of roll-up of the coarse grid solution. 

\subsubsection{Basic mesh refinement study}\label{sec-RTI-mesh-refinement}

Next, we perform a basic mesh refinement study to analyze the behavior of our multiscale algorithm as both 
the underlying mesh, as well as the interface discretization, is refined. More precisely, we shall consider 
grids with $8 \times 32$, or $10 \times 40$ or $12 \times 48$ cells, and the interface discretized with 
either $N=64$ or $N=128$ or $N=256$. We shall keep the artificial viscosity parameters 
$\beta$ and $\tilde{\delta}$ fixed, with $\beta=5$ and $\tilde{\delta}=1.5$. The time-step $\delta t$ and 
artificial viscosity parameter ${\mu}$ are allowed to vary as both the underlying mesh resolution, as well 
as the interface resolution, are varied. The exact choices for these parameters, as well as 
the corresponding runtimes $T_{\mathrm{CPU}}$ and speed-up factors $\Lambda$ compared to the 
high resolution
reference solution calculation, are presented in \Cref{table:RMI-parameters} in Appendix 
\ref{sec-appendices-b}, 
but we note here that the values of ${\mu}$ are almost the same (${\mu} \approx 0.001$) for all 
of the runs. 

The results for the mesh refinement study are shown in \Cref{fig:RT_LiWe2003_z_compare} in 
Appendix \ref{sec-appendices-b}. As $N$ increases with the number of cells fixed, the computed interface positions 
are roughly the same, except in the KHI roll-up regions, in which the interfaces computed with larger $N$ show 
significantly more roll-up (compare, for instance, \Cref{fig:RT_LiWe2003_z_8x32_64-compare} with 
\Cref{fig:RT_LiWe2003_z_8x32_256-compare}). 
This is in line with the observations in \Cref{z-model-numerical-study} 
for numerical simulations using the incompressible and irrotational $z$-model, and is due to the scaling of 
the parameter $\tilde{\delta}$. 

On the other hand, if the number of cells contained in the underlying coarse mesh is increased with $N$ held fixed, 
the resulting solutions for the interface do not have more roll-up, but instead appear to converge to the 
reference solution away from the roll-up region (compare, for instance, 
\Cref{fig:RT_LiWe2003_z_8x32_128-compare} with \Cref{fig:RT_LiWe2003_z_12x48_128-compare}). 
This ``convergence'' is particularly noticeable in the ``pits'' of the mushroom shape.

\subsection{A single-mode RMI problem}

We next consider the single-mode RMI with the following problem setup. The domain is 
$(x_1,x_2) \in [-1/6,+1/6] \times [-1,1]$, the gravitational constant is $g=0.5$, and the initial data is given by 
\begin{equation}\label{RM-initialdata}
\begin{bmatrix}
\rho_0 \\ (\rho u)_0 \\ E_0 
\end{bmatrix}
=
\begin{bmatrix}
1 \\ 0 \\ f/(\gamma-1) 
\end{bmatrix}
\mathbbm{1}_{[-1,\eta_0)}(x_2)
+
\begin{bmatrix}
2 \\ 0 \\ f/(\gamma-1)
\end{bmatrix}
\mathbbm{1}_{[\eta_0,0.8)}(x_2) 
+
\begin{bmatrix}
4.857143  \\ 0 \\  36.6666+ f/(\gamma-1)
\end{bmatrix}
\mathbbm{1}_{[0.8,1]}(x_2)  \,, 
\end{equation}
where the initial interface $\Gamma_0$ is parameterized by $(x_1,\eta_0(x_1))$ with $\eta_0(x_1)=0.5 + 0.1\cos(6 \pi x_1)$, and $f$ is 
the function defined as
\begin{equation*}\label{initial-pressure-RMI-hydrostatic}
 f(x_1,x_2) = \begin{cases}
 	   2.4 + g \, (\eta_0(x_1) - x_2) + 2g\,(1-\eta_0(x_1)) &, \text{ if } x_2 < \eta_0(x_1) \\
 	   2.4 +2g \, (1 - x_2) &, \text{ if } x_2 \geq \eta_0(x_1) \\	
 	   \end{cases} \,,
\end{equation*}
Periodic conditions are applied in the horizontal direction $x_1$, while free-flow and solid-wall conditions are 
applied at the bottom and top boundaries, respectively. 

A high resolution reference solution is computed using the $C$-method on a fine grid with $50 \times 400$ cells. The 
time-step is fixed as $\delta t = 0.0001$, which results in $\mathrm{CFL} \in (0.14,0.2)$. Again, we have 
found that this relatively small CFL number is required to prevent the occurrence of high-frequency noise 
in the computed solution. 

\subsubsection{The multiscale algorithm applied to the RMI}
We apply the multiscale algorithm to the RMI problem \eqref{RM-initialdata} with the following parameter 
choices: the coarse mesh for $v$ contains $10 \times 80$ cells, the fine mesh for $w$ uses $N=64$, and the 
time-step is $\delta t = 0.00125$, which yields $\mathrm{CFL} \in (0.30,0.45)$. The artificial viscosity 
parameters in \eqref{C-method-viscosity-parameters} are set as $\beta=1$, $\beta_s=1$, while 
$\tilde{\delta}=1.25$ and ${\mu}=0.005$. The Atwood number is set as $A=1/3$, and the initial data 
is given by \eqref{RM-initialdata} (with $u$ replaced by $v$), together with
\begin{align*}\label{RM_LiWe2003_initialdata}
 z_1(\alpha,0) &= \alpha  \,, \\
z_2(\alpha,0) &= 0.5 +  0.1\cos(6 \pi \alpha) \,, \\
\varpi(\alpha,0) &= 0\,. 
\end{align*} 

\begin{figure}[h]
\centering
\subfigure[Multiscale $z$]{\label{fig:RM_LiWe2003_z_10x80_64}\includegraphics[width=35mm]{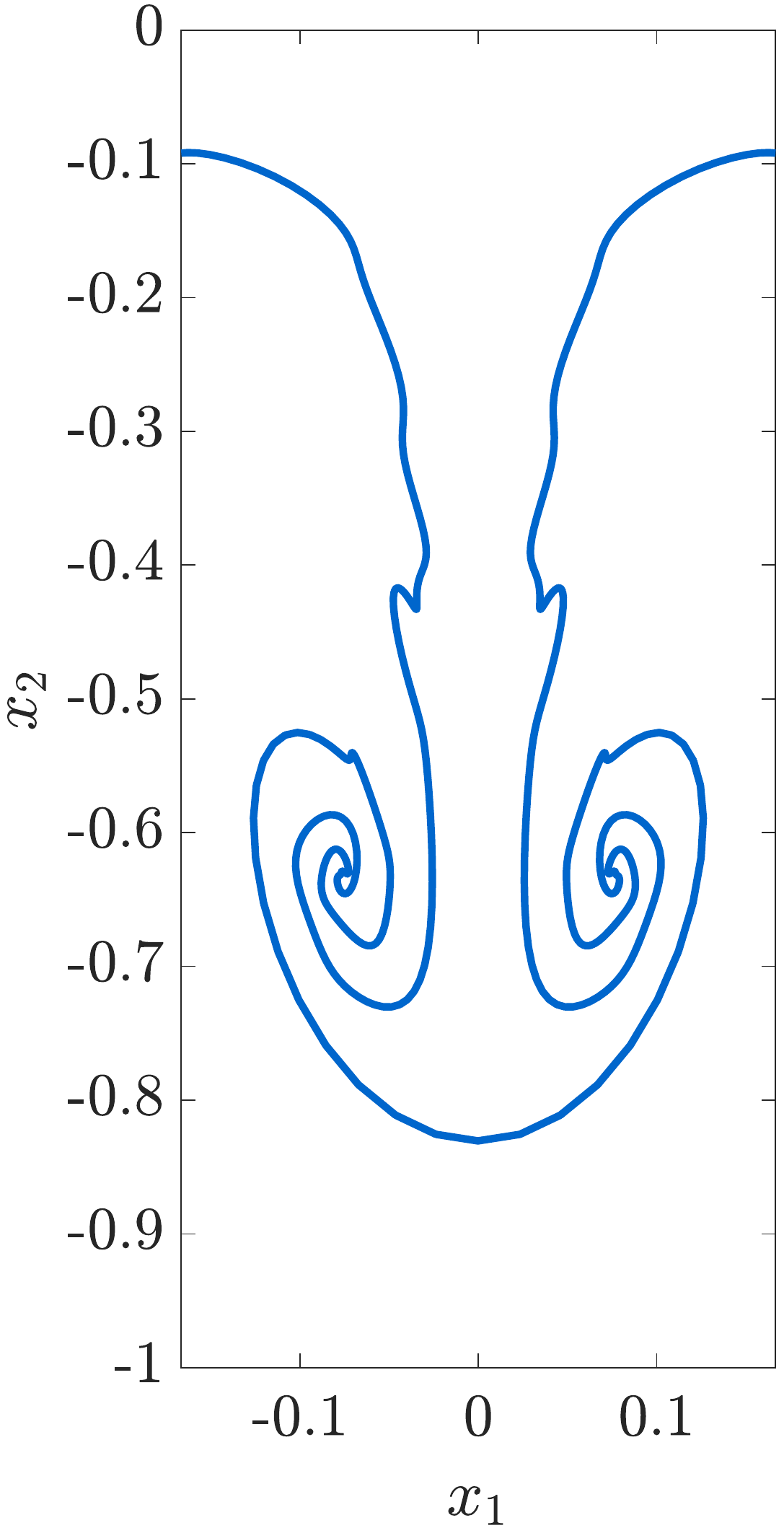}}
\hspace{2em}
\subfigure[High-res.]{\label{fig:RM_LiWe2003_z_highres}\includegraphics[width=27.175mm]{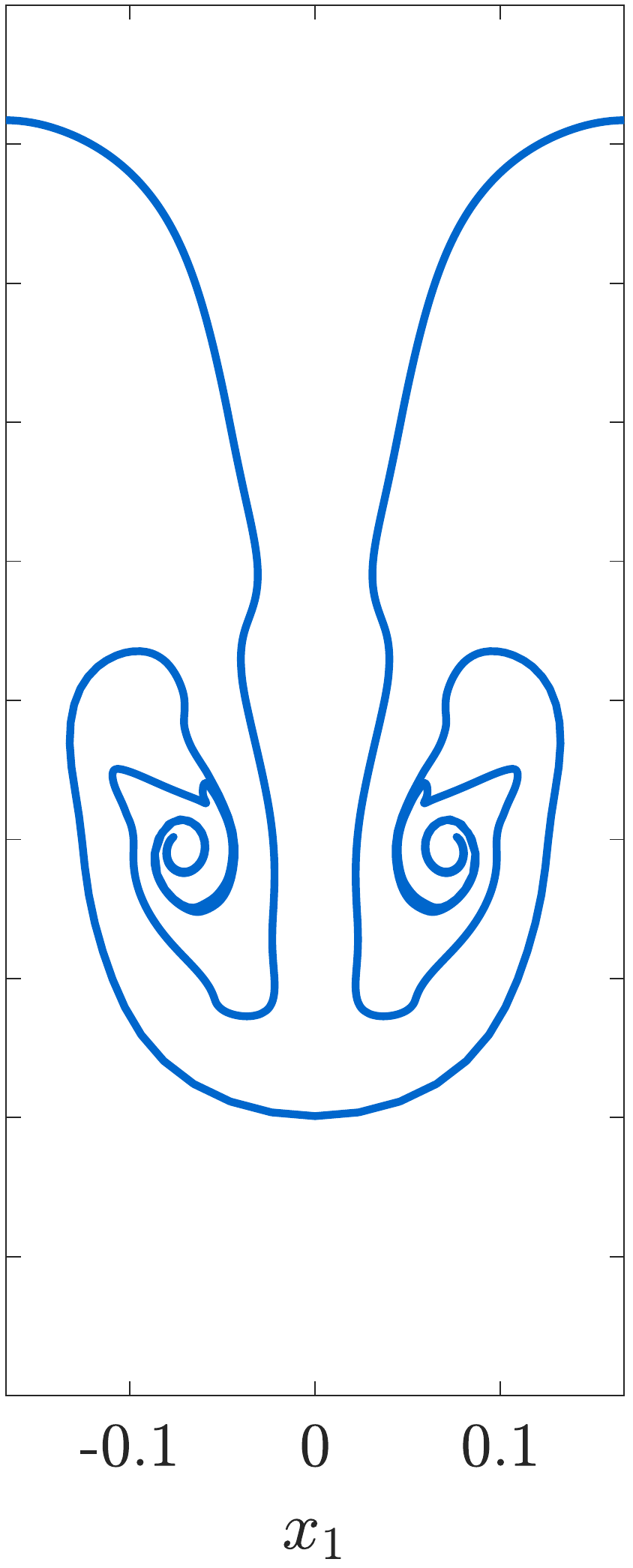}}
\hspace{2em}
\subfigure[Comparison]{\label{fig:RM_LiWe2003_z_compare_10x80_64}\includegraphics[width=27.175mm]{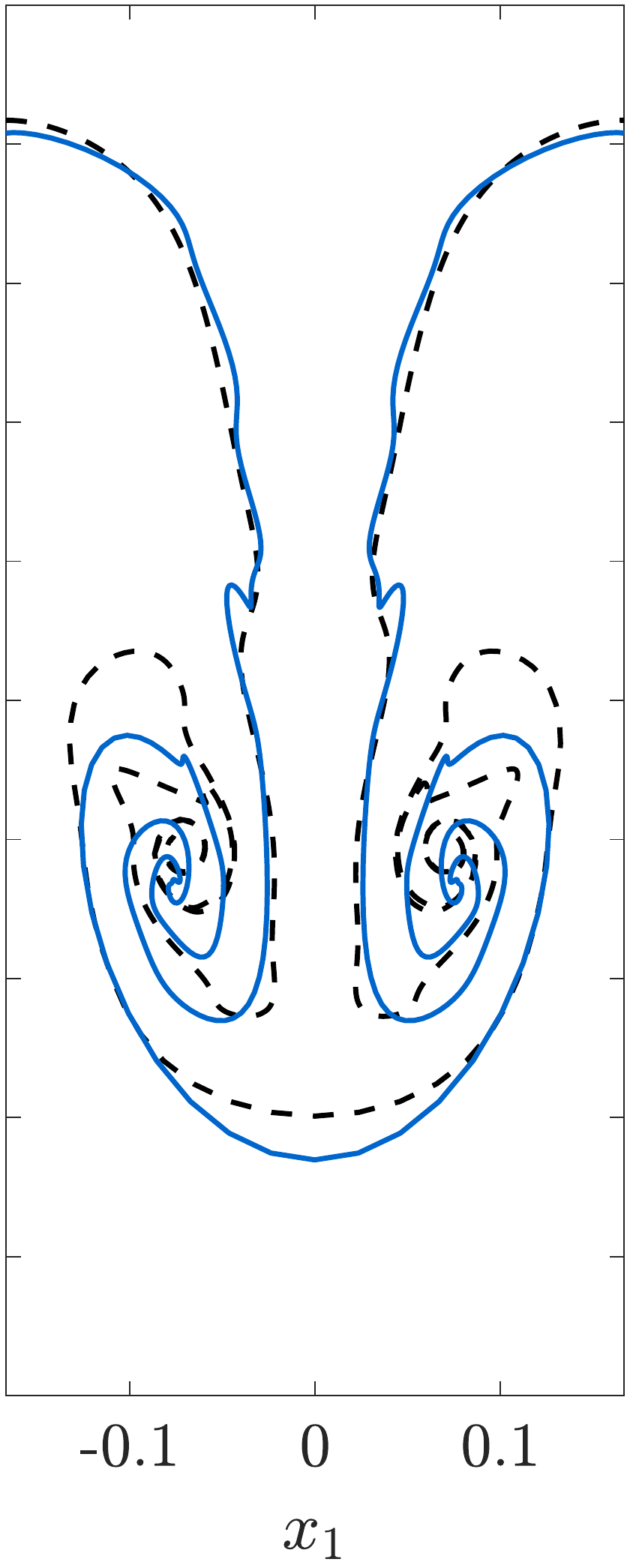}}
\caption{Results for the multiscale algorithm applied to single-mode RMI test, with the 
underlying grid containing $10 \times 80$ cells and the interface discretized with $N=64$. Shown are (a)  
computed interface parametrization $z$, (b) benchmark interface position, and (c) 
the computed interface $z$ (blue) overlaying the benchmark solution (dashed black)
at the final time $t=1.6$.}
\label{fig:RM_LiWe2003_10x80_64}
\end{figure}

\begin{figure}[h]
\centering
\subfigure[$t=0.1$]{\label{fig:RM_LiWe2003_z_10x80_64_t=10}\includegraphics[width=26.475mm]{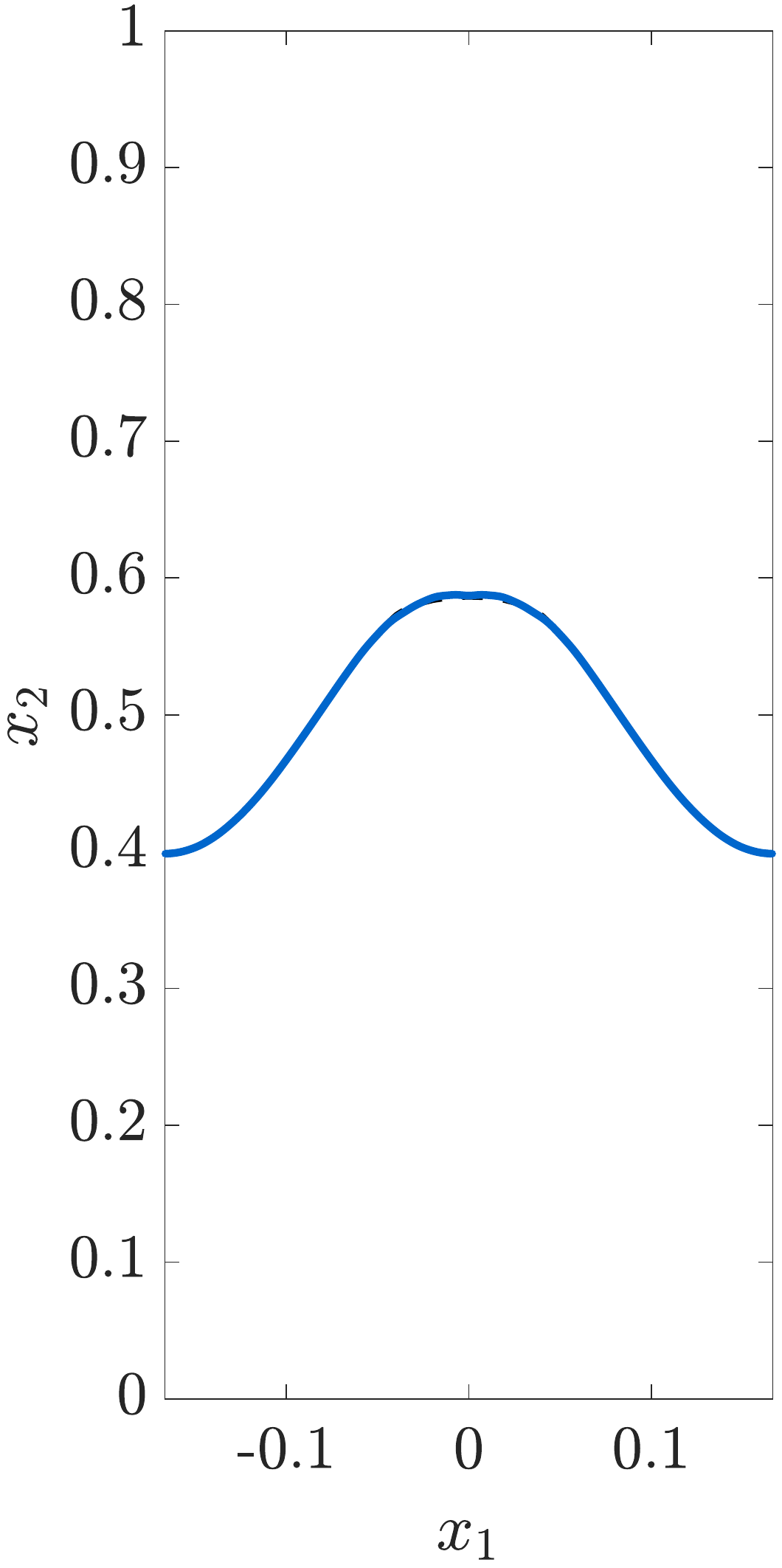}}
\hspace{1em}
\subfigure[$t=0.4$]{\label{fig:RM_LiWe2003_z_10x80_64_t=40}\includegraphics[width=25.0mm]{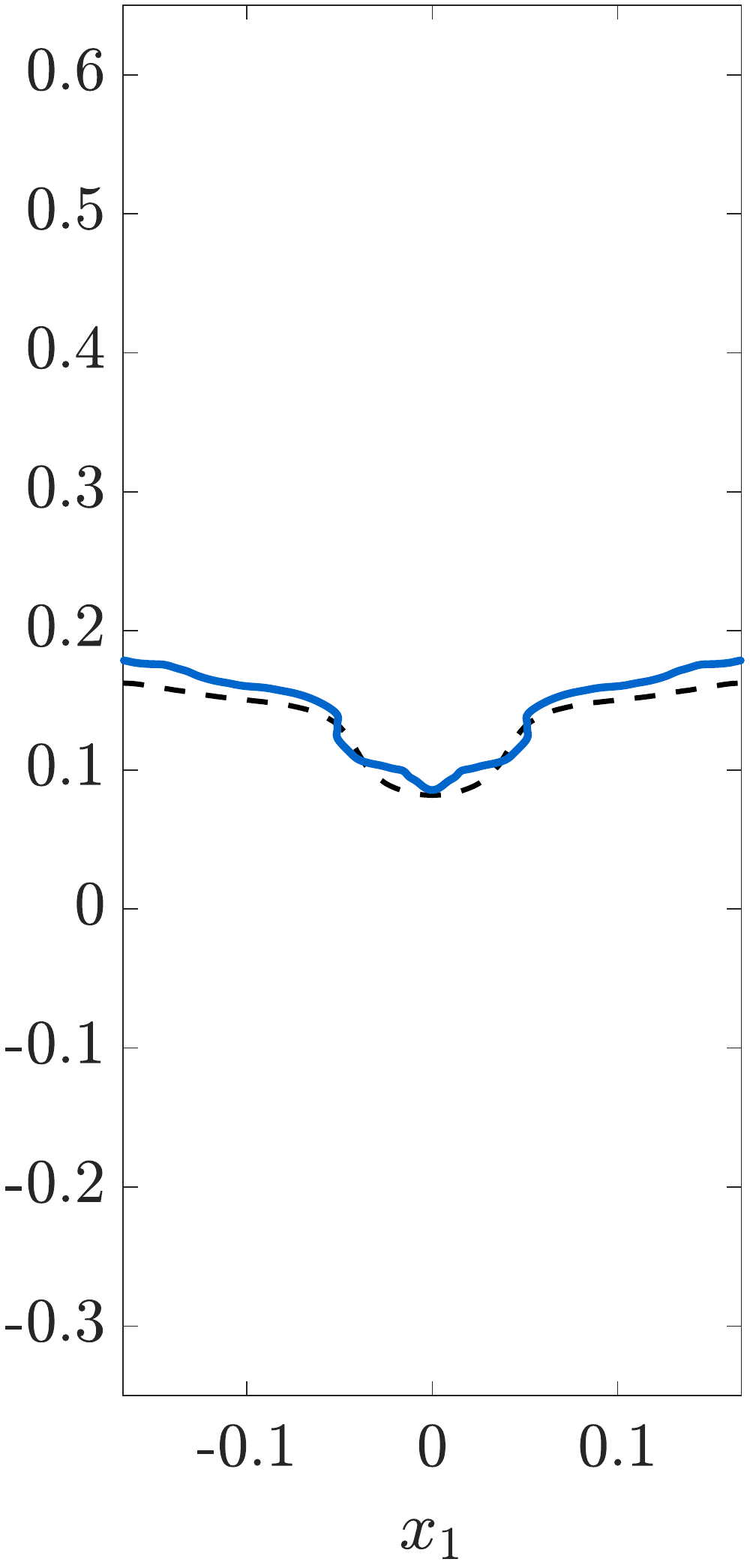}}
\hspace{1em}
\subfigure[$t=0.8$]{\label{fig:RM_LiWe2003_z_10x80_64_t=80}\includegraphics[width=25.0mm]{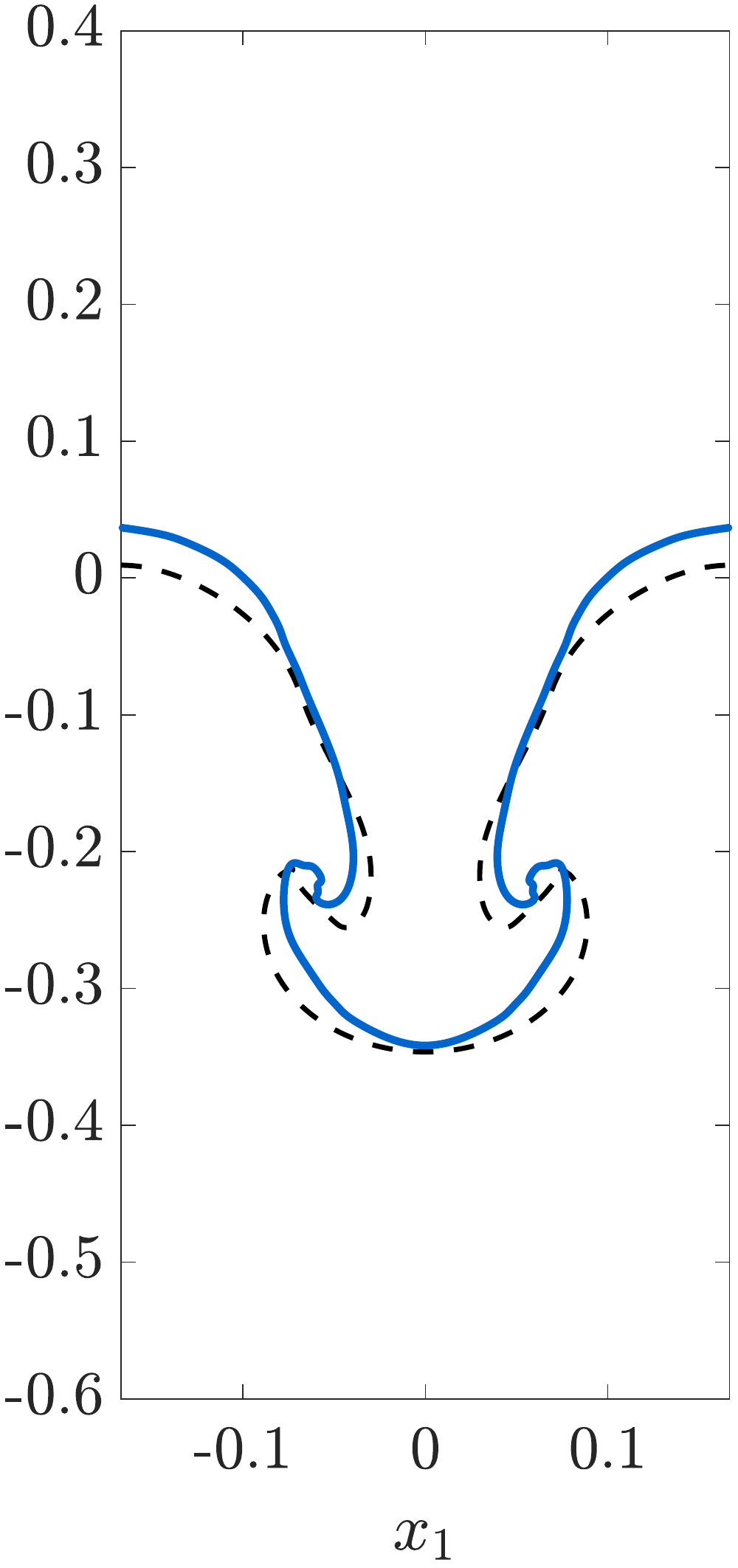}}
\hspace{1em}
\subfigure[$t=1.2$]{\label{fig:RM_LiWe2003_z_10x80_64_t=120}\includegraphics[width=25.0mm]{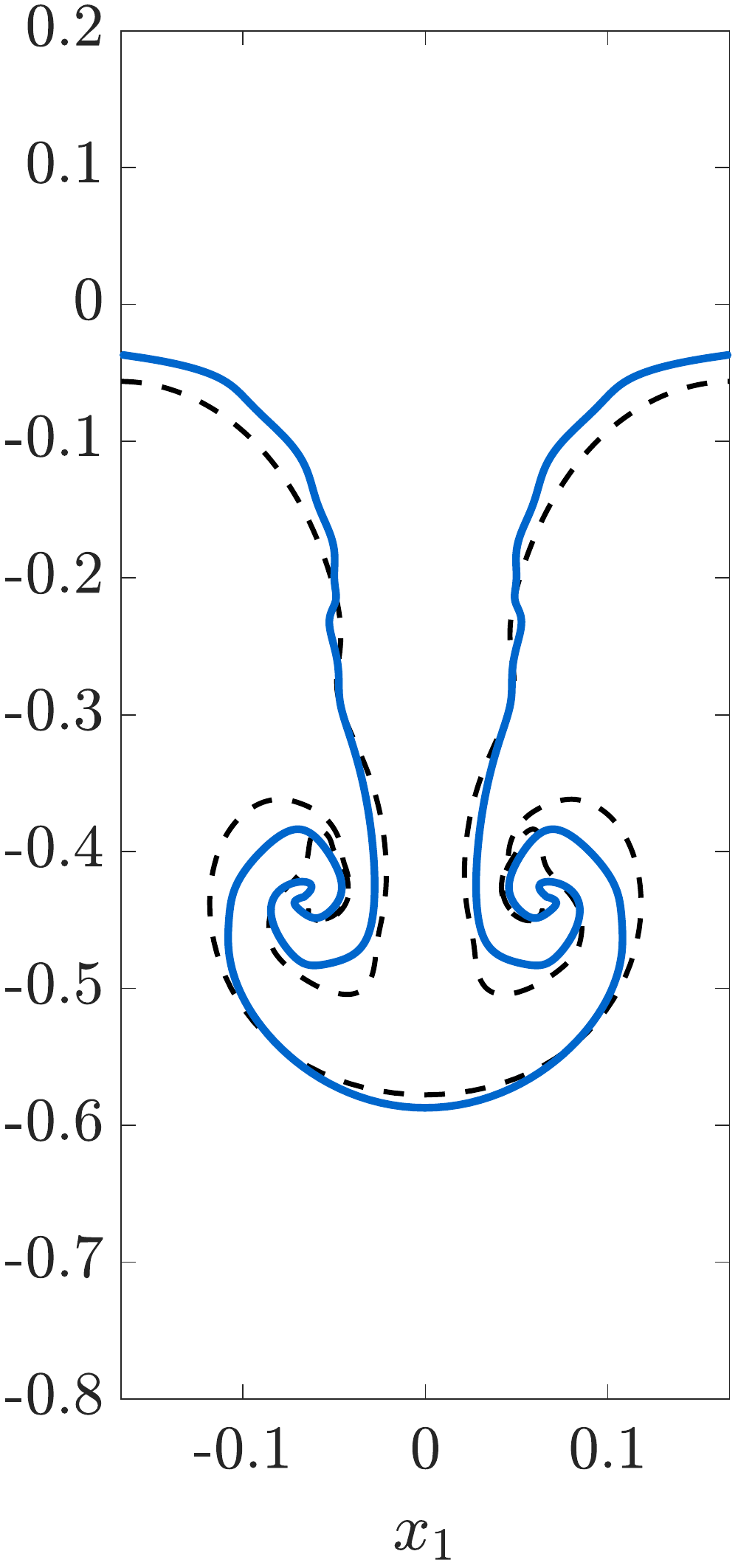}}
\hspace{1em}
\subfigure[$t=1.6$]{\label{fig:RM_LiWe2003_z_10x80_64_t=160}\includegraphics[width=25.0mm]{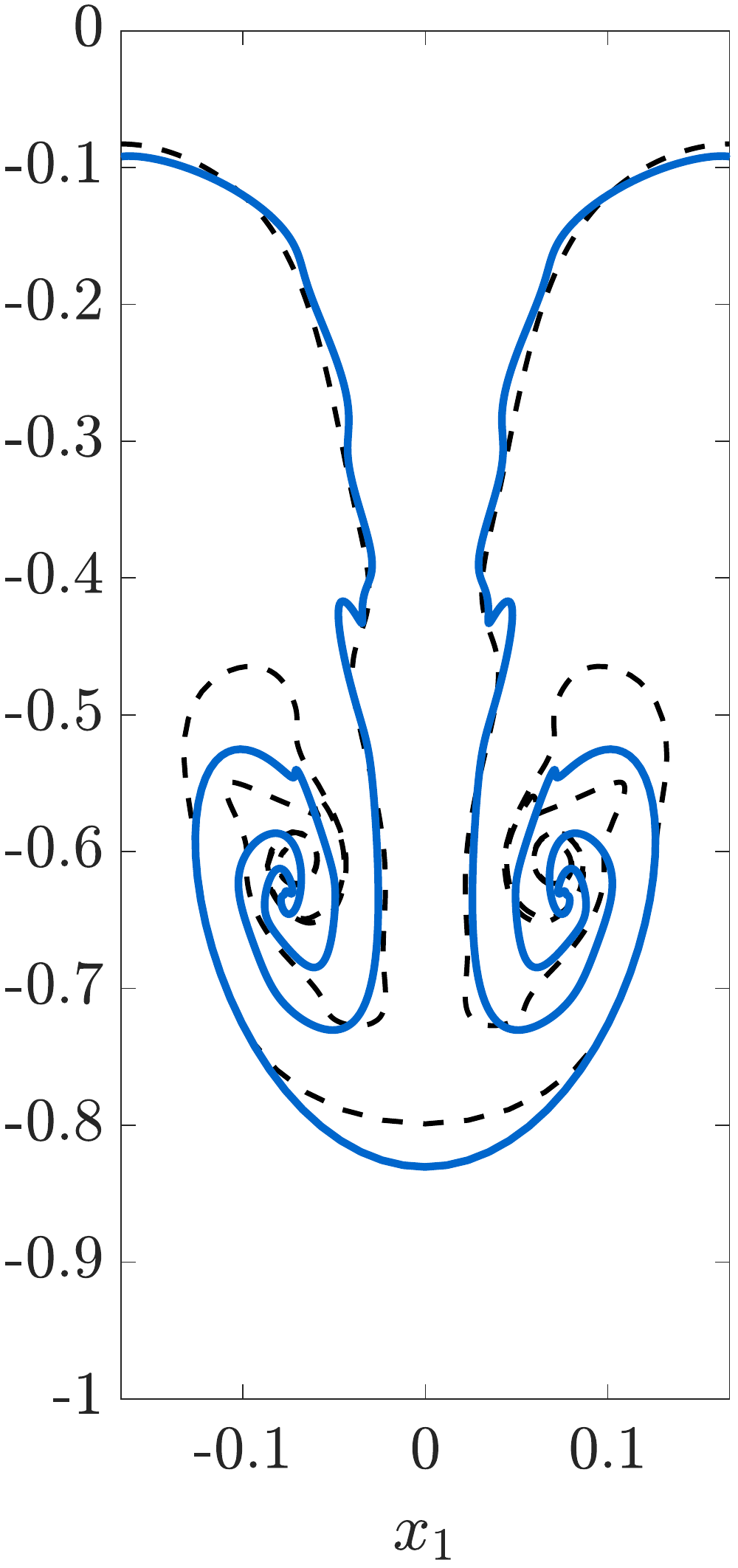}}
\hspace{1em}
\caption{Evolution over time $t$ of the interface for the single-mode RMI. 
Here $z$ is computed using the multiscale algorithm on a mesh with $10 \times 80$ cells and an 
interface discretized with $N=64$. The blue curve is the computed $z(\alpha,t)$, and the dashed black 
curve is the reference solution.}
\label{fig:RM_LiWe2003_10x80_64_time_evolution}
\end{figure}

We present results for the computed interface parametrization at the final time $t=1.6$ in 
\Cref{fig:RM_LiWe2003_z_10x80_64}. Plots showing the evolution of the interface over time are shown in 
\Cref{fig:RM_LiWe2003_10x80_64_time_evolution}. 
The computed interface position $z$ agrees well with the reference 
solution for the duration of simulation, and both interfaces display similar amounts of roll-up at the 
final time.The positions of the bubbles coincide for the two 
solution, while the spike tip positions are also in good agreement. 
We also note that in the high resolution reference solution
 there is slight ``kink'' in the ``stem'' of the mushroom; this ``kink'' is also displayed in the interface 
 computed using the multiscale algorithm. 
 
The approximate runtime of the reference solution calculation is 
$T_{\mathrm{CPU}} \approx 1653$ s, whereas our multiscale algorithm computation had a runtime of only 
$T_{\mathrm{CPU}} \approx 15$ s, yielding a speed-up factor of almost 110 times. As with the multiscale algorithm applied to the 
RTI, we are able to accurately predict large scale quantities and small scale structure with minimal 
computational expense. 

\subsubsection{The effects of the baroclinic term}
\begin{figure}[h]
\centering
\subfigure[$t=0.1$]{\label{fig:RM_LiWe2003_zbaro_10x80_64_t=10}\includegraphics[width=26.475mm]{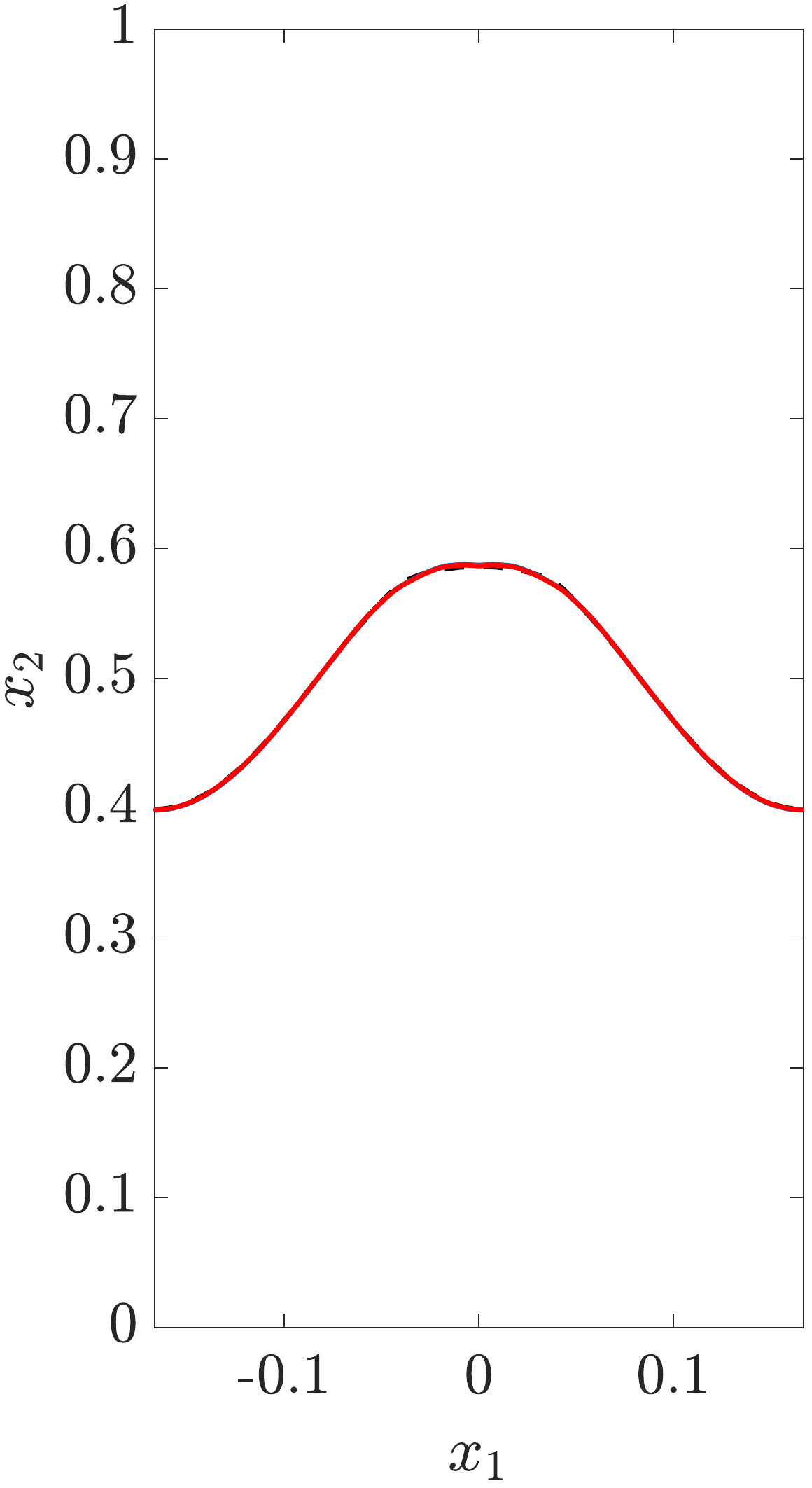}}
\hspace{1em}
\subfigure[$t=0.4$]{\label{fig:RM_LiWe2003_zbaro_10x80_64_t=40}\includegraphics[width=25.0mm]{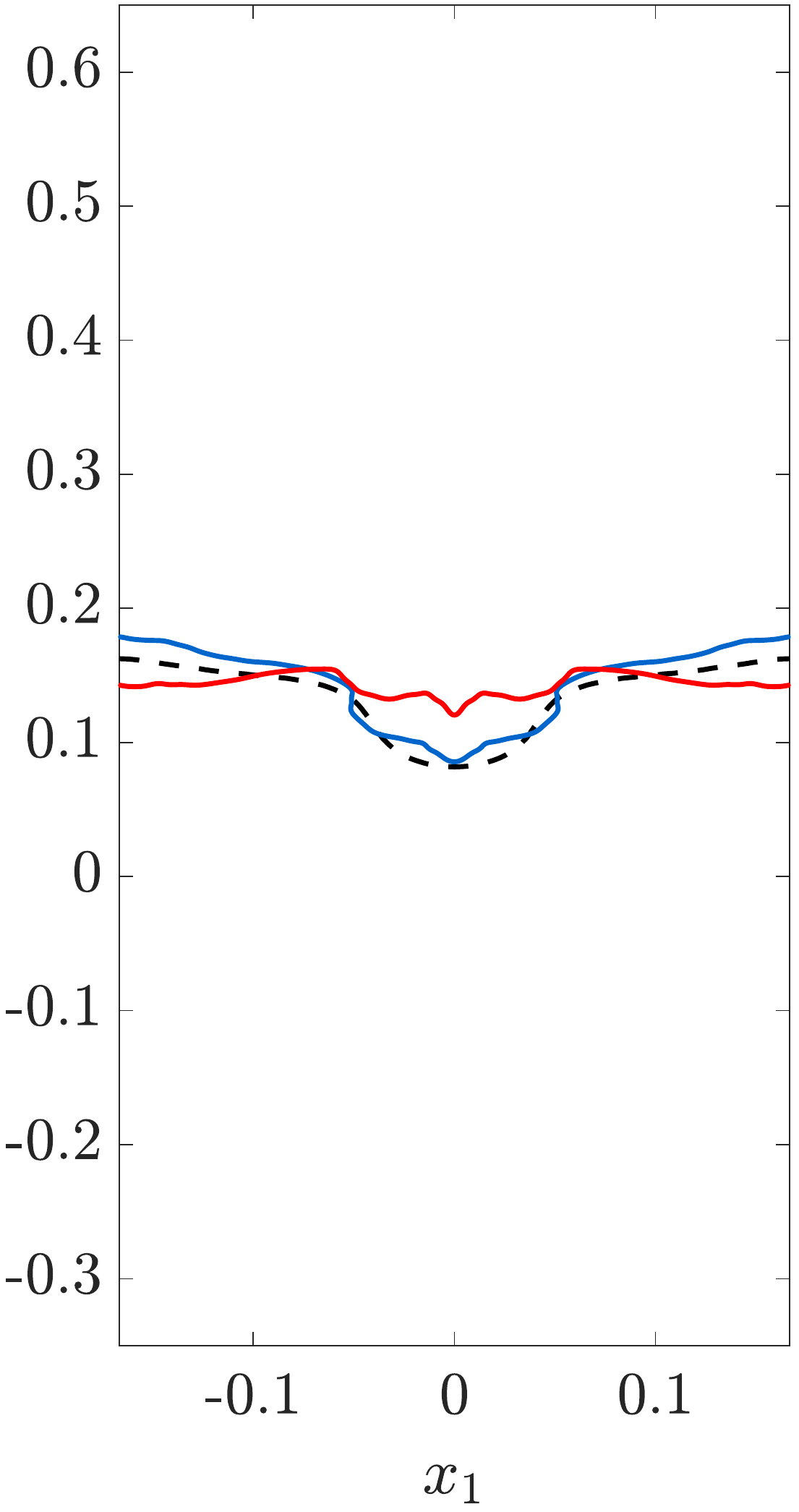}}
\hspace{1em}
\subfigure[$t=0.8$]{\label{fig:RM_LiWe2003_zbaro_10x80_64_t=80}\includegraphics[width=25.0mm]{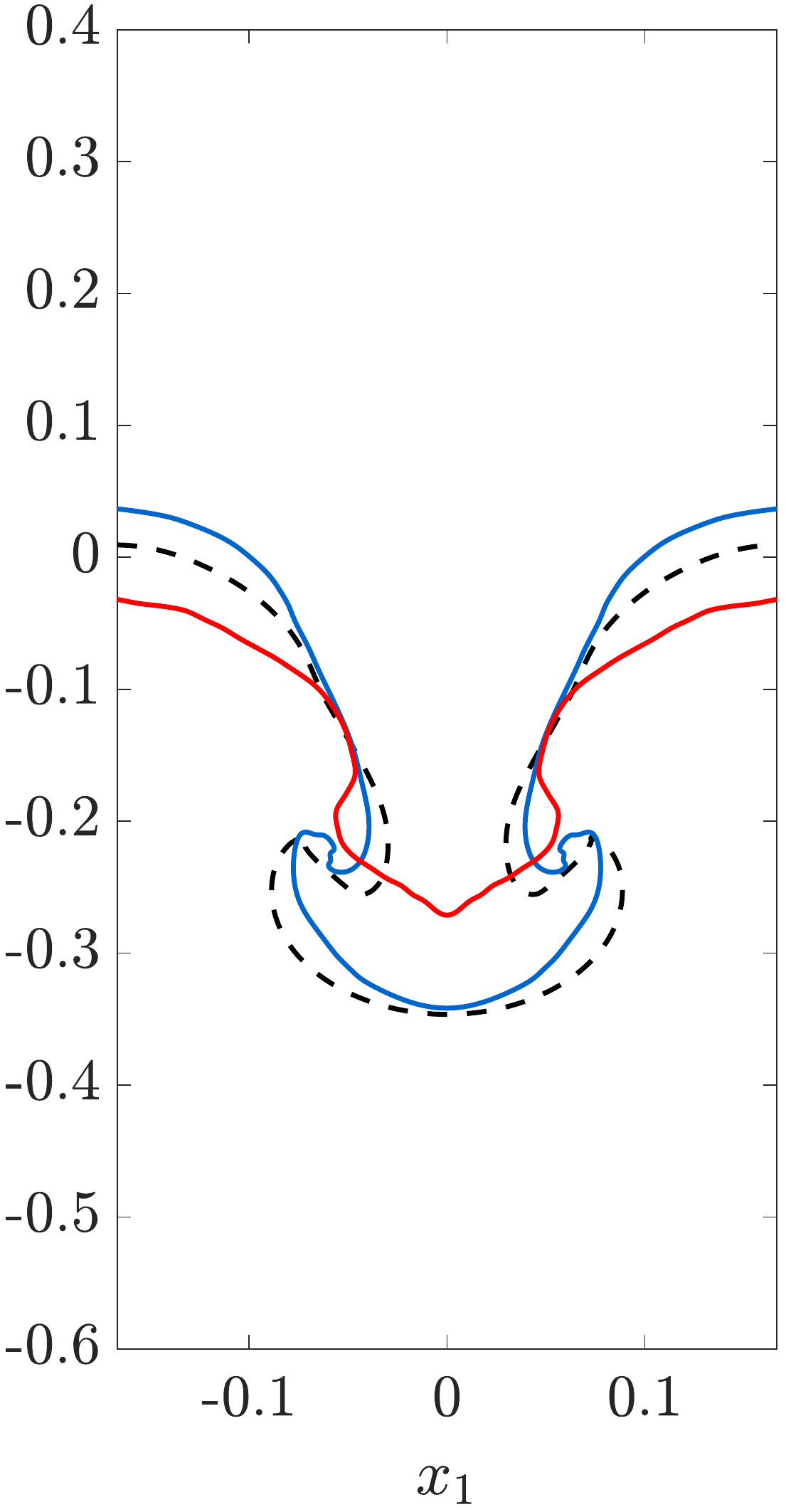}}
\hspace{1em}
\subfigure[$t=1.2$]{\label{fig:RM_LiWe2003_zbaro_10x80_64_t=120}\includegraphics[width=25.0mm]{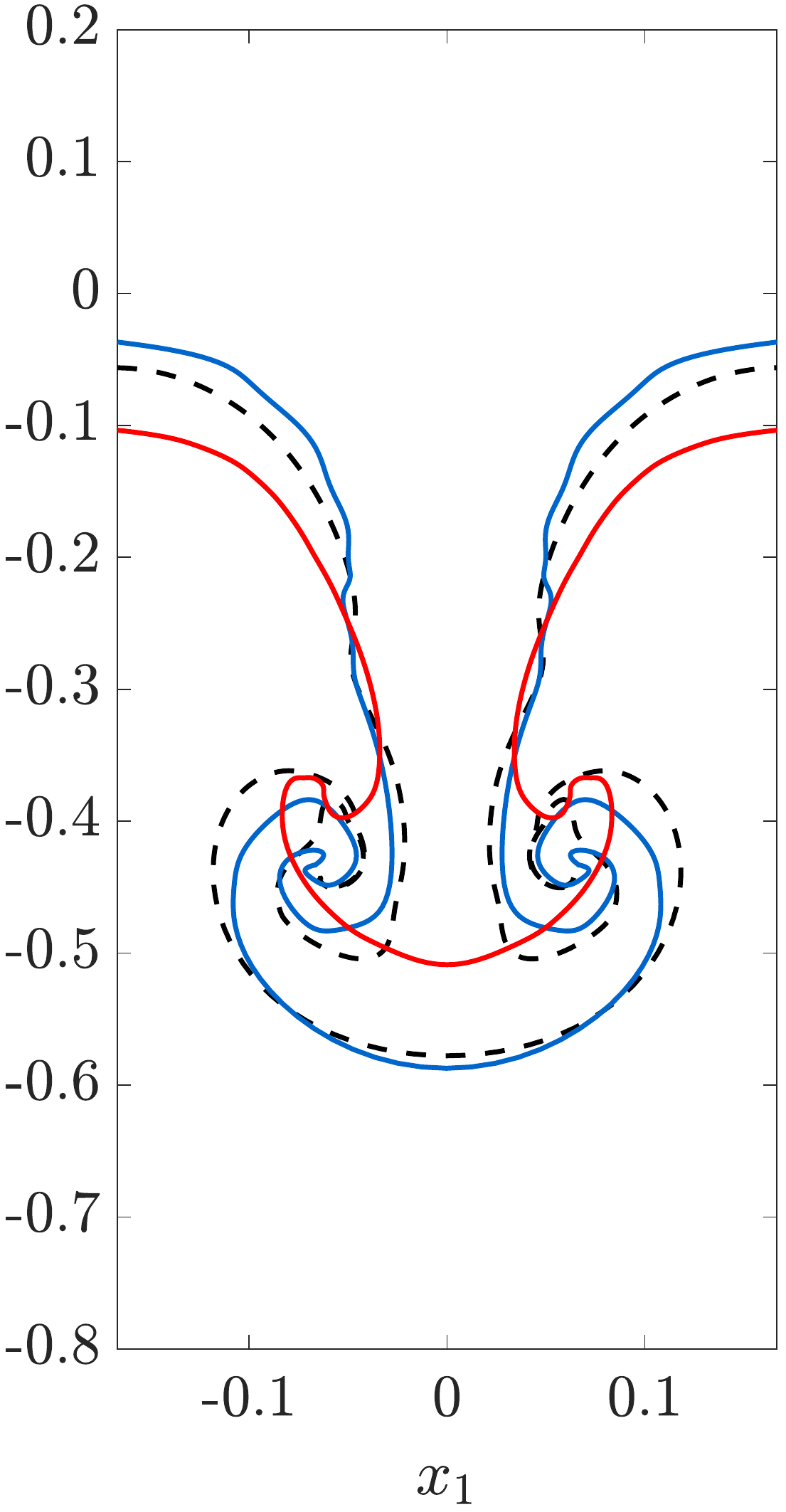}}
\hspace{1em}
\subfigure[$t=1.6$]{\label{fig:RM_LiWe2003_zbaro_8x46_64_t=160}\includegraphics[width=25.0mm]{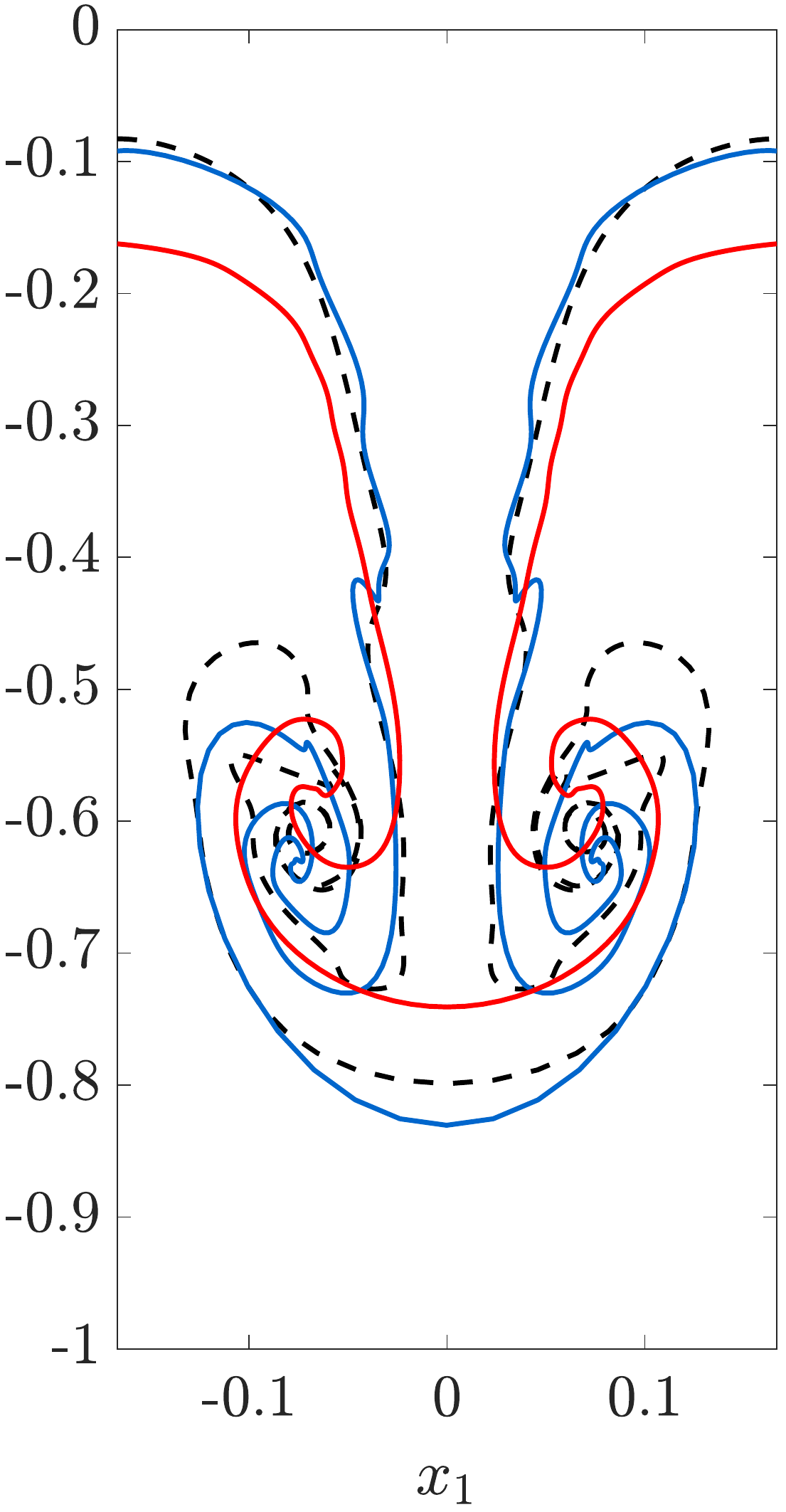}}
\hspace{1em}
\caption{Evolution over time $t$ of the interface for the single-mode RMI. 
Here, the red curve is $z$ computed using the multiscale algorithm, 
but without the use of the baroclinic term in the 
$\varpi$-equation, on a mesh with $10 \times 80$ cells and an 
interface discretized with $N=64$. The blue curve is $z$ computed using the complete 
multiscale algorithm, while dashed black curve is the reference solution.}
\label{fig:RM_LiWe2003_10x80_64_baro_time_evolution}
\end{figure}

In this section, we briefly discuss the importance of 
including the baroclinic term in the modified $\varpi$-equation 
\eqref{zmodelRM-b}. This term is crucial in ensuring that the interface $z$ is advected 
by the correct velocity.  Without 
this term, information about the baroclinic deposition of vorticity by the shock 
on the interface is not transmitted to 
the incompressible portion of the multiscale algorithm. 
We show in \Cref{fig:RM_LiWe2003_10x80_64_baro_time_evolution} 
the evolution of a multiscale solution,  computed  without the inclusion of the baroclinic term,  but
 otherwise identical to the RMI multiscale algorithm. This solution (displayed as a red curve) is compared 
against the solution obtained using the actual RMI multiscale algorithm (displayed as a blue curve), as well 
as the high resolution reference solution (displayed as a dashed black curve). It is clear that the omission of the baroclinic term leads to a solution with an 
incorrect interface position, and with significantly less KH roll-up.

\begin{figure}[h]
\centering
\subfigure[$t=0.05$]{\label{fig:RM_LiWe2003_wbar_t=5}\includegraphics[width=29mm]{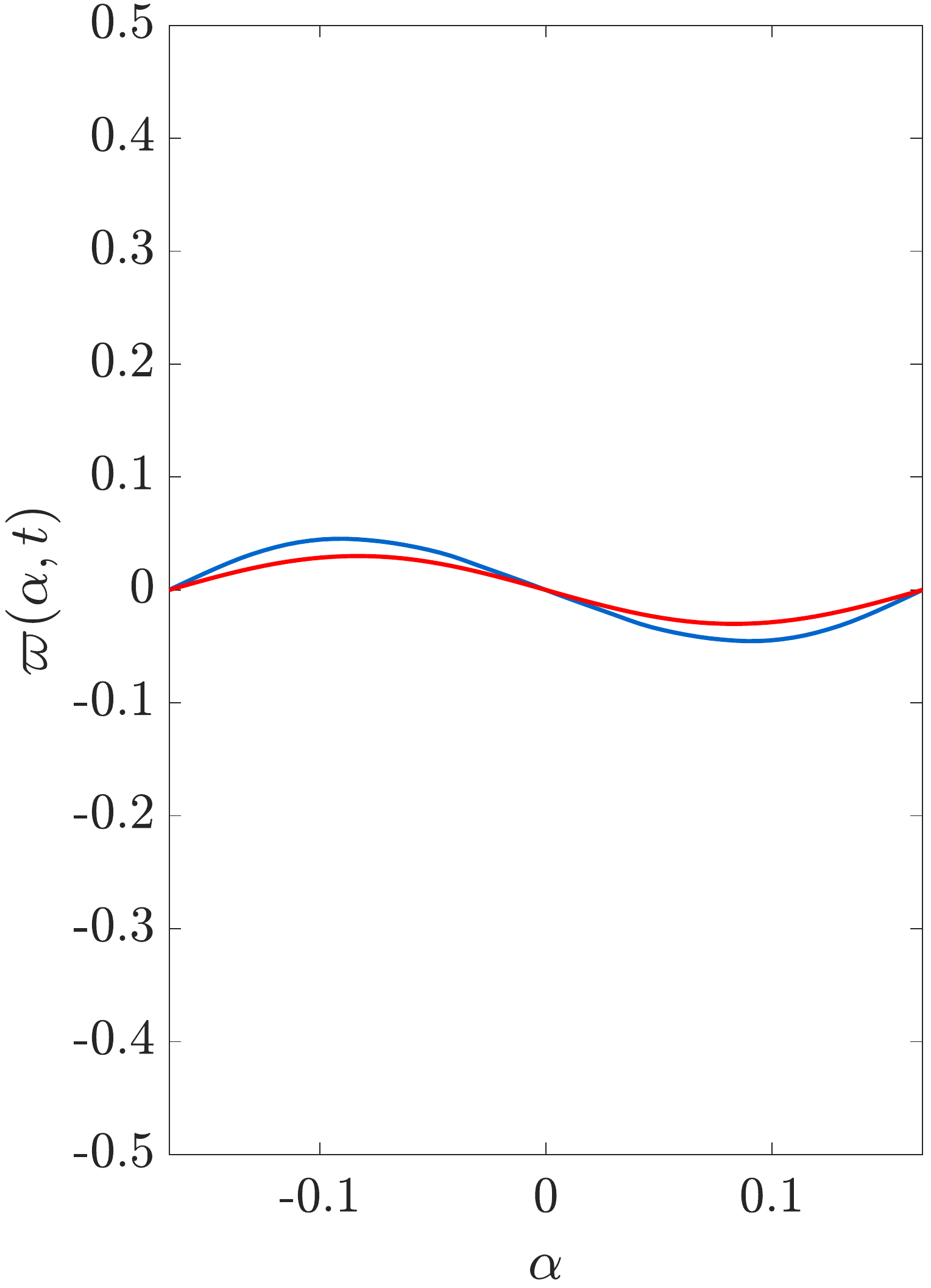}}
\hspace{0.2em}
\subfigure[$t=0.10$]{\label{fig:RM_LiWe2003_wbar_t=10}\includegraphics[width=24mm]{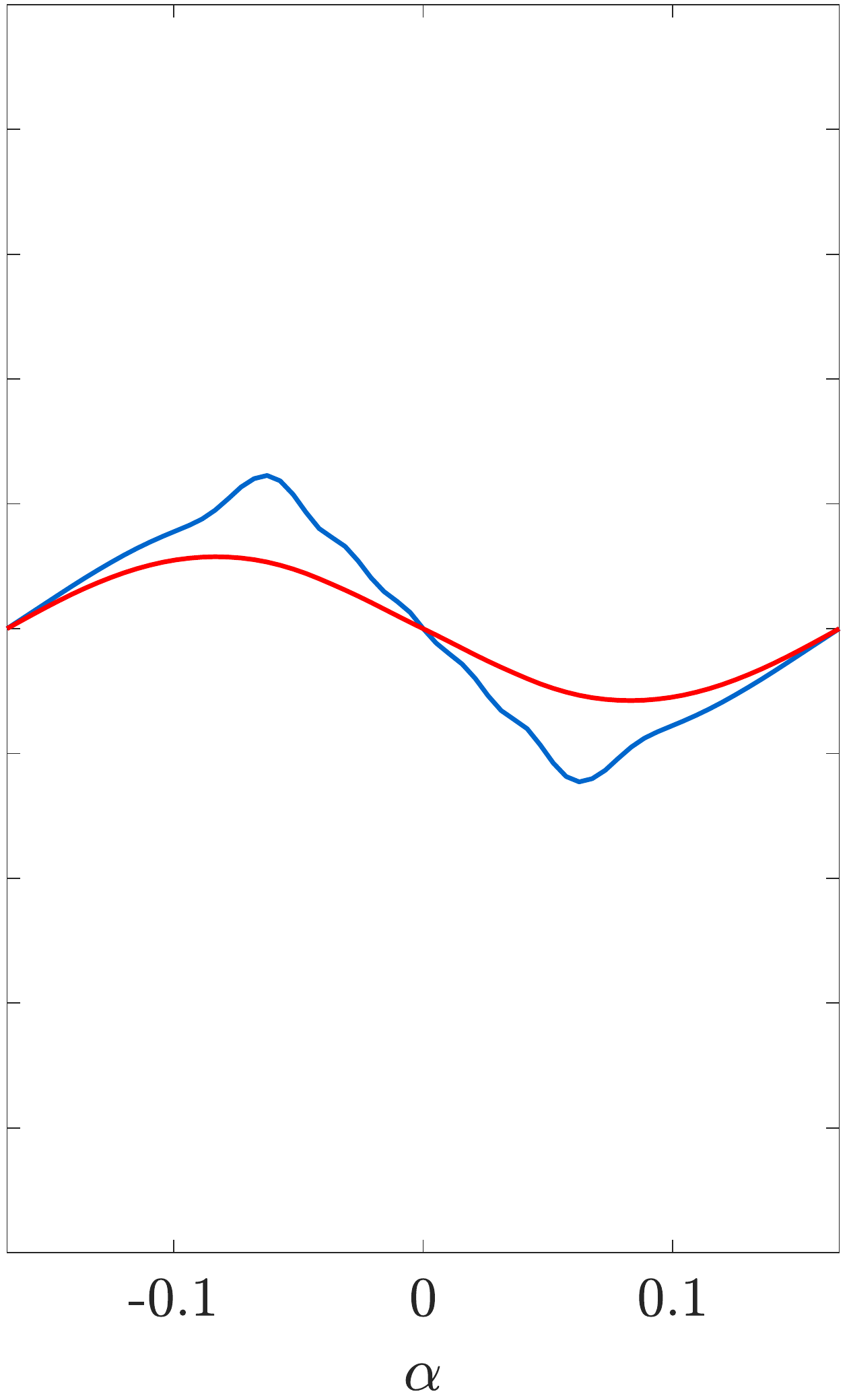}}
\hspace{0.2em}
\subfigure[$t=0.12$]{\label{fig:RM_LiWe2003_wbar_t=12}\includegraphics[width=24mm]{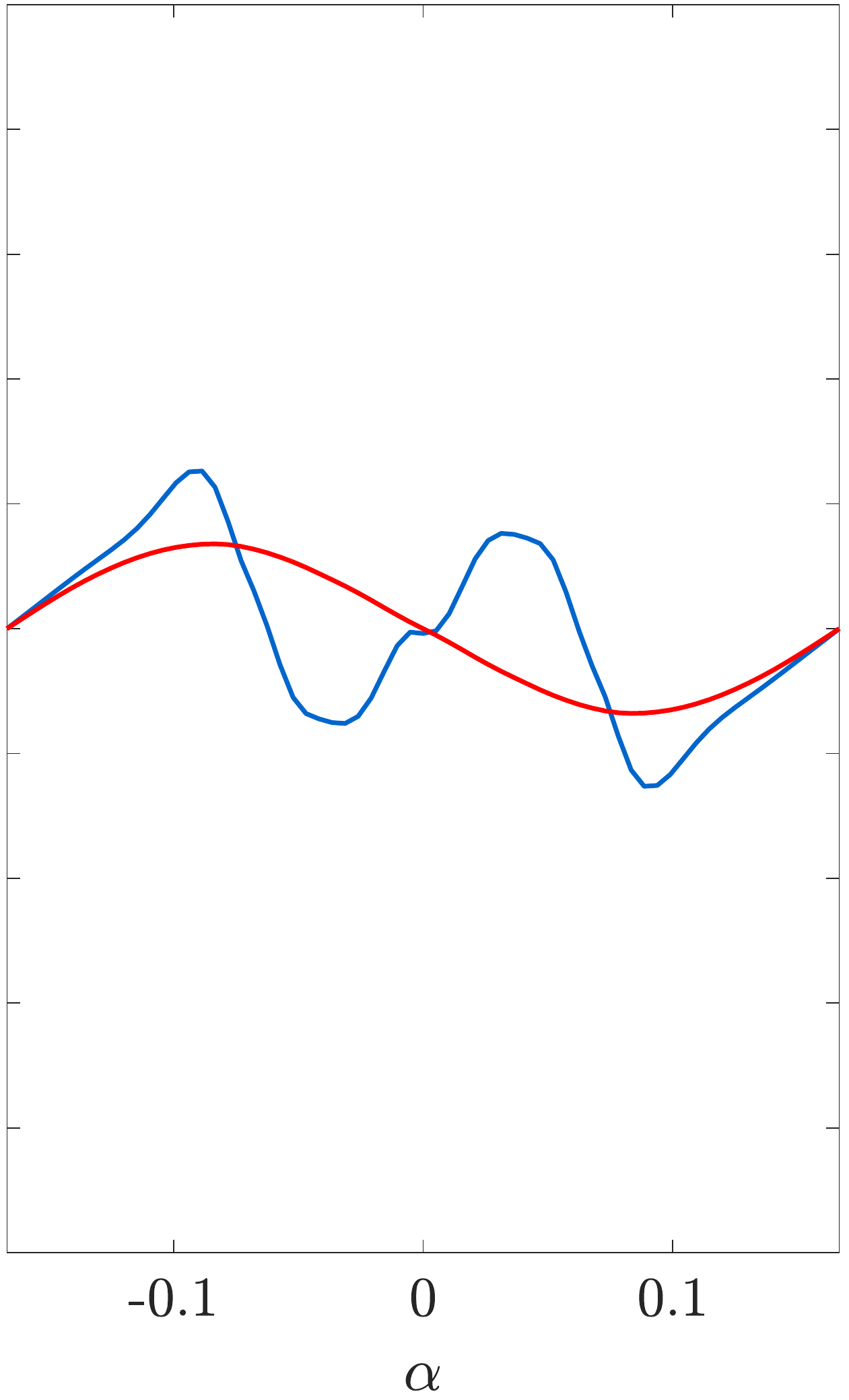}}
\hspace{0.2em}
\subfigure[$t=0.15$]{\label{fig:RM_LiWe2003_wbar_t=15}\includegraphics[width=24mm]{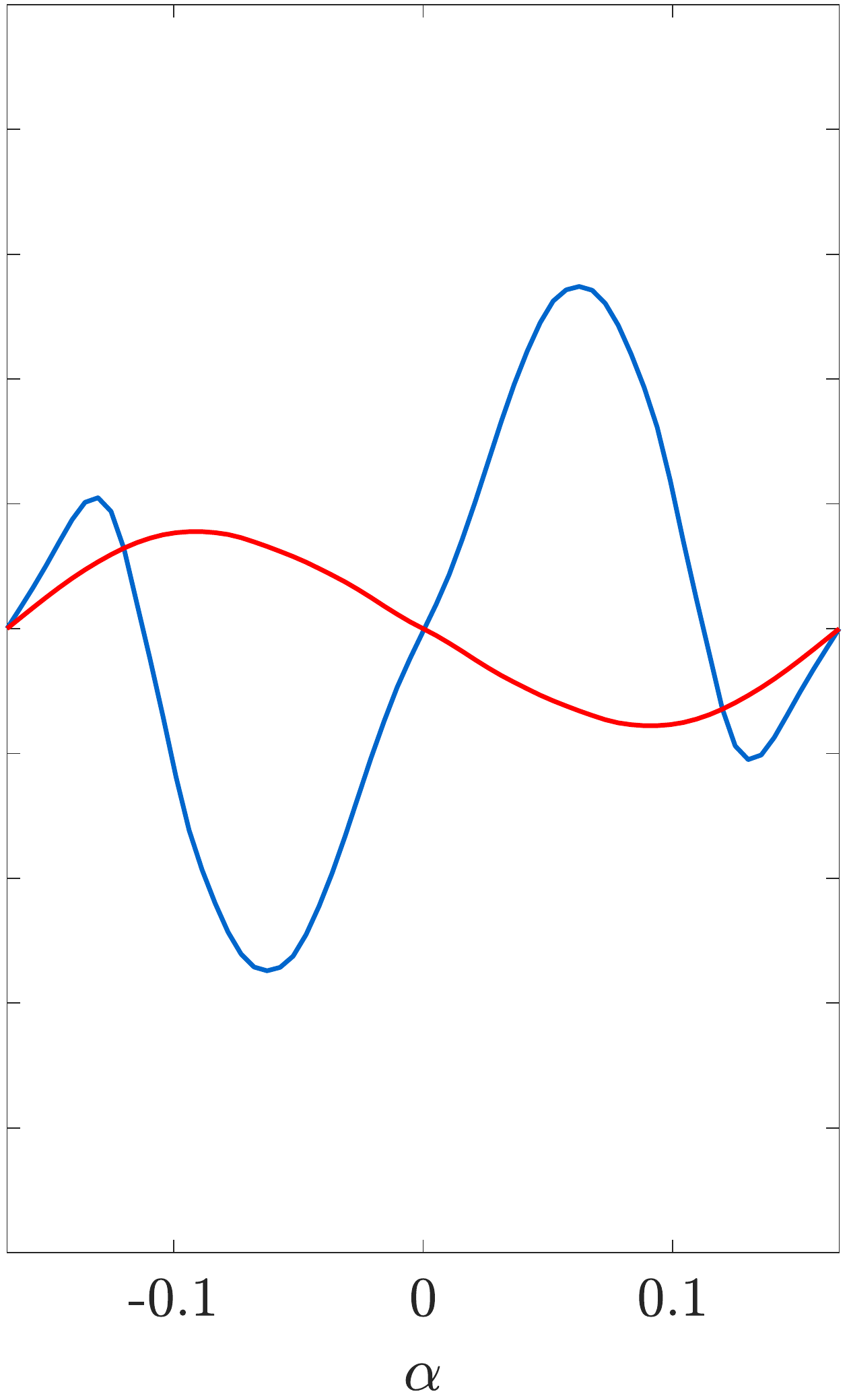}}
\hspace{0.2em}
\subfigure[$t=0.20$]{\label{fig:RM_LiWe2003_wbar_t=20}\includegraphics[width=24mm]{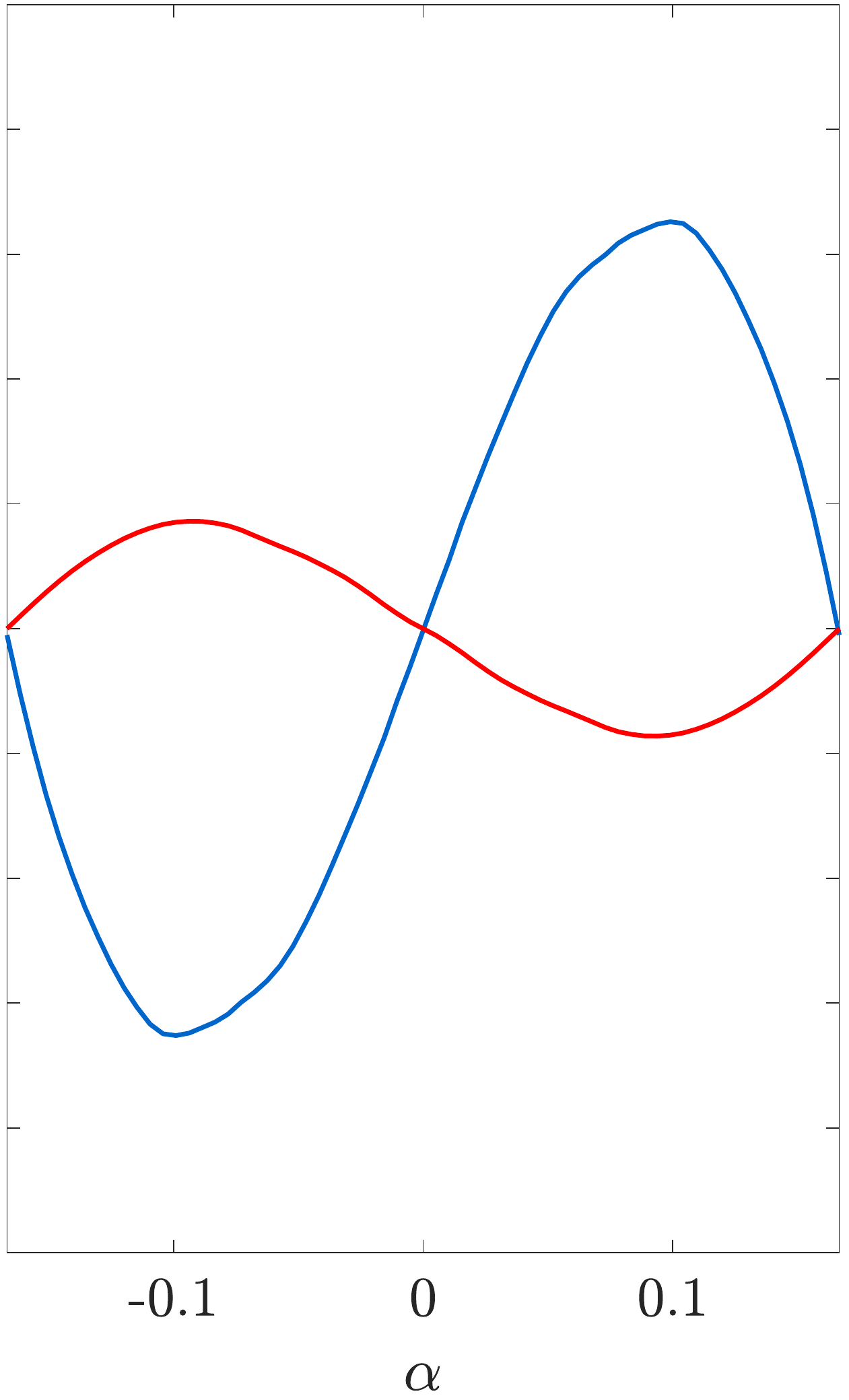}}
\hspace{0.2em}
\subfigure[$t=0.70$]{\label{fig:RM_LiWe2003_wbar_t=70}\includegraphics[width=24mm]{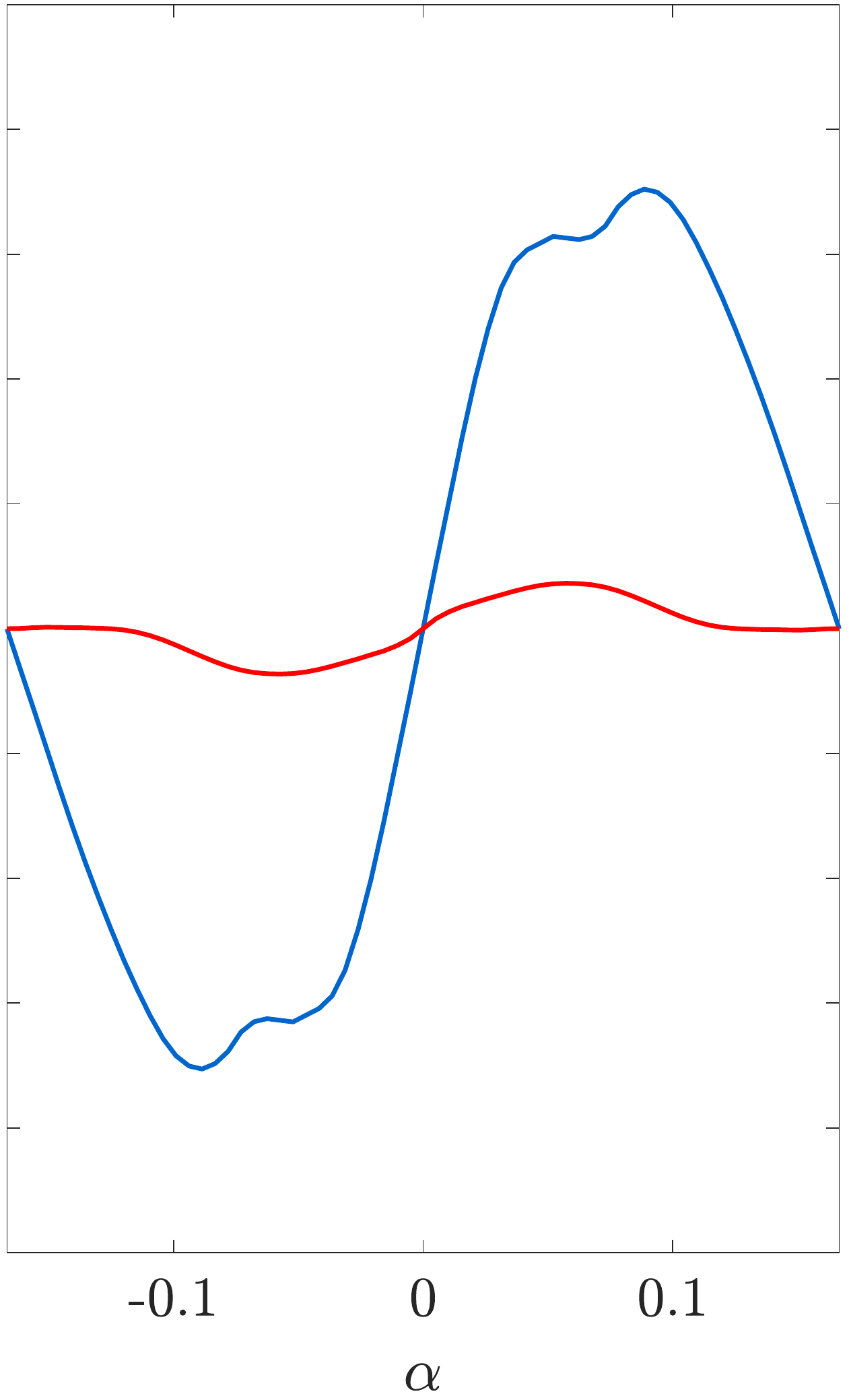}}
\hspace{0.2em}
\caption{Evolution over time $t$ of the amplitude of vorticity $\varpi(\alpha,t)$ 
versus $\alpha$ for the single-mode RMI. 
Here, the red curve is $\varpi$ 
computed using the multiscale algorithm, but without the use of the baroclinic term in the 
$\varpi$-equation, on a mesh with $10 \times 80$ cells and an 
interface discretized with $N=64$. The blue curve is $\varpi$ computed using the complete 
multiscale algorithm.}
\label{fig:RM_LiWe2003_wbar_time_evolution}
\end{figure}

In the absence of this baroclinic term, 
while the coarse scale shock velocity $v$ still advects the 
interface upon shock-contact collision, the computed amplitude of vorticity $\varpi$
(shown in \Cref{fig:RM_LiWe2003_wbar_time_evolution} as a red curve) has the wrong sign after the shock-contact 
collision at $t \approx 0.12$, which 
yields an incorrect calculation of the fine scale velocity $w$, and subsequently an incorrect interface position 
$z$.

We compare this incorrect $\varpi$ with the
 amplitude of vorticity $\varpi$ computed using the complete RMI multiscale algorithm (shown in 
\Cref{fig:RM_LiWe2003_wbar_time_evolution} as a blue curve);
without the baroclinic term, $\varpi$ has the wrong sign until time $t=0.70$, at which point 
the advection of the interface by the coarse scale velocity $v$ 
\emph{forces} $\varpi$ to have the correct sign. This is in contrast to our RMI
multiscale algorithm, in which the baroclinic term forces $\varpi$ to switch sign after the shock-contact 
interaction, which results in the correct computation of the fine grid velocity $w$ and, consequently, the correct interface position.


\subsection{The RMI test of Nourgaliev et al.}\label{subsec:RM_TeTr2009}

We next consider the RMI problem introduced in \cite{NoDiTh2006}, and later considered in 
\cite{TeTr2009,NoMoTeObFu2012,WoLe2017}. A heavy fluid of density $\rho^{+}=5.04$ 
lies below a light fluid of density $\rho^{-}=1$, and a planar Mach 1.24 shock travels vertically 
downards through the light fluid and eventually collides with the interface separating the two fluids, resulting 
in a transmitted shock and a reflected shock. The instability is generated by the subsequent acceleration of 
the light fluid into the heavy fluid. 
The domain is $(x_1,x_2) \in [-0.5,0.5] \times [0,4]$, the 
adiabatic constant is $\gamma = 1.276$, gravity is assumed to be negligible (i.e. $g=0$), and the initial data is 
defined as follows. The initial interface $\Gamma_0$ is parametrized by 
$(x_1,\eta_0(x_1))$, with $\eta_0(x_1)=2.9+0.1 \cos(2 \pi x_1)$, and the 
initial shock position is at $x_2=3.2$. The initial 
horizontal velocity vanishes $u_1(x,0)=0$, and the initial vertical velocity satisfies 
$u_2(x,0) = -0.550368 \cdot \mathbbm{1}_{x_2 \geq 3.2}(x)$. The initial pressure is 
$p_0(x) = \mathbbm{1}_{x_2 < 3.2}(x) + 1.628 \cdot \mathbbm{1}_{x_2 \geq 3.2}(x)$. As in the 
numerical experiments in \Cref{subsec:RT_CASTRO}, the initial density is smoothed over a length 
scale $h$ using a $\tanh$ profile:
\begin{equation}\label{RM_TeTr2009_initial-rho}
\rho_0(x_1,x_2) = \rho^{-} + \frac{\rho^{+}-\rho^{-}}{2} \left[  1+ \tanh \left(\frac{\psi(x_1) - x_2}{h} \right) \right]\,. 
\end{equation}
For our high resolution reference 
solution computed using the $C$-method on a fine mesh, we use $h=0.005$. 
Periodic boundary conditions are applied in the 
$x_1$-direction, and free-flow conditions are imposed in the $x_2$-direction. 

\begin{figure}[h]
\centering
\includegraphics[width=28mm]{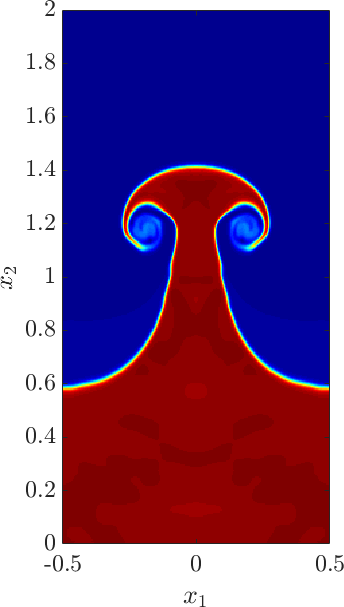}
\caption{The density profile at time $t=8.0$ for the RMI test of \citet{NoDiTh2006}. This high-resolution solution is 
computed using the $C$-method on a mesh with $100 \times 800$ cells.}
\label{fig:RM_TeTr2009_rho_highres}
\end{figure}

Our high resolution reference solution is computed on a mesh with $100 \times 800$ cells and a time-step 
of $\delta t = 6.25 \times 10^{-4}$, which gives $\mathrm{CFL} \approx 0.4$. The computed density is 
shown in \Cref{fig:RM_TeTr2009_rho_highres} at the final time $t=8$, and can be compared with 
Figures 10, 11, 23, and 23 in \cite{NoDiTh2006}, \cite{TeTr2009}, \cite{NoMoTeObFu2012}, and 
\cite{WoLe2017}, respectively. 

\subsubsection{The multiscale algorithm applied to the RMI}
Next, we apply the multiscale RMI algorithm to the problem described above. The relevant parameters are 
chosen as follows. The coarse mesh for $v$ contains $10 \times 80$ cells, the fine mesh for $w$ 
uses $N=128$, and the time-step is set as $\delta t = 6.25 \times 10^{-3}$, giving $\mathrm{CFL} \approx 0.4$.
The initial velocities are $v_1(x,0)=0$ and $v_2(x,0)= -0.550368 \cdot \mathbbm{1}_{x_2 \geq 3.2}(x)$, and
the initial pressure is 
$p_0(x) = \mathbbm{1}_{x_2 < 3.2}(x) + 1.628 \cdot \mathbbm{1}_{x_2 \geq 3.2}(x)$.
The initial density is smoothed using \eqref{RM_TeTr2009_initial-rho} with $h=0.02$. The 
Atwood number is $A \approx 0.67$, and the initial data for the interface calculation is 
\begin{align*}
 z_1(\alpha,0) &= \alpha  \,, \\
z_2(\alpha,0) &= 2.9 +  0.1\cos(2 \pi \alpha) \,, \\
\varpi(\alpha,0) &= 0\,. 
\end{align*} 
The artificial 
viscosity parameters are set as $\beta=1$, $\beta_s=0$, $\tilde{\delta}=1$, and $\varpi=1 \times 10^{-4}$. 

\begin{figure}[h]
\centering
\subfigure[Multiscale $z$]{\label{fig:RM_TeTr2009_z_10x80_128}\includegraphics[width=28mm]{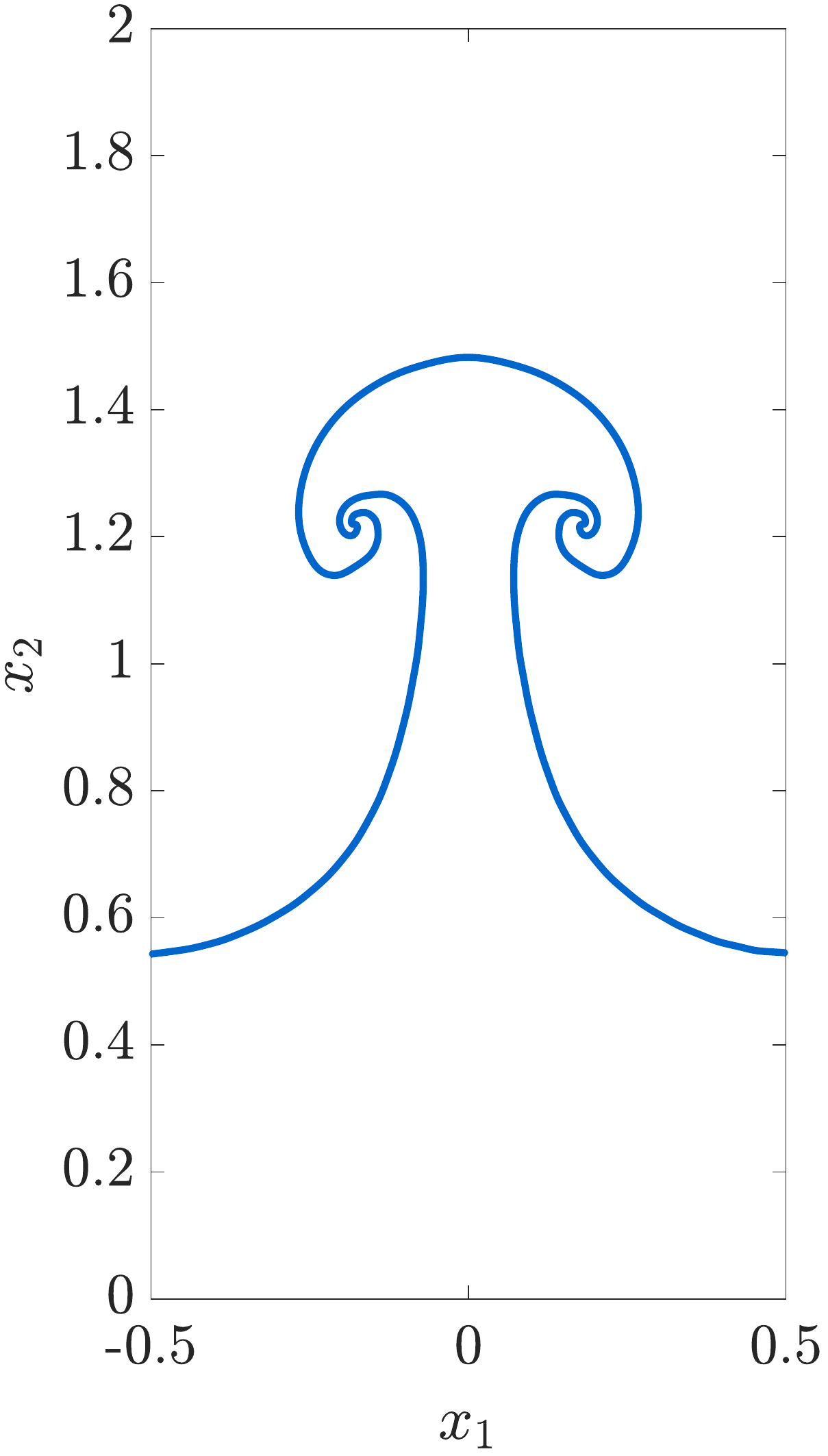}}
\hspace{2em}
\subfigure[High-res.]{\label{fig:RM_TeTr2009_z_highres}\includegraphics[width=24.54mm]{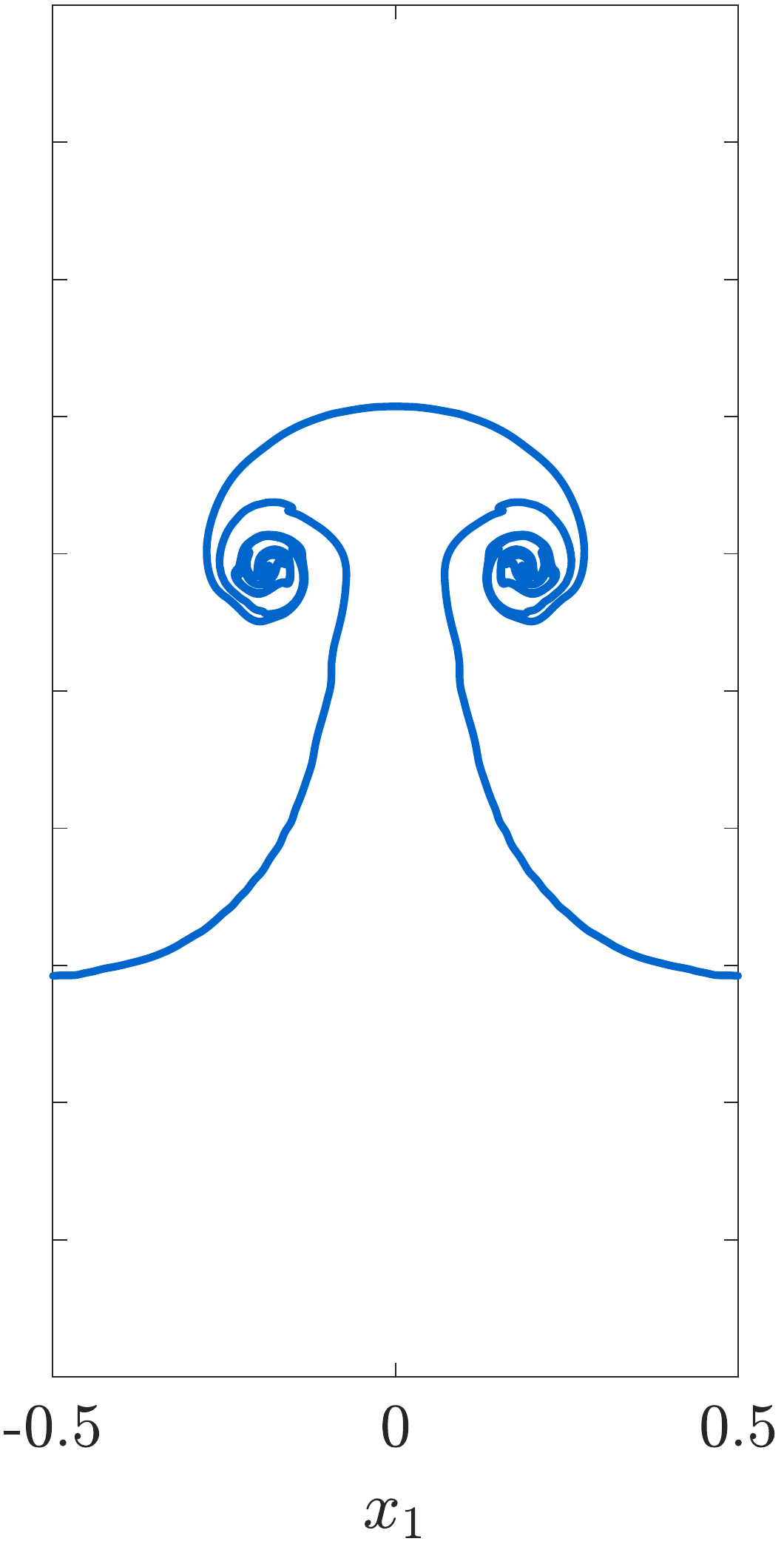}}
\hspace{2em}
\subfigure[Comparison]{\label{fig:RM_TeTr2009_z_compare_10x80_128}\includegraphics[width=24.54mm]{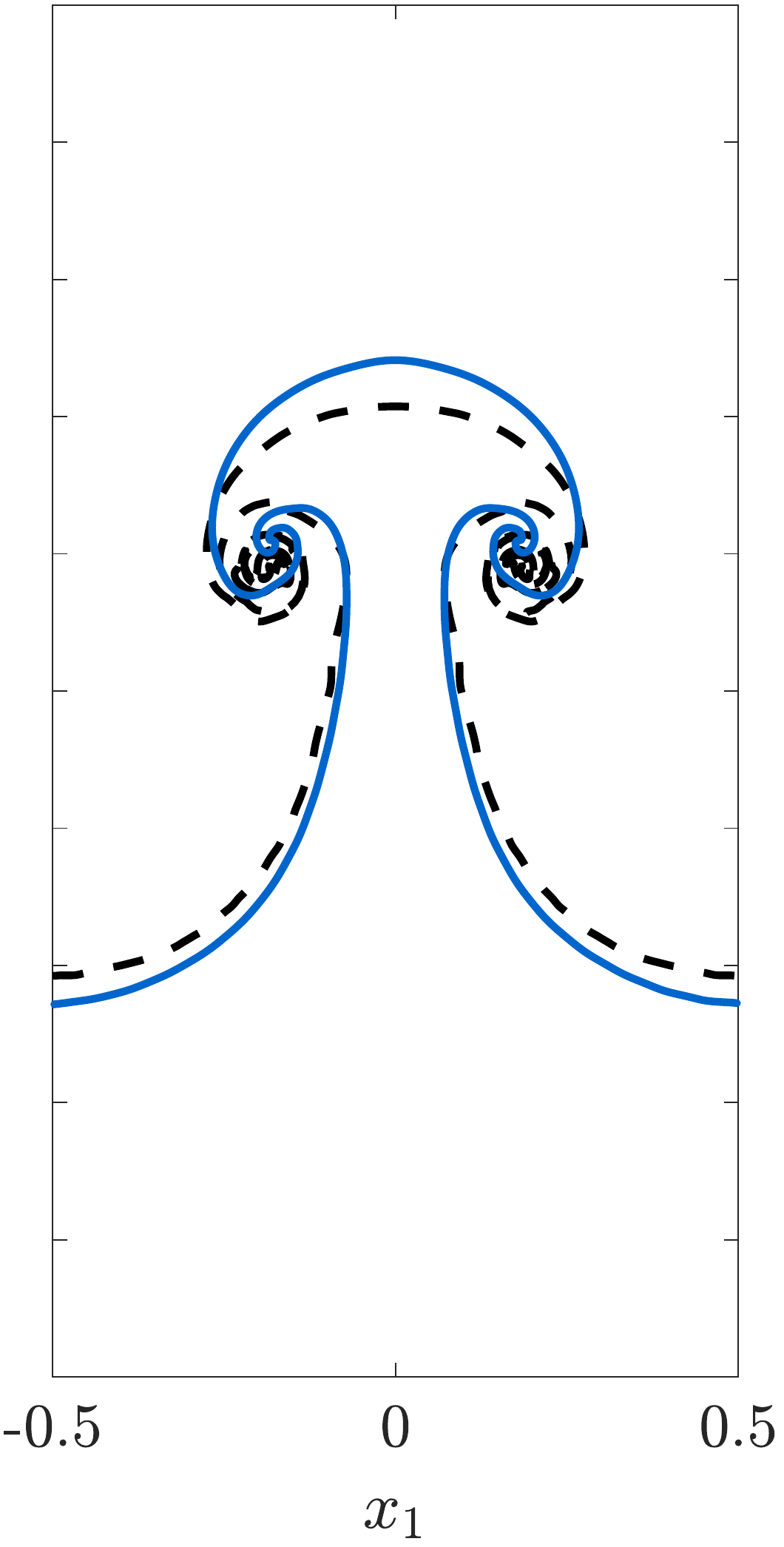}}
\caption{Results for the multiscale algorithm applied to single-mode RMI test of 
\citet{NoDiTh2006}, with the 
underlying grid containing $10 \times 80$ cells and the interface discretized with $N=128$. Shown are (a) 
computed interface parametrization $z$, (b) benchmark interface position, and (c) 
the computed interface $z$ (blue) overlaying the benchmark solution (dashed black)
at the final time $t=8.0$.}
\label{fig:RM_TeTr2009_10x80_128}
\end{figure}

The computed interface is shown  
in \Cref{fig:RM_TeTr2009_z_10x80_128}, the high resolution reference solution is shown in 
\Cref{fig:RM_TeTr2009_z_highres}, and a comparison of the two solutions is made in 
\Cref{fig:RM_TeTr2009_z_compare_10x80_128}. The multiscale algorithm is able to simulate both 
the transport of the contact as well as the KHI roll up; moreover,  
\Cref{fig:RM_TeTr2009_z_compare_10x80_128} shows 
that the bubble tip and spike tip positions compare well with the high resolution reference solution. The 
runtime for the high resolution simulation was $T_{\mathrm{CPU}} \approx 6298$ s, 
whereas the runtime for the 
multiscale simulation was only $T_{\mathrm{CPU}} \approx 24$ s, giving a speed up factor of approximately 
260 times. 

\begin{figure}[h]
\centering
\subfigure[Multiscale $z$]{\label{fig:RM_TeTr2009_z_20x160_256}\includegraphics[width=28mm]{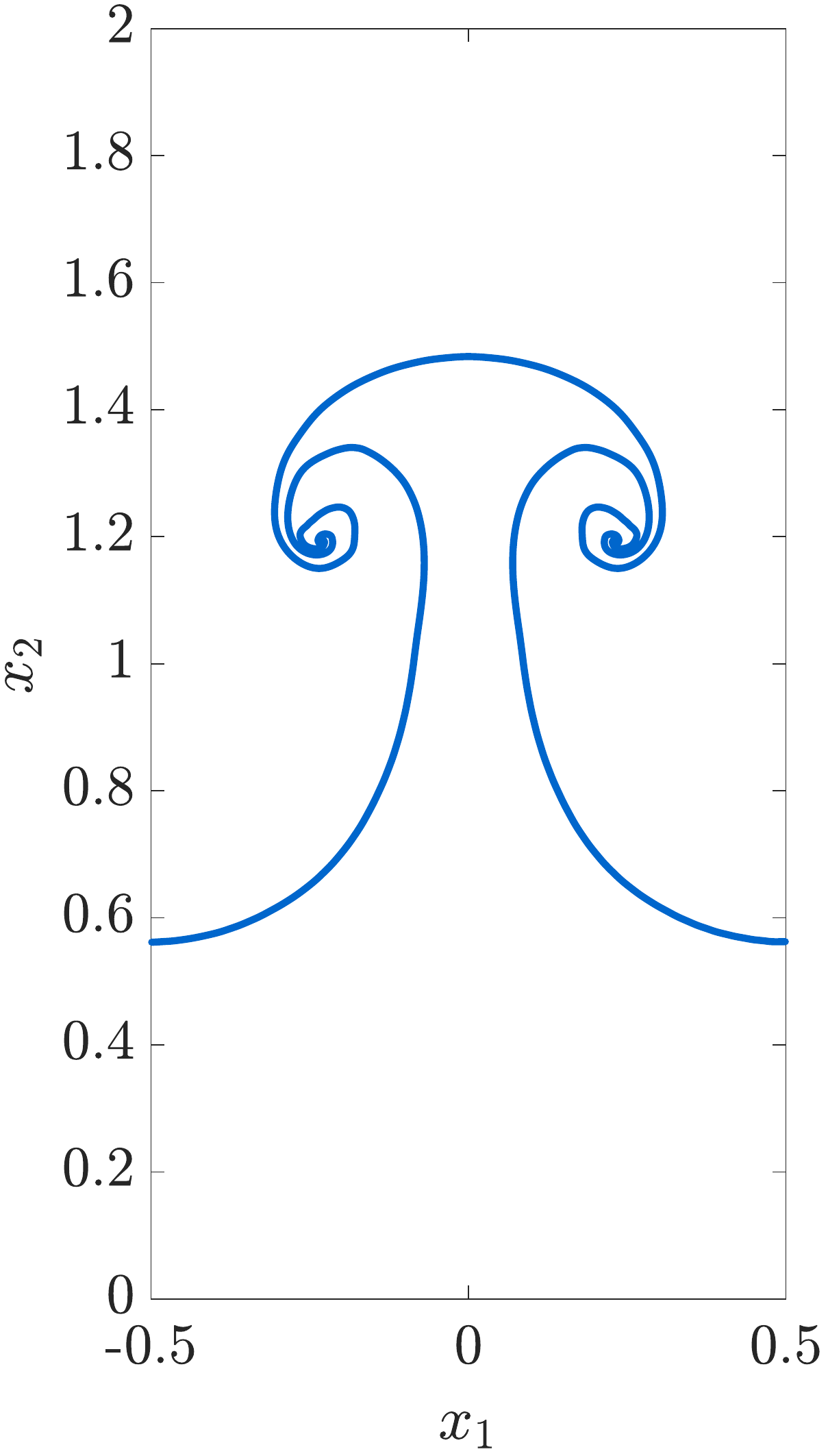}}
\hspace{2em}
\subfigure[High-res.]{\label{fig:RM_TeTr2009_z_highres_2}\includegraphics[width=24.54mm]{RM_TeTr2009_z_highres.pdf}}
\hspace{2em}
\subfigure[Comparison]{\label{fig:RM_TeTr2009_z_compare_20x160_256}\includegraphics[width=24.54mm]{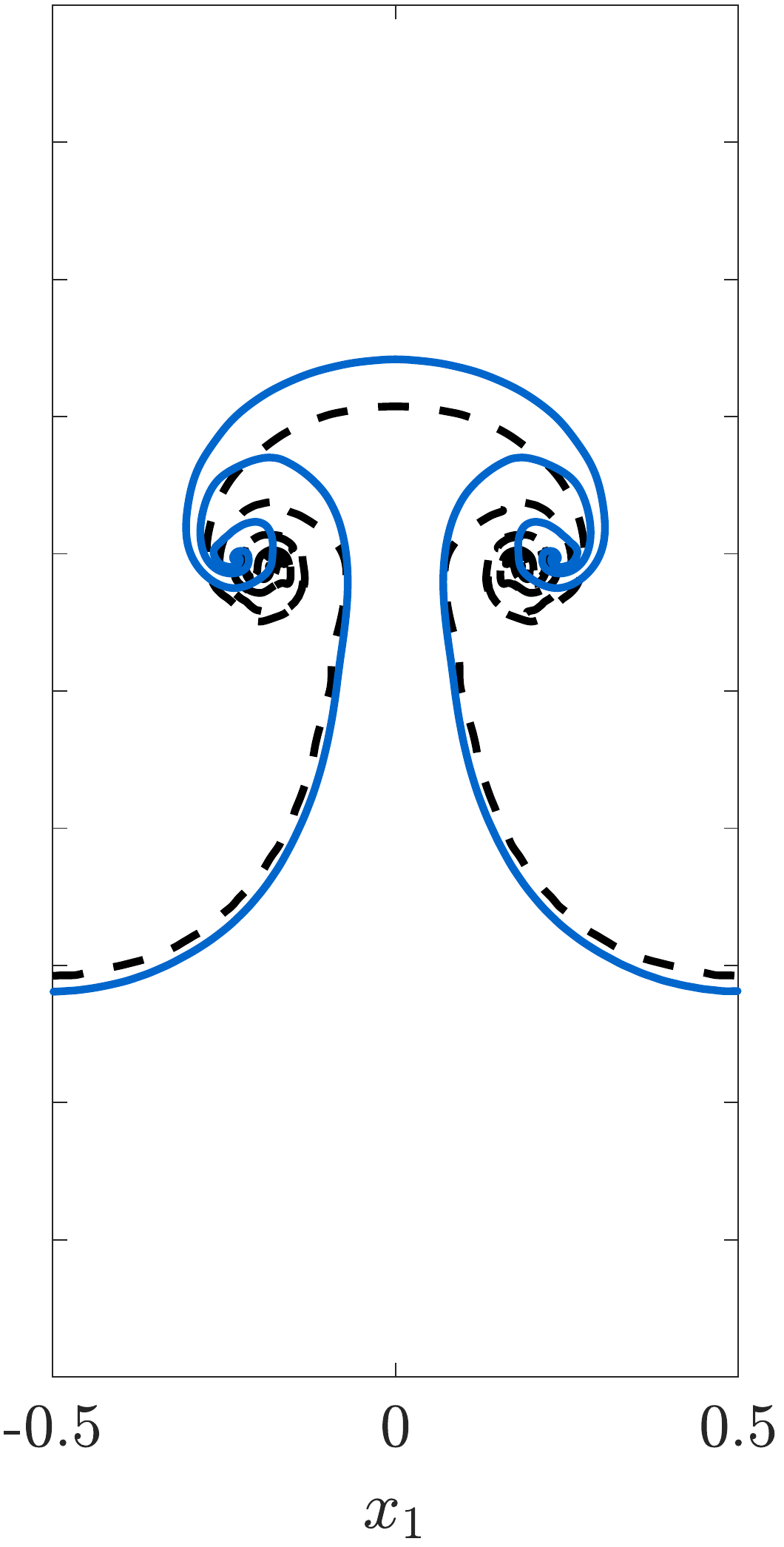}}
\caption{Results for the multiscale algorithm applied to single-mode RMI test of 
\citet{NoDiTh2006}, with the 
underlying grid containing $20 \times 160$ cells and the interface discretized with $N=256$. Shown are (a)  
computed interface parametrization $z$, (b) benchmark interface position, and (c) 
the computed interface $z$ (blue) overlaying the benchmark solution (dashed black)
at the final time $t=8.0$.}
\label{fig:RM_TeTr2009_20x160_256}
\end{figure}

Next, we double the resolution of both the coarse mesh for $v$ as well as the fine mesh for $w$ i.e. the grid 
now contains $20 \times 160$ cells and $N=256$. The time-step is halved, $\delta t = 3.125 \times 10^{-3}$, 
so that the CFL number is still approximately 0.4. All the other 
parameters from the previous run are kept fixed, and 
we employ the multiscale algorithm; the resulting solution is shown in \Cref{fig:RM_TeTr2009_z_20x160_256}. 
We see that there is more roll-up of the vortex sheet, and that the ``cap'' of the mushroom is ``flattened'', which 
is in qualitative agreement with the reference solution. The runtime for this simulation is 
$T_{\mathrm{CPU}} \approx 265$ s, giving a speed up factor of approximately 24 times over the 
high resolution simulation.  

\subsubsection{The effects of the Taylor hypothesis}

We now briefly discuss the effects of 
the Taylor "frozen turbulence" hypothesis in our multiscale RMI algorithm. We recall that the hypothesis 
asserts that the fine grid velocity $w$ is transported by  the coarse grid velocity $v$ (over small time intervals).   
The primary effect of the Taylor hypothesis is that the dynamics of the coarse grid compressible velocity
$v$ are evolved  using \eqref{v-eqn-b'} in place of \eqref{v-eqn-b}. Without the use of the hypothesis, 
the additional terms in \eqref{v-eqn-b} lead to an incorrect calculation of the velocity field $v$, which in turn 
leads to an incorrect update of the interface position $z$ (i.e. Steps \ref{RM-step1e}, 
\ref{RM-step1f} and \ref{RM-step4b}, \ref{RM-step4c} of the RMI algorithm). 

\begin{figure}[h]
\centering
\subfigure[$t=0.15$]{\label{fig:RM_TeTr2009_taylor1}\includegraphics[width=28mm]{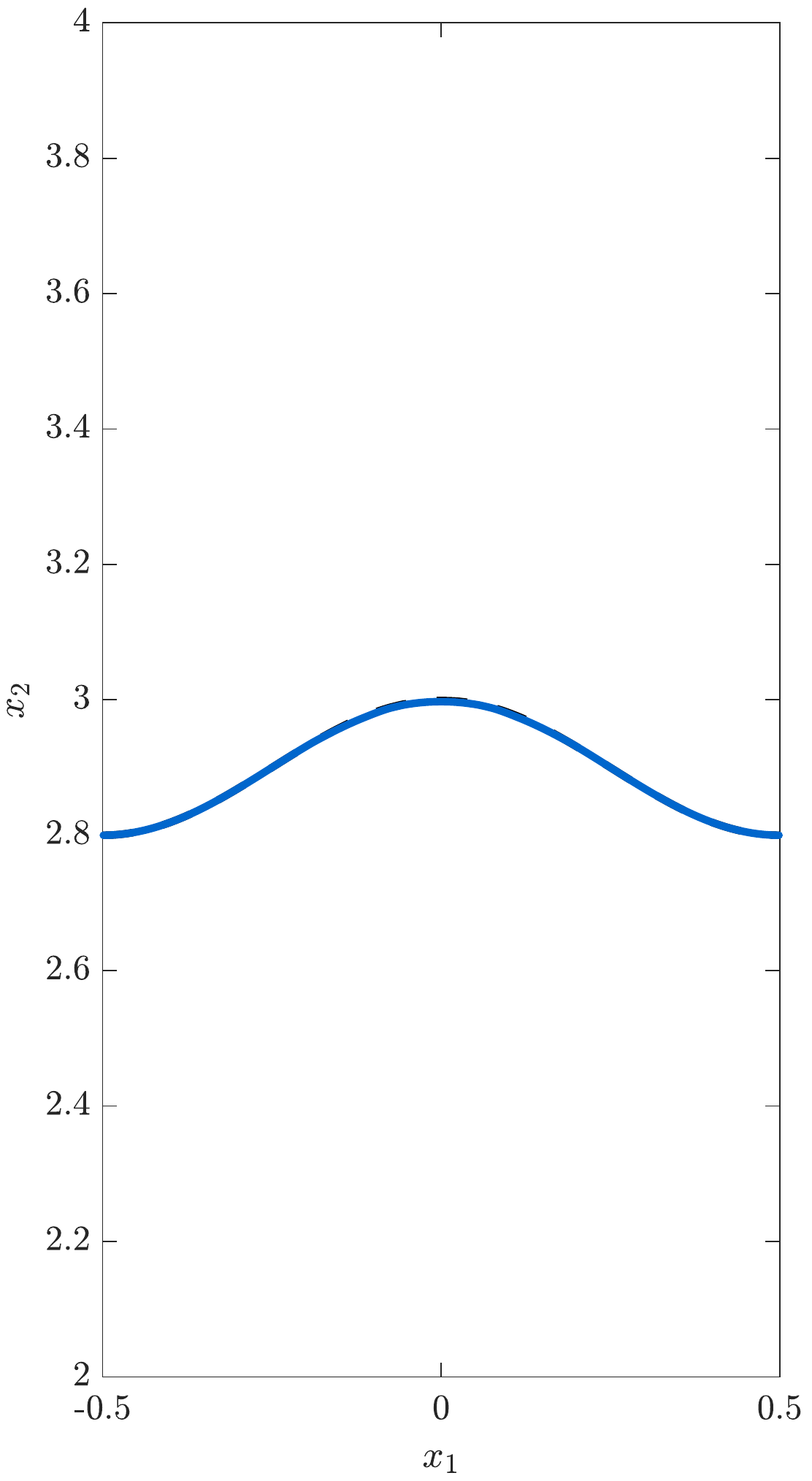}}
\hspace{2em}
\subfigure[$t=1.0$]{\label{fig:RM_TeTr2009_taylor2}\includegraphics[width=25.55mm]{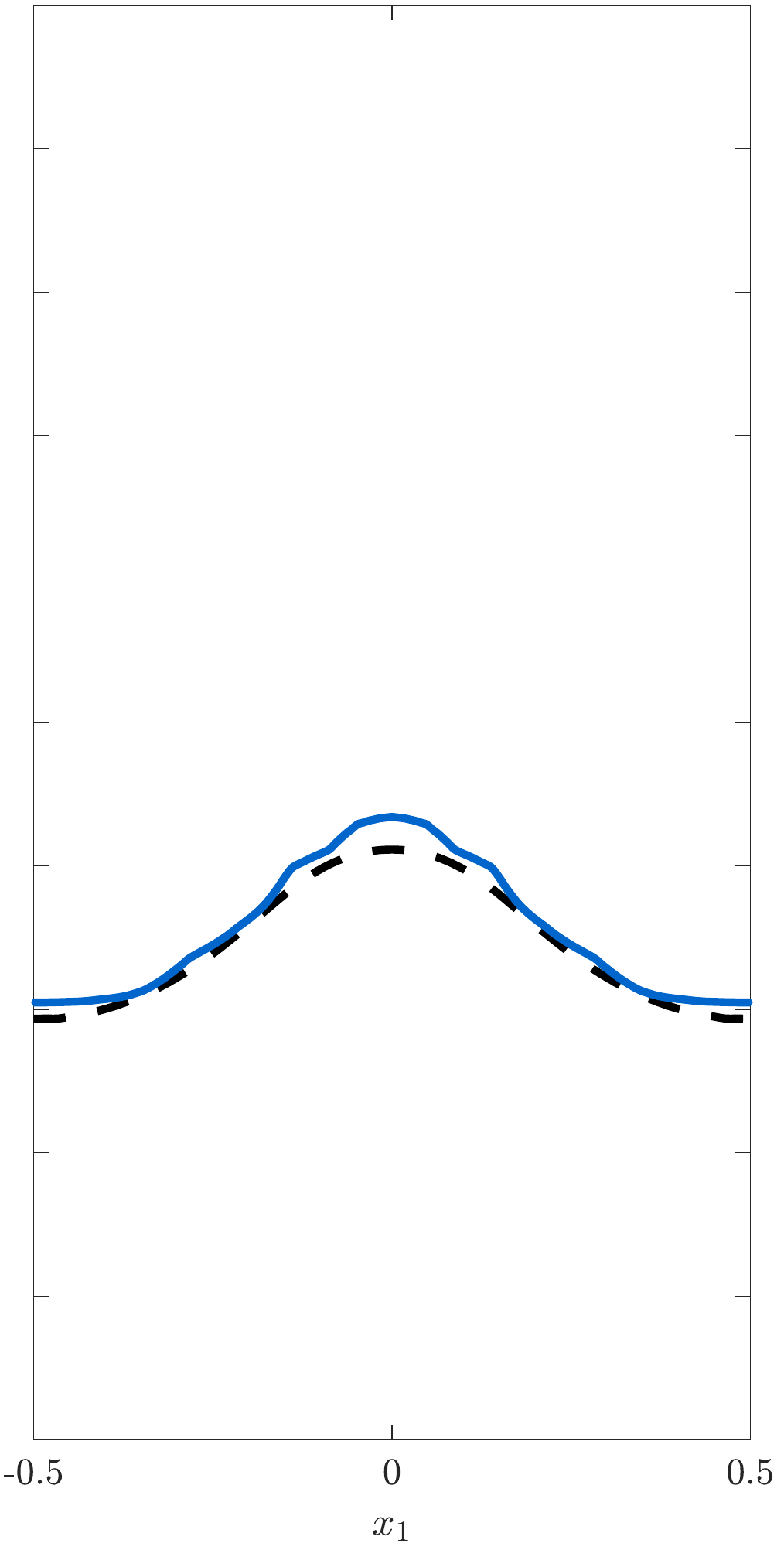}}
\hspace{2em}
\subfigure[$t=1.25$]{\label{fig:RM_TeTr2009_taylor3}\includegraphics[width=25.55mm]{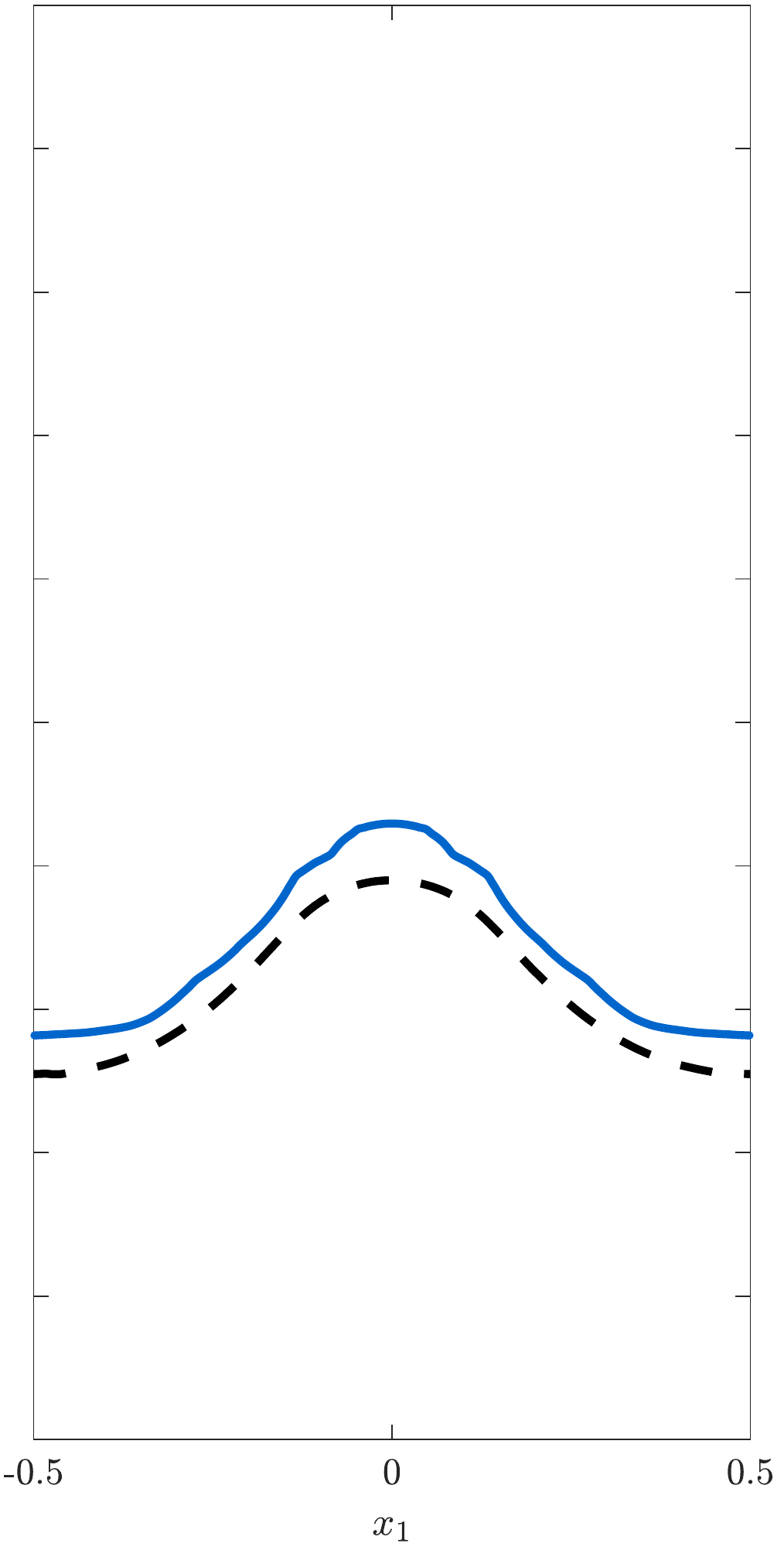}}
\hspace{2em}
\subfigure[$t=2.0$]{\label{fig:RM_TeTr2009_taylor4}\includegraphics[width=25.55mm]{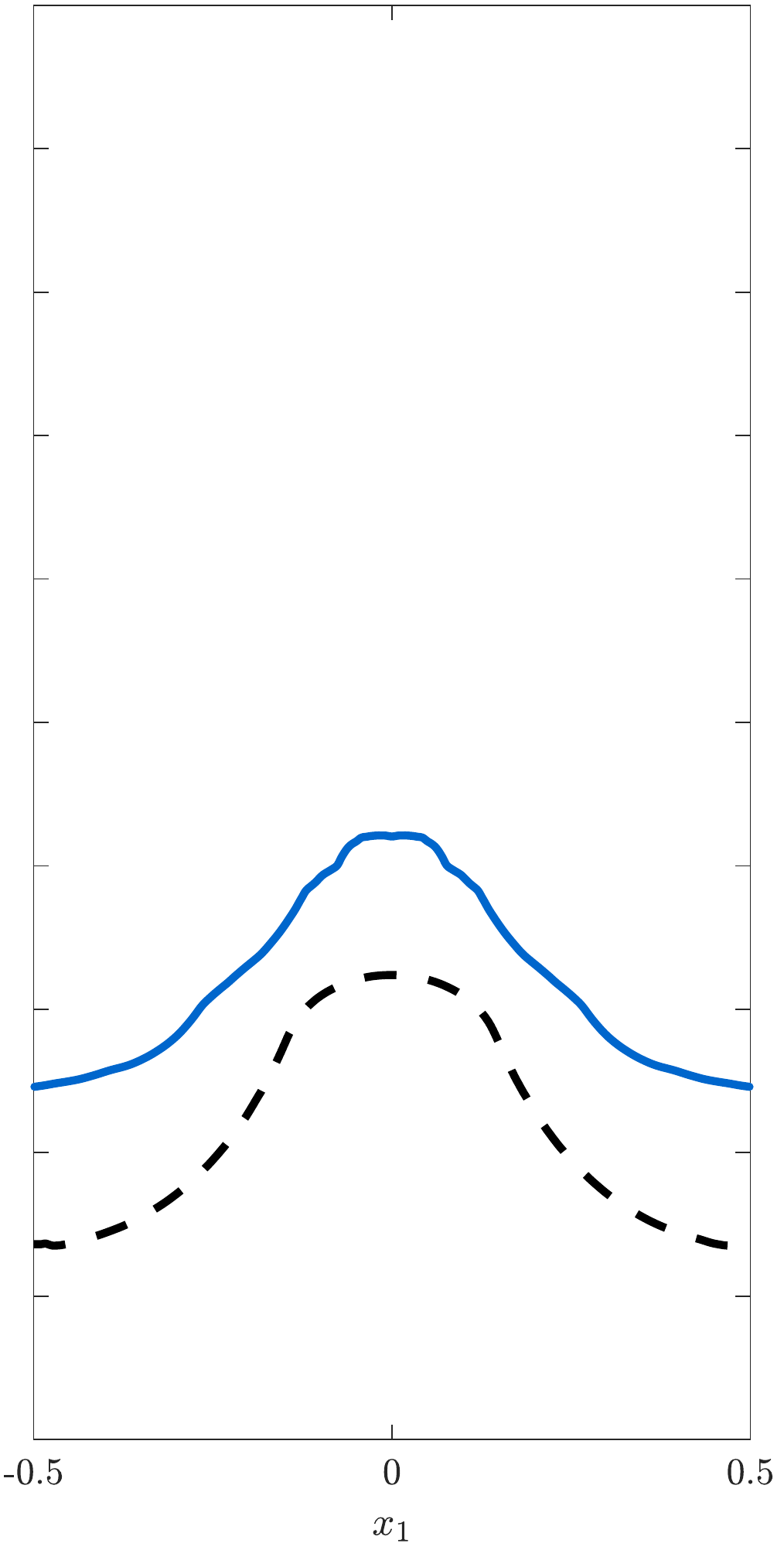}}
\caption{Results for the modified multiscale algorithm without the Taylor ``frozen turbulence'' hypothesis 
applied to single-mode RMI test of \citet{NoDiTh2006}, with all relevant 
parameters unchanged from the simulation presented in \Cref{fig:RM_TeTr2009_10x80_128}. Shown are 
the computed interface $z$ (blue) and reference solution (dashed black) at various times $t$.}
\label{fig:RM_TeTr2009_taylor}
\end{figure}

We demonstrate this in \Cref{fig:RM_TeTr2009_taylor}, in which  
we show results of a simulation performed using a modified 
version of the RMI multiscale algorithm in which the Taylor hypothesis is not employed, but is 
otherwise identical to the RMI  multiscale algorithm in \Cref{subsec:algorithms}. The solution is 
computed using the same parameters as those used for the simulation presented in 
\Cref{fig:RM_TeTr2009_z_10x80_128}. 

The shock collides with the interface at time $t=0.15$, shown in 
\Cref{fig:RM_TeTr2009_taylor1}; at this time, vorticity is deposited on the interface by the shock. 
 The interface is transported with reasonable accuracy for $t < 1.0$ (as 
demonstrated in \Cref{fig:RM_TeTr2009_taylor2}),  because the velocity $w$ is very small at early times, 
and thus does not have a noticeable effect on the interface evolution; for early times, the evolution of the 
 interface is mainly due to the ``transport'' coarse grid velocity $v$. However, as the magnitude of the 
  vorticity on the interface 
 grows with time, $w$ is no longer negligible, and affects the evolution of the interface. The 
 extra terms appearing on the right-hand side of \eqref{v-eqn-b} (which do not appear in 
 \eqref{v-eqn-b'}) lead to an incorrect calculation of the coarse scale velocity $v$. 
 Since the fine grid velocity $w$ is not being transported by 
 $v$, the coarse scale compression and vorticity information is not correctly conveyed to the interfacial 
 algorithm, which leads to a deceleration of the interface (shown in 
 \Cref{fig:RM_TeTr2009_taylor3}). The solution then quickly degrades, with the interface position 
 completely incorrect by time $t=2.0$, as shown in \Cref{fig:RM_TeTr2009_taylor4}. 
 For problems in which an unstable interface is transported as the instability develops, the use of the Taylor 
 hypothesis ensures that coarse scale information is accurately conveyed to the small scale calculations, 
 thereby producing accurate solutions with correct interface positions. 
 
 \section{Concluding remarks} \label{sec:conclusion}
 
 This paper introduces a novel multiscale model describing the evolution of contact discontinuities in 
 compressible fluid flow. The multiscale model is based on a decomposition of the 
 velocity field $u=v+w$. While the velocity 
 $w$ is discontinuous, it is also incompressible and irrotational, and can be solved efficiently on fine meshes 
 using a new asymptotic (high-order) $z$-model to approximate   the full
  Birkhoff-Rott system of singular integral equations. The velocity $v$ is compressible and rotational, but is 
 smooth near the contact discontinuity and can thus be computed efficiently on relatively coarse meshes. 
 
 We have proposed an extremely simple numerical implementation of the incompressible $z$-model, and 
 have presented numerical results for the $z$-model which simulate  classical RTI experiments. These results 
 show excellent qualitative and quantitative 
 agreement of our computed solutions with both experimental predictions, as well 
 as ``reference'' solution calculations performed using the complete Birkhoff-Rott system. In the latter case, 
 we have demonstrated that our $z$-model algorithm 
 is at least 75 times faster than a standard numerical algorithm for the 
 Birkhoff-Rott system. 
 \textcolor{black}{
 We have additionally compared our simple numerical implementation with more sophisticated methods using 
 higher order regularizations, and found good agreement (as in previous numerical studies 
 \cite{BaPh2006,Sohn2014}) between all three methods in predicting both the 
 bubble and spike tip locations, as well as the radii and location of the spiral roll up structures.
 } 
 While our simple numerical method leads to a fast-running algorithm, in the future, more 
 sophisticated numerical implementations of the $z$-model will be considered, including implementations 
 of a fast summation method \cite{GrRo1987}, a point-insertion procedure \cite{Krasny1987}, 
 non-oscillatory shock-capturing, and space-time smooth artificial viscosity \cite{RaReSh2019a}. 
 
 We have also developed multiscale models and algorithms for compressible flows with vorticity undergoing
  RTI and RMI.  Our algorithms 
 couple the fine scale velocity $w$, which controls the small scale structure of the interface, with the 
 coarse grid velocity $v$, which controls the bulk compression and vorticity of the fluid.
 The RMI multiscale algorithm includes the effects of vorticity deposition on the 
 interface during shock-contact interaction, and also enforces a version of Taylor's frozen turbulence 
 hypothesis, which asserts that small scale velocity fluctuations are transported by the mean flow. 
 We have presented numerical results for both the RTI and RMI, and demonstrated 
 that the computed solutions are in good agreement, both qualitatively and quantitively, with high 
 order gas dynamics simulations, while having computational run times that are at least two orders of 
 magnitude smaller. In particular, the multiscale solutions exhibit KHI roll up regions 
 of the contact discontinuity that are in good agreement with the roll up regions observed in solutions
 obtained from high resolution calculations. 
 Such roll up of the contact is in general 
 not observed in low resolution Euler simulations; however, the coupling of the 
 fine scale velocity $w$ to the coarse scale velocity $v$  through our multiscale model leads to simulations (on coarse meshes) which
 exhibit  roll-up regions similar to those in high-resolution simulations. 
 
 In future work, we shall generalize our models to three space dimensions, and 
 consider its applications to the numerical simulation of other physical problems, such as
 RTI and RMI with multimode initial perturbations, 
 rising bubbles, and shock-bubble interaction. The mathematical analysis of the $z$-model and the 
 multiscale models will also be considered in future work. 
 \textcolor{black}{
 In particular, we shall consider a detailed 
 theoretical and numerical study of the convergence 
 behavior of our multiscale solutions, as the interfacial and planar meshes 
 are refined, and artificial viscosity parameters converge to zero.
 }

\subsubsection*{Acknowledgements} \addcontentsline{toc}{section}{Acknowledgements}
SS was partially supported by DTRA HDTRA11810022. 
\textcolor{black}{We would like to express our gratitude to the anonymous referee for their numerous 
suggestions that have greatly improved the manuscript.}

\begin{appendices}\label{sec-appendices}

\section{Mesh refinement for the multiscale algorithm applied to the RTI}\label{sec-appendices-b}

\begin{figure}[p]
\centering
\subfigure[\tiny{$8 \times 32$, $N=64$}]{\label{fig:RT_LiWe2003_z_8x32_64-compare}\includegraphics[width=25mm]{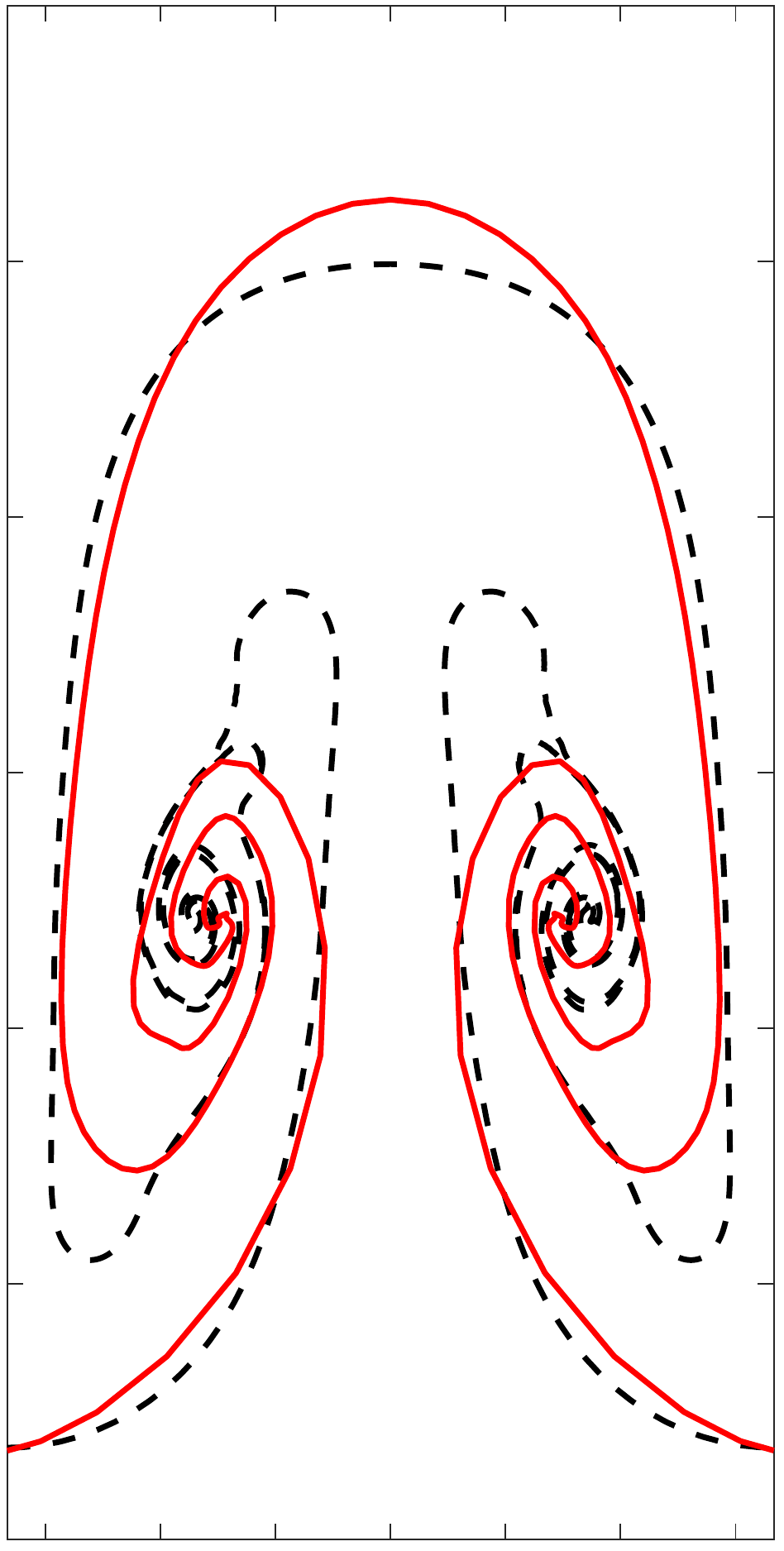}}
\hspace{1em}
\subfigure[\tiny{$8 \times 32$, $N=128$}]{\label{fig:RT_LiWe2003_z_8x32_128-compare}\includegraphics[width=25mm]{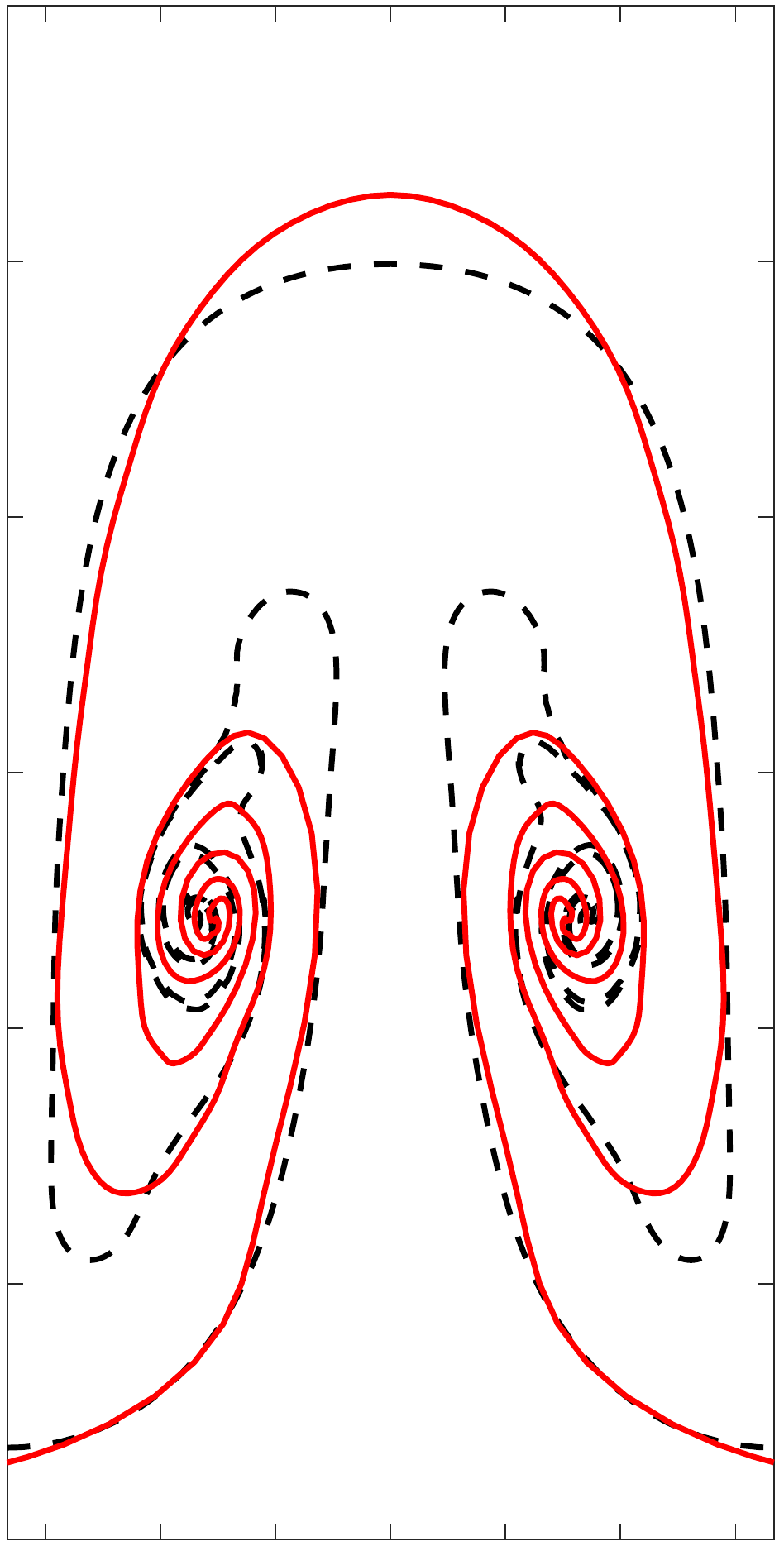}}
\hspace{1em}
\subfigure[\tiny{$8 \times 32$, $N=256$}]{\label{fig:RT_LiWe2003_z_8x32_256-compare}\includegraphics[width=25mm]{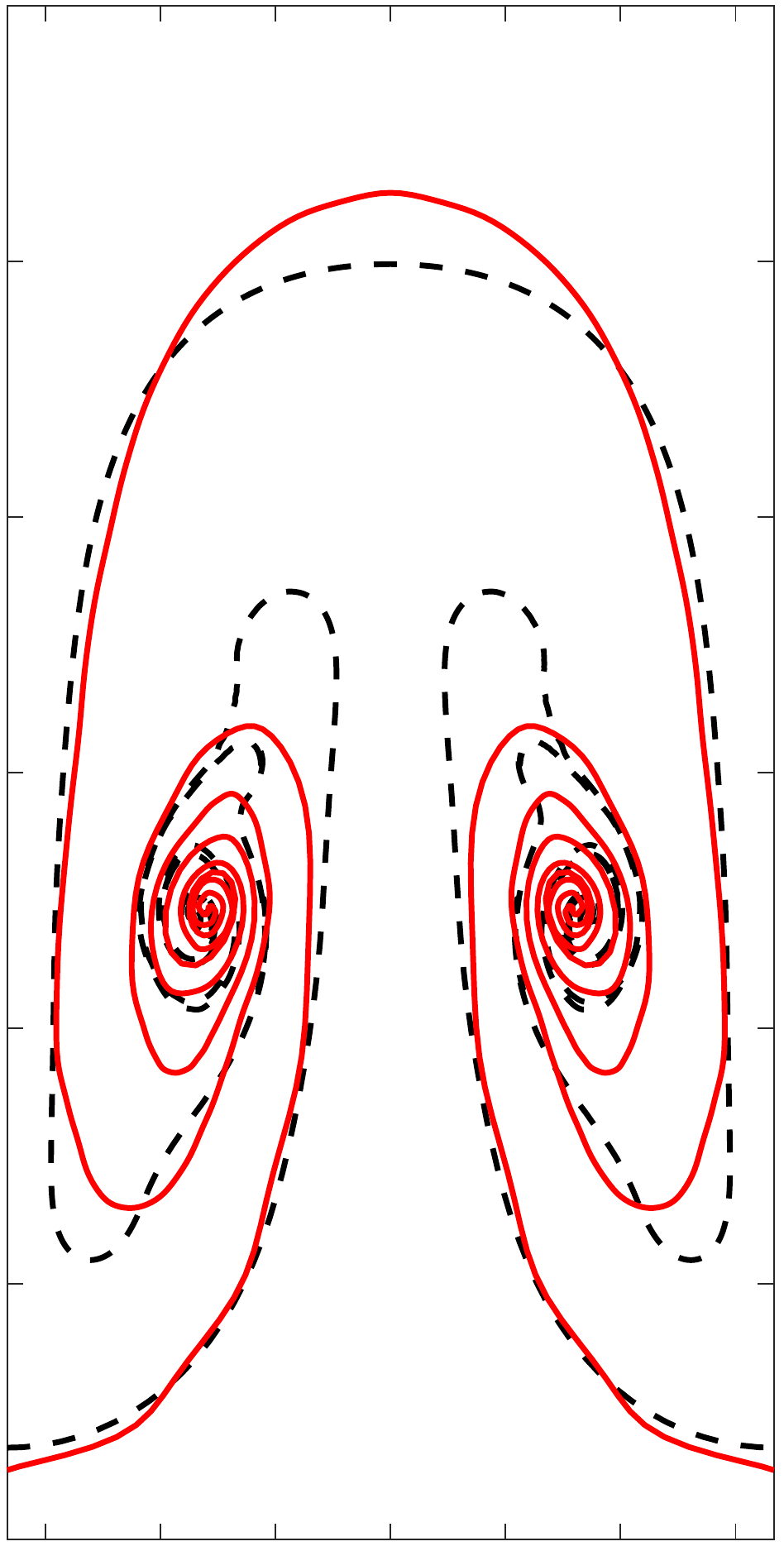}}
\par
\hspace{48em}
\subfigure[\tiny{$10 \times 40$, $N=64$}]{\label{fig:RT_LiWe2003_z_10x40_64-compare}\includegraphics[width=25mm]{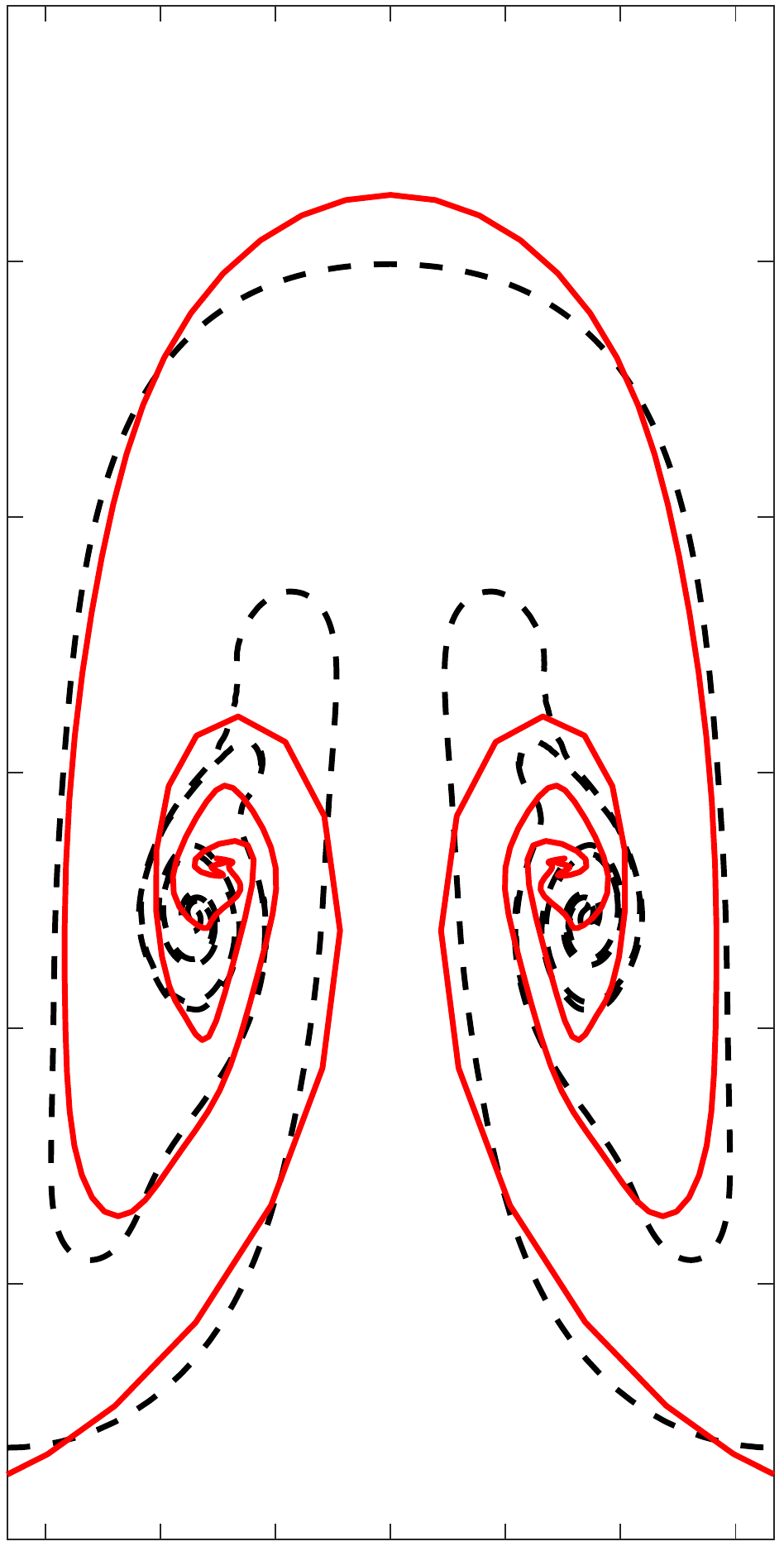}}
\hspace{1em}
\subfigure[\tiny{$10 \times 40$, $N=128$}]{\label{fig:RT_LiWe2003_z_10x40_128-compare}\includegraphics[width=25mm]{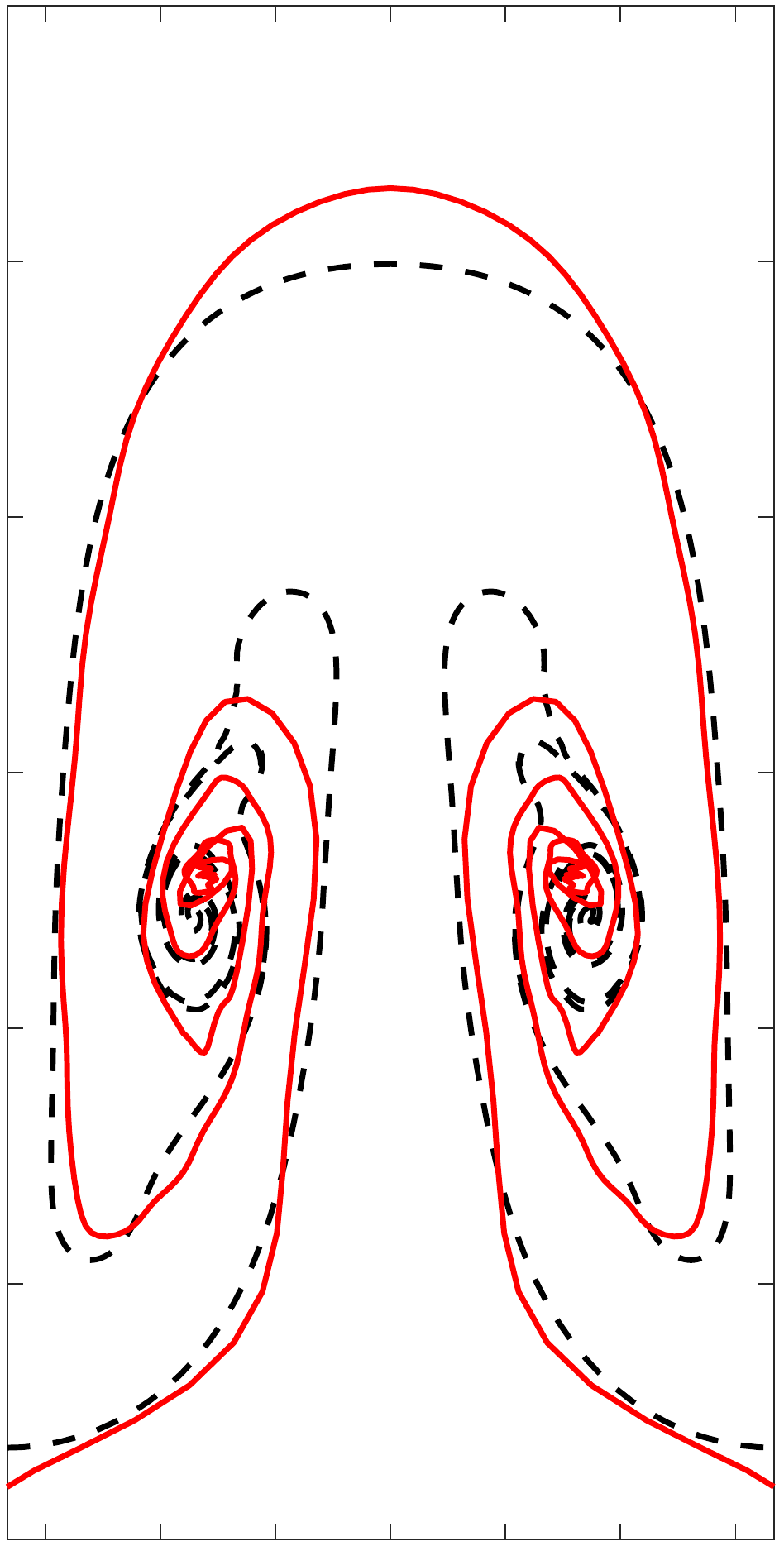}}
\hspace{1em}
\subfigure[\tiny{$10 \times 40$, $N=256$}]{\label{fig:RT_LiWe2003_z_10x40_256-compare}\includegraphics[width=25mm]{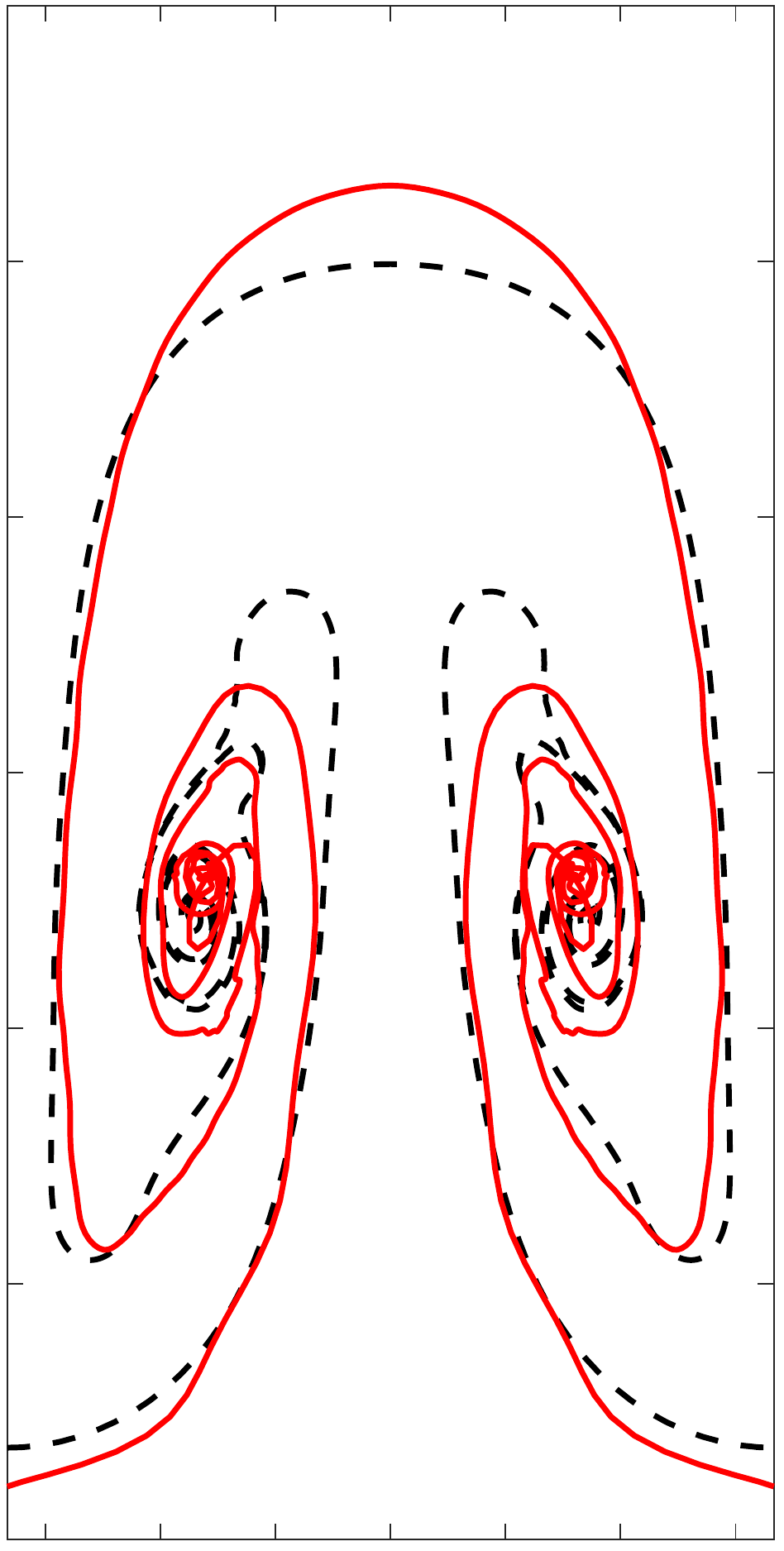}}
\par
\subfigure[\tiny{$12 \times 48$, $N=64$}]{\label{fig:RT_LiWe2003_z_12x48_64-compare}\includegraphics[width=25mm]{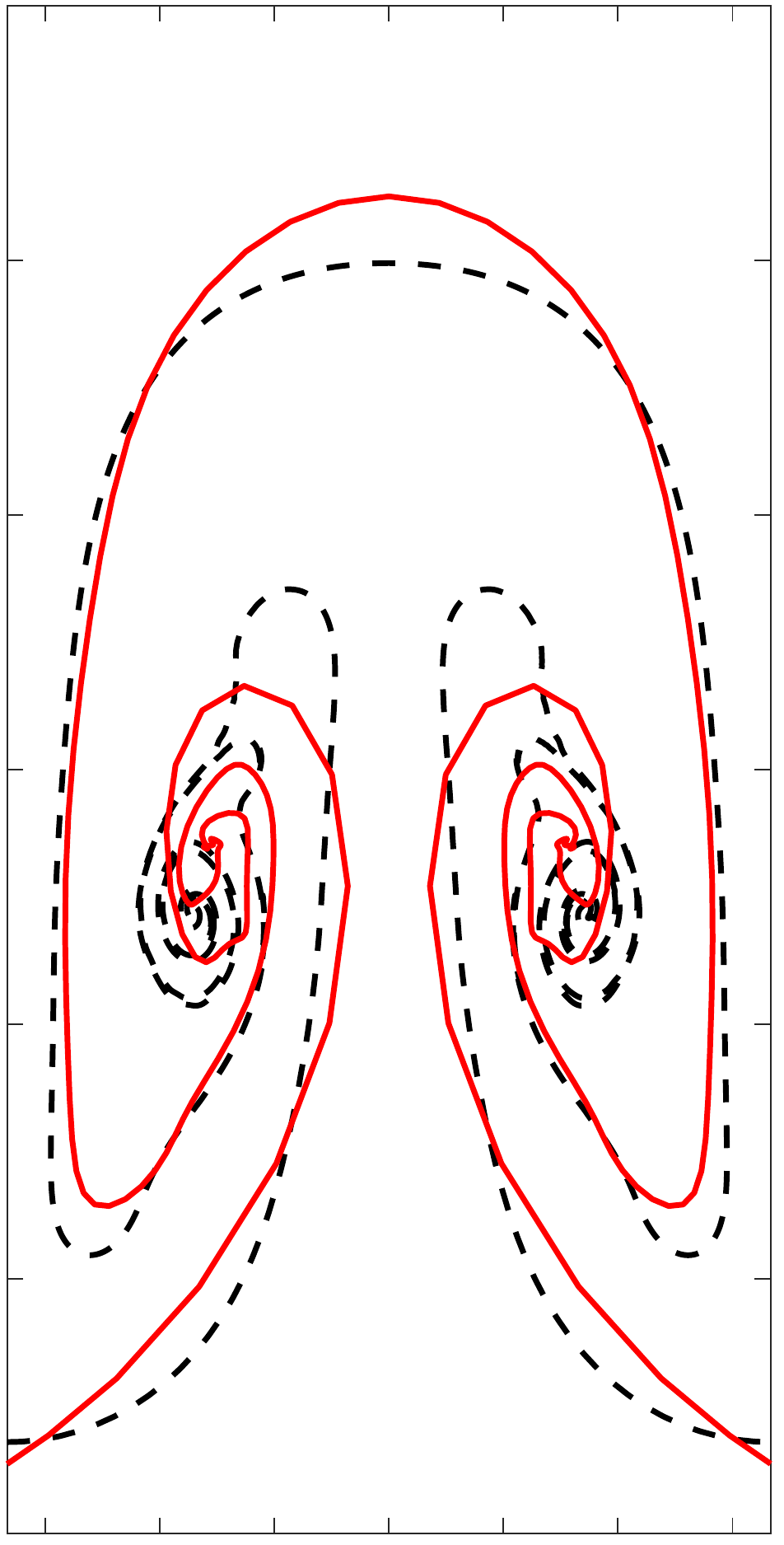}}
\hspace{1em}
\subfigure[\tiny{$12 \times 48$, $N=128$}]{\label{fig:RT_LiWe2003_z_12x48_128-compare}\includegraphics[width=25mm]{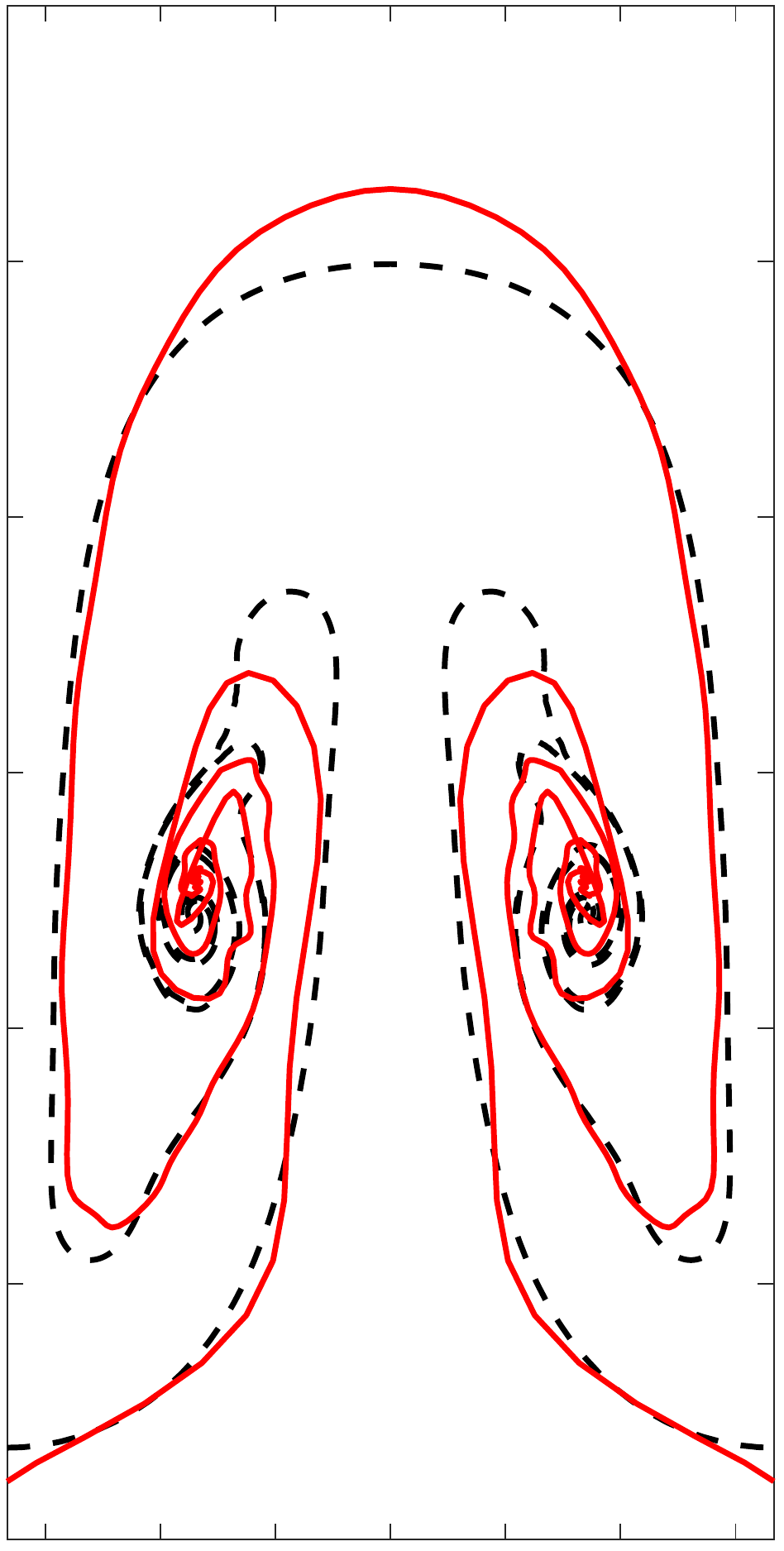}}
\hspace{1em}
\subfigure[\tiny{$12 \times 48$, $N=256$}]{\label{fig:RT_LiWe2003_z_12x48_256-compare}\includegraphics[width=25mm]{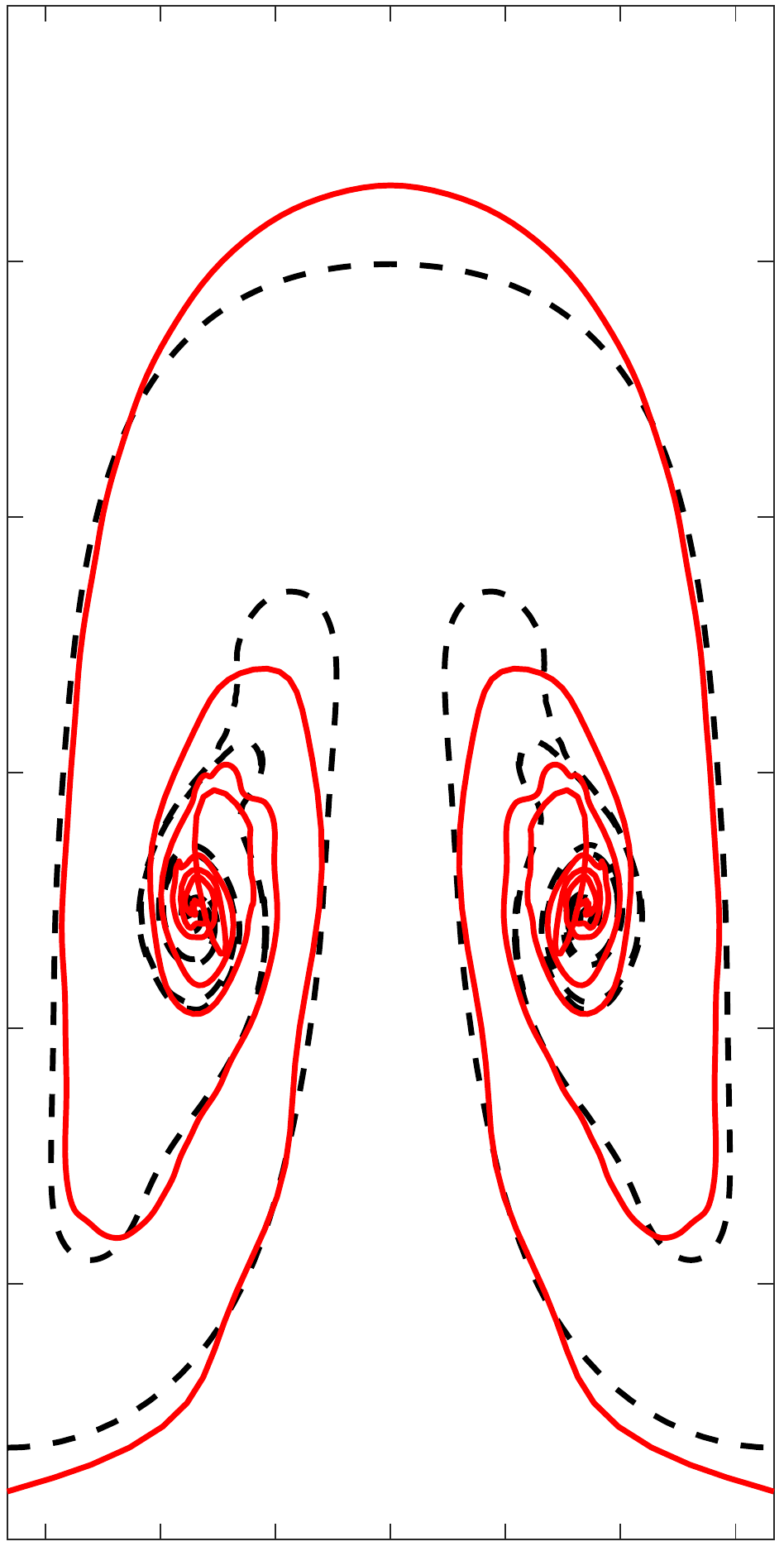}}
\caption{Plots of the interface computed using the multiscale algorithm applied to the 
single-mode compressible RTI test of \citet{LiWe2003} 
for different grid and interface resolutions. The red curve is the computed 
interface, and the dashed black curve is the reference solution.}
\label{fig:RT_LiWe2003_z_compare}
\end{figure}

In this section, we present the results for the mesh refinement study of the multiscale algorithm applied to the 
single-mode compressible RTI test of Liska \& Wendroff (see \Cref{sec-RTI-mesh-refinement}). 
\Cref{table:RMI-parameters} contains a list of the choices for the parameters $\delta t$ and $\mu$, as well as 
the runtimes $T_{\mathrm{CPU}}$ and relative speed up factors $\Lambda$
for the RTI mesh refinement study. The results from this study are presented in 
\Cref{fig:RT_LiWe2003_z_compare}, which show the reference solution (dashed black curve) overlaid by the 
computed interface parametrization $z$ (red curve). 

\begin{table}[h]
\small
\centering
\begin{tabular}{|c|c|c|c|}
\toprule
 \backslashbox{Cells}{$N$} & 64 & 128 & 256 \\
 \midrule
\multirow{4}{*}{$8 \times 32$} &  $\delta t = 2.5\e3$   &  $\delta t =2.5\e3$   &   $\delta t =6.25\e4$ \\[0.1em]		
					      &  ${\mu}= 1\e3$   &   ${\mu}= 9\e4$   &   ${\mu}= 8.75\e4$ \\	[0.1em]
					      &  $T_\mathrm{CPU}= 14$ s  &  $T_\mathrm{CPU}=28$ s   &  $T_\mathrm{CPU}=244$ s \\[0.1em]	
					      &  $\Lambda=203$  &  $\Lambda=105$    &  $\Lambda=12$  \\[0.1em]
\midrule					      
\multirow{4}{*}{$10 \times 40$} &  $\delta t = 2.5\e3$   &  $\delta t =2.5\e3$   &   $\delta t =6.25\e4$ \\[0.1em]		
					      &  ${\mu}= 1\e3$   &   ${\mu}= 8.75\e4$   &   ${\mu}= 9\e4$ \\[0.1em]	
					      &  $T_\mathrm{CPU}=19$ s  &  $T_\mathrm{CPU}=37$ s   &  $T_\mathrm{CPU}=304$ s \\[0.1em]	
					      &  $\Lambda=149$  &  $\Lambda=78$    &  $\Lambda=10$  \\[0.1em]					    \midrule
\multirow{4}{*}{$12 \times 48$} &  $\delta t = 1.67\e3$   &  $\delta t =1.67\e3$   &   $\delta t =4.55\e4$ \\[0.1em]		
					      &  ${\mu}= 1.5\e3$   &   ${\mu}= 1.25\e3$   &   ${\mu}= 1.1\e3$ \\[0.1em]	
					      &  $T_\mathrm{CPU}=38$ s  &  $T_\mathrm{CPU}=69$ s   &  $T_\mathrm{CPU}=578$ s \\[0.1em]	
					      &  $\Lambda=76$  &  $\Lambda=42$    &  $\Lambda=5$  \\[0.1em]					      			  
\bottomrule
\end{tabular}
\caption{Time-step $\delta t$ and artificial viscosity parameter ${\mu}$ choices for the 
compressible single-mode RTI mesh refinement study described in \Cref{sec-RTI-mesh-refinement}. 
Shown also are the runtimes $T_{\mathrm{CPU}}$ and speed-up factors $\Lambda$.}
\label{table:RMI-parameters}
\end{table}

\section{The $C$-method for space-time smooth artificial viscosity}\label{sec-appendices-a}

The presence of jump discontinuities in the solution $\bf{U}$ to \eqref{Euler-2d} poses a significant challenge for numerical 
schemes attempting to approximate such solutions, due to the occurrence of small-scale oscillations 
(or Gibbs phenomenon). A variety of high-order discretization techniques have been developed to combat this
issue, such as MUSCL \cite{VanLeer1979,Colella1985,Huynh1995}, 
PPM \cite{CoWo1984}, WENO \cite{LiOsCh1994,JiSh1996,Shu2003}, and its predecessor, 
ENO \cite{HaEnOsCh1997,ShOs1988,ShOs1989}. These methods rely on a careful reconstruction of the 
numerical flux; centered numerical fluxes, such as the Lax-Friedrichs flux \cite{Lax1954}, 
add dissipation implicitly to preserve stability and monotonicity, while upwinding methods based upon 
exact or approximate Riemann solvers tend to be complex and computationally costly. We refer the 
reader to \cite{RaReSh2019a} and the references therein for further details. 

Explicit artificial viscosity methods provide a simple way to stabilize shock fronts and contact discontinuities. 
These methods regularize solutions by introducing diffusion terms to the equations of motion; 
discontinuities are \emph{smeared} over a small region in space, which stabilize the solution and 
prevent the occurrence of spurious oscillations, while high-order accuracy is maintained in smooth regions of 
the flow. We next describe a method for adding localized, space-time smooth artificial viscosity to the system 
\eqref{Euler-2d}, which we call the $C$-method \cite{ReSeSh2012,RaReSh2019a,RaReSh2019b}. 

The $C$-method is a variant of the original classical artificial viscosity method of 
\citet{VonNeRi1950}, and
 couples the Euler equations \eqref{Euler-2d} to a set of scalar reaction-diffusion equations, whose solutions
 act as space-time smooth artificial viscosity indicators. The $C$-method tracks the geometry of the
evolving fronts, which allows for the implementation of both directionally isotropic and anisotropic artificial 
viscosity schemes. The latter is important for the capture of the roll-up of vortex sheets subject to the 
Kelvin-Helmholtz instability. 

We introduce the following Euler-$C$-$\hat{C}$ system:
\begin{subequations}
\label{Euler-C-Chat}
\begin{align}
\partial_t \rho + \nabla \cdot (\rho u) &= 0 \,, \label{Euler-C-Chat1} \\
\partial_t (\rho u) + \nabla \cdot (\rho u \otimes u) + \nabla p + \rho g e_2 &= \partial_{x_i} \left( \tilde{\beta} \rho \hat{C} C^{\tau_i} C^{\tau_j} \partial_{x_j} u \right) + \nabla \cdot (\tilde{\beta_s} \rho C \nabla u )\,, \label{Euler-C-Chat2}\\ 
\partial_t E + \nabla \cdot (u(E+p)) + \rho g u_2 &=0 \,, \label{Euler-C-Chat3} \\
\partial_t C - \mathscr{L}[C\,; \varepsilon,\kappa] &= \frac{S(u)}{\varepsilon |\delta x|} G_{\rho}\,, \label{Euler-C-Chat4} \\
\partial_t \hat{C} - \mathscr{L}[\hat{C}\,; \varepsilon,\kappa] &= \frac{S(u)}{\varepsilon |\delta x|} \hat{G}_{\rho} \,, \label{Euler-C-Chat5}\\ 
\partial_t C^{\tau_i} - \mathscr{L}[C^{\tau_i}\,; \varepsilon,\kappa] &= \frac{S(u)}{\varepsilon |\delta x|} \hat{\tau}_i \,, \text{ for } i=1,2\,. \label{Euler-C-Chat6}
\end{align}
\end{subequations}
Here, the operator $\mathscr{L}$ is defined by 
\begin{equation}
\mathscr{L}[C\,; \varepsilon,\kappa] = -\frac{S(u)}{\varepsilon |\delta x|} C + \kappa S(u) \Delta C\,,
\end{equation}
where $\varepsilon$ and $\kappa$ are parameters controlling the support and smoothness of the function $C$, 
respectively, $\delta x = (\delta x_1\,,\delta x_2)$ is the grid spacing, and 
$\Delta = \partial_{x_1}^2 + \partial_{x_2}^2$ is the Laplacian operator. The forcing functions to the 
$C$-equations are
\begin{subequations}
\label{C-forcings}
\begin{align}
G_{\rho} &= \mathbbm{1}_{(-\infty,0)}(\nabla \cdot u) [1-\mathbbm{1}_{(\infty,0)}(\partial_{n} \rho \partial_{n} e)]|\nabla \rho| \,, \label{G-rho} \\
\hat{G}_{\rho} &= \mathbbm{1}_{(-\infty,0)}(\partial_{n} \rho \partial_{n} e) |\nabla \rho|\,, \label{G-hat-rho} \\ 
\hat{\tau}_1 &= -\mathbbm{1}_{(-\infty,0)}(\partial_{n} \rho \partial_{n} e) \partial_y \rho \,, \label{tau-1-hat} \\
\hat{\tau}_2 &= \mathbbm{1}_{(-\infty,0)}(\partial_{n} \rho \partial_{n} e) \partial_x \rho \,, \label{tau-2-hat}
\end{align}
\end{subequations}
where $\partial_{n} = n \cdot \nabla$ is the normal derivative operator and the function 
$\mathbbm{1}_{(-\infty,0)}(\xi)$ is a {compression} or {expansion} switch defined by 
$$
\mathbbm{1}_{(-\infty,0)}(\xi) = \begin{cases}
						1\,, & \text{if } \xi < 0\,, \\
						0\,, & \text{if } \xi \geq 0\,.
						\end{cases} 
$$
The artificial viscosity parameters $\tilde{\beta}$ and $\tilde{\alpha}$ are defined by 
\begin{equation}\label{C-method-viscosity-parameters}
\tilde{\beta} = |\delta x|^2 \cdot \frac{\max_x | \nabla u|}{\mu^2 \max_x \hat{C}}\beta \quad \text{ and } \quad \tilde{\beta_s} =  |\delta x|^2 \cdot \frac{\max_x | \nabla u|}{\max_x C}\beta_s\,,
\end{equation}
with $\mu = \max_x \left\{ |C^{\tau_1}| \,, |C^{\tau_2}| \right\}$ a normalization constant, and $\beta$ and $\alpha$
constant positive numbers.  

In writing the system \eqref{Euler-C-Chat}, we have utilized the summation convention which specifies that 
a repeated free index in the same term implies summation over all values of that index.

The two artificial viscosity terms in \eqref{Euler-C-Chat} are 
\begin{subequations}\label{artificial-visc-Euler-C-Chat}
\begin{gather}
\partial_{x_i} \left( \tilde{\beta} \rho \hat{C} C^{\tau_i} C^{\tau_j} \partial_{x_j} u \right)\,, \label{anisotropic-visc} \\
\nabla \cdot (\tilde{\beta_s} \rho C \nabla u )\,. \label{isotropic-visc}
\end{gather}
\end{subequations}
\eqref{anisotropic-visc} is an \emph{anisotropic} artificial viscosity term that adds dissipation only in directions
tangential to the front. This artificial viscosity term is localized to the vortex sheet through the use of the 
$C$-functions; the expansion switch $\mathbbm{1}_{(-\infty,0)}(\partial_{n} e \partial_{n} \rho)$ in the 
forcing functions $\hat{G}_{\rho}$ and $\hat{\tau}_i$ vanishes at shock fronts, but is active at contact 
discontinuities. The \emph{isotropic} artificial viscosity operator \eqref{isotropic-visc}, on the other other hand, 
adds dissipation in all directions; this artificial viscosity term is localized to the shock fronts through the 
compression switch $\mathbbm{1}_{(-\infty,0)}(\nabla \cdot u)$ in the forcing function $G_{\rho}$. For the 
vortex sheet, it is important that dissipation is added only in direction tangential to the sheet; therefore, the 
switch $1 - \mathbbm{1}_{(-\infty,0)}(\partial_{n} e \partial_{n} \rho)$ ``turns off'' the isotropic dissipation in the 
regions where a shock front intersects with a vortex sheet.

\begin{remark}
If there are no shock fronts present in a solution, as is the case for the classical Rayleigh-Taylor problems, then 
the isotropic diffusion term \eqref{isotropic-visc} in \eqref{Euler-C-Chat} is omitted, and consequently so is 
equation \eqref{Euler-C-Chat4} for the $C$-function localized to shock waves. 
\end{remark}

\begin{remark}
For problems which are symmetric about $x_1=0$, we compute the solution only for $x_1 \geq 0$ and then 
use reflection to obtain the solution for $x_1 < 0$. A similar reflection procedure is used in the 
numerical implementation of the $z$-model. 
\end{remark}

For the purposes of brevity, we have omitted some of the details and refer the reader to \cite{RaReSh2019b} for 
further discussion of the $C$-method in 2-$D$. 

\end{appendices}

\end{document}